\documentclass{article}

\usepackage{PRIMEarxiv}
\usepackage{amsmath}
\usepackage{bm}
\usepackage{color}
\usepackage{amssymb}
\usepackage{graphicx}
\usepackage{epstopdf}
\usepackage{caption}
\usepackage{subcaption}
\usepackage{multirow}
\usepackage{color}
\usepackage{xcolor}
\usepackage[final]{pdfpages}
\usepackage{lineno}
\usepackage{setspace}
\modulolinenumbers[5]

\usepackage{placeins}

\let\Oldsection\section
\renewcommand{\section}{\FloatBarrier\Oldsection}

\let\Oldsubsection\subsection
\renewcommand{\subsection}{\FloatBarrier\Oldsubsection}

\let\Oldsubsubsection\subsubsection
\renewcommand{\subsubsection}{\FloatBarrier\Oldsubsubsection}

\usepackage{tikz}
\usetikzlibrary{tikzmark,calc,arrows,shapes,decorations.pathreplacing}
\usetikzlibrary{positioning}
\tikzset{every picture/.style={remember picture}}
\definecolor{Red}{rgb}{1,0,0}

\newcommand{\btxt}[1]{\textcolor{black}{#1}} 

\newcommand{\fref}[1]{Figure (\ref{#1})}
\newcommand{\tref}[1]{Table \ref{#1}}
\newcommand{\eref}[1]{Eq.(\ref{#1})}
\newcommand{\cref}[1]{Chapter~\ref{#1}}

\newcommand{\subf}[2]{%
  {\small\begin{tabular}[t]{@{}c@{}}
  #1\\#2
  \end{tabular}}%
}

\renewcommand{\vec}[1]{\mathbf{#1}}

\newcommand{\x}{\vec{x}}

\usepackage[utf8]{inputenc}

\definecolor{pink1}{RGB}{219, 48, 122}
\definecolor{asparagus}{rgb}{0.53, 0.66, 0.42}
\definecolor{amber}{rgb}{1.0, 0.35, 0.0}
\definecolor{applegreen}{rgb}{0.55, 0.71, 0.0}
\definecolor{nodiffusion}{rgb}{0.6, 0.0, 0.0}
\definecolor{diffusion}{rgb}{0.0 0.6, 0.0}
\definecolor{br}{rgb}{0.0, 0.0, 1.0}

\title{Probabilistic Prediction of Coalescence Flutter Using Measurements: Application to the Flutter Margin Method
}

\author{
Sandip Chajjed\\
Department of Civil and Environmental Engineering\\
  Carleton University\\
  Ottawa,Ontario\\
  Canada.\\
   \And 
Mohammad Khalil\thanks{\textit{Sandia National Laboratories is a multimission laboratory managed and operated by National Technology and Engineering Solutions of Sandia, LLC., a wholly owned subsidiary of Honeywell International, Inc., for the U.S. Department of Energy’s National Nuclear Security Administration under contract DE-NA-0003525.}} \\ 
Quantitative Modeling \& Analysis Department\\
Sandia National Laboratories\\
Livermore, CA\\
United States\\  
   \And
Dominique Poirel\\
Department of Mechanical and Aerospace Engineering\\
Royal Military College of Canada, Kingston, Ontario\\
Canada\\
   \And
Chris Pettit\\
Department of Aerospace Engineering\\
United States Naval Academy, Annapolis, MD\\
 United States\\
  \And
   Abhijit Sarkar\\
   Department of Civil and Environmental Engineering\\
  Carleton University\\
  Ottawa,Ontario\\
  Canada.\\
 \texttt{abhijit.sarkar@carleton.ca} \\
}

\begin{document}
\maketitle

\begin{abstract}
{\btxt{Zimmerman and Weissenburger's flutter margin method is widely used to estimate
the aeroelastic coalescence flutter speed. In contrast to aeroelastic decay rates, the flutter margin exhibits monotonic decay with respect to airspeed redering it effective in
extrapolating the flutter speed using flight test data conducted at pre-flutter airspeeds. 
This paper reports the generalization of the Bayesian formulation of the flutter margin
 method by Khalil {\it et al.} developed to tackle measurement and modeling uncertainties. 
This paper improves the predictive performance of the previous algorithm by incorporating the 
joint prior of aeroelastic modal frequencies and decay rates among airspeeds in order to better estimate the joint posterior of modal parameters using observational data. The modal 
parameter prior is constructed using the classical  two-degree-of-freedom pitch-plunge
aeroelastic model whose system matrices (e.g. structural stiffness and damping matrices) vary randomly. Such joint modal parameter prior  enforces statistical dependence among posteriors
of modal parameters and the associated flutter margins across airspeeds. Numerical studies demonstrate a considerable reduction of uncertainties on the predicted flutter speed obtained
from the generalized Bayesian flutter margin method. This improved algorithm can cut cost by
reducing the number of flight tests and better assess the uncertainty against aeroelastic flutter.
}}
\end{abstract}


{
\noindent
{\bf Abbreviations}\\
\textbf{AMMCMC} \,\, \textbf{a}daptive \textbf{M}etropolis  \textbf{M}arkov \textbf{C}hain \textbf{M}onte \textbf{C}arlo \\
\textbf{COV} \,\, \textbf{c}oefficient \textbf{o}f \textbf{v}ariation\\ 
\textbf{CG} \,\, \textbf{c}enter of \textbf{g}ravity\\
\textbf{dof} \,\, \textbf{d}egree \textbf{o}f \textbf{f}reedom\\
\textbf{EA} \,\, \textbf{e}lastic \textbf{a}xis\\
\textbf{FM} \,\, \textbf{f}lutter \textbf{m}argin\\
\textbf{MCS} \,\, \textbf{M}onte \textbf{C}arlo \textbf{S}imulation\\
\textbf{MH} \,\, \textbf{M}etropolis \textbf{H}asting \\
\textbf{MCMC} \,\, \textbf{M}arkov \textbf{C}hain \textbf{M}onte \textbf{C}arlo \\
\textbf{pdf}   \,\, \textbf{p}robability \textbf{d}ensity \textbf{f}unction\\
\textbf{MAP}   \,\, Maximum {\it a posteriori} estimate\\
}

{
\noindent
{\bf Symbols}\\
$b$ \,\, Semi-chord \\
$c$ \,\, Chord  \\
$m$ \,\, Mass \\
${I_{EA}}$ \,\, Mass moment of inertia about the elastic axis \\
$U$ \,\, True airspeed  \\
$L$ \,\, Lift \\
$M_{EA}$ \,\, Aerodynamic moment about the elastic axis \\
$h$ \,\, Heave displacement \\
$\alpha$ \,\, Angle of attack \\
$\omega_i^{(j)}$ \,\,  Frequency of $i^{th}$ mode and $j^{th}$ test point  \\
$\beta_i^{(j)}$ \,\, Decay rate of $i^{th}$ mode and $j^{th}$ test point  \\
$\rho$ \,\, Free stream fluid density \\
$k_h$ \,\, heave stiffness \\
$k_{\alpha}$ \,\, pitch stiffness \\
$x_{\alpha}$ \,\, Normalized static imbalance \\
$a_h$ \,\, Normalized distance between EA and mid-chord \\
$\xi_1$ \,\, Damping ratio (mode 1) \\
$\xi_2$ \,\, Damping ratio (mode 2)  \\
} 

\section{Introduction}
{The aeroelastic flutter instability can be predicted tracking damping against airspeed or dynamic pressure. At the critical airspeed at which the flutter instability occurs, the damping of the unstable aeroelastic 
mode becomes zero. However, this prediction approach experiences difficulties as the damping may not exhibit monotonic 
decay against airspeed. Furthermore, the damping estimates obtained using noisy and sparse observational
data exhibit large scatter or uncertainties, compared to the estimated aeroelastic frequencies. For the
coalescence flutter for which two aeroelastic modes participate in the instability, Zimmerman and Weissenburger~\cite{weissenburger1964prediction} defines the flutter margin using the aeroelastic modal frequencies  $(\omega_1,  \omega_2)$ and  decay rates $(\beta_1, \beta_2)$ as:
\begin{eqnarray}\label{eq:fm}
F(U)&=\left[ \left(\frac{\omega_2^2-\omega_1^2}{2}\right)+\left(\frac{\beta_2^2-\beta_1^2}{2}\right)\right]^2\\\nonumber
& + 4\beta_1\beta_2\left[ \left(\frac{\omega_2^2+\omega_1^2}{2}\right)+2\left(\frac{\beta_2+\beta_1}{2}\right)^2\right]\\\nonumber
&  -\left[ \left(\frac{\beta_2-\beta_1}{\beta_2+\beta_1}\right) \left(\frac{\omega_2^2-\omega_1^2}{2}\right)+2\left(\frac{\beta_2+\beta_1}{2}\right)^2\right]^2
\end{eqnarray}
The flutter margin, a stability indicator, tends to zero when  the system becomes unstable at the flutter speed. In contrast to the decay rate, the flutter margin monotonically decays to zero at the critical (flutter) airspeed. By nature, the aeroelastic modal
frequencies have more influence on coalescence flutter that decay rates.
They also generally exhibit less scatter in their estimates than the decay rates.
Note from ~\eref{eq:fm}, the expression of flutter margin includes the modal frequencies $\omega_1$ and  $\omega_2$ which leads to its better performance (than just decay rates) in predicting flutter speed. Under certain practical assumptions (e.g. quasi-steady aerodynamics, negligible structural damping etc.), 
the flutter margin approximately becomes a  function of airspeed $U$ as (e.g. \cite{khalil2015jsv}):
\begin{equation}\label{eq:fm2}
F(U)=B_1U^4+B_2U^2+B_3
\end{equation}
Using the above equation, the flutter speed can be estimated using the flight test data obtained from
subcritical (pre-flutter) airspeeds. This process involves the following steps. At a given airspeed $U$,
the aeroelastic modal frequencies  ($\omega_1, \omega_2$) and  decay rates ($\beta_1, \beta_2$) are estimated using noisy and sparse flight test data in order to estimate the flutter margin in  ~\eref{eq:fm}.
 Using the estimates of the flutter margin at a number of subcritical airspeeds, the parameters $B_1$, $B_2$ and $B_3$ can be estimated using the method of  least-squares. Finally, the flutter speed $U_f$ can be 
computed as the positive real root of the equation:
\begin{equation}\label{eq:fm3}
B_1U_f^4+B_2U_f^2+B_3=0
\end{equation}
The flutter margin method  proposed by Zimmerman and Weissenburger \cite{weissenburger1964prediction} has had significant impact on flutter flight test.  Earlier use of the flutter margin method to wind tunnel data on aircraft wing is presented by Bennett \cite{bennett1982application}.   Zimmerman and Weissenburger implemented Routh's stability criteria to derive a flutter margin equation for 2D airfoil with 
quasi-steady aerodynamics \cite{weissenburger1964prediction,khalil2015jsv,khalil2013bayesian}.  
The flutter margin equation constitutes of frequencies $(\omega_1, \omega_2)$, and the decay rates $(\beta_1, \beta_2)$ of the aeroelastic modes participating in the flutter  at a given airspeed $U$~\cite{weissenburger1964prediction}. 
The positive flutter margin indicates a stable aeroelastic system. 
The flutter margin's dependence on airspeed is given by \eref{eq:fm2}.  
The airspeed at which $F=0$ is the flutter speed \cite{weissenburger1964prediction}.\\

Various extension of the original flutter margin method~\cite{weissenburger1964prediction} have been 
proposed. Price and Lee~\cite{price1993evaluation} extended the flutter margin method to the cases where
the flutter mechanism involves three aeroelastic modes, leading to the so-called trinary flutter.
Torii and Matsuzaki~\cite{torii1997flutter} extended the flutter margin method to the discrete-time systems using Jury's stability analysis. Dimitriadis and Cooper \cite{dimitriadis2001flutter} compared the predictive performance of the flutter margin method against several other methods, based on damping fit,
response envelope function, Nissim and Gilyard method and autoregressive moving average method.
The presence of considerable uncertainties in the modal parameter estimates prompted Poirel  
{\it et al.}~\cite{poirel2005flutter} to introduce probabilistic flutter margin method. Due to the nonlinear
relationships among the modal parameters, flutter margin and flutter speed, the Gaussian approximation of modal parameters translates to non-Gaussian pdfs of flutter margins and flutter speed as demostrated using
F18 flight data~\cite{poirel2005flutter}. Heege~\cite{heeg2007stochastic} futher examined this method to compute the flutter speed within the framework of maximum likelihood estimation. Abbasi and Cooper~\cite{Abbasi2009} concluded that the flutter margin pdfs become non-Gaussian as the aeroelastic 
frequencies come closer, being a typical feature of coalescence flutter. 

Khalil {\it et al.}~\cite{khalil2015jsv,khalil2013bayesian,khalil2009application,khalil2010bayes,khalil2011application,khalil2011probabilistic,khalil2010application}
formulated the probabilistic flutter margin method by Poirel {\it et al.}~\cite{poirel2005flutter} within the
framework of Bayesian inference. Using the Markov Chain Monte Carlo (MCMC) method, the joint samples 
of the modal parameters are generated first from the noisy measurement data at pre-flutter airspeeds which are finally used to construct the pdf of flutter speed. However, several simplifying assumptions are made in this approach: (a) the flat (uniform) priors with positive supports are assumed at each airspeed, (b) while modal damping and frequencies are statistically correlated at each airspeed, they are statistically independent at different airspeed. Consequently, the flutter margin are also statistically independent at different airspeeds. The current investigation generalizes the Bayesian flutter margin method by Khalil {\it at al.}~\cite{khalil2015jsv,khalil2013bayesian,khalil2009application,khalil2010bayes,
khalil2011application,khalil2011probabilistic,khalil2010application} by relaxing these
simplifying assumptions. First, the prior pdfs of modal parameters are constructed using the parametric uncertainty quantification method for a classical pitch-plunge airfoil whose system properties (e.g. structural  stiffness and damping) exhibit random variabilities. 
Note that no (aeroelastic) model was used in the previous investigation.
This joint modal prior (i.e. joint prior pdf of modal frequencies and decay rates for airspeeds) considers statistical dependence among  modal parameters within and across airspeeds. The  posterior MCMC samples of the modal parameters are generated using this joint modal parameter prior and the likelihood function obtained from measurement data at various airspeeds. Consequently, the  pdf of flutter speed is estimated using  the joint statistics of flutter margins for all airspeeds at which tests are conducted.  In this case, the uncertainty is significantly reduced in the  flutter speed prediction compared to the case of non-informative modal parameter prior and the statistical  independence of
margins across airspeeds.

\section{Dynamics of  a  two-dimensional  airfoil with quasi-steady  aerodynamics}
 
In this section, the dynamics of a simple two-dimensional (2D) pitch-plunge airfoil is described as shown in Fig.(\ref{2dairfoil}). The formulation of the quasi-steady aerodynamics is adapted from \cite{bisplinghoff2013aeroelasticity,fung2008,hodges2002introduction}. The parametric uncertainty analysis will be conducted
on this system in order to construct the prior of  modal frequencies and 
decay rates. 

An extensive body of literature exists on  the parametric uncertainty quantification in  aeroelastic systems. 
Pettit~\cite{pettit2003effects,pettit2004uncertainty} reported the effect of parametric uncertainty on the aeroelastic behavior of an airfoil undergoing limit cycle oscillations. Kurdi~\cite{kurdi2007uncertainty} studied Goland wing to estimate flutter boundaries using Monte Carlo simulation (MCS) due to random variabilities in structural parameters. Castravete~\cite{castravete2008effect} used a cantilever airfoil
to highlight the influence of uncertain stiffness on aeroelastic flutter analysis.    
Ueda~\cite{ueda2005aeroelastic} studied the effect of uncertainty in the location of the center of gravity and elastic axis of an aeroelastic system. Borello~\cite{borello2010structural} investigated the influence of  random  Young's modulus, shear modulus and mass on  flutter speed prediction.  Khodaparasi \cite{khodaparasi2009estimation}  used probabilistic and fuzzy logic methods to quantify uncertainty in flutter boundaries in the presence of uncertain structural parameters. Poirel and Price~\cite{dp-airfoil}
conducted a probabilistic analysis to study the effect of random airspeed on the flutter analysis of a typical section.

\begin{figure}[htbp]
\begin{center}
\includegraphics[scale= 0.8]{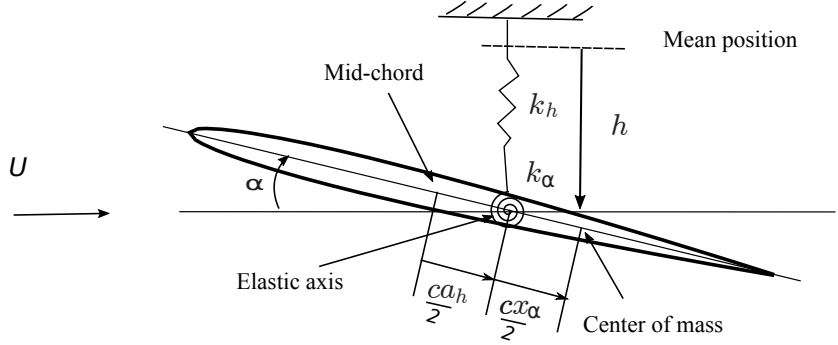}
\caption{symmetric airfoil (adapted from \cite{dp-airfoil,airfoil-plotter})}
\label{2dairfoil}
\end{center}
\end{figure}

\subsection{Equations of motion}
Next we briefly review the dynamics  of the classical two-degree-of-freedom airfoil undergoing pitch and plunge (heave) motions .
The quasi-steady thin-airfoil theory aerodynamic model is used for the system. 
The equations of motion of the pitch $\alpha$ and plunge $h$ motions shown below are adopted from \cite{bisplinghoff2013aeroelasticity,fung2008,hodges2002introduction,matachniouk2013parametric}.
\begin{equation}\label{eq:1}
m\frac{d^2h}{dt^2}+\frac{mcx_\alpha}{2}\frac{d^2\alpha}{dt^2}+B_1\frac{dh}{dt}+B_2\frac{d\alpha}{dt}+k_h h =-L(t)
\end{equation}
\begin{equation}\label{eq:2}
I_{EA}\frac{d^2\alpha}{dt^2}+\frac{mcx_\alpha}{2}\frac{d^2h}{dt^2}+B_3\frac{dh}{dt}+B_4\frac{d\alpha}{dt}+k_\alpha \alpha =M_{EA}(t)
\end{equation}
where
\begin{eqnarray*}\label{eq:3}
L(t)=\frac{\rho U^2}{2}c2\pi\left(\alpha+\frac{\dot{h}}{U}+\frac{\dot{\alpha c}(\frac{1}{2}-a_h)}{2U}\right),
\end{eqnarray*}
\begin{eqnarray*}\label{eq:4}
M_{EA}&=\frac{\rho U^2}{2}c2\pi\left(\alpha\frac{c(\frac{1}{2}+a_h)}{2}+\dot{h}\frac{c(\frac{1}{2}+a_h)}{2U}\right.\\\nonumber&+\left. \dot{\alpha}\left(\frac{c^2(\frac{1}{2}-a_h)(\frac{1}{2}+a_h)}{4U}-\frac{c^2}{16U}\right) \right)
\end{eqnarray*}

In the matrix-vector notation, the above equations can be expressed 
as follows:
\begin{eqnarray}\label{eq:7.1}
\underbrace{\left[\begin{array}{cc} m & \frac{mcx_{\alpha}}{2}\\ \frac{mcx_{\alpha}}{2} & I_{EA}\end{array}\right]}_{\text{Structural mass matrix}}\left\{\begin{array}{c}\ddot{h}\\ \ddot{\alpha}\end{array}\right\}+\underbrace{\left[\begin{array}{cc}B_1 &B_2\\ B_3 & B_4\end{array}\right]}_{\text{Structural damping matrix}}\left\{\begin{array}{c} {\dot{h}}\\ {\dot{\alpha}}\end{array}\right\}\\\nonumber+\underbrace{\left[\begin{array}{cc} k_h & 0\\ 0 & k_{\alpha}\end{array}\right]}_{\text{Structural stiffness matrix}}\left\{\begin{array}{c} {h}\\ {\alpha}\end{array}\right\}=\underbrace{\left\{\begin{array}{c} {-L(t)}\\ {M_{EA}}\end{array}\right\}}_{\text{Aerodynamic load vector}}
\end{eqnarray}
where
\begin{equation}\label{eq:7.2}
\left\{\begin{array}{c} {-L(t)}\\ {M_{EA}}\end{array}\right\}=\frac{\rho U^2}{2}c2\pi\left\{\begin{array}{c}
\underbrace{\left[\begin{array}{cc} 0 & -1\\ 0 & c/2(\frac{1}{2}+a_h)\end{array}\right]}_{\text{Aerodynamic stiffness matrix}}\left\{\begin{array}{c}{h}\\ {\alpha}\end{array}\right\} \\+\frac{1}{U}\underbrace{\left[\begin{array}{cc} -1 & -c/2(\frac{1}{2}-a_h)\\ c/2(\frac{1}{2}+a_h) & \frac{c^2(\frac{1}{2}-a_h)(\frac{1}{2}+a_h)}{4}-\frac{c^2}{16}\end{array}\right]}_{\text{Aerodynamic damping matrix}}\left\{\begin{array}{c} \dot{h}\\ \dot{\alpha}\end{array}\right\}\end{array}\right\}
\end{equation}

\noindent
The proportional Rayleigh damping   is set with respect to the structural mass and stiffness matrices given by the following equation (e.g.~\cite{humar2012dynamics}):\begin{equation}
[B_{str}]=\alpha_0[M_{str}]+\alpha_1[K_{str}]
\end{equation}
The coefficients $\alpha_0$ and $\alpha_1$ are evaluated using prescribed damping ratios $\xi_i$ and $\xi_j$ using the following relation (e.g. ~\cite{humar2012dynamics}):
\begin{equation}
{\frac{1}{2}\left[\begin{array}{cc}\frac{1}{\omega_i} &\omega_i\\\omega_j & \omega_j\end{array}\right]}\left\{\begin{array}{c} {\alpha_0}\\ {{\alpha_1}}\end{array}\right\}=\left\{\begin{array}{c} {\xi_i}\\ {{\xi_j}}\end{array}\right\}
\end{equation}
where $\omega_i$ and $\omega_j$ are natural frequencies of the two modes at $U=0$. When ~\eref{eq:7.1} is converted into the state-space form: $\dot{x}=Ax$ (where $x$ being the state-vector containing heave and pitch displacements and velocities) , the state matrix $A$  is given 
by (e.g.~\cite{matachniouk2013parametric}):
\begin{equation}\label{eq:7}
A=\left[\begin{array}{cccccc} &  &  &  &  &  \\ & [0]_{2\times 2} &  &  & [I]_{2\times 2} &  \\ &  &  &  &  &  \\  &   &   &   &   &   \\  & [M^{-1}K]_{2\times 2} &   &   & [M^{-1}C]_{2\times 2} &   \\  &   &   &   &   &  \end{array}\right]
\end{equation}
where
\begin{equation}\label{eq:7.2}
[M]=\left[\begin{array}{cc} m & \frac{mcx_{\alpha}}{2}\\ \frac{mcx_{\alpha}}{2} & I_{EA}\end{array}\right],
\end{equation}
\begin{equation}
[K]=\left[\begin{array}{cc} k_h & \rho U^2c\pi\\ 0 & k_{\alpha}-\frac{\rho U^2c^2\pi}{2}(\frac{1}{2}+a_h)\end{array}\right],
\end{equation}
\begin{equation}
[C]=\left[\begin{array}{cc} B_1+\frac{\rho U}{2}c2\pi & B_2+\frac{\rho U}{2}c^2\pi(\frac{1}{2}-a_h)\\ B_3-\frac{\rho U}{2}c^2\pi(\frac{1}{2}+a_h) &B_4-\frac{\rho U}{2}c2\pi\left( \frac{c^2(\frac{1}{2}-a_h)(\frac{1}{2}+a_h)}{4}-\frac{c^2}{16}\right)\end{array}\right].
\end{equation}
Solving the  eigenvalue problem, two pairs of complex conjugate roots  are obtained as~\cite{hoen2005engineering}
\begin{equation}\label{eq:8}
s_i=\beta_i\pm i\omega_i
\end{equation}
\noindent
where $\beta_i$ and $\omega_i$ are the negative decay rate and the frequency respectively.

\section{Application of Bayesian Inference in flutter speed estimation}
The free decay response (heave or pitch) of the aeroelastic system described in ~\eref{eq:7.1} can be expressed as~\cite{khalil2013bayesian,hoen2005engineering}:
\begin{equation}\label{eq:11}
u_k={\hat u}_k +\eta_k,\,\,\,\, k=1,...,N
\end{equation}
and
\begin{equation}\label{eq:111}
{\hat u}_k =a_1e^{-\beta_1 t_k}\cos(\omega_1 t_k+b_1)+a_2e^{-\beta_2 t_k}\cos(\omega_2 t_k+b_2)
\end{equation}
where $\eta_k$ is the Gaussian noise with zero mean and variance $\gamma$; and it can account for both measurement and model errors (e.g. due to the effects of higher aeroelastic modes neglected in the 2-dof pitch-plunge model and the influence of unsteady aerodynamics~\cite{khalil2015jsv}).   While we have considered a white noise process to model the measurement noise (leading to statistically independent 
${\{\eta_k\}}$s, i.e. $E(\eta_i \eta_j)=0,\, i\ne j$ where $E$ denotes the expectation operator)  as the initial step, further studies  are needed to investigate  the benefits  of colored, non-stationary and  non-Gaussian noise models (e.g. \cite{philippe2022}).

The likelihood function of these unknown parameters\\ 
$\alpha_p=\{a_1,a_2,b_1,b_2,\beta_1,\beta_2,\omega_1,\omega_2,\gamma \}$ is~\cite{khalil2015jsv}:
\begin{equation}\label{eq:12}
p( u_{k=1:N}|\alpha_p) =\frac{1}{(2\pi\gamma)^{N/2}} 
\exp{\sum_{k=1}^{N}-\frac{1}{2\gamma}(u_k-{\hat u}_k (\alpha_p))^2}
\end{equation}\\

Using Bayesian formulation, the posterior pdf of the parameter  is given by:
\begin{equation}\label{eq:13}
p(\alpha_p^{n_u}|u_{k=1:N}^{n_u})\propto p( u_{k=1:N}^{n_u}|\alpha_p^{n_u})p(\alpha_p^{n_u})
\end{equation}
where $n_u$ is the index of airspeed at which experiments are conducted,
 and the prior pdf $p(\alpha_p^{n_u})$ is estimated using the parametric uncertainty quantification method through MCS \cite{matachniouk2013parametric}. Samples of modal parameters 
${\hat \alpha}_q=(\beta_1^{n_u},\beta_2^{n_u},\omega_1^{n_u},\omega_2^{n_u})$ in~\eref{eq:8} are generated using Monte Carlo simulation by solving the eigenvalue problem. In Khalil {\it et\, al.}~\cite{khalil2015jsv}, the noise variance $\gamma$ was estimated to account for both model and measurement errors as the unsteady aerodynamic model was used to generate the synthetic  data (corrupted by a Gaussian measurement noise) while the quasi-steady aerodynamic model was used for the inference. In the current case, the quasi-steady aerodynamic model is used for both generating synthetic data and inference. In this case, $\gamma$ denotes only measurement noise  which is assumed to be known (hence $\gamma$ is not estimated). The flat (non-informative) priors are used  for $a_1$, $a_2$, $b_1$ and $b_2$.

The prior pdf $p({\hat \alpha}_q)$ of the modal parameters  is approximated to be Gaussian as given by \cite{khalil2013bayesian,silverman1986density}: 
\begin{equation}\label{eq:13.1}
p({\hat \alpha}_q^{i})=\frac{1}{\sqrt{(2\pi)^{q}|\sum_i|}}\exp{ \left(-\frac{1}{2}({\hat \alpha}_q^{i}-\mu_i)^T{\sum}_i^{-1}({\hat \alpha}_q^{i}-\mu_i)\right)},\,\,\, i=1,...,n_u
\end{equation}
where $\mu_i$ and  $\sum_i$ are  the mean vector and  covariance matrix  of ${\hat \alpha}_q^{i}$ respectively (where  $q=4$, being the dimension of the modal parameter vector). Although the Gaussian approximation of the modal prior is adequate (as explained later) for this investigation, the proposed method is general and can handle any non-Gaussian prior. 

For the independent Gaussian measurement noise, 
the joint parameter posterior for airspeeds at  the $n_u$ points   is given by: 
\begin{equation}\label{eq:13.3}
p(\alpha_p^1,\ldots , \alpha_p^{n_u}|u_{k=1:N}^1, \ldots , u_{k=1:N}^{n_u})\propto p(\alpha_p^1,\ldots , \alpha_p^{n_u})\prod_{i=1}^{n_u}p(u_{k=1:N}^i|\alpha_p^i).
\end{equation}

where the prior $p(\alpha_p^1,\cdots , \alpha_p^{n_u})$ is estimated using parametric uncertainty quantification \cite{matachniouk2013parametric}.
Using \eref{eq:13.3}, the posterior samples of $(\alpha_p^1,\cdots , \alpha_p^{n_u})$
are generated  using the AMMCMC (e.g.~\cite{gilks1996introducing}) sampling technique. 
The implementational details of the AMMCMC are available~\cite{sc_thesis}.
Using these samples, the corresponding samples of flutter margin are obtain using~\eref{eq:fm} and subsequently the joint pdf of flutter margin $p(F|B)$ is generated.
The joint pdf of flutter margin  $p(F|B)$   is used to estimate unknown coefficients $B=\{B_1,B_2,B_3\}$ from \eref{eq:fm2} using the Bayesian formulation as \cite{khalil2015jsv,khalil2010application}:
\begin{equation}\label{eq:14}
p(B|F)\propto p(F|B)p(B)
\end{equation}
where
\begin{equation}
p(F|B)=p(F^1,\ldots ,F^{n_u}|{\alpha^{1}_p},\ldots , {\alpha^{n_u}_p} ).
\end{equation}
In~\eref{eq:fm2}, the flutter margin is very sensitive to the accuracy of $B_1$ term \cite{weissenburger1964prediction,khalil2015jsv,bennett1982application}. To tackle the issue, it is preferred to use the following second order polynomial as suggested by Zimmerman and Weissenburger \cite{weissenburger1964prediction} as well as  Bennett \cite{bennett1982application}:  
\begin{equation}\label{eq:23}
F=B_2U^2+B_3
\end{equation}
where its positive real root provides the flutter speed.
Moreover, the prior of $p(B)$ is given by \cite{khalil2015jsv, khalil2010application}
\begin{equation}\label{eq:prior}
p(B)\propto\begin{cases}
1, & \text{if $ B_2, B_3$ satisfy conditions(1) and (2)}. \\
0, & \text{otherwise}.
\end{cases}
\end{equation}
where\\
 \hspace{1.5cm} 
 condition(1): $B_2<0$\\
  \hspace{1.5cm}
 condition(2): $B_3>0$\\
Condition 1 reflects that the flutter speed has real roots and condition 2 ensures the flutter margin is positive at zero airspeed \cite{khalil2015jsv,khalil2010application}.\\

In order to highlight the benefit of the current formulation compared to previous investigations \cite{khalil2013bayesian,khalil2009application,khalil2010bayes,khalil2011application,khalil2011probabilistic,
khalil2010application}, the following cases are considered:

\begin{enumerate}
\item {\it \underline{Flat prior}}: At each airspeed, the prior of model parameter vector is considered to be uniformly distributed with a large  support (also positivity enforced for modal parameters)   as in Khalil {\it et al.} \cite{khalil2013bayesian,khalil2009application,khalil2010bayes,khalil2011application,khalil2011probabilistic,
khalil2010application}:

\begin{equation}
p(\alpha_p^{i})\propto 1, \,\,\, i=1,...,n_u.
\end{equation}

The model parameter vectors and flutter margins are  statistically independent at different airspeeds, leading to:
\begin{equation}\label{eq:prior_ind}
p(\alpha_p^1,\ldots, \alpha_p^{n_u})=\prod_{i=1}^{n_u}p(\alpha_p^i)
\end{equation}
and 
\begin{equation}\label{eq:fm_ind}
  p(F^1,\ldots, F^{n_u})=\prod_{i=1}^{n_u}p(F^{i})
\end{equation}

This case relates to the previous work by  Khalil {\it et al.} \cite{khalil2013bayesian,khalil2009application,khalil2010bayes,khalil2011application,khalil2011probabilistic,
khalil2010application}

\item {\it \underline{Independent prior}}: While the statistical dependence among the model parameters are now considered at a given airspeed, the  model parameter priors and flutter margins are still assumed to be  independent from one airspeed to another as described in ~\eref{eq:prior_ind} and ~\eref{eq:fm_ind}. The results of this case are new contributions of this paper.

\item {\it \underline{Joint prior}}: The  statistical dependence is now consider among modal parameter vectors and  flutter margins at all airspeeds. The results of this  general case are new contributions.  

\end{enumerate}

\section{Deterministic baseline}
The nominal values of the system parameters in~\tref{Case10cov} (values in the middle column)  are 
used for deterministic simulation.  The  results in this section provide true estimate of flutter speed which is used to compare the results obtained from the Bayesian method. 
The calculated  flutter speed (FS)  is $54.01\ m/s$ as shown in~\fref{FM}. The flutter margin calculated using
~\eref{eq:fm} is plotted using the blue line solid line (FM: method1). 
Five flutter margin values are plotted in red dots at airspeeds ranging from 50\% to 70\% of the flutter speed. 
Using these five data points, the flutter margin is estimated  by the least square fit (dashed line, FM: method2) using ~\eref{eq:23}. The inaccuracy in the least square estimate of flutter speed using ~\eref{eq:23} is evident in~\fref{FM}. The aeroelastic (modal) frequencies and decay rates versus airspeed are plotted in~\fref{Freq} and~~\fref{Decay} respectively. As evident from these figures, the two frequencies approaches closer as the decay rate on the unstable mode become negative.
\begin{table}[ht!]
\caption{System parameter values}
\begin{center}\begin{tabular}{|c|c|c|}\hline parameter & Value & $\sigma$ for 10\% COV \\\hline mass & 50 kg& 5 \\\hline span & 1 m&  \\\hline $I_{EA}$ & 0.25 kg-m$^2$& 0.025\\\hline c & 0.2 & \\\hline $k_h$ & 3000 N-m$^2$& 300\\\hline $k_{\alpha}$ & 150 N-m$^2$& 15\\\hline $x_{\alpha}$ & 0.25 & 0.025 \\\hline $a_h$ & -0.45 & 0.045  \\\hline $\rho$ & 1.19 kg/m$^3$& \\\hline $\xi_1$ & 0.02 & \\\hline $\xi_2$ & 0.02 & \\\hline \end{tabular} 
\end{center}
\label{Case10cov}
\end{table}

\begin{figure}[htbp!]
\begin{center}
\includegraphics[scale= 0.3]{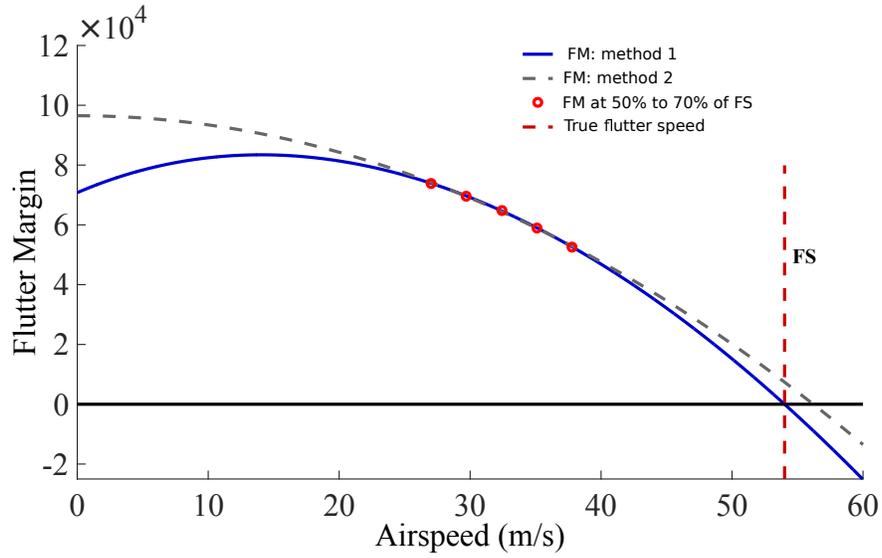}
\caption{Flutter Margin versus airspeed}
\label{FM}
\end{center}
\end{figure}

\begin{figure}[htbp!]
\begin{center}
\includegraphics[scale= 0.3]{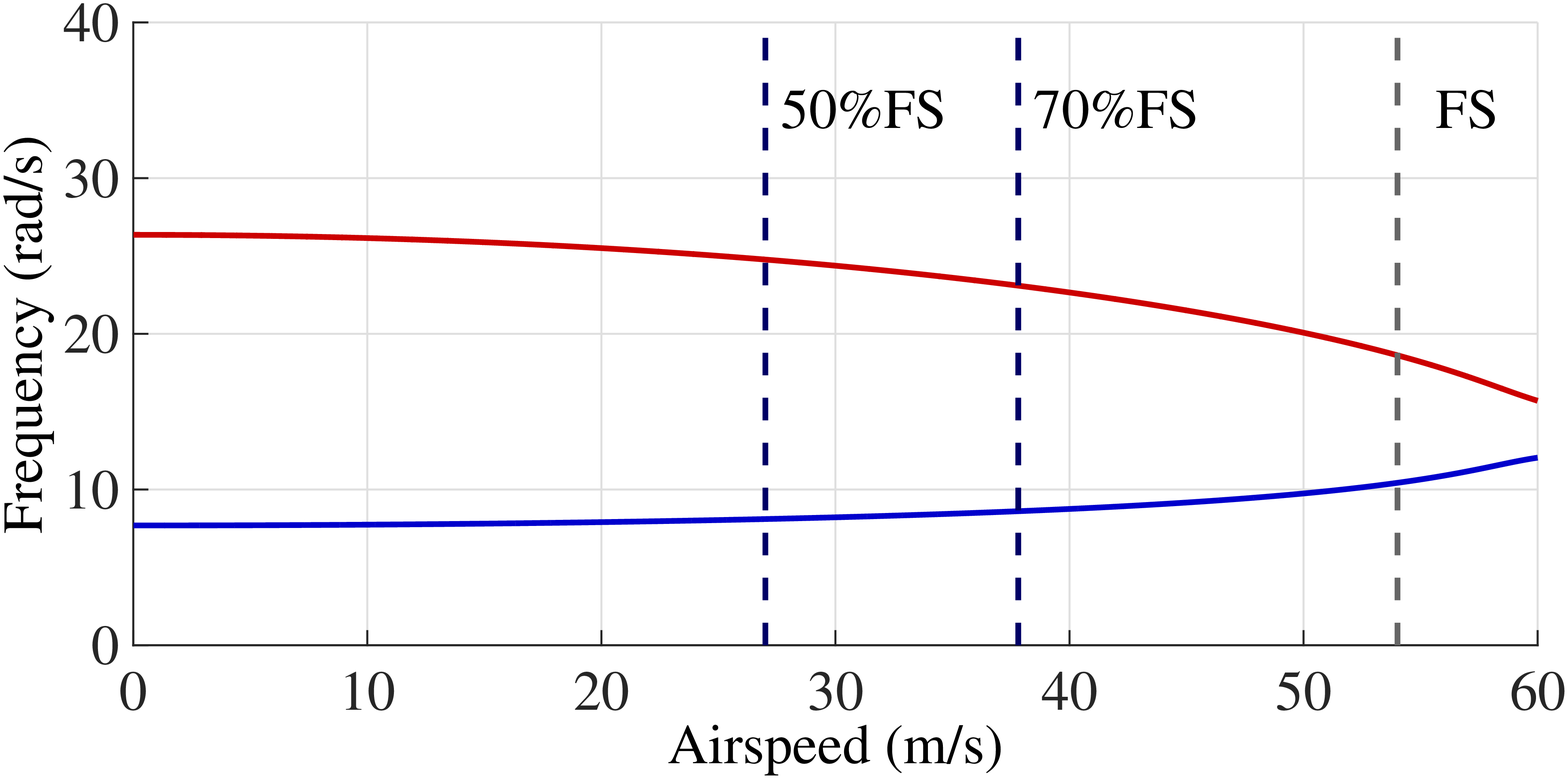}
\caption{Frequency versus airspeed}
\label{Freq}
\includegraphics[scale= 0.3]{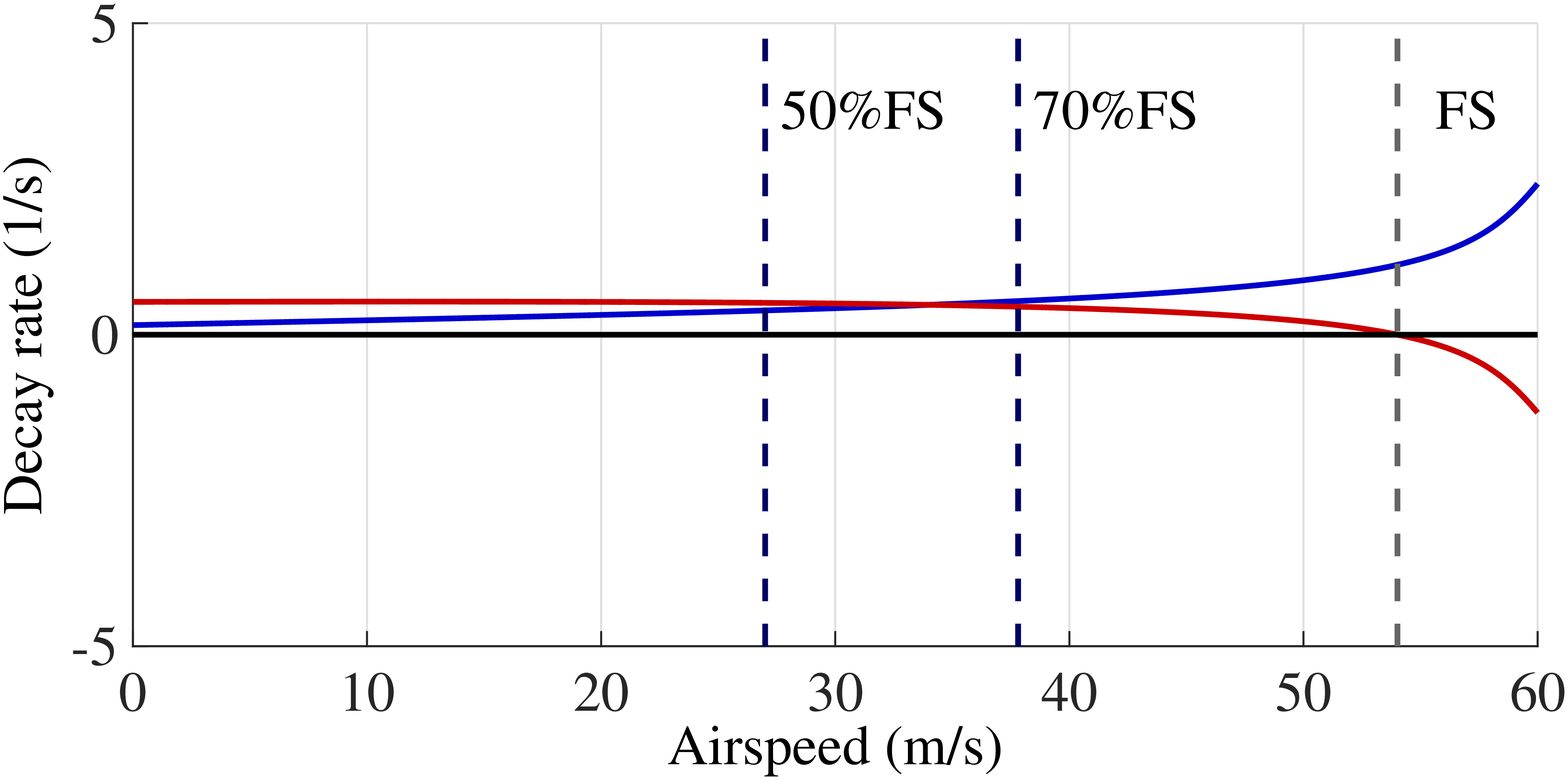}
\caption{Decay versus airspeed}
\label{Decay}
\end{center}
\end{figure}

\section{Bayesian inference results}
\fref{3Uresponse1}, ~\fref{3Uresponse2}  and ~\fref{3Uresponse3} show the noisy measurement data and the associated true free-decay  response at the airspeeds 
$U=27.00$ $m/s$, $U=32.40$ and $U=37.80$ $m/s$,  computed at the 50\%, 60\%  and 70\% of the flutter speed, respectively. The standard deviation for measurement noise is 12\% of the root mean squared (rms) value of the true signal. These noisy measurement data are used for the Bayesian inference as described next.

To construct parameter  priors, 10\% COV  is considered in the Gaussian distributions of mass ($m$), 
mass moment of inertia about the elastic axis (${I_{EA}}$), heave and pitch stiffnesses ($k_h$  and $k_{\alpha}$), the normalized distance between the elastic axis and mid-chord ($a_h$)  and the normalized static imbalance ($x_{\alpha}$) as described in ~\tref{Case10cov}. For the small COV, we ignore the fact that negetive values may be produced in the strictly positive system parameters due to Gaussian assumptions.
The  marginal prior pdfs of modal parameters are shown in \fref{fig3Uprior10cov}. The assumption of Gaussian approximation for the modal prior pdfs in~\eref{eq:13.1} can be validated from~\fref{fig3Uprior10cov} which shows marginal pdfs overlapped with the corresponding Gaussian distributions (in green color). 

 In ~\fref{fig3ujointprior10cva} and 
~\fref{fig3ujointprior10cvb}, the two-dimensional joint pdfs of modal parameters are plotted at each airspeed and between two airspeeds. Most of these modal parameters show significant correlations as exhibited by their tilted elliptic shapes in the subplots of  ~\fref{fig3ujointprior10cva} and ~\fref{fig3ujointprior10cvb}. Note that the mass, stiffness and
damping matrices nonlinearly map to aeroelastic modal parameters (frequencies and decay rates).
Hence the statistically independent Gaussian randomness in the mass, stiffness and damping matrices can 
introduce correlations among modal parameters, including non-Gaussian effects in their pdfs due to nonlinear mappings. However, for small uncertainty (i.e. small COVs), the nonlinear mappings
among input parameters (i.e. mass, stiffness and damping matrices) and modal parameters
can be locally approximated to be linear maps (functions) for the probabilistic analysis (related to transformations of random vectors, see \cite{matachniouk2013parametric} for more details). In these cases, the Gaussian randomness in the mass, stiffness and damping matrices will also lead to the pdfs of modal parameters which
are nearly Gaussian.

The marginal posterior  pdfs of modal parameters are plotted in 
~\fref{3u10cvomega11} --~\fref{3u10cvbeta32} which are generated using AMMCMC samples. These figures  compare the posterior pdfs obtained using the flat prior, independent prior and joint prior of modal parameters.  The corresponding modal priors (shown in blue color) are also plotted which are obtained from MCS.\\
The modal frequencies and decay rates obtained using the flat priors exhibit significant bias and uncertainty in their estimates. Note that the posteriors of modal parameters  are primarily governed by measurement data alone for the case of flat priors. In  general, the  uncertainties in the estimated modal frequencies are lower that those in the decay rates.
The modal parameter posteriors are sharper (narrower) for the independent priors than the flat priors.
Hence the modal frequencies and decay rates are better captured with the independent priors than the flat priors. The uncertainties in the estimated decay rates are reduced with the independent priors compared to the flat priors.

In ~\fref{3u10cvomega11} --~\fref{3u10cvbeta32} , the  marginal posterior pdfs of modal parameters  obtained using the joint priors demonstrate that the accuracy of the estimates is significantly improved when statistical dependence of the modal parameters are  accounted for among airspeeds.
The modal parameter posteriors are the sharpest (narrowest) for the case of the joint priors 
compared to the flat and independent priors. It means that the best accuracy in the modal parameter estimates is achieved  when the statistical dependence among these parameters across the airspeeds are enfored through the joint priors.
The bias in the estimated modal parameters using the independent priors and the joint priors is due to the finite information about the system encoded in the noisy and sparse observational data.

~\fref{3u10cvjointpostflat},~\fref{3u10cvjointpostindependent} and~\fref{3u10cvjointpostjoint}  show joint posterior pdfs of modal parameters  among airspeeds estimated using   the flat prior, independent prior and joint prior respectively. In ~\fref{3u10cvjointpostjoint}, some joint posterior pdfs estimated using the joint prior show the statistical dependence (reflected through the tilted elliptic shapes) among modal parameters across the airspeeds, while joint posterior pdfs estimated using the flat prior (in ~\fref{3u10cvjointpostflat}) and independent prior (in ~\fref{3u10cvjointpostindependent})   can not account for the underlying statistical dependence  among modal parameters across airspeeds.

~\fref{3u10cvjointpostfm} shows joint posterior pdfs of flutter margins for the airspeeds for the cases of the flat prior, independent prior and joint prior.  The joint  pdf  estimated using the joint prior can predict the inherent statistical dependence of flutter margins between the airspeeds as reflected in the tilted elliptic shape (see the third column in ~\fref{3u10cvjointpostfm}). The joint pdf of flutter margin  estimated using the flat prior is non-Gaussian as shown in the first column of ~\fref{3u10cvjointpostfm}. The symmetric 
elliptic shape (with respect to the horizontal and vertical axes) of the joint posterior pdf in the second column of ~\fref{3u10cvjointpostfm} points out the statistical independence of the flutter margins among airspeeds due to the use of the independent prior.

~\fref{3u10cvjointpostB1B2} shows the joint posterior pdfs of parameters $B_2$ and $B_3$ from  ~\eref{eq:23} used to estimate the flutter speed. The best accuracy of the estimates of $B_2$ and $B_3$ are obtained using the joint prior (shown in the third panel), having the most compact elliptic shape of the joint posteriors compared to the cases of the flat and independent priors (shown in  the first and second panels).

~\fref{Fs3U10cov} summarizes the main results of this paper. It shows the comparison of the posterior of flutter speed estimated using the flat prior, independent prior, joint prior and prior alone (i.e. flutter speed pdf constructed just with prior information alone without measurement data) with the true flutter speed estimated using the deterministic  calculation and least square fit using the true flutter margin values computed  at three airspeeds.
Note that the uncertainty  in the estimated flutter
speed (reflected by the width of pdfs) decrease while using independent prior and joint prior, compared to  
the flat prior. The uncertainty in the estimated flutter speed is minimum for  the case of the joint
prior as it contains the maximum prior information about the modal parameters among the airspeeds.
For the flat prior, independent prior and joint prior,
~\fref{Fsboxplot3U10cov} represents the bias and uncertainty through  the box plots of flutter speed estimates corresponding to  ~\fref{Fs3U10cov}. In \fref{Fs3U10cov} and ~\fref{Fsboxplot3U10cov},  the corresponding colors match  for the flat prior, independent prior and joint prior.
 For each prior in ~\fref{Fsboxplot3U10cov}, the middle horizontal line in the box  provides the MAP estimate of flutter speed using the Bayesian inference.  On the other hand,  the top and bottom horizontal lines above and below each box indicate    the MAP plus three standard deviation and the MAP minus three standard deviation respectively. The distance between the MAP estimate and true 
flutter speed is minimum for the joint prior (hence having the least bias). Futhermore the uncertainty in the estimated flutter speed is also minimum  for the
case of the joint prior.
\tref{Case3FScv} shows the COVs of the estimated flutter speed for the flat prior, independent prior and joint prior. As expected, the COV is minimum for the joint prior and maximum for the 
flat prior.
\begin{figure}[htbp]
\begin{center}
\includegraphics[width=120mm,height=65mm]{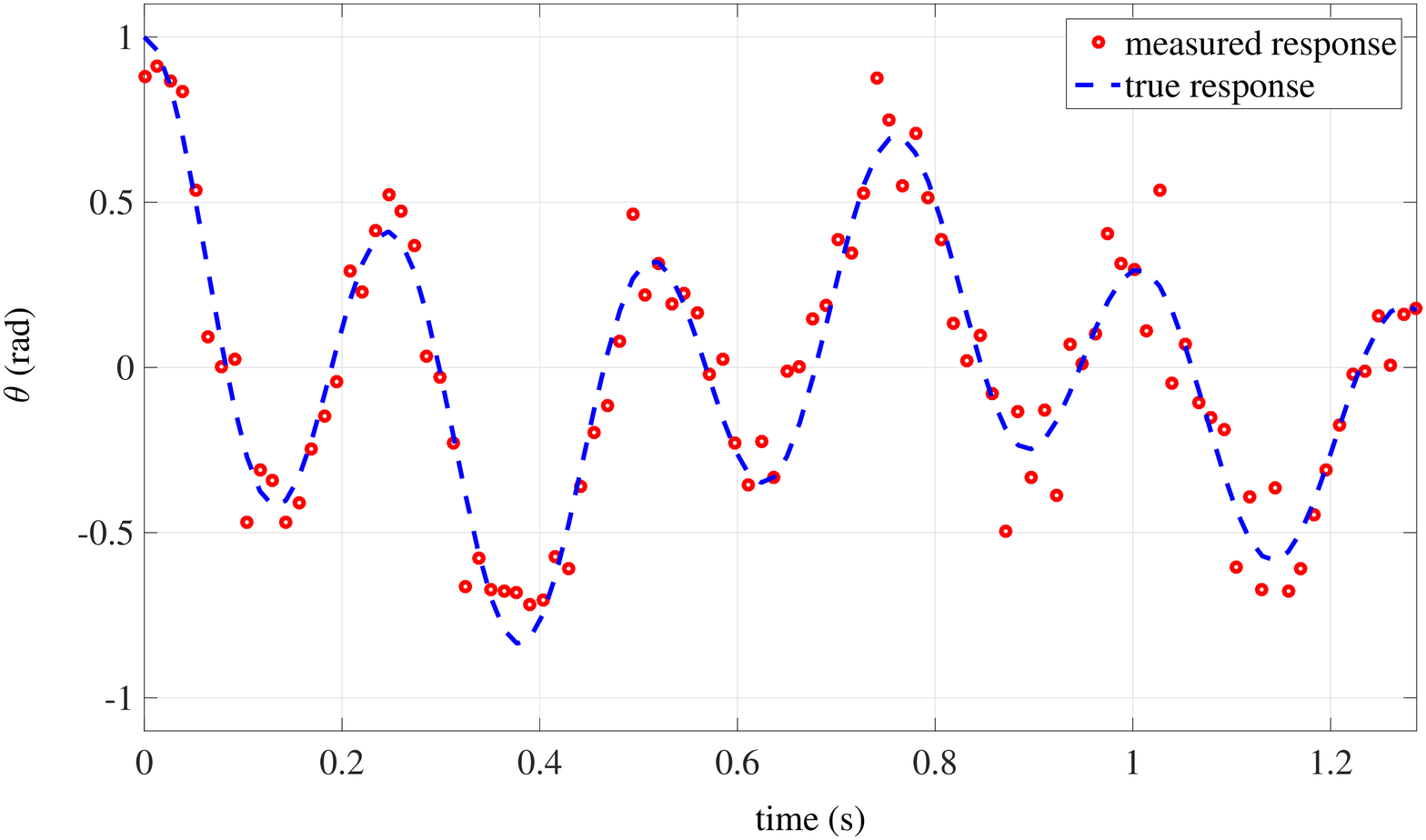}
\caption{Measurement at first airspeed $U=27.00$ $m/s$}
\label{3Uresponse1}
\end{center}
\end{figure}
\begin{figure}[htbp]
\begin{center}
\includegraphics[width=120mm,height=65mm]{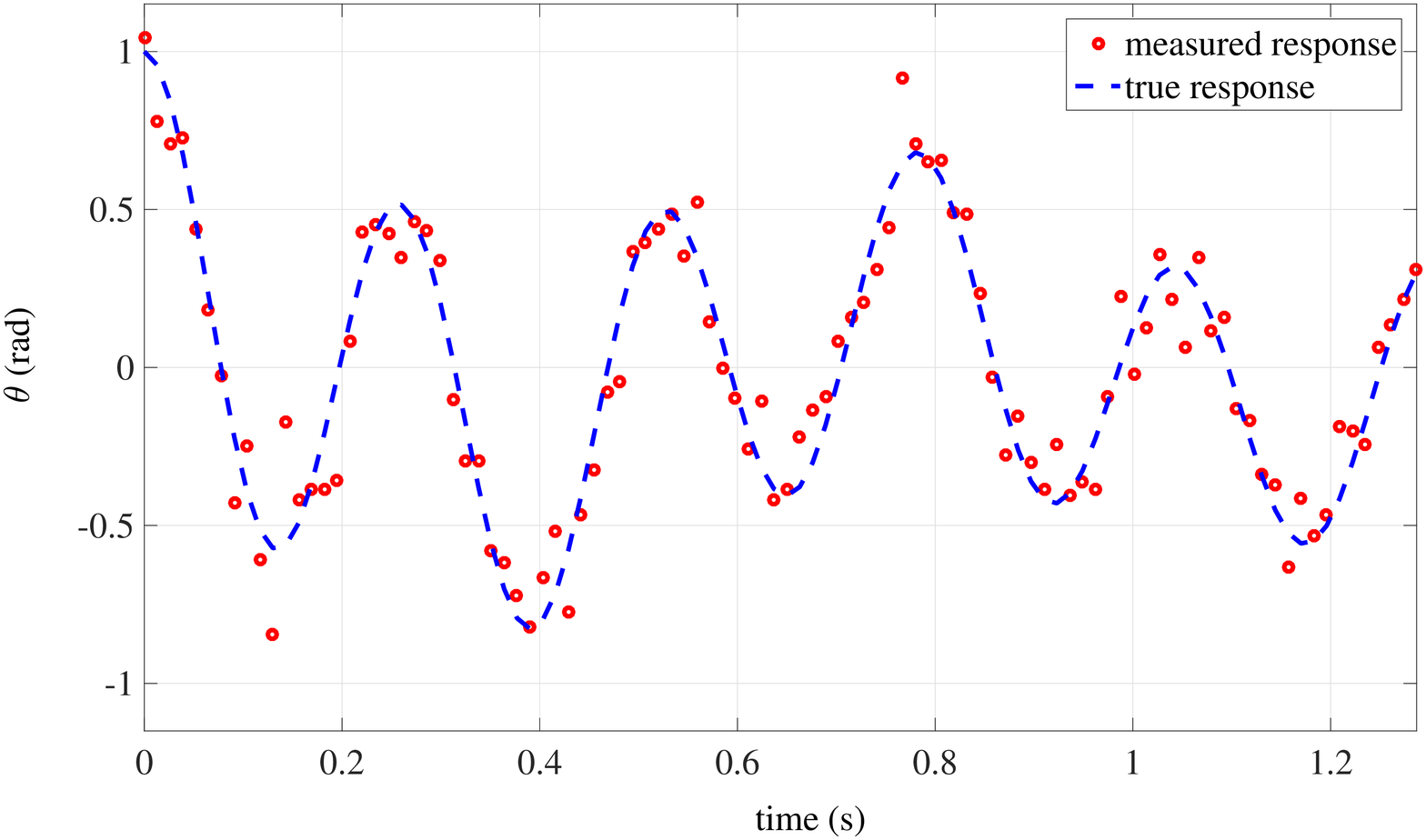}
\caption{Measurement at second airspeed $U=32.40$}
\label{3Uresponse2}
\end{center}
\end{figure}
\begin{figure}[htbp]
\begin{center}
\includegraphics[width=120mm,height=65mm]{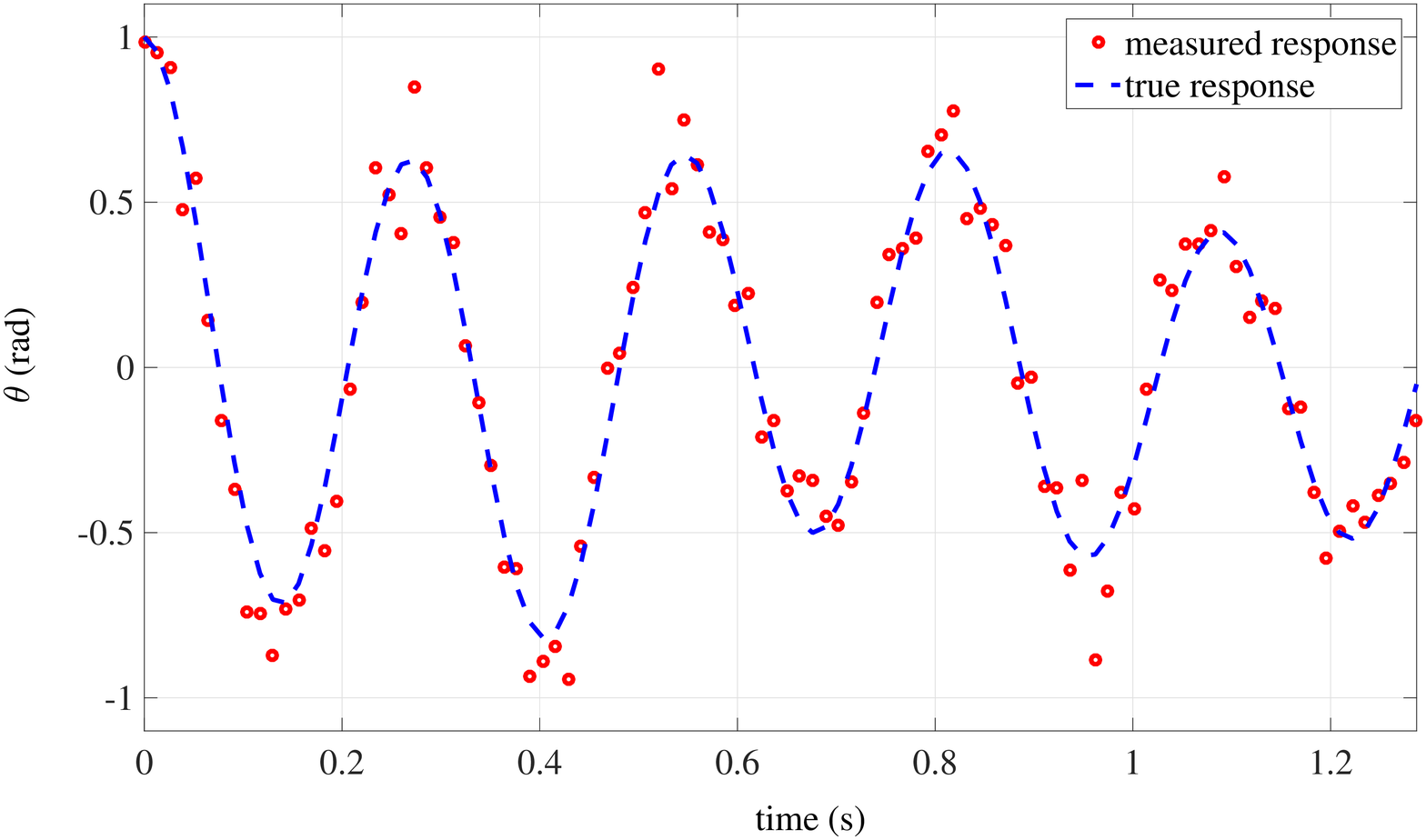}
\caption{Measurement at third airspeed $U=37.80$ $m/s$}
\label{3Uresponse3}
\end{center}
\end{figure}

\begin{figure}[h!]
\centering
\begin{tabular}{ccc}
\subf{\includegraphics[width=45mm]{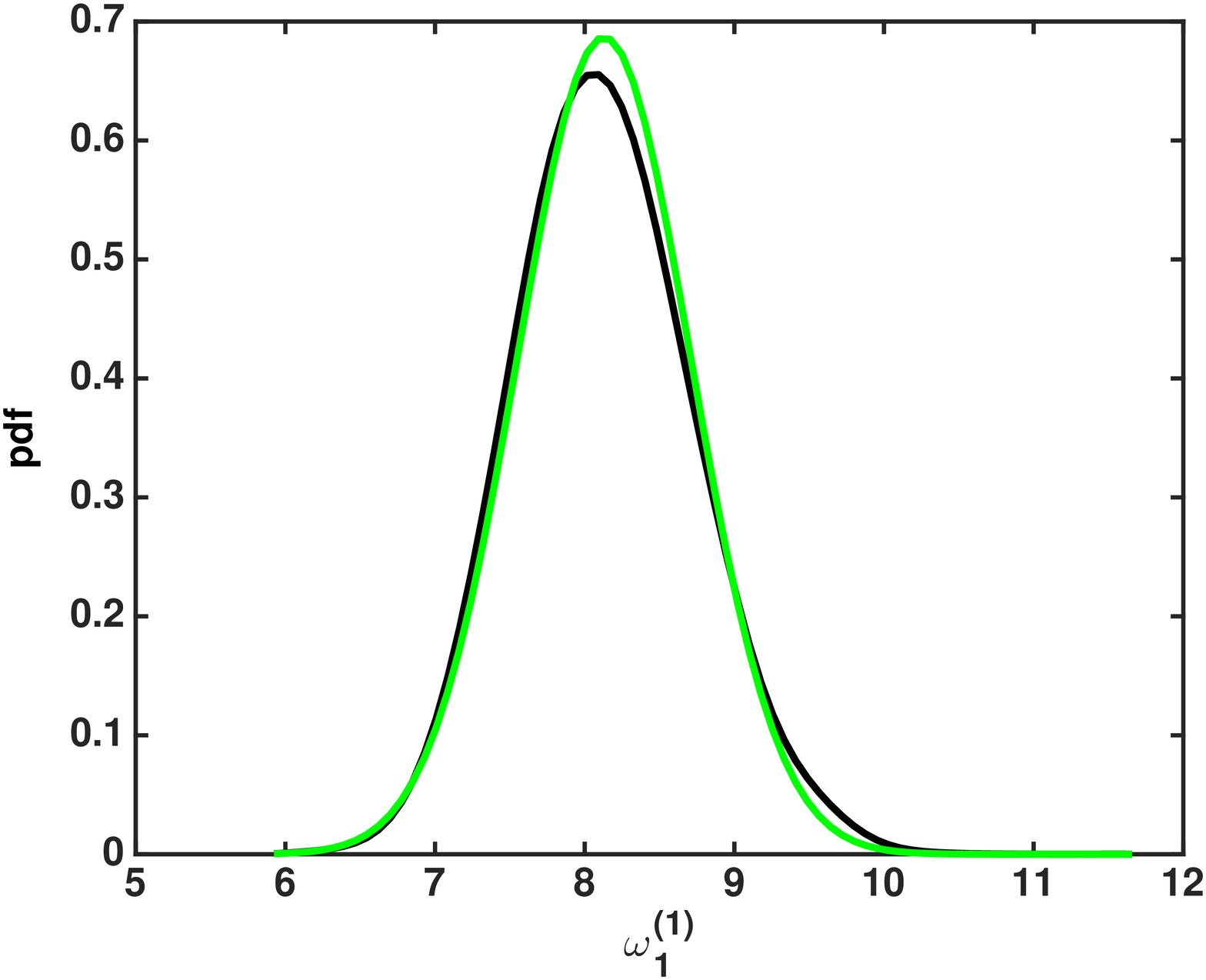}}
     {$$}
&
\subf{\includegraphics[width=45mm]{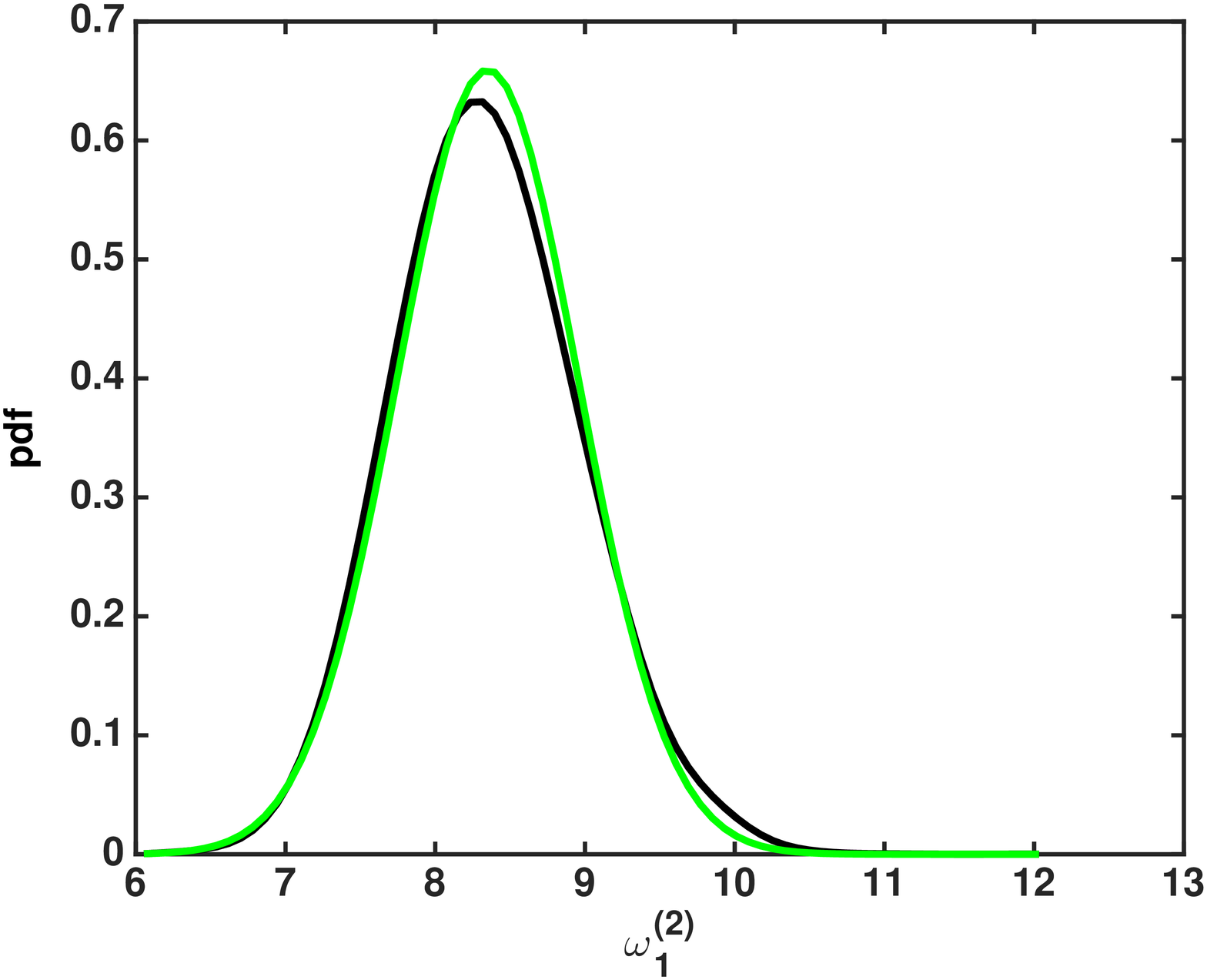}}
     {$$}
&
\subf{\includegraphics[width=45mm]{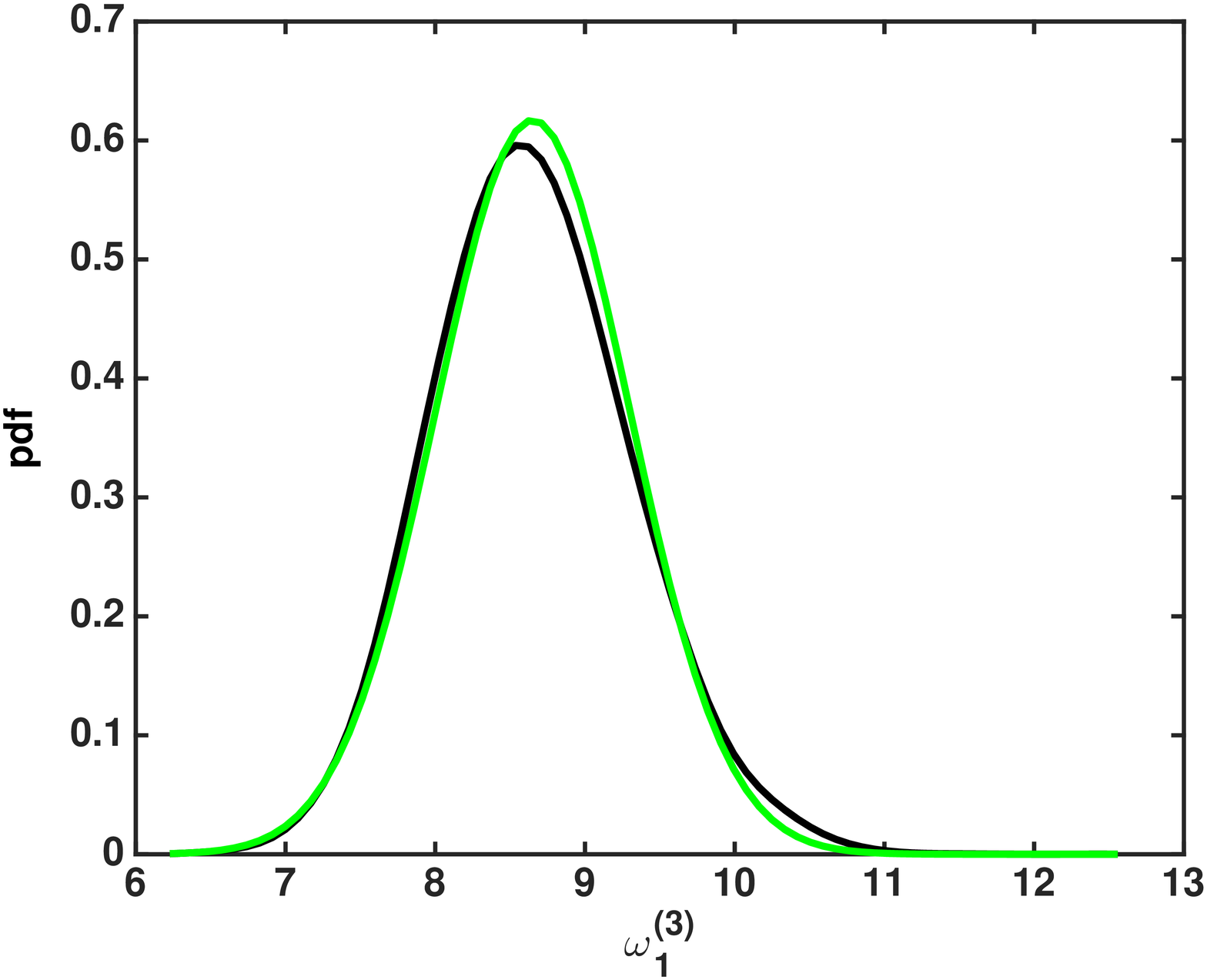}}
     {$$}      
\\
\subf{\includegraphics[width=45mm]{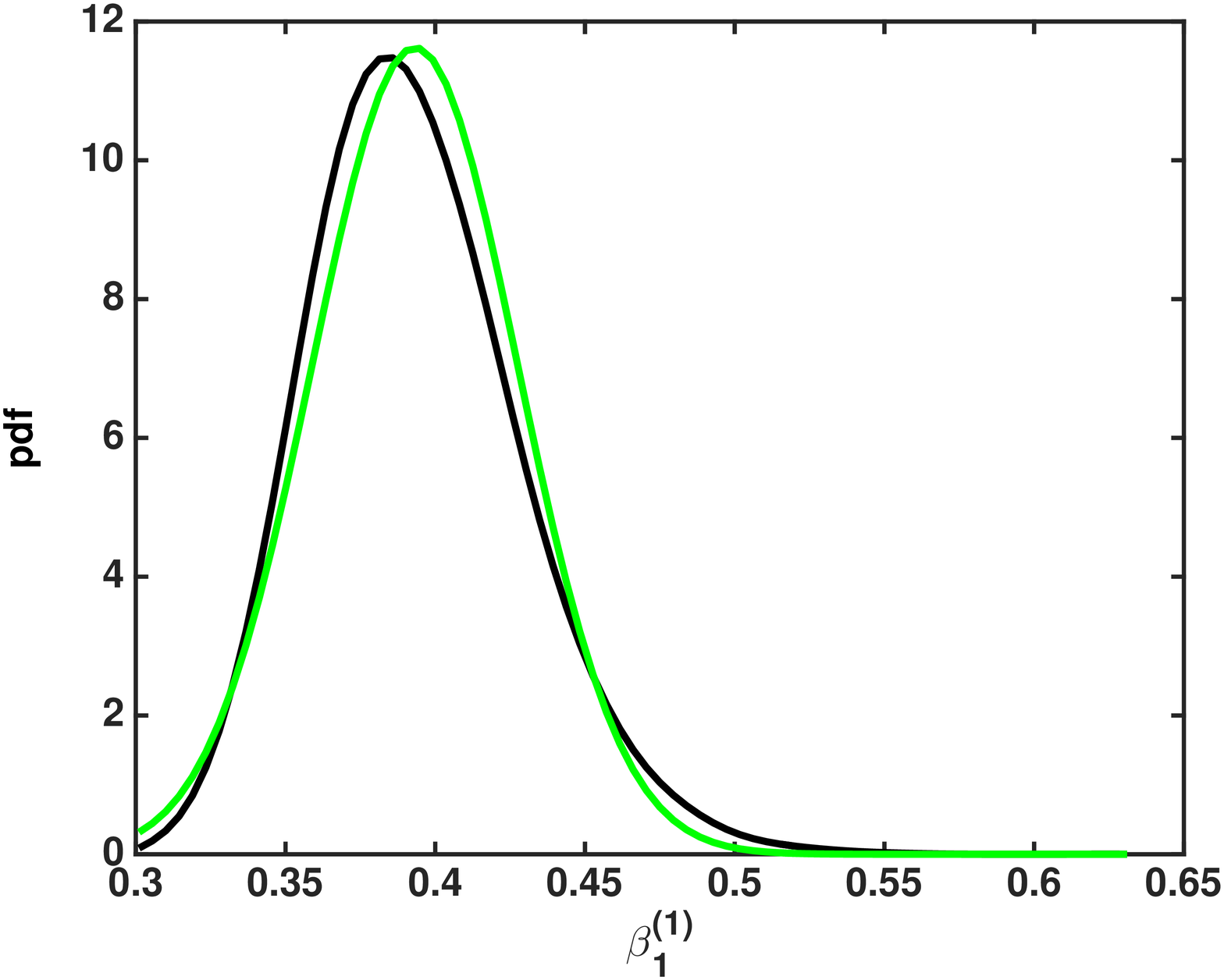}}
     {$$}
&
\subf{\includegraphics[width=45mm]{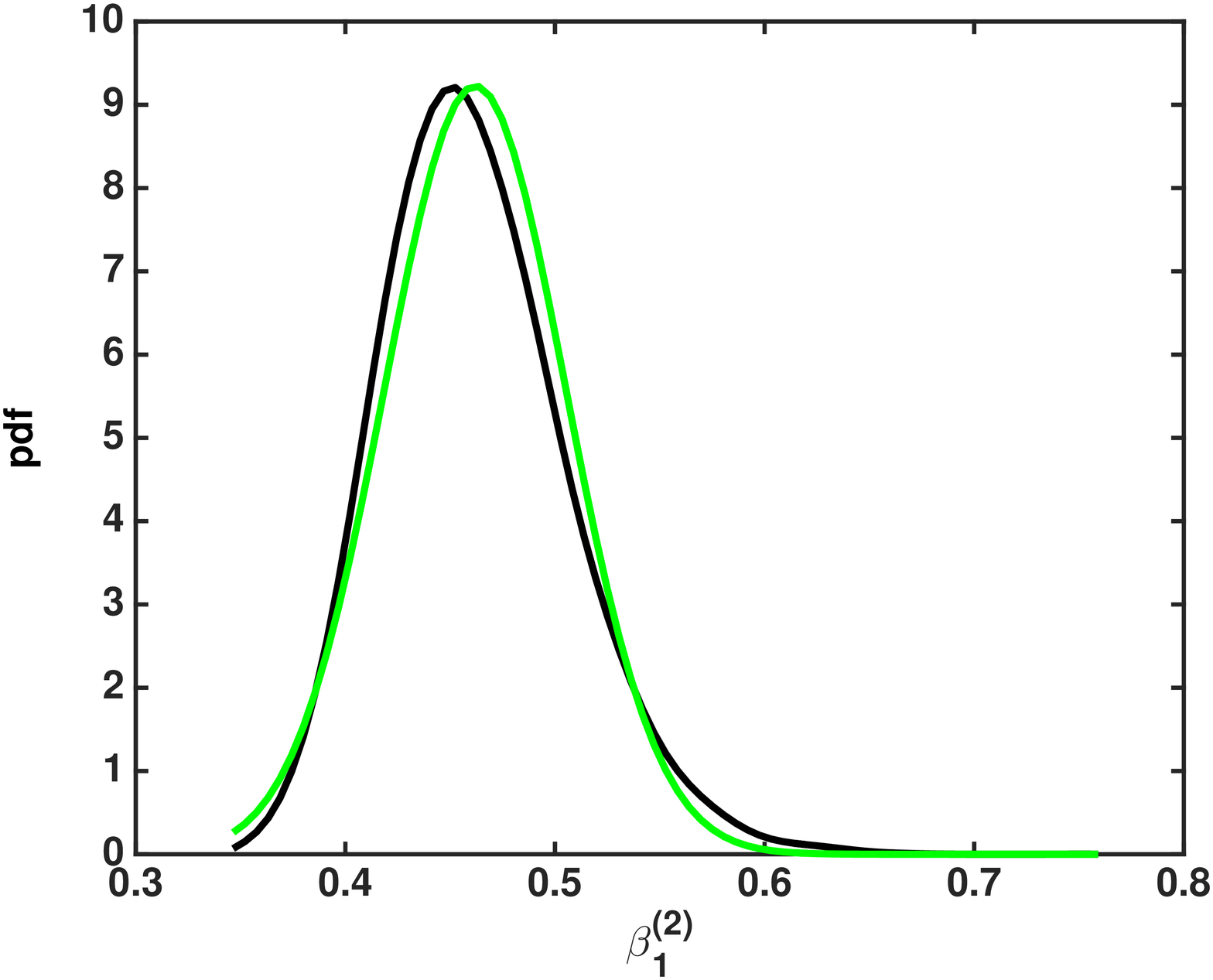}}
     {$$}
&
\subf{\includegraphics[width=45mm]{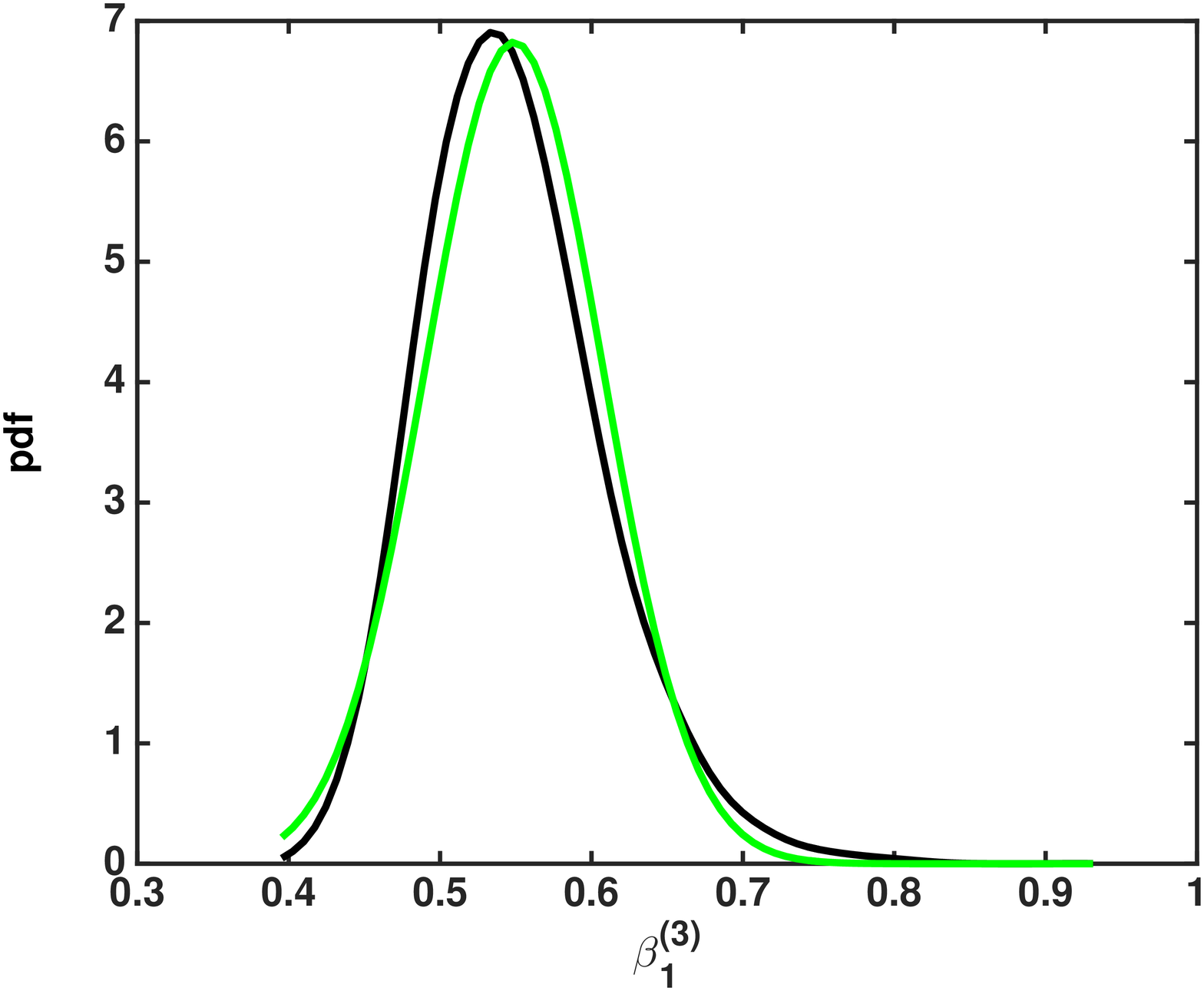}}
     {$$}
\\
\subf{\includegraphics[width=45mm]{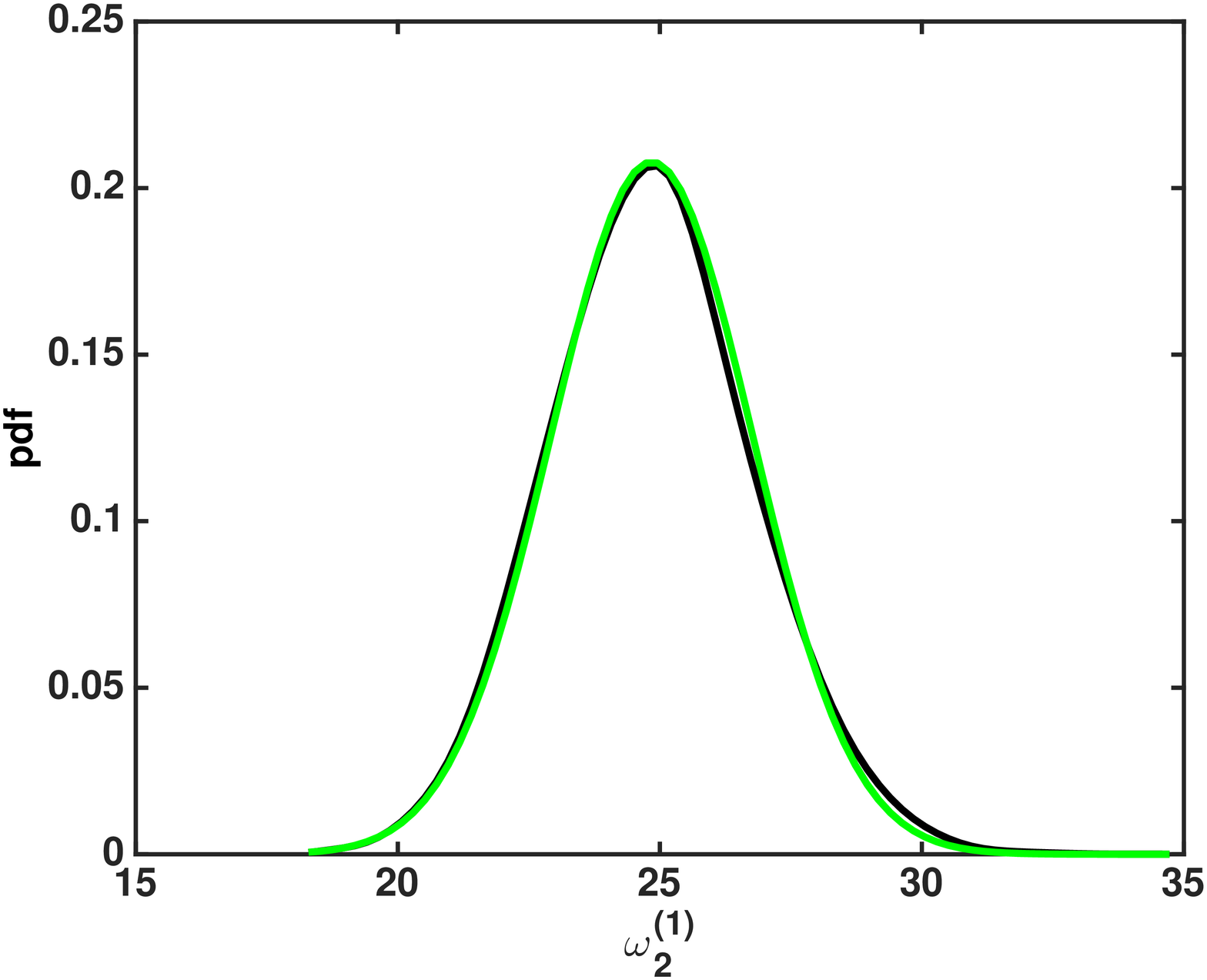}}
     {$$}
&
\subf{\includegraphics[width=45mm]{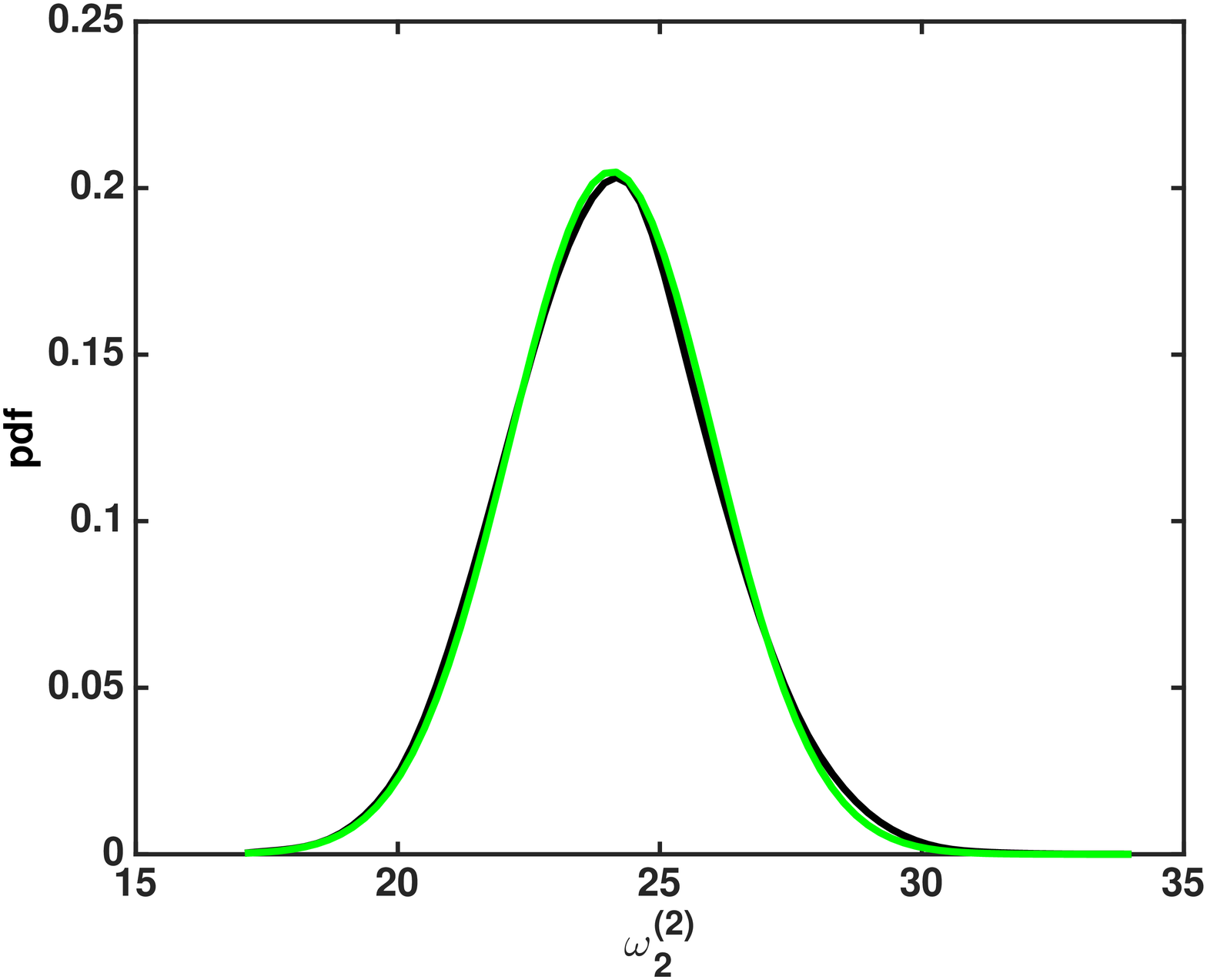}}
     {$$}
&
\subf{\includegraphics[width=45mm]{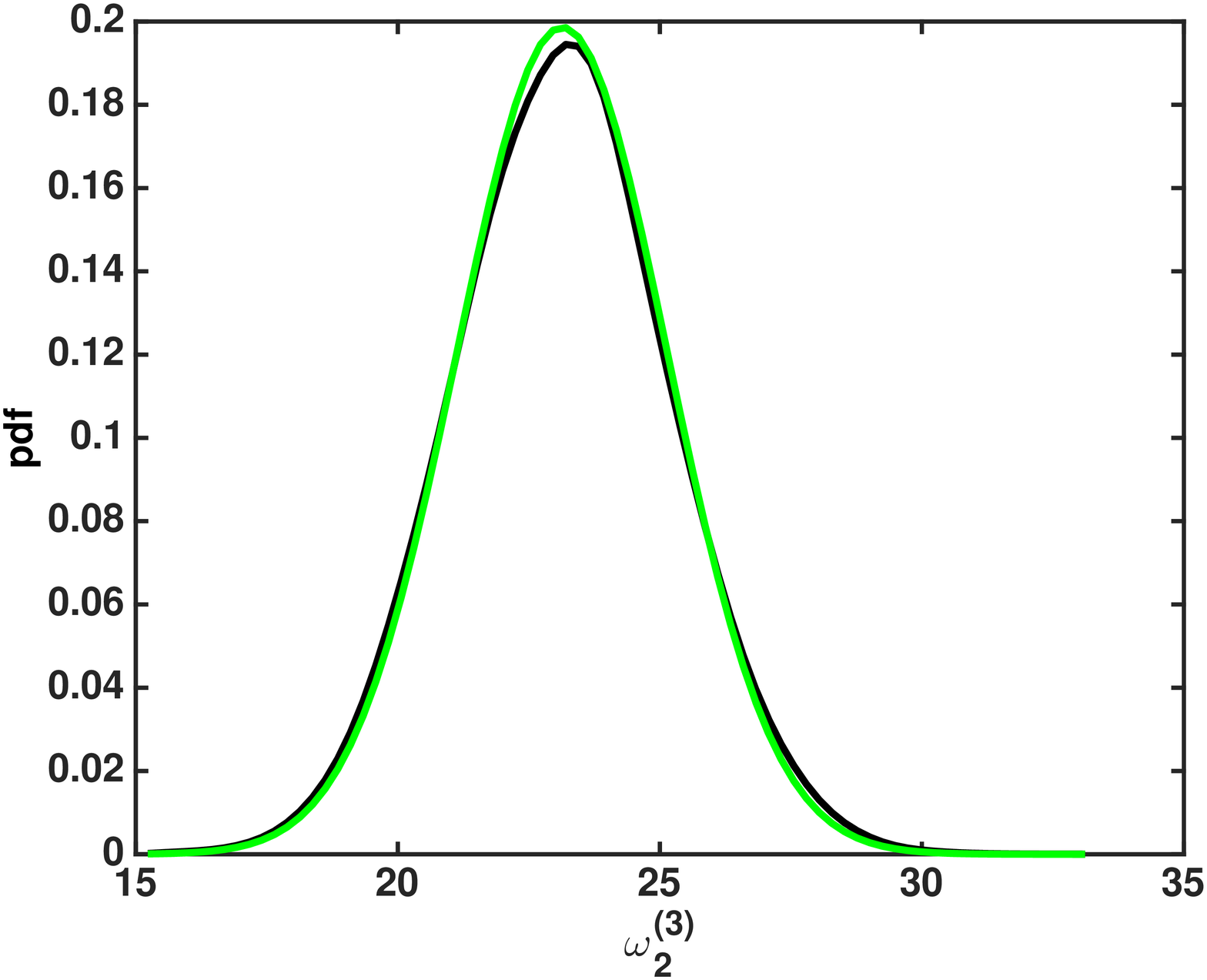}}
     {$$}
\\
\subf{\includegraphics[width=45mm]{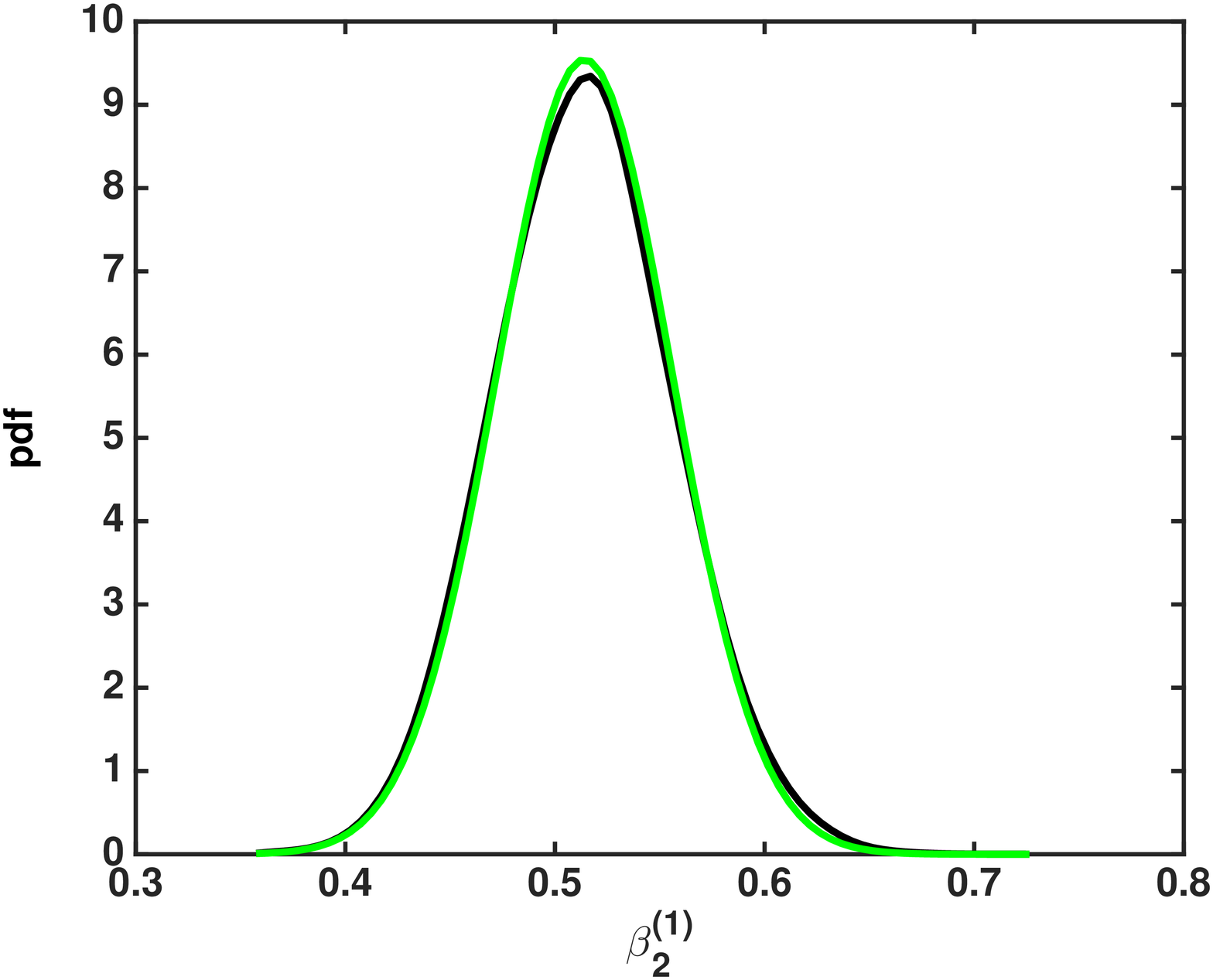}}
     {$$}
&
\subf{\includegraphics[width=45mm]{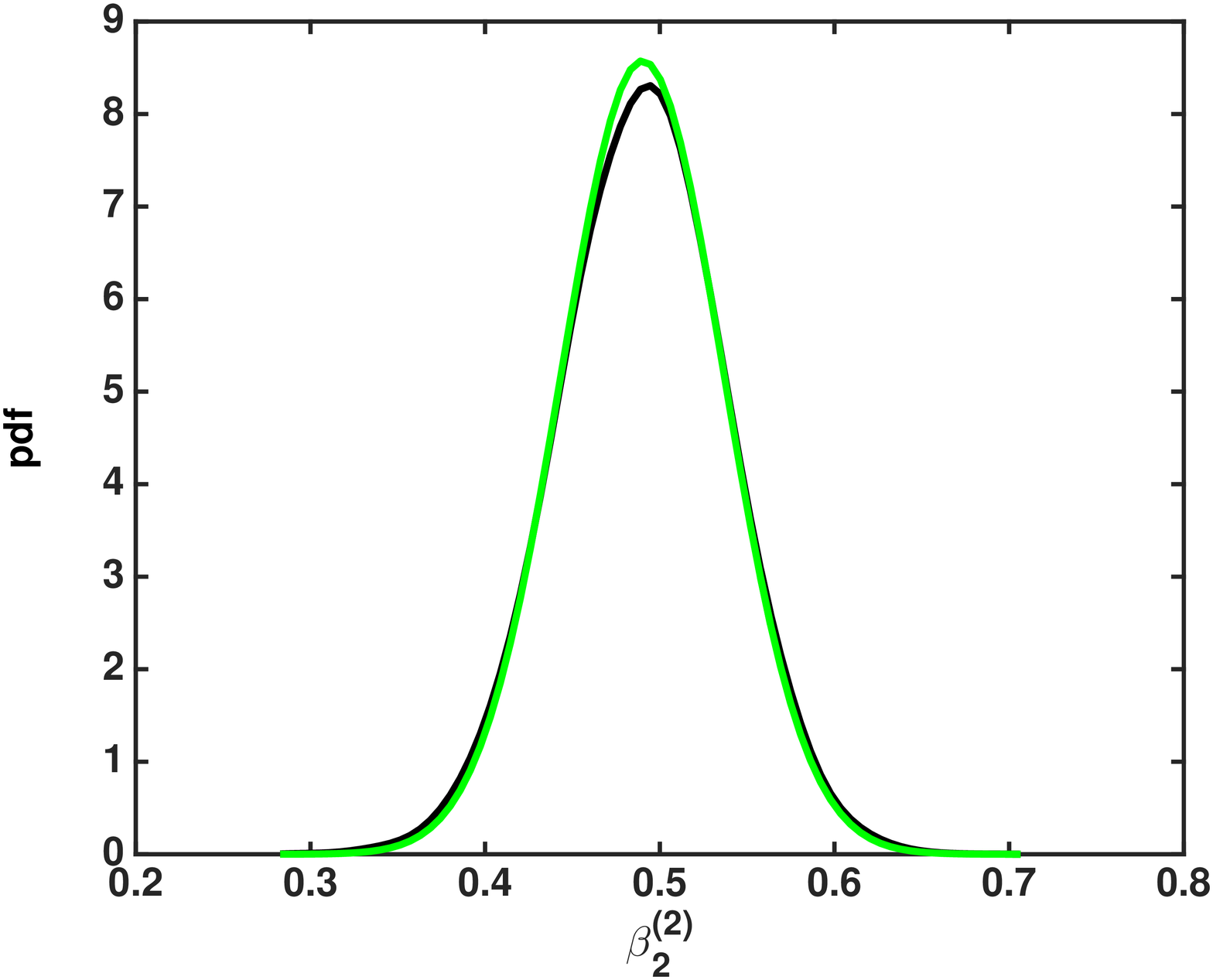}}
     {$$}
&
\subf{\includegraphics[width=45mm]{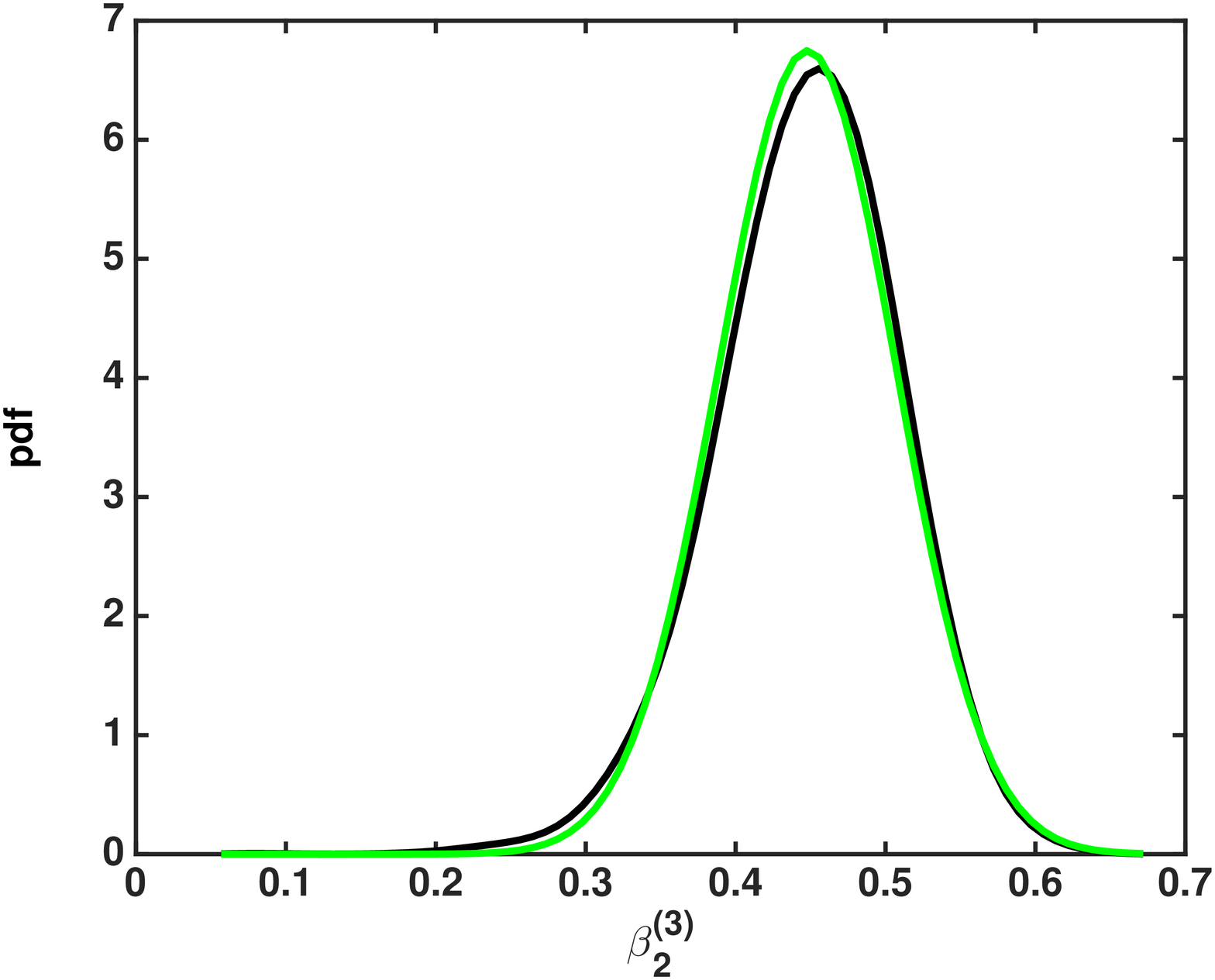}}
     {$$}
\\
\end{tabular}
\caption{Marginal  prior pdfs of parameters $\omega_1, \beta_1, \omega_2$ and $\beta_2$ at three airspeeds for 10\% input COV. MCS: black,  equivalent Gaussian: green}
\label{fig3Uprior10cov}
\end{figure}


\begin{figure}[h!]
\centering
\begin{tabular}{ccc}
\subf{\includegraphics[width=45mm]{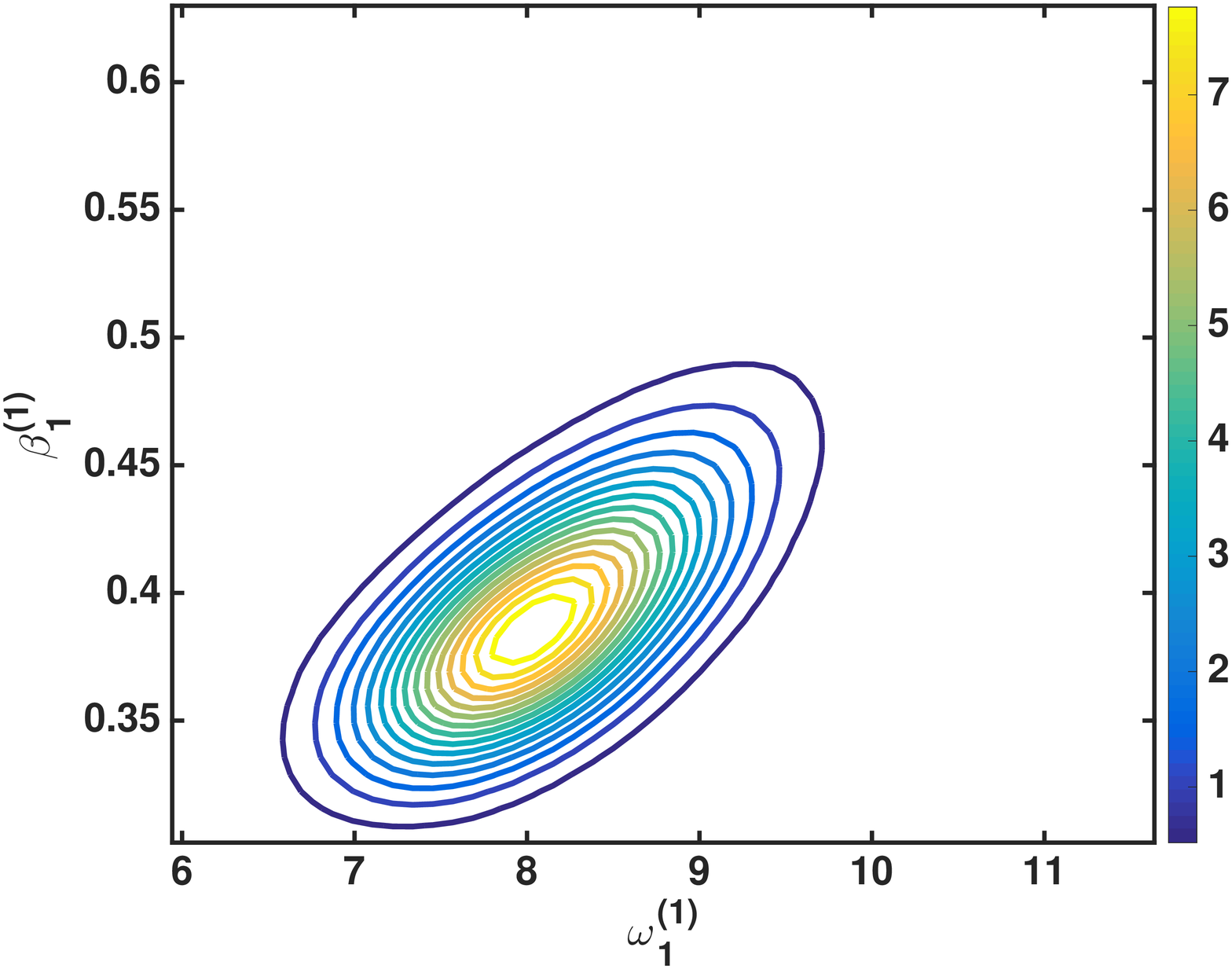}}
     {$$}
&
\subf{\includegraphics[width=45mm]{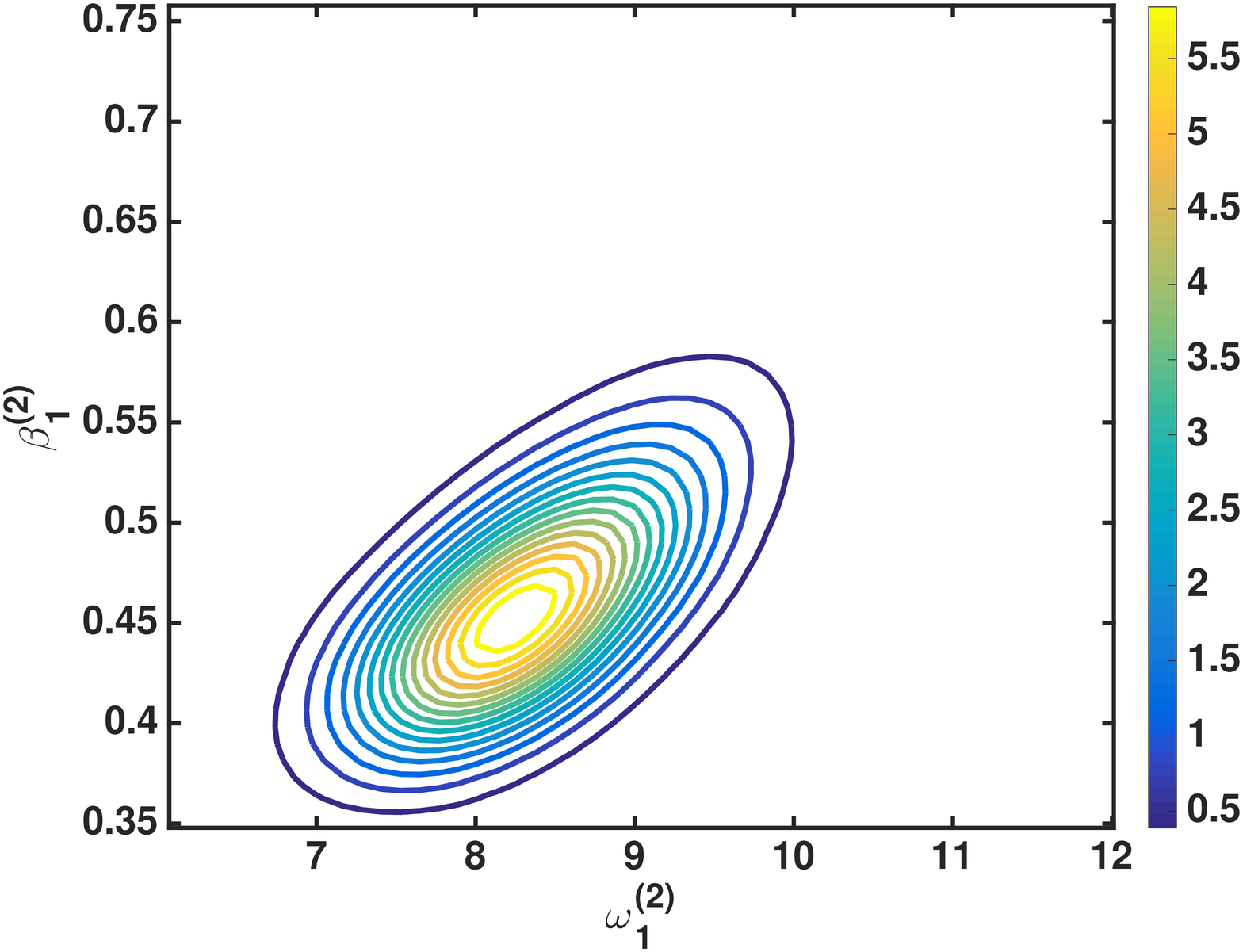}}
     {$$}
&
\subf{\includegraphics[width=45mm]{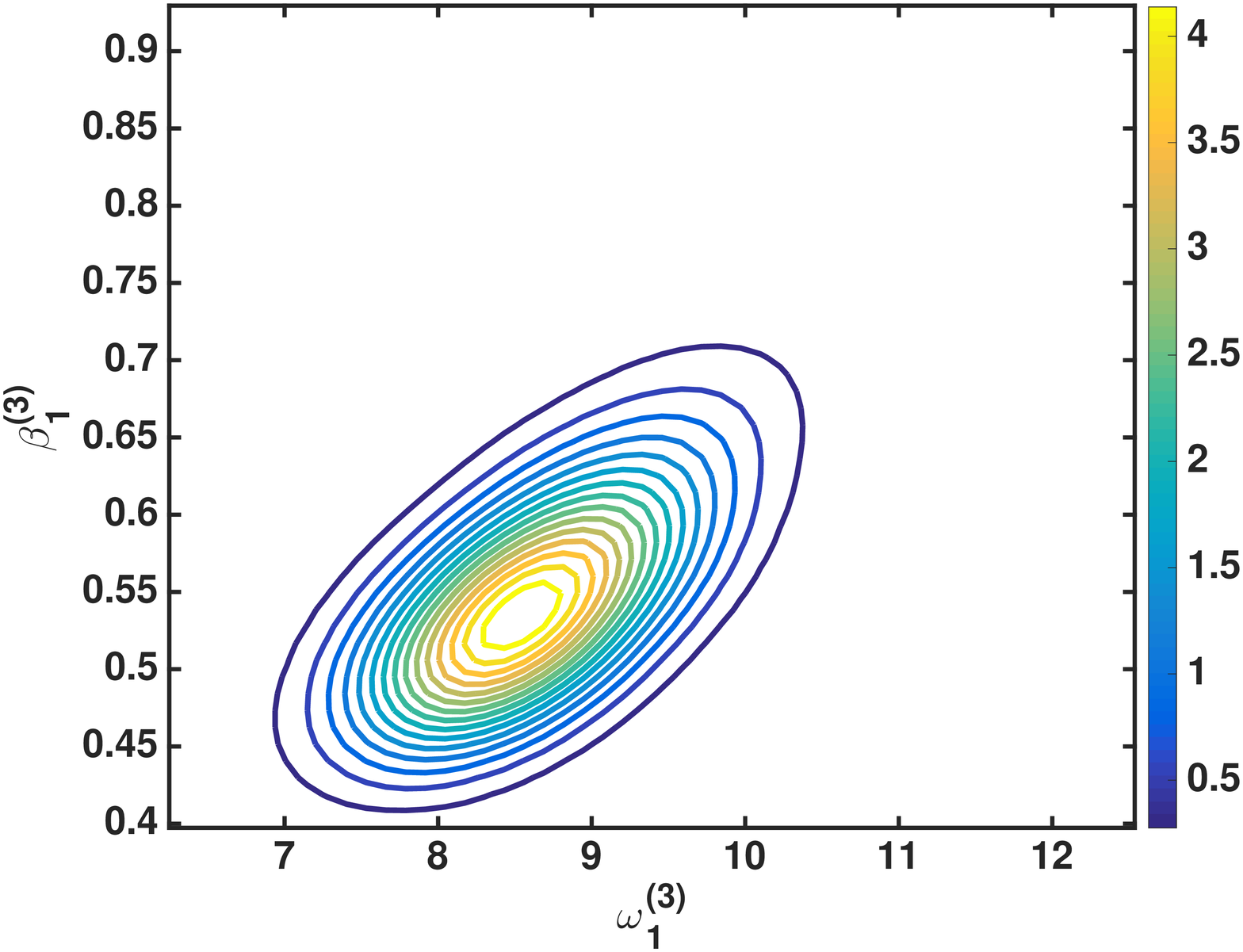}}
     {$$}     
\\
\subf{\includegraphics[width=45mm]{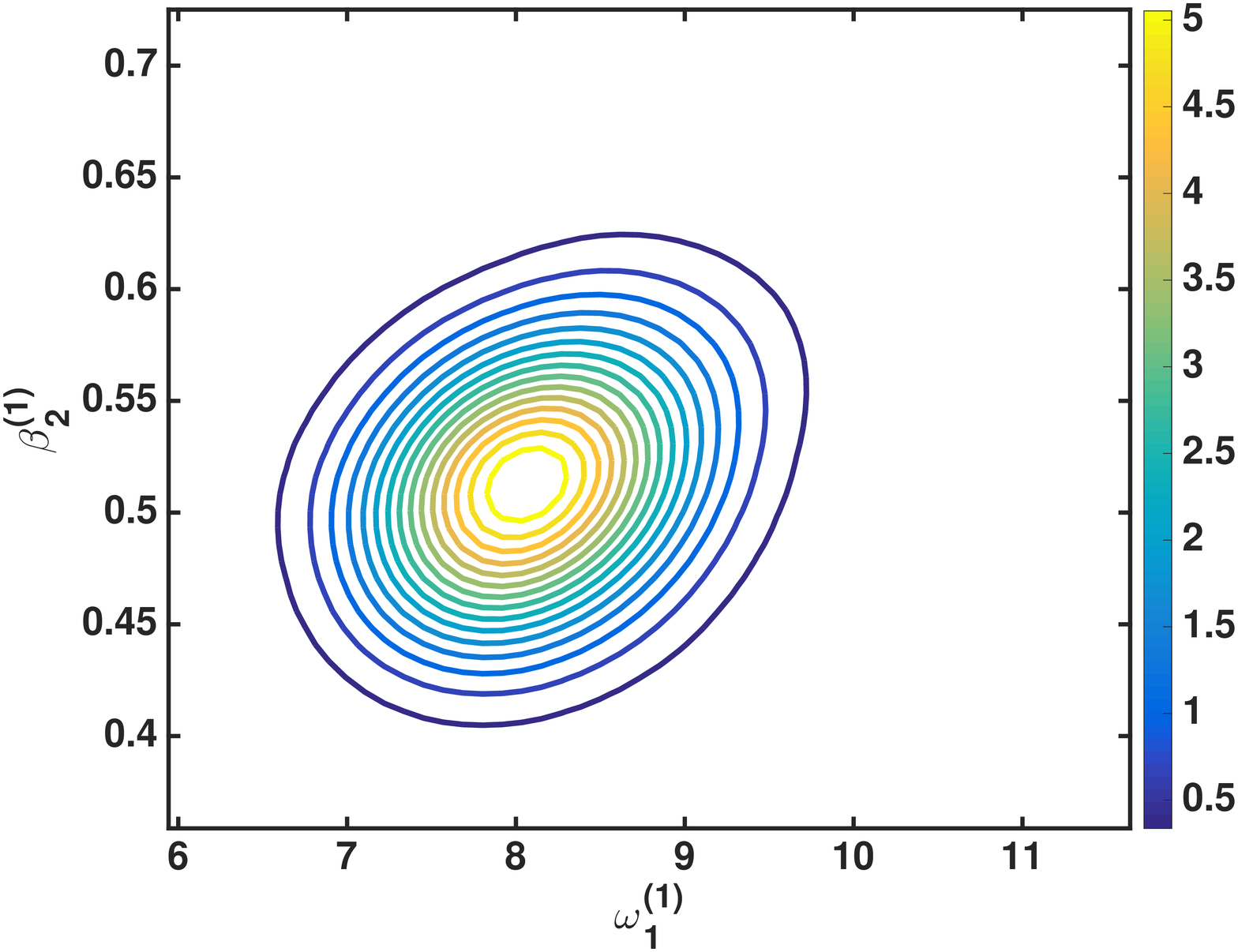}}
     {$$}
&
\subf{\includegraphics[width=45mm]{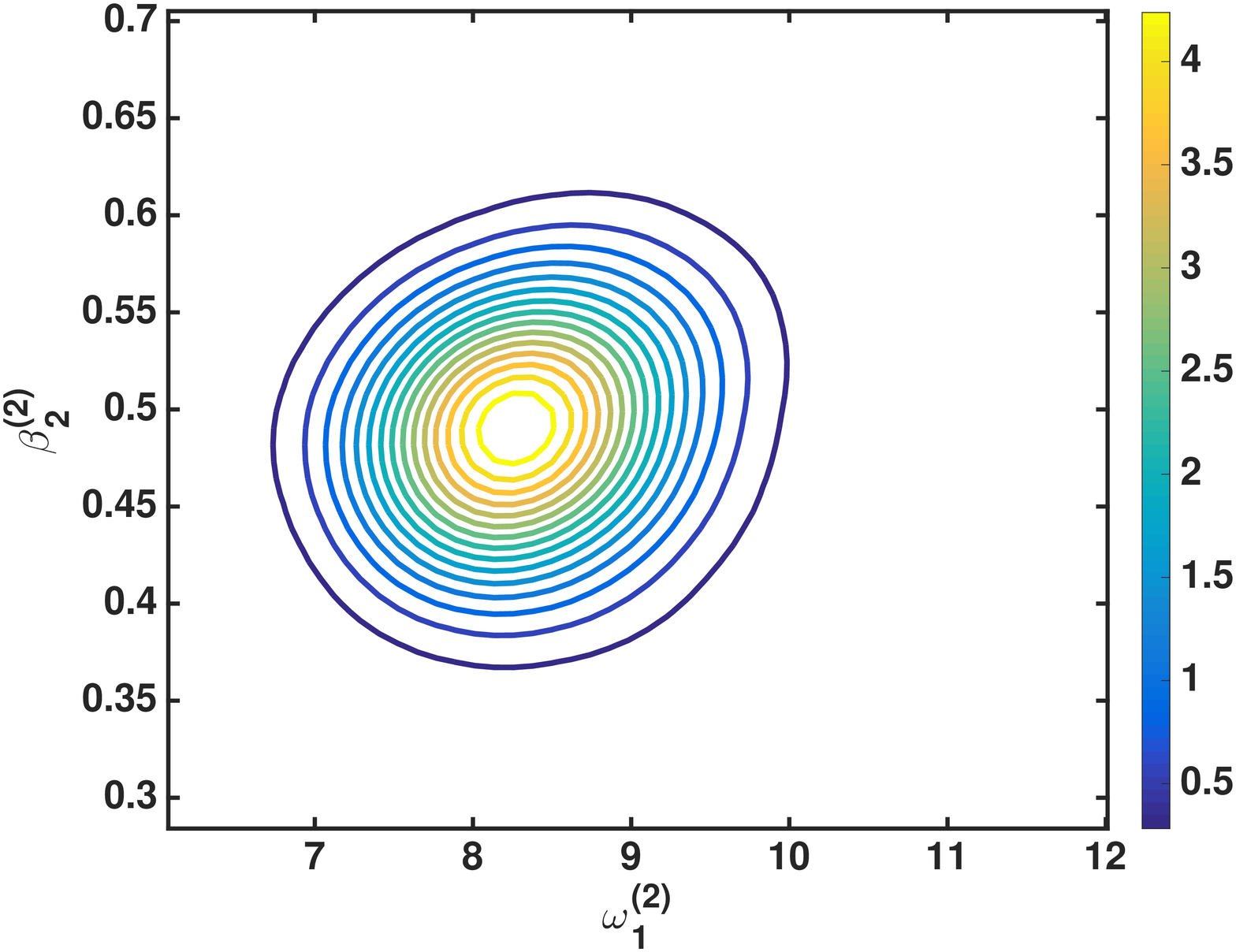}}
     {$$}
&
\subf{\includegraphics[width=45mm]{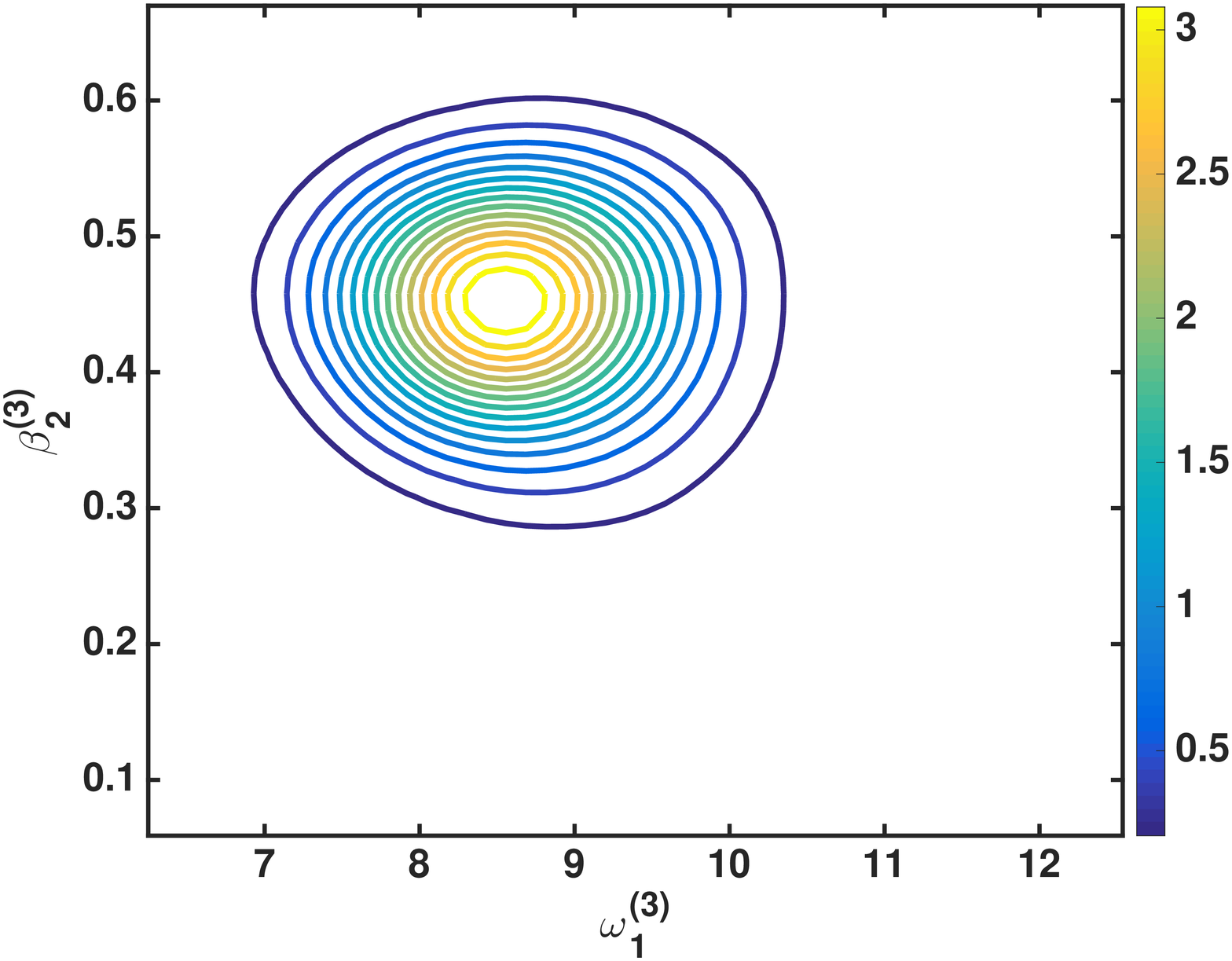}}
     {$$}
\\
\subf{\includegraphics[width=45mm]{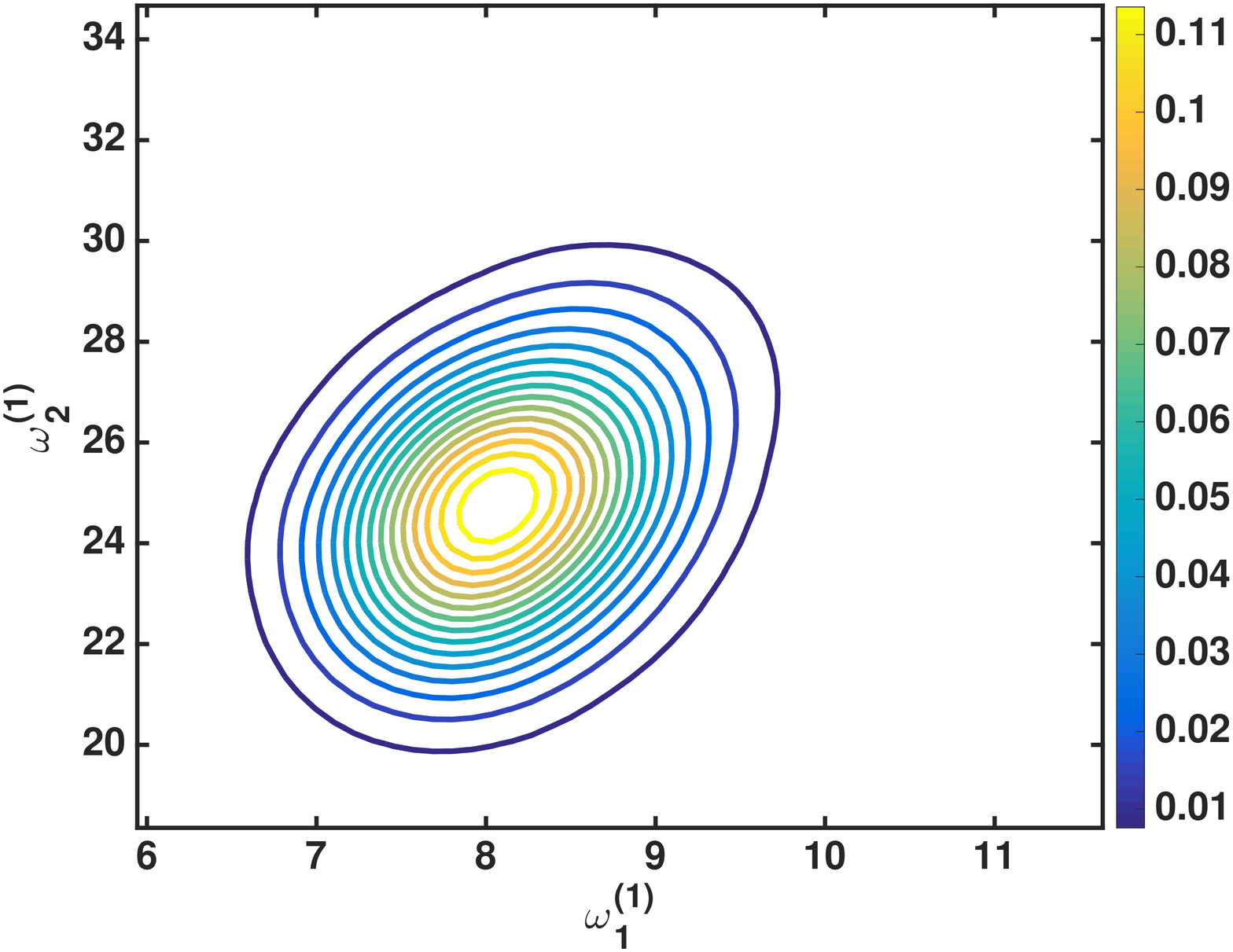}}
     {$$}
&
\subf{\includegraphics[width=45mm]{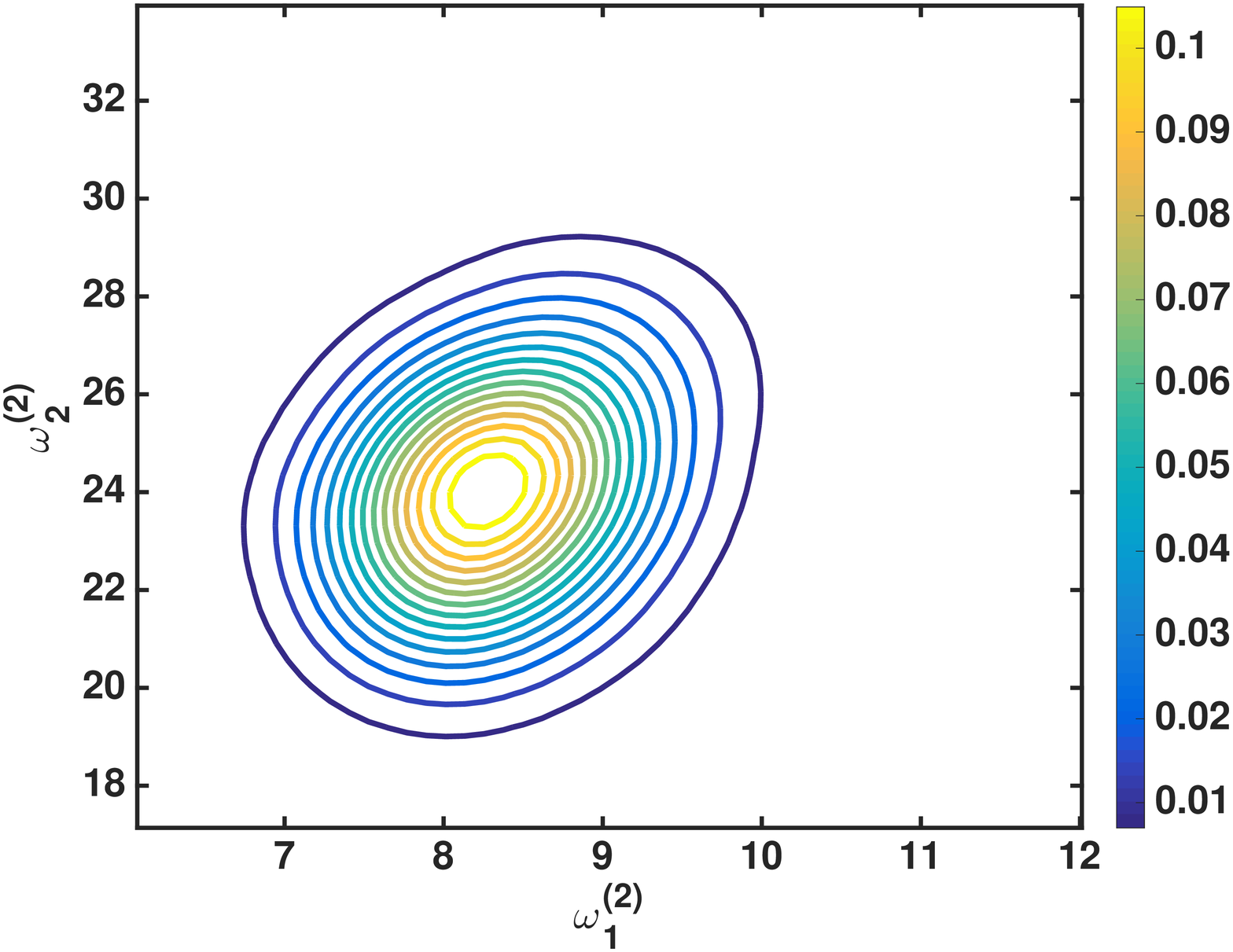}}
     {$$}
&
\subf{\includegraphics[width=45mm]{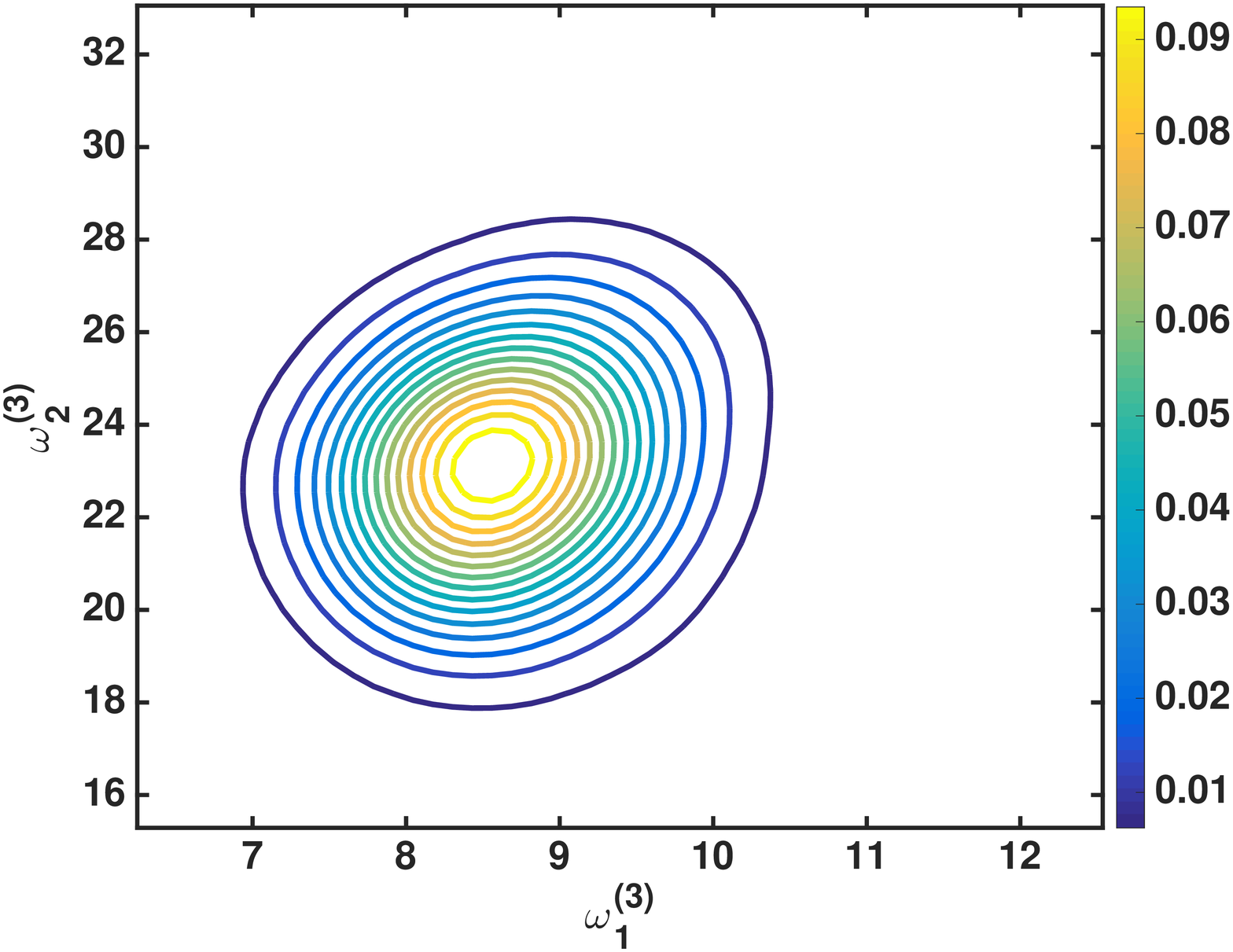}}
     {$$}
\\
\subf{\includegraphics[width=45mm]{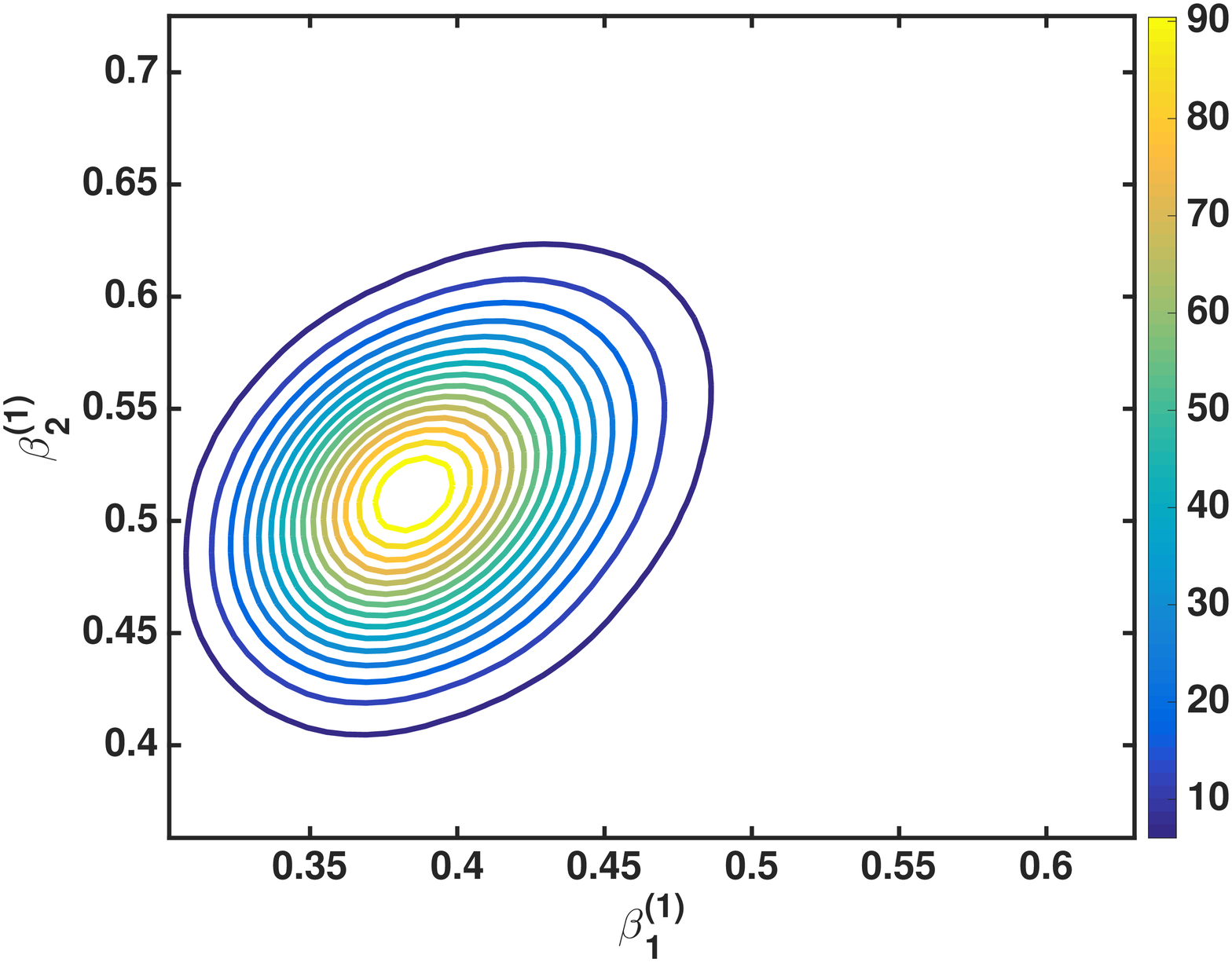}}
     {$$}
&
\subf{\includegraphics[width=45mm]{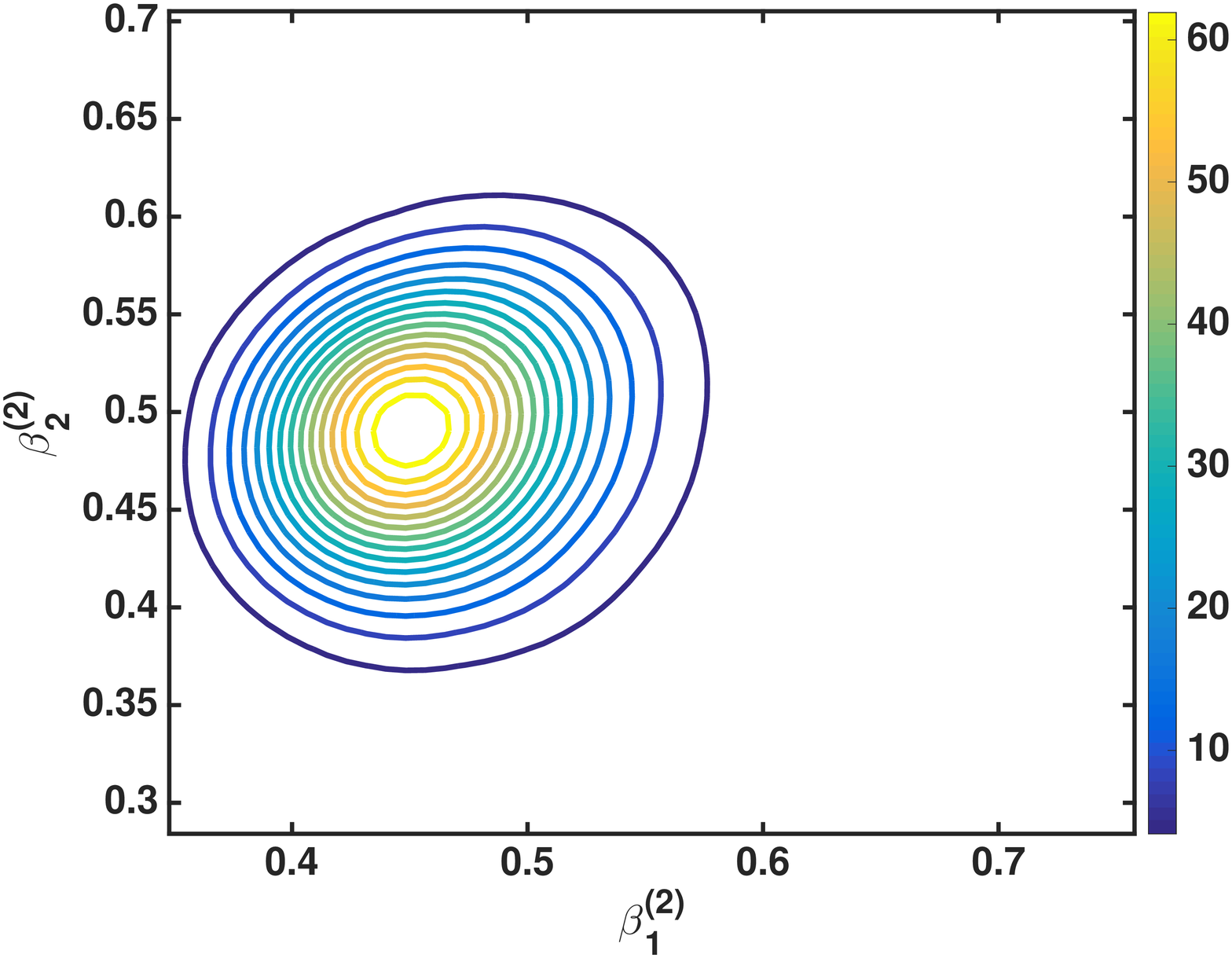}}
     {$$}
&
\subf{\includegraphics[width=45mm]{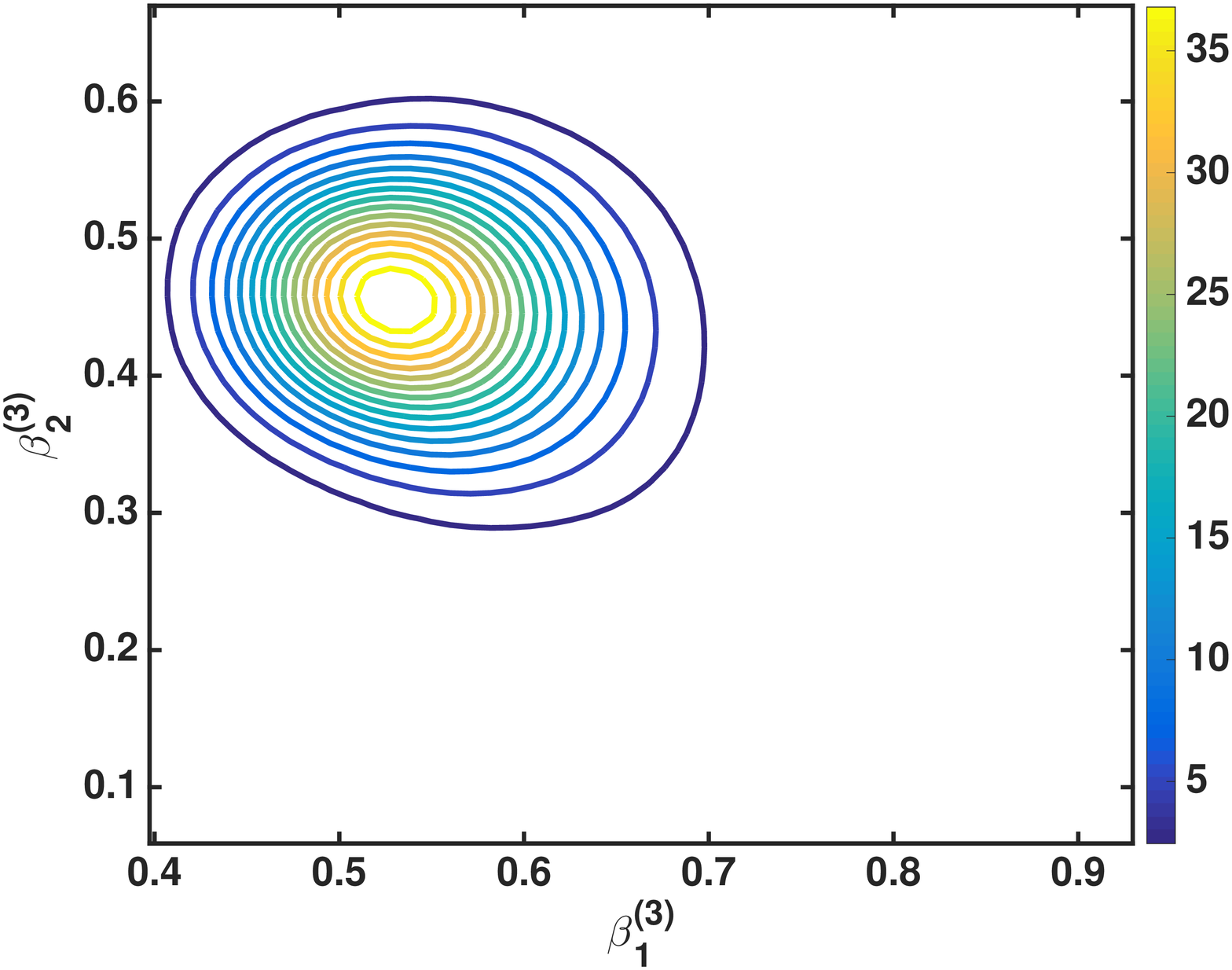}}
     {$$}
\\
\subf{\includegraphics[width=45mm]{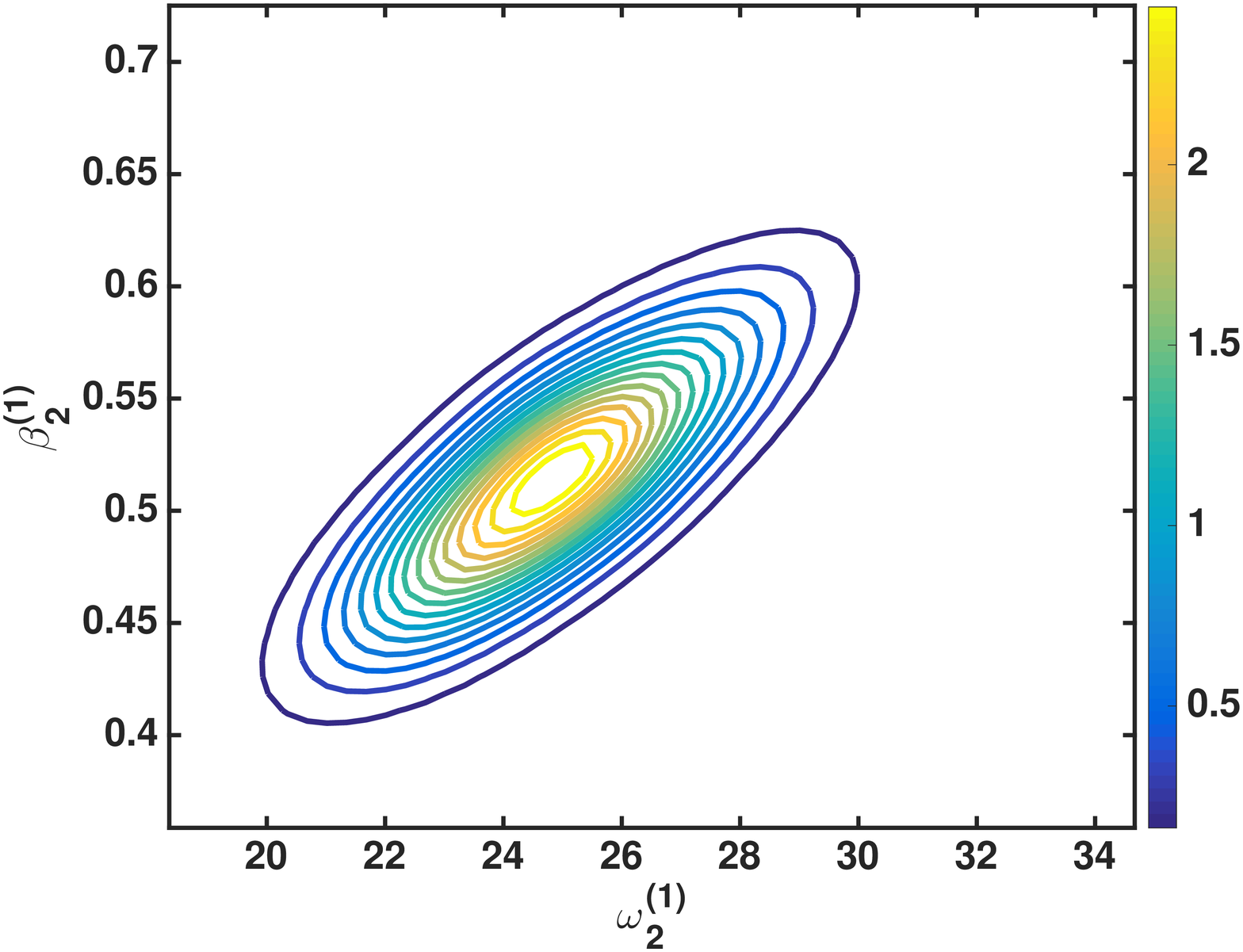}}
     {$$}
&
\subf{\includegraphics[width=45mm]{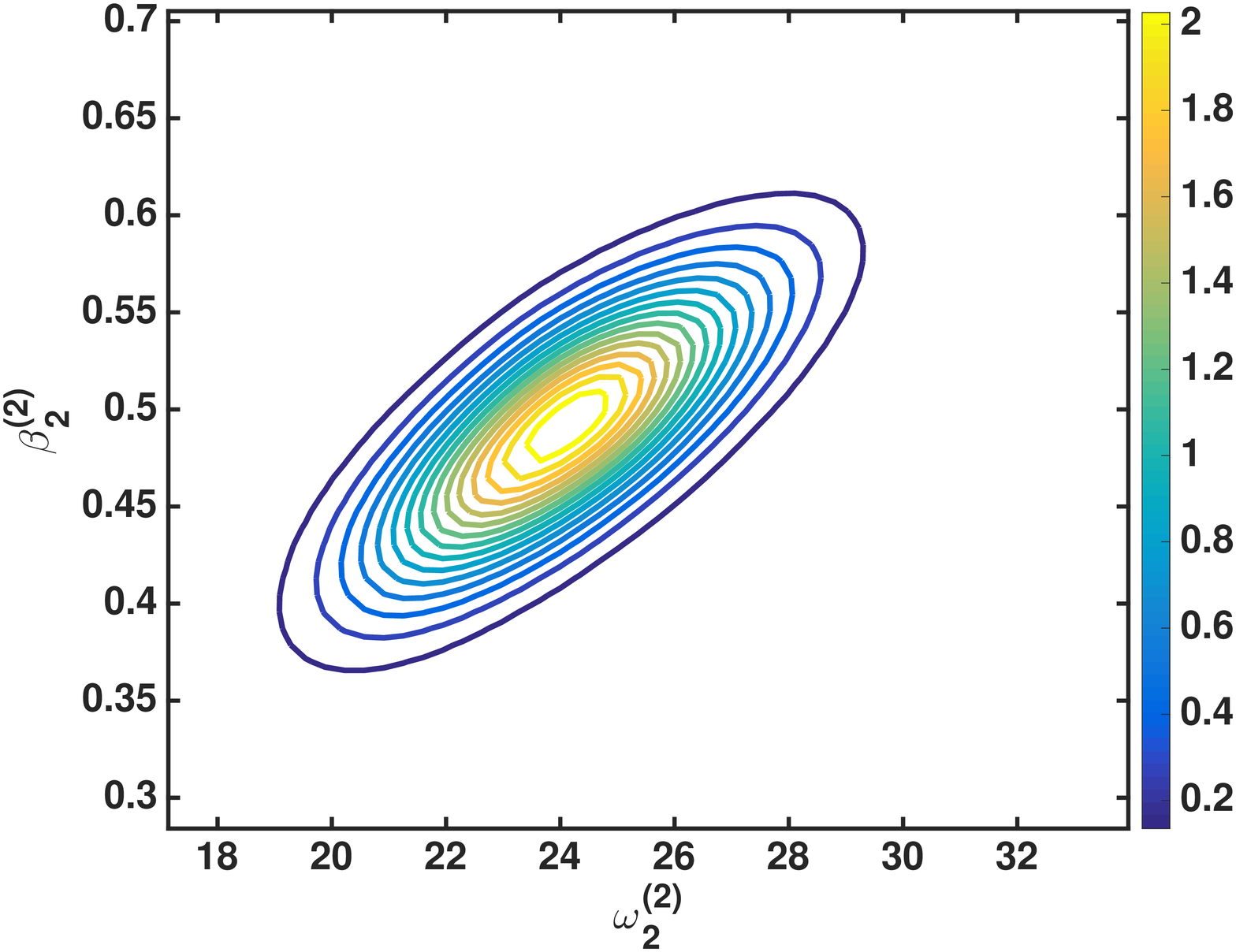}}
     {$$}
&
\subf{\includegraphics[width=45mm]{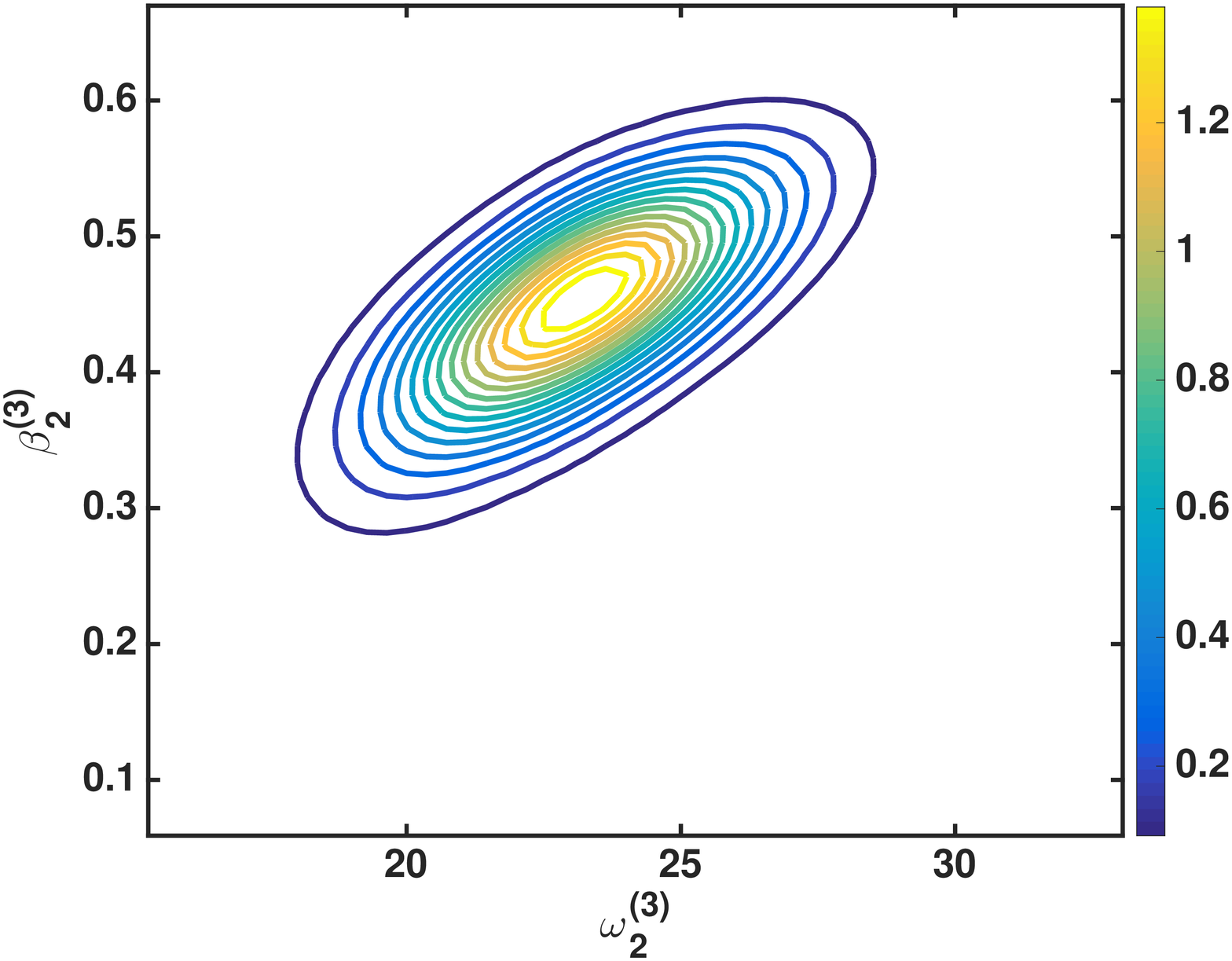}}
     {$$}     
\\
\end{tabular}
\caption{Joint prior pdfs of parameters $\omega_1, \beta_1, \omega_2$ and $\beta_2$ at three airspeeds}
\label{fig3ujointprior10cva}
\end{figure}

\begin{figure}[h!]
\centering
\begin{tabular}{ccc}
\subf{\includegraphics[width=45mm]{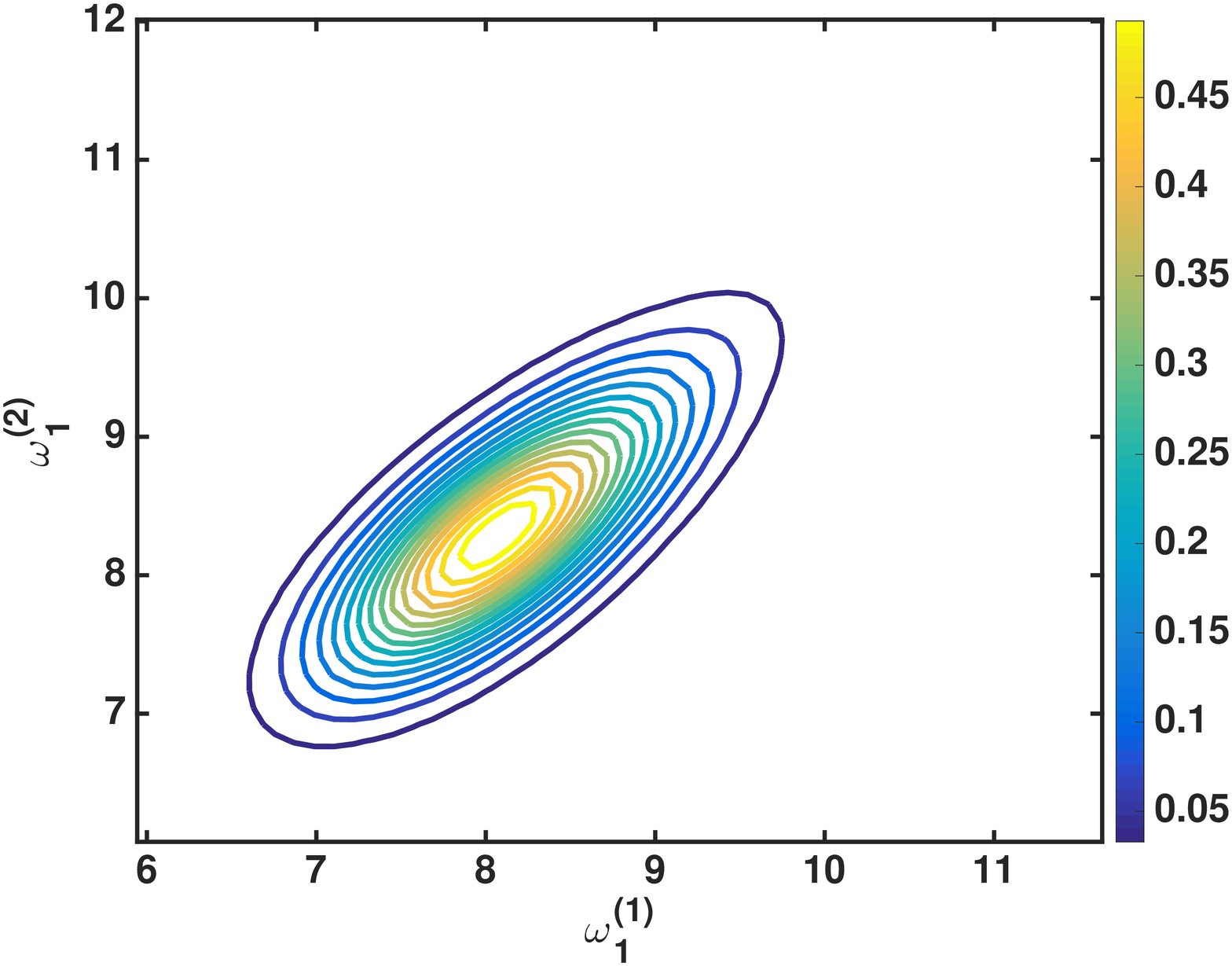}}
     {$$}
&
\subf{\includegraphics[width=45mm]{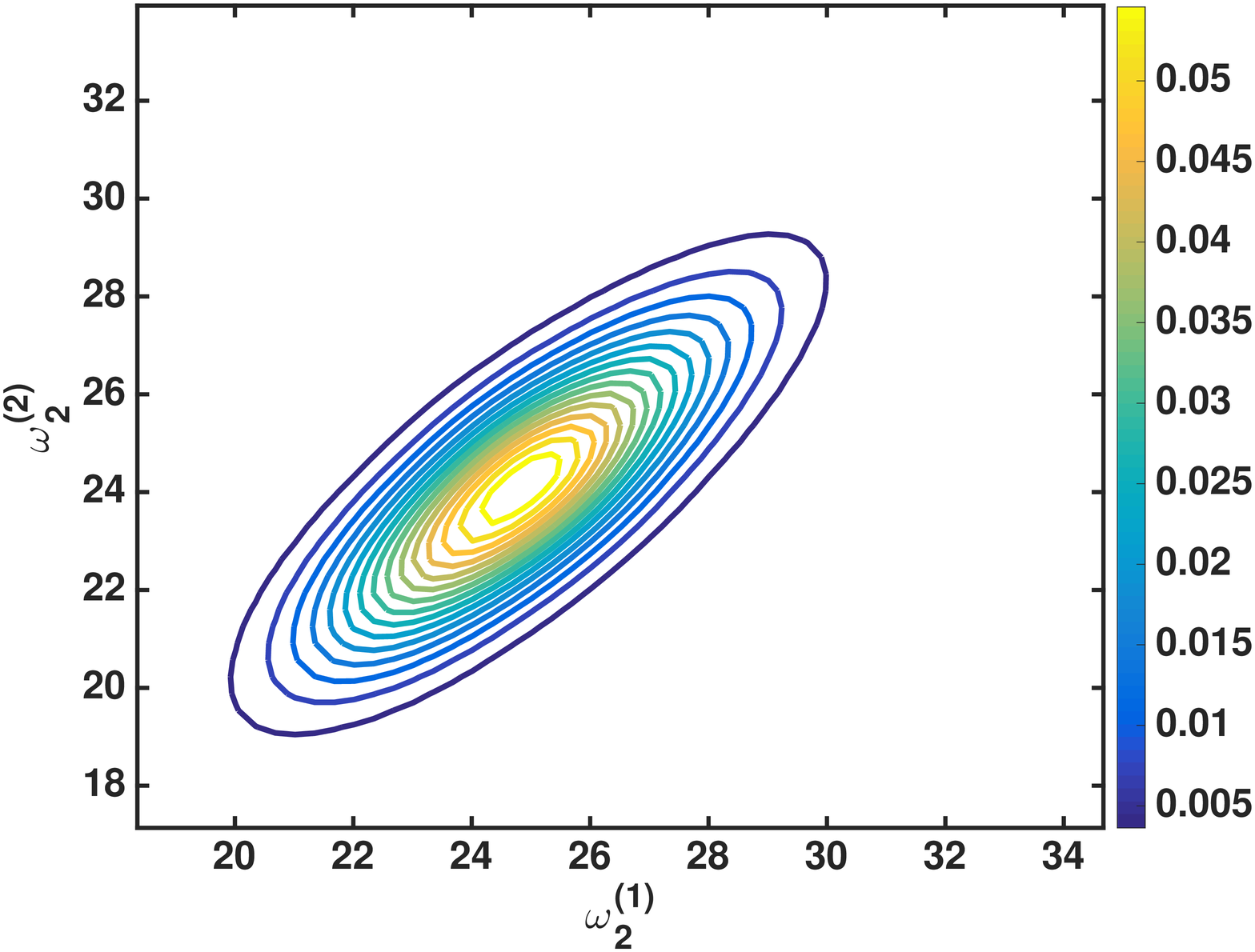}}
     {$$}
&
\subf{\includegraphics[width=45mm]{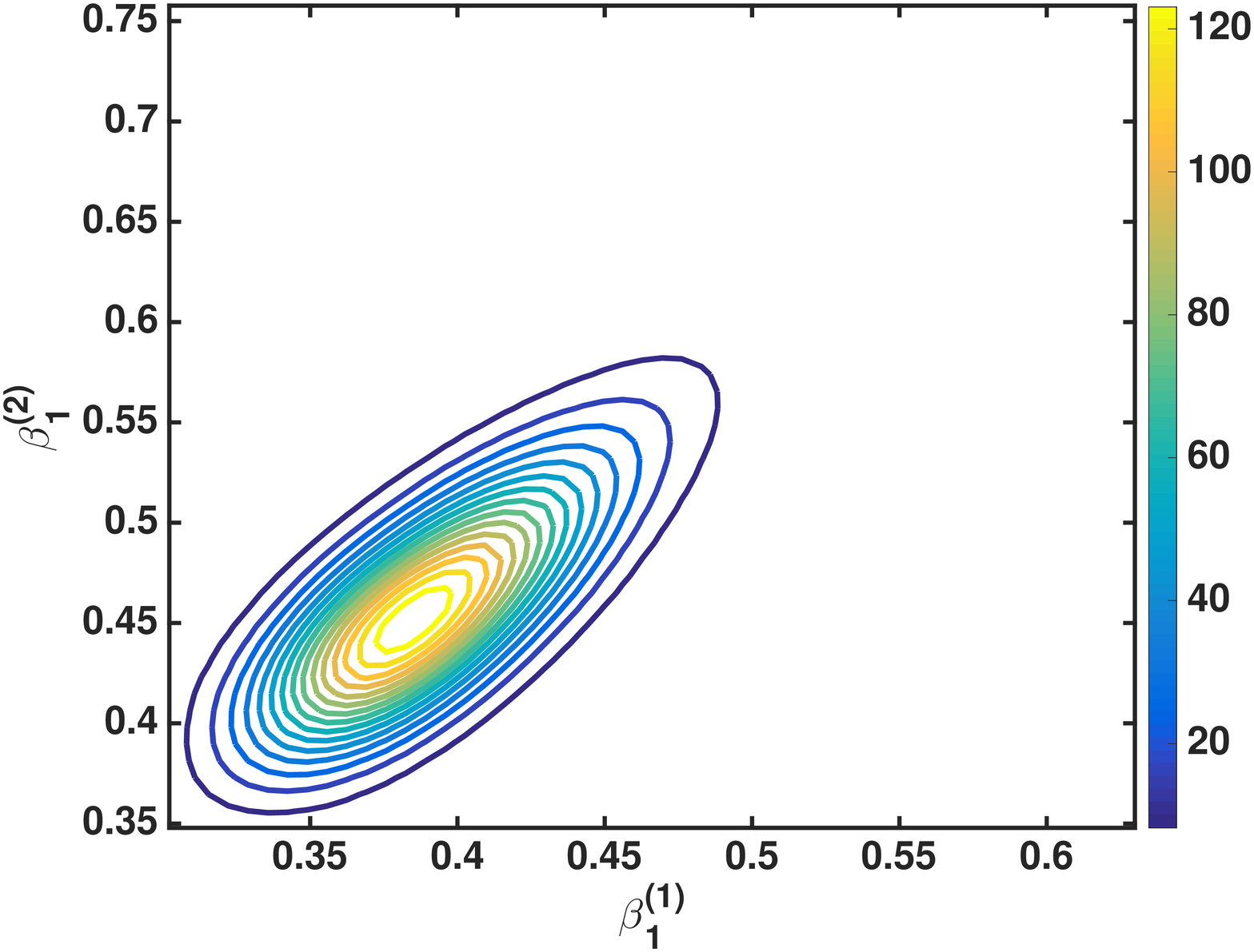}}
     {$$}
\\
\subf{\includegraphics[width=45mm]{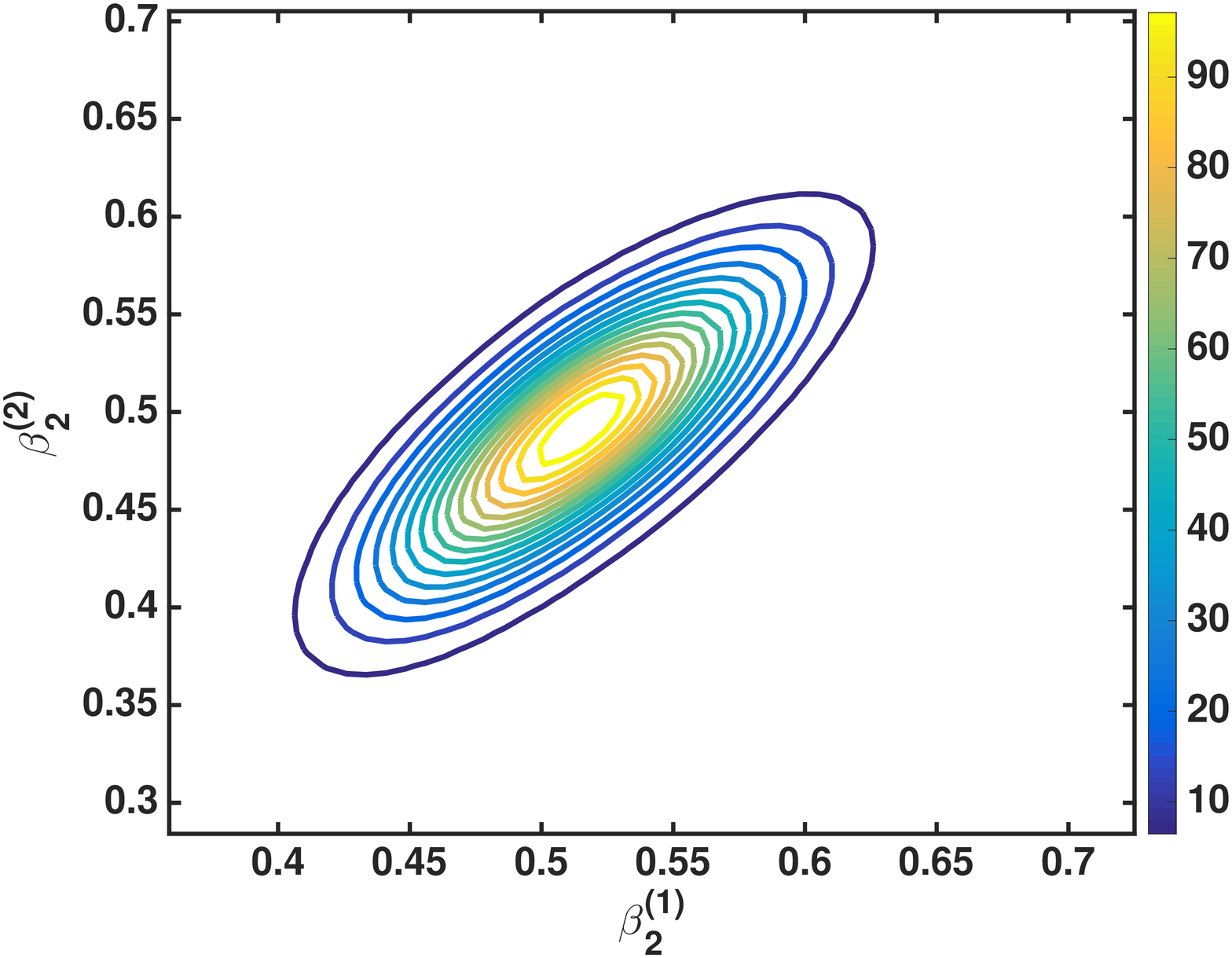}}
     {$$}
&
\subf{\includegraphics[width=45mm]{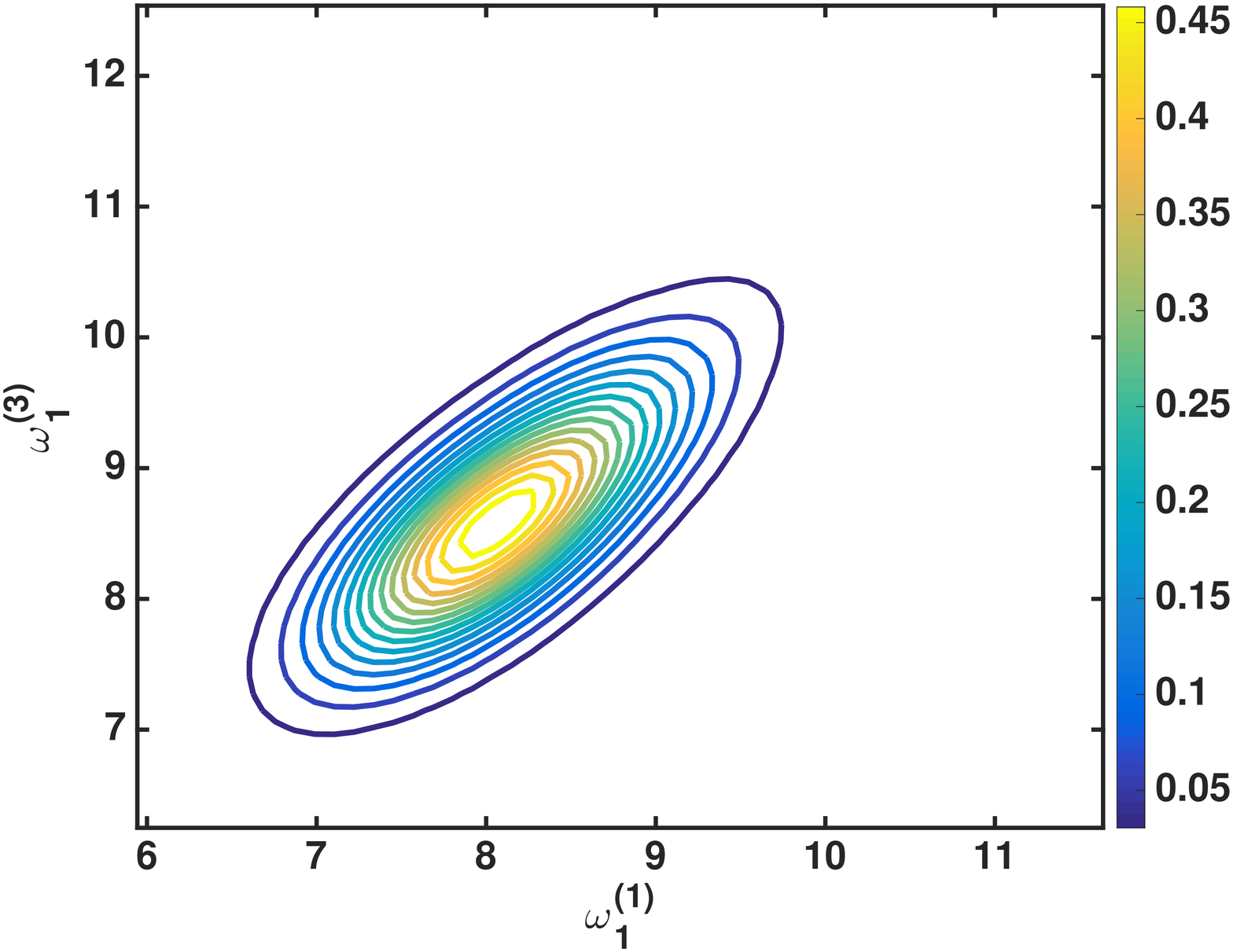}}
     {$$}
&
\subf{\includegraphics[width=45mm]{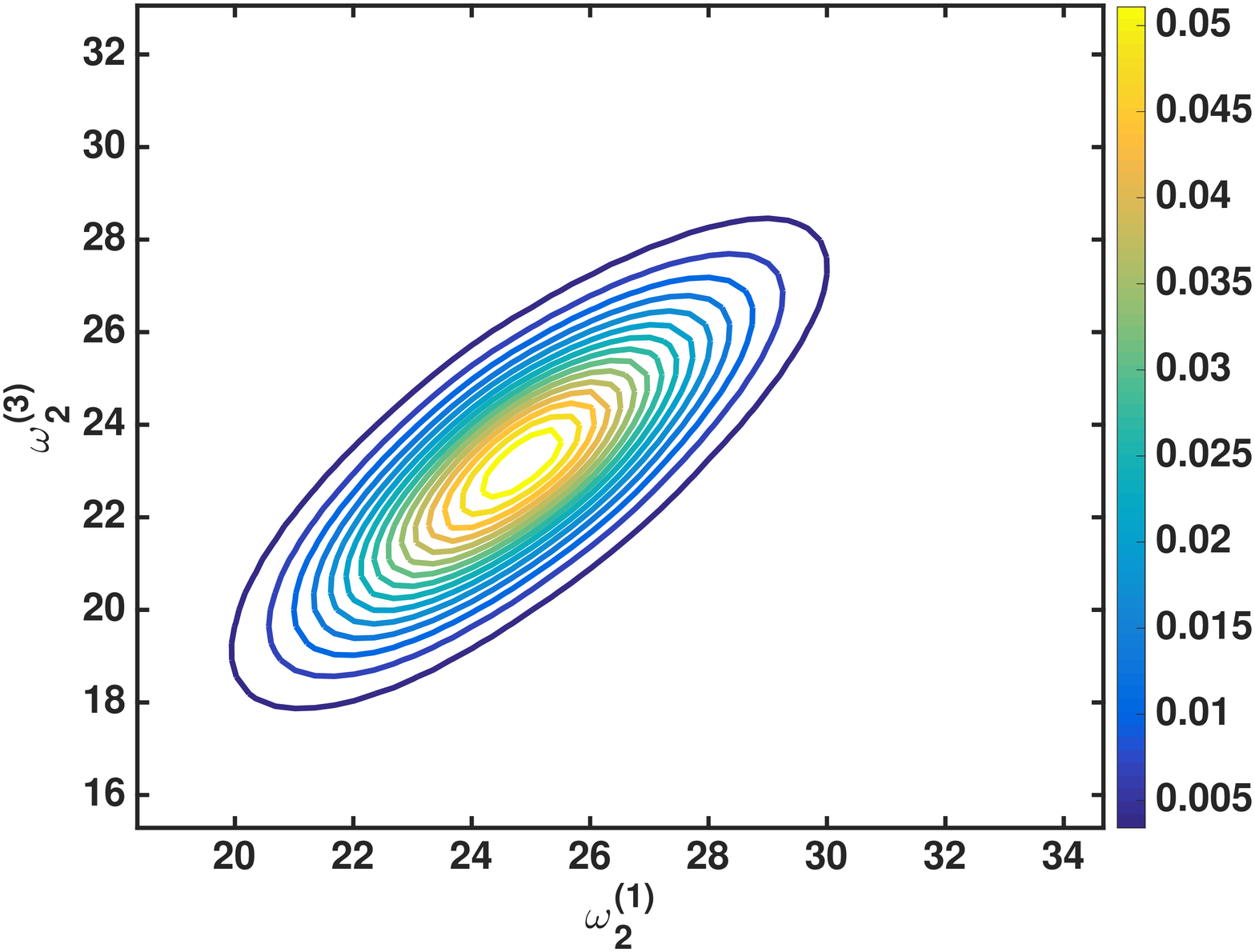}}
     {$$}
\\
\subf{\includegraphics[width=45mm]{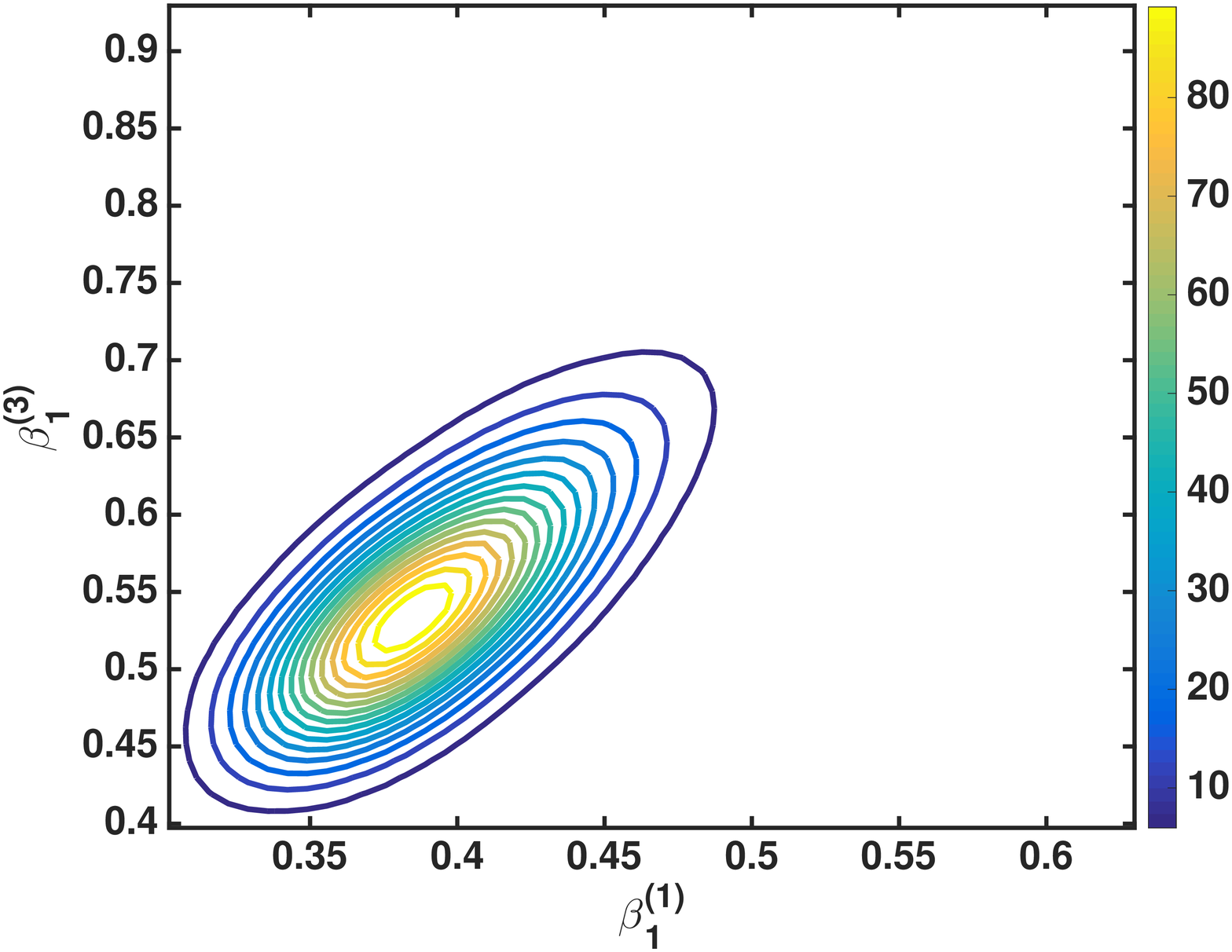}}
     {$$}
&
\subf{\includegraphics[width=45mm]{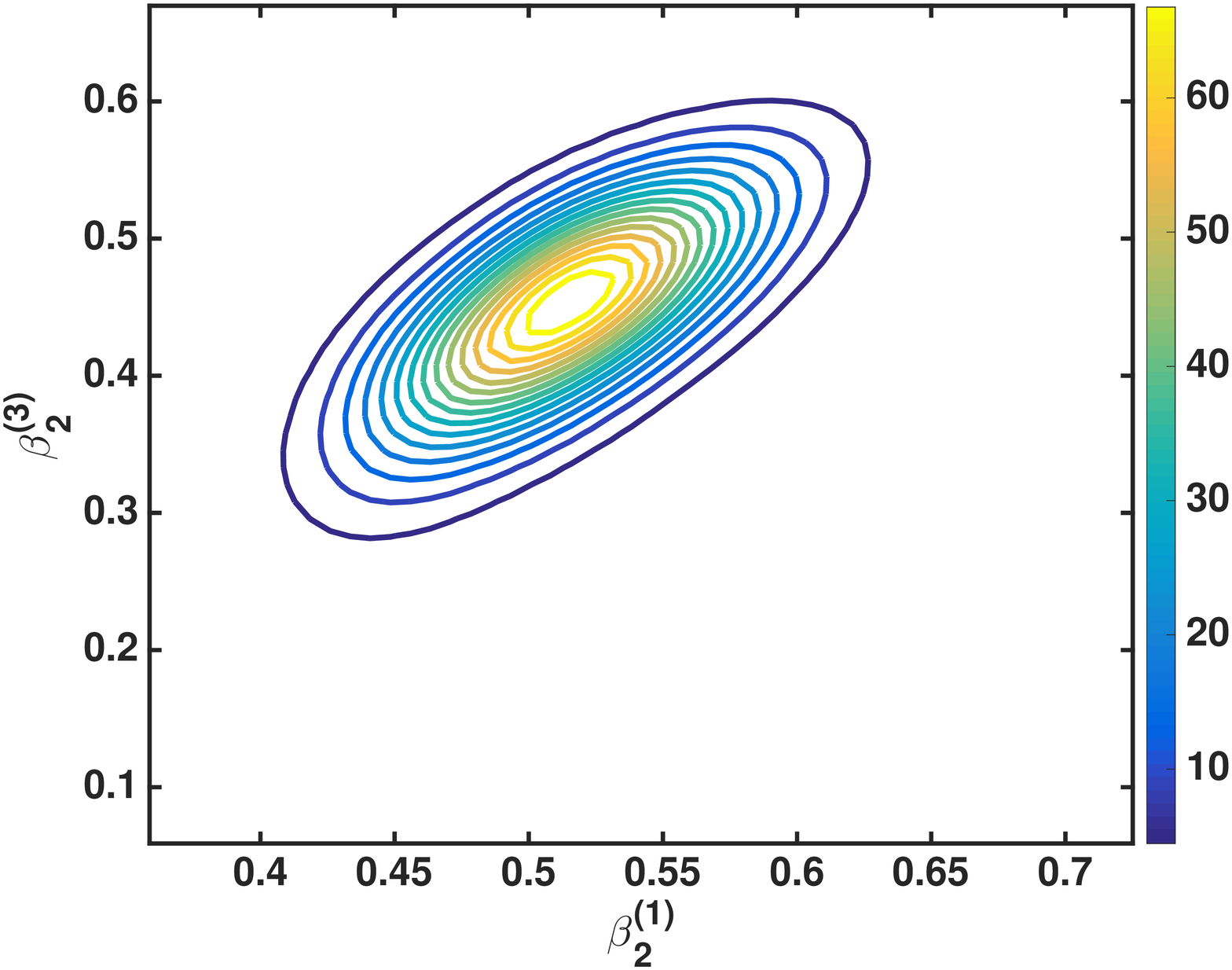}}
     {$$}
&
\subf{\includegraphics[width=45mm]{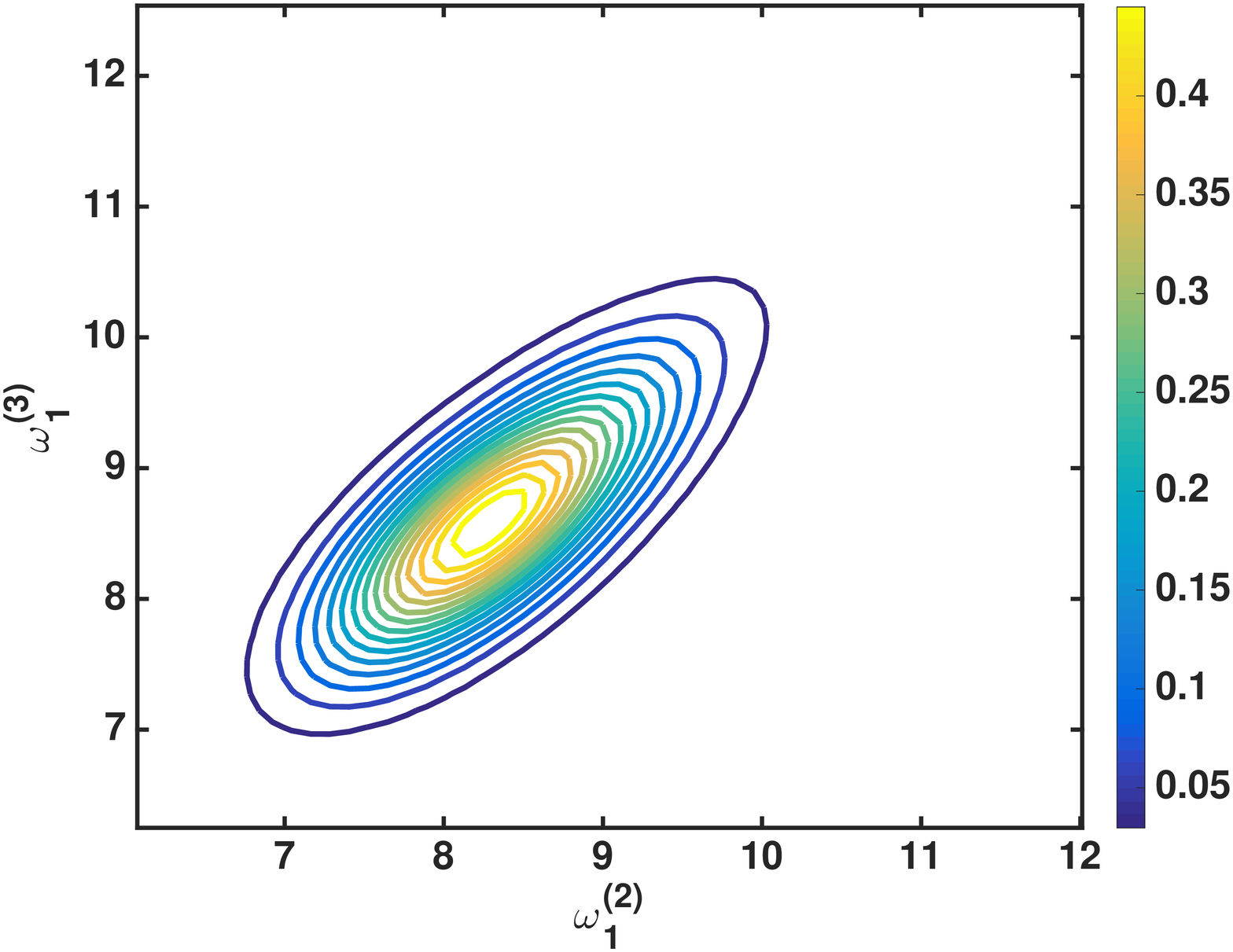}}
     {$$}
\\
\subf{\includegraphics[width=45mm]{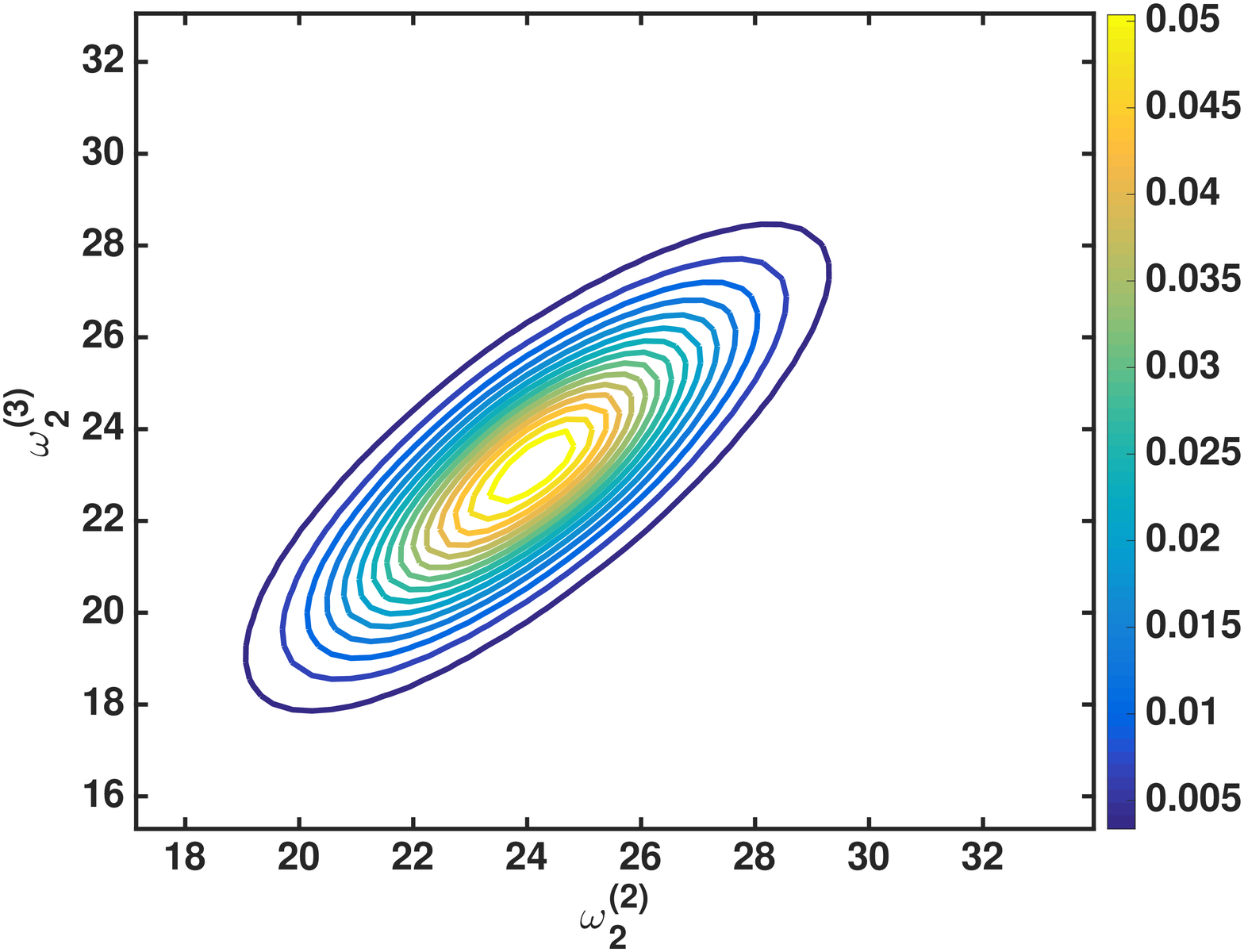}}
     {$$}
&
\subf{\includegraphics[width=45mm]{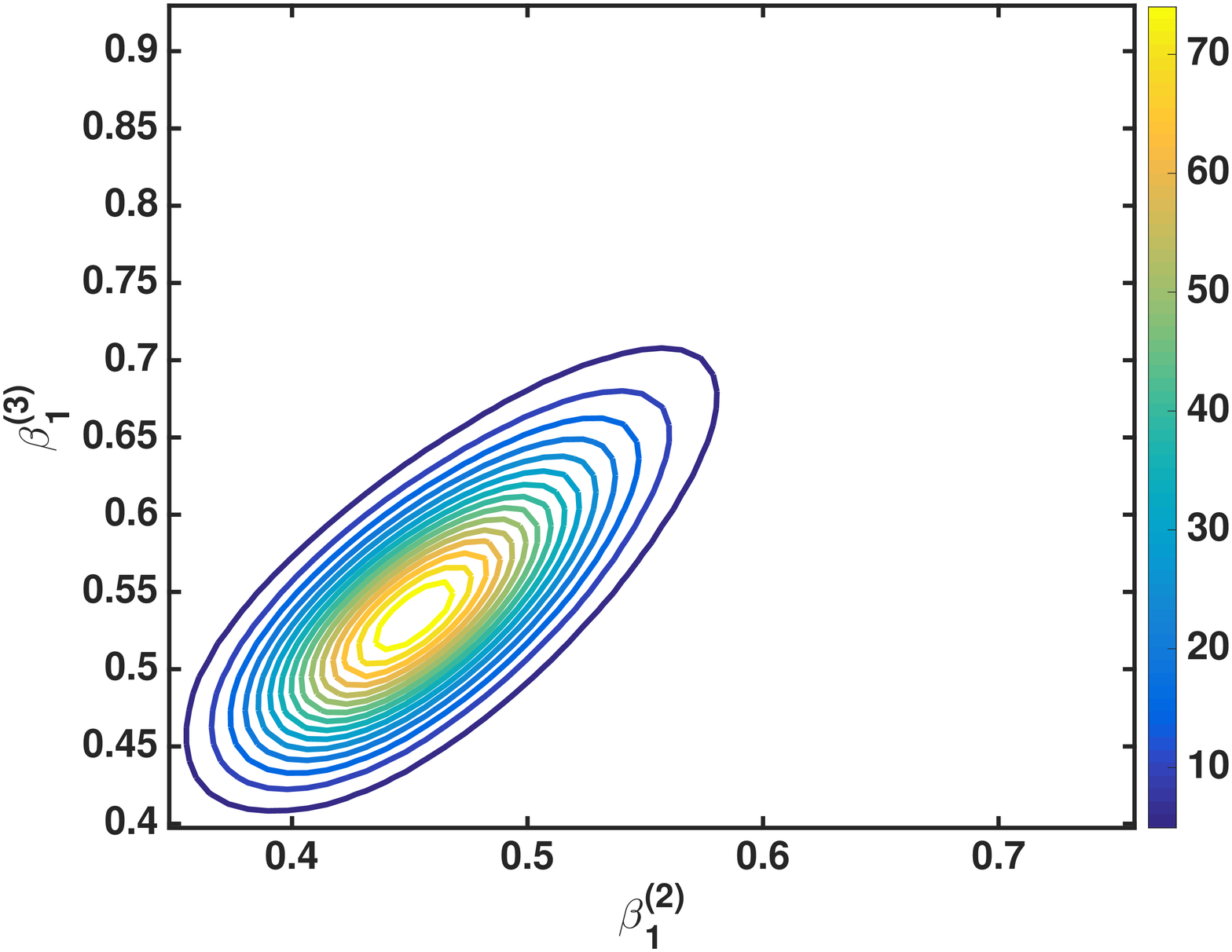}}
     {$$}
&
\subf{\includegraphics[width=45mm]{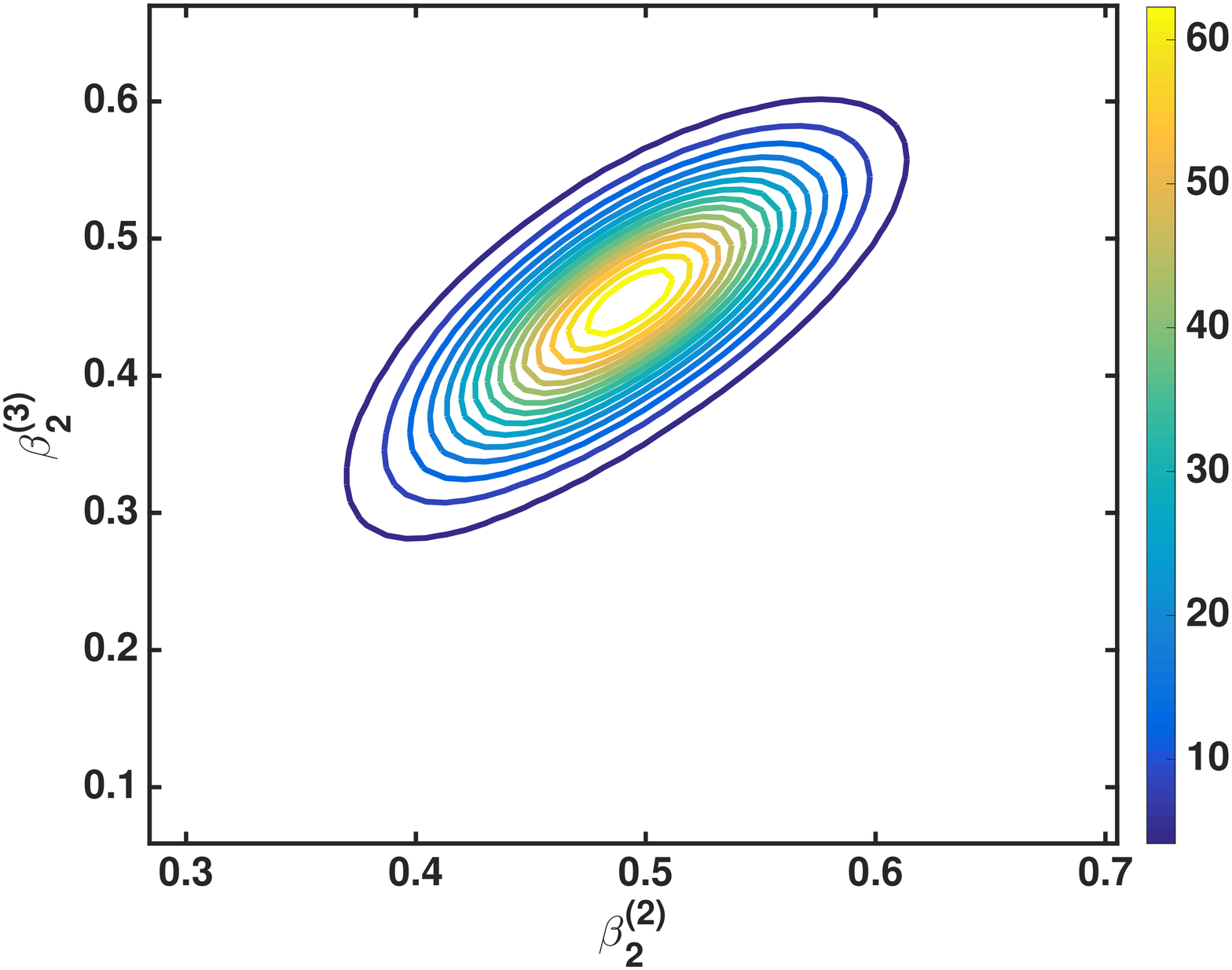}}
     {$$}
\\
\end{tabular}
\caption{Joint prior pdfs of parameters $\omega_1, \beta_1, \omega_2$ and $\beta_2$ among three airspeeds}
\label{fig3ujointprior10cvb}
\end{figure}

\begin{figure}[htbp]
\begin{center}
\includegraphics[width=120mm,height=80mm]{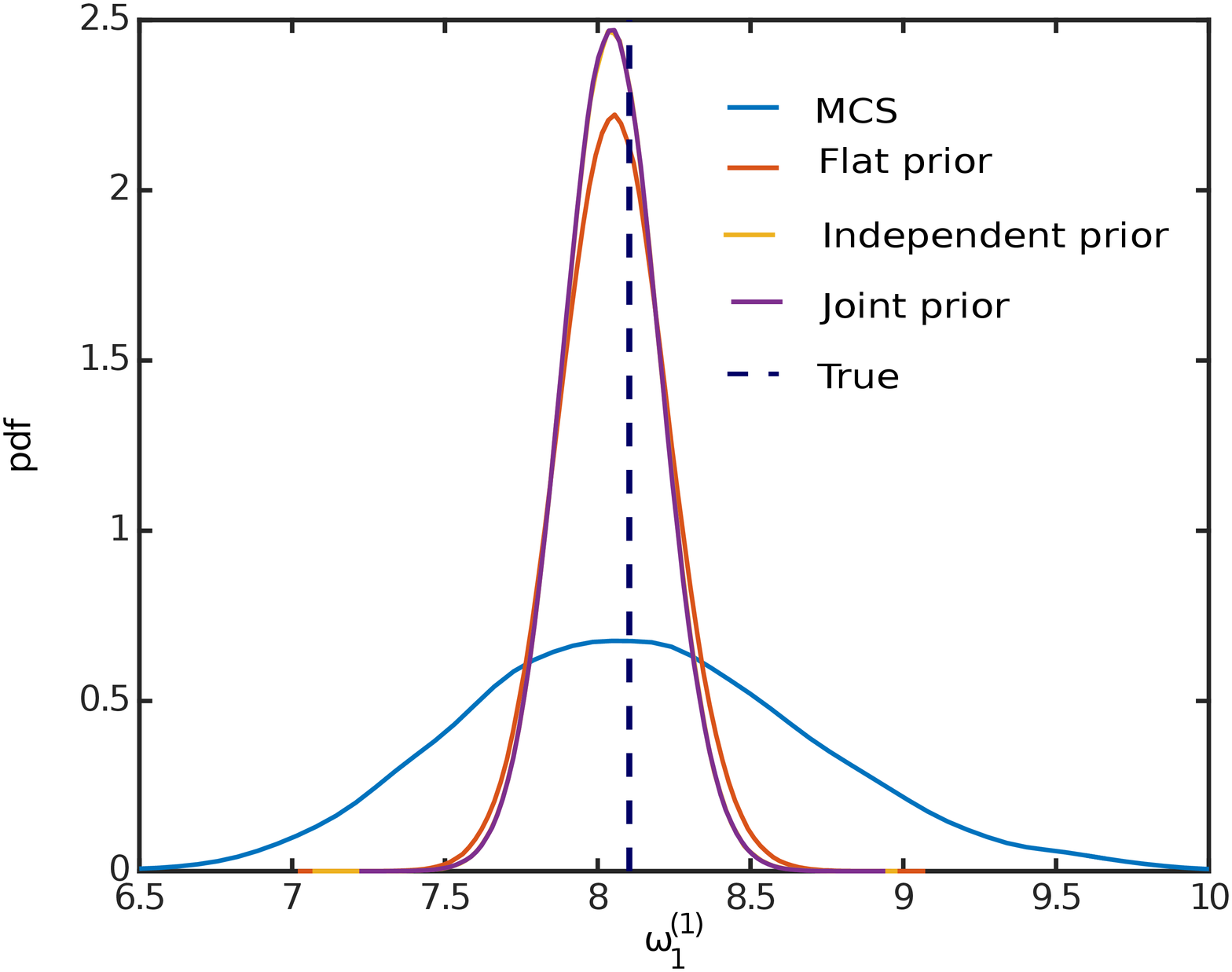}
\caption{Comparison of posterior of $\omega_1^{(1)}$ for different priors }
\label{3u10cvomega11}
\end{center}
\end{figure}
\begin{figure}[htbp]
\begin{center}
\includegraphics[width=120mm,height=80mm]{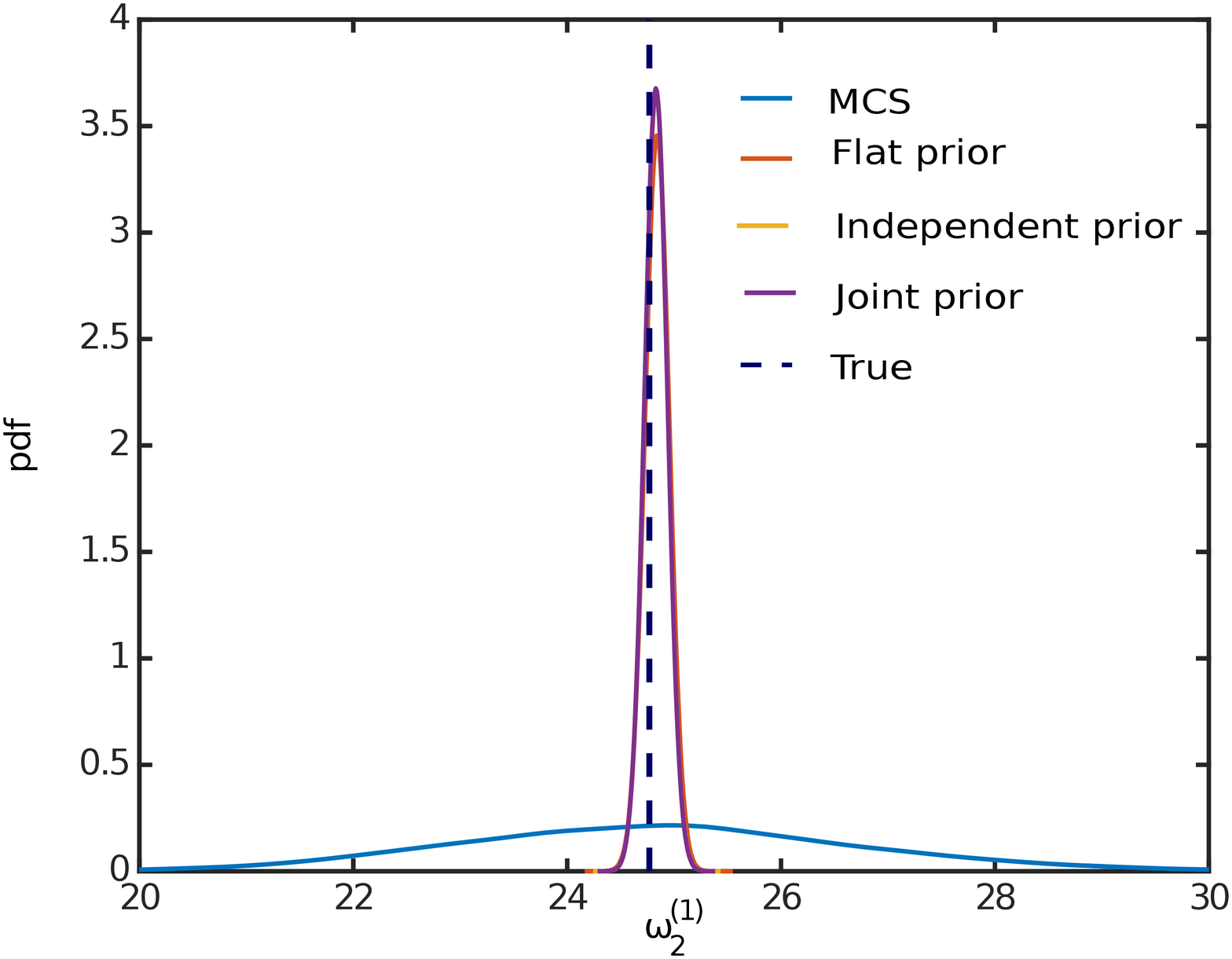}
\caption{Comparison of posterior of $\omega_2^{(1)}$ for different priors }
\label{3u10cvomega12}
\end{center}
\end{figure}
\begin{figure}[htbp]
\begin{center}
\includegraphics[width=120mm,height=80mm]{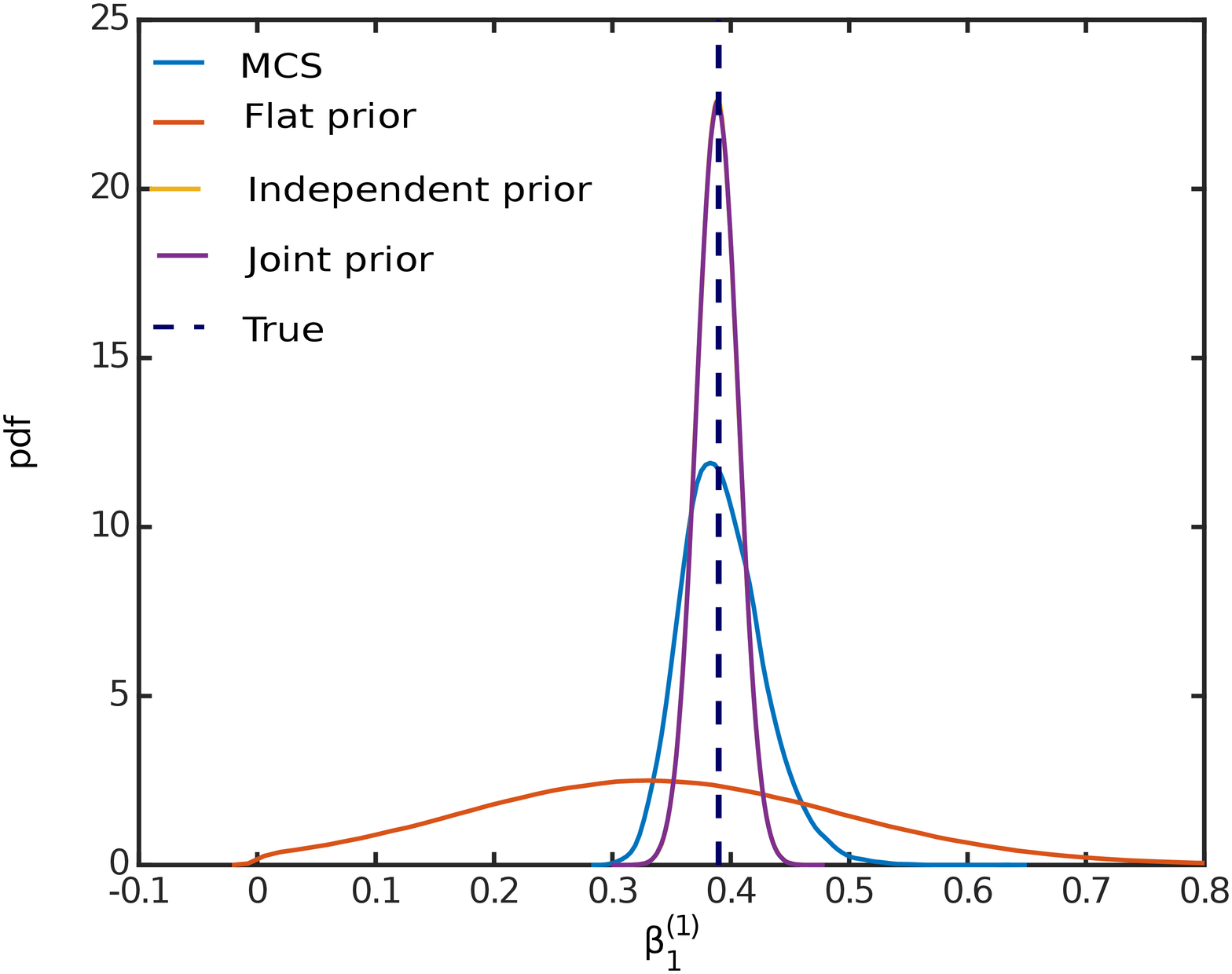}
\caption{Comparison of posterior of $\beta_1^{(1)}$ for different priors }
\label{3u10cvbeta11}
\end{center}
\end{figure}
\begin{figure}[htbp]
\begin{center}
\includegraphics[width=120mm,height=80mm]{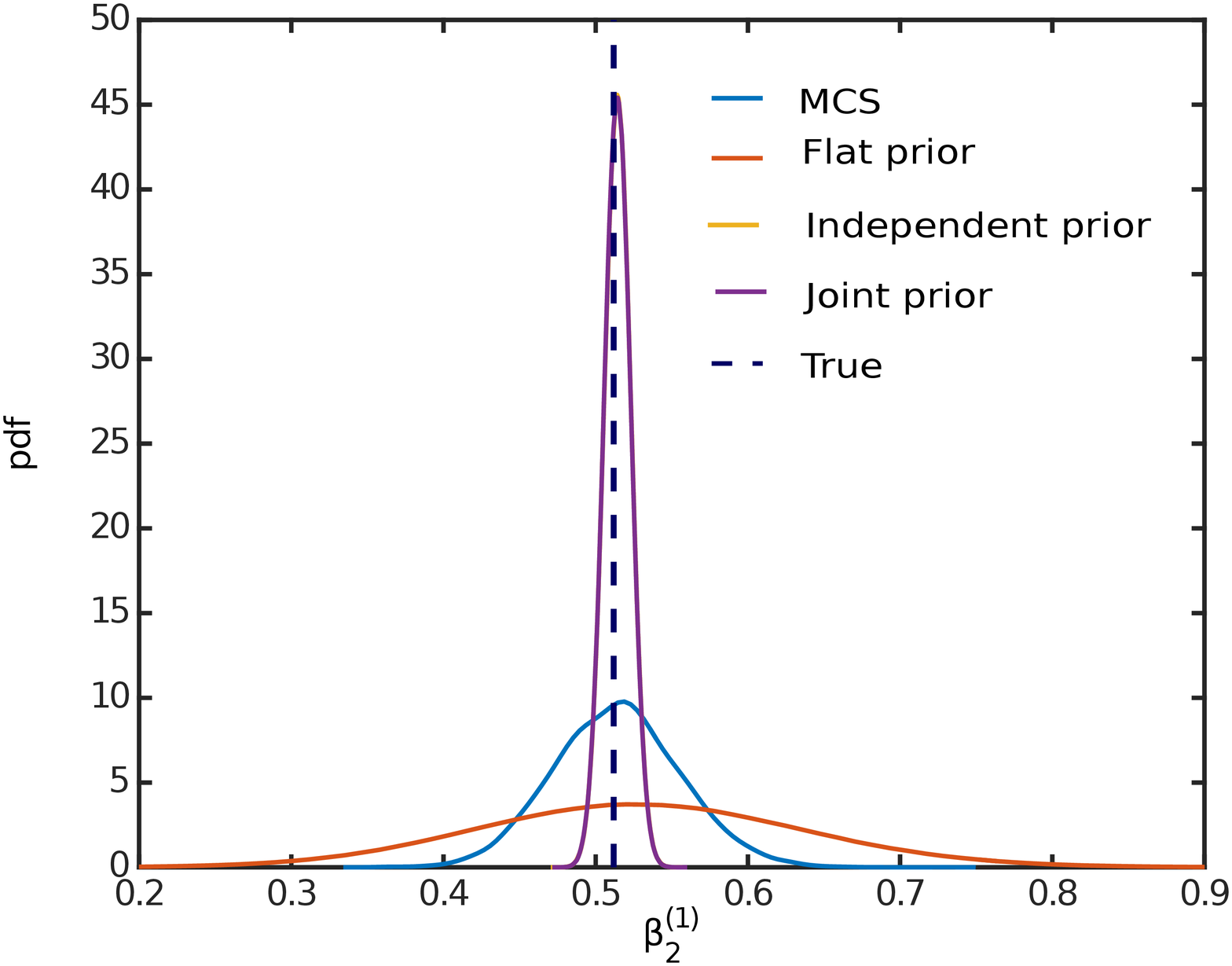}
\caption{Comparison of posterior of $\beta_2^{(1)}$ for different priors }
\label{3u10cvbeta12}
\end{center}
\end{figure}
\begin{figure}[htbp]
\begin{center}
\includegraphics[width=120mm,height=80mm]{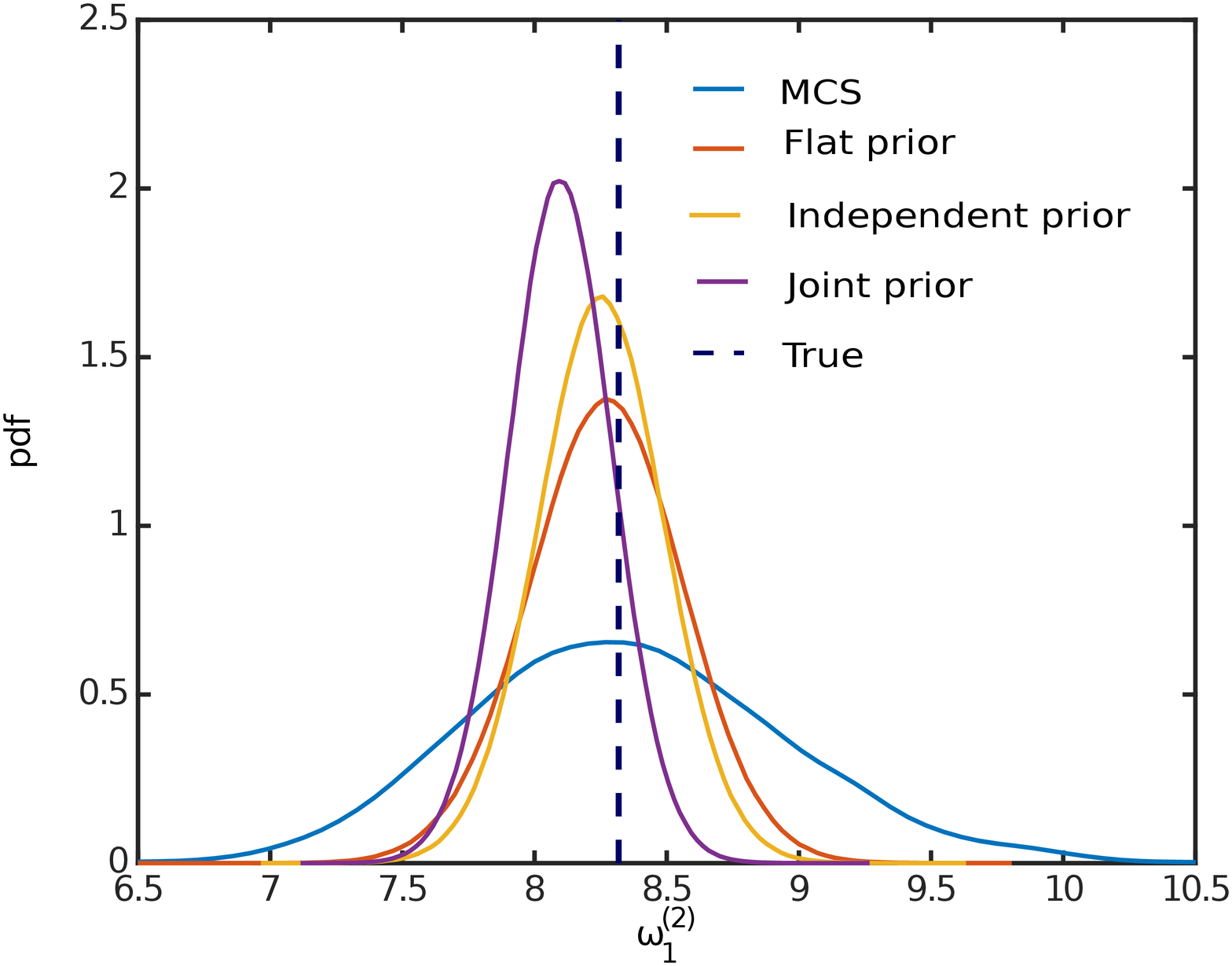}
\caption{Comparison of posterior of $\omega_1^{(2)}$ for different priors }
\label{3u10cvomega21}
\end{center}
\end{figure}
\begin{figure}[htbp]
\begin{center}
\includegraphics[width=120mm,height=80mm]{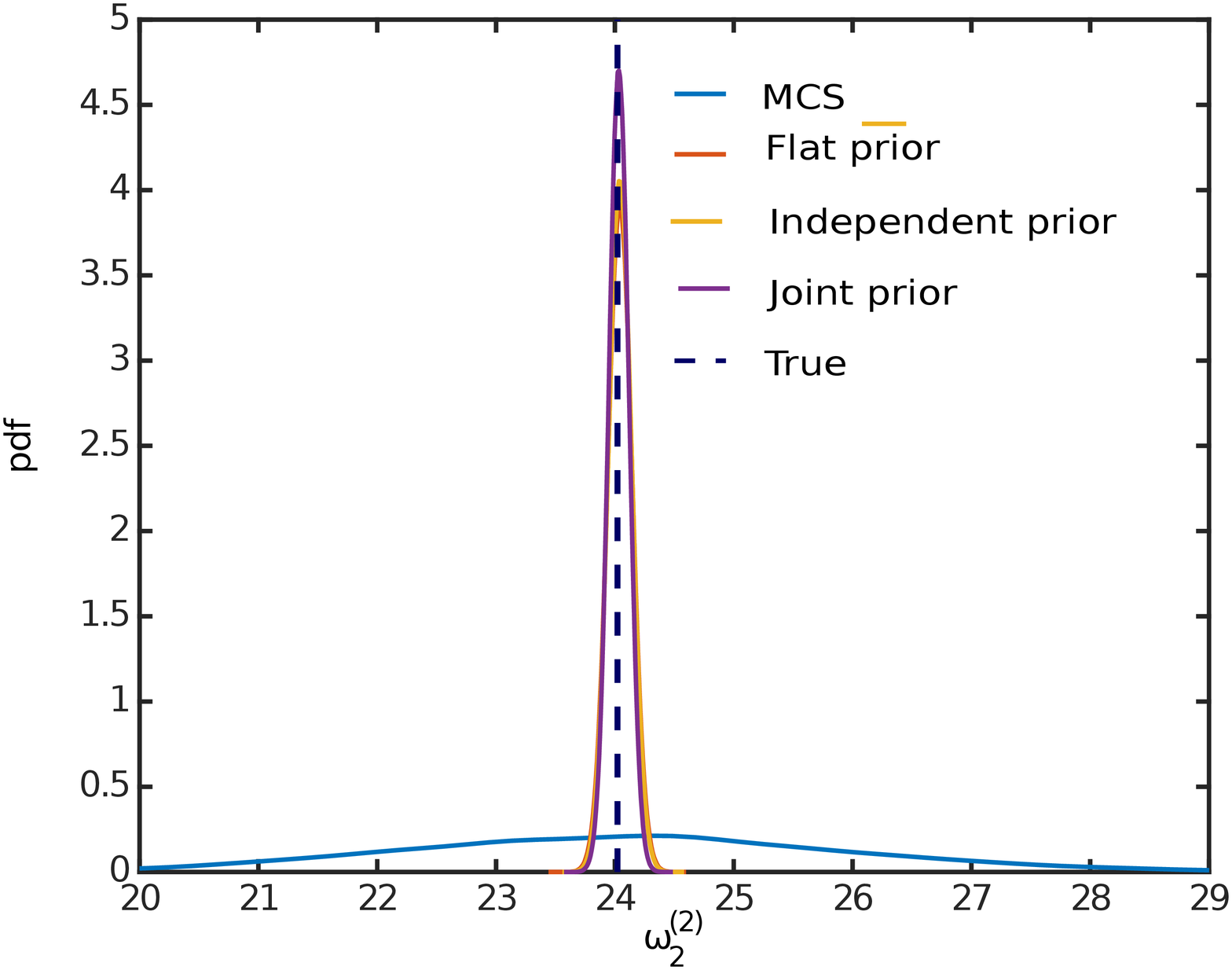}
\caption{Comparison of posterior of $\omega_2^{(2)}$ for different priors }
\label{3u10cvomega22}
\end{center}
\end{figure}
\begin{figure}[htbp]
\begin{center}
\includegraphics[width=120mm,height=80mm]{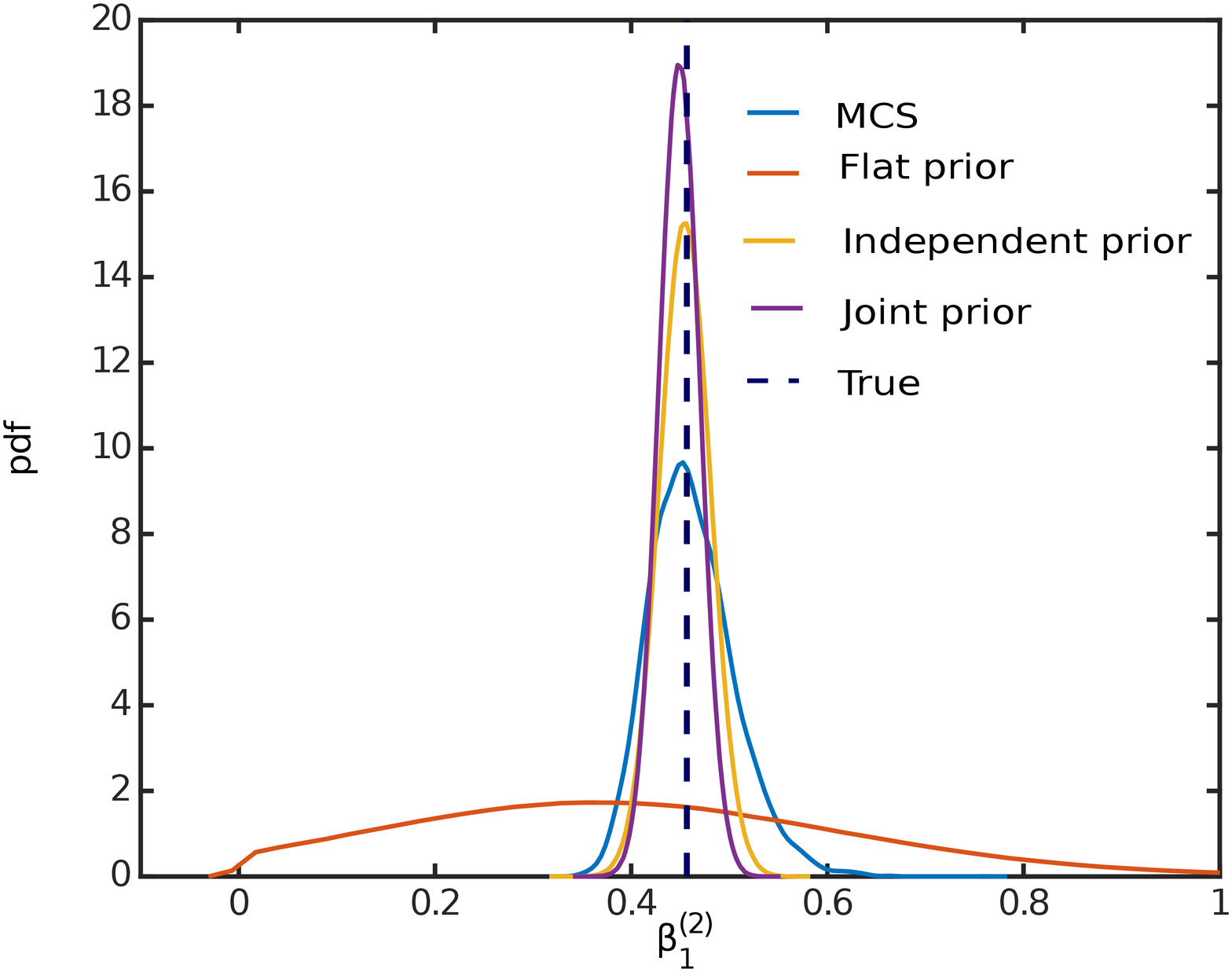}
\caption{Comparison of  posterior of $\beta_1^{(2)}$ for different priors }
\label{3u10cvbeta21}
\end{center}
\end{figure}
\begin{figure}[htbp]
\begin{center}
\includegraphics[width=120mm,height=80mm]{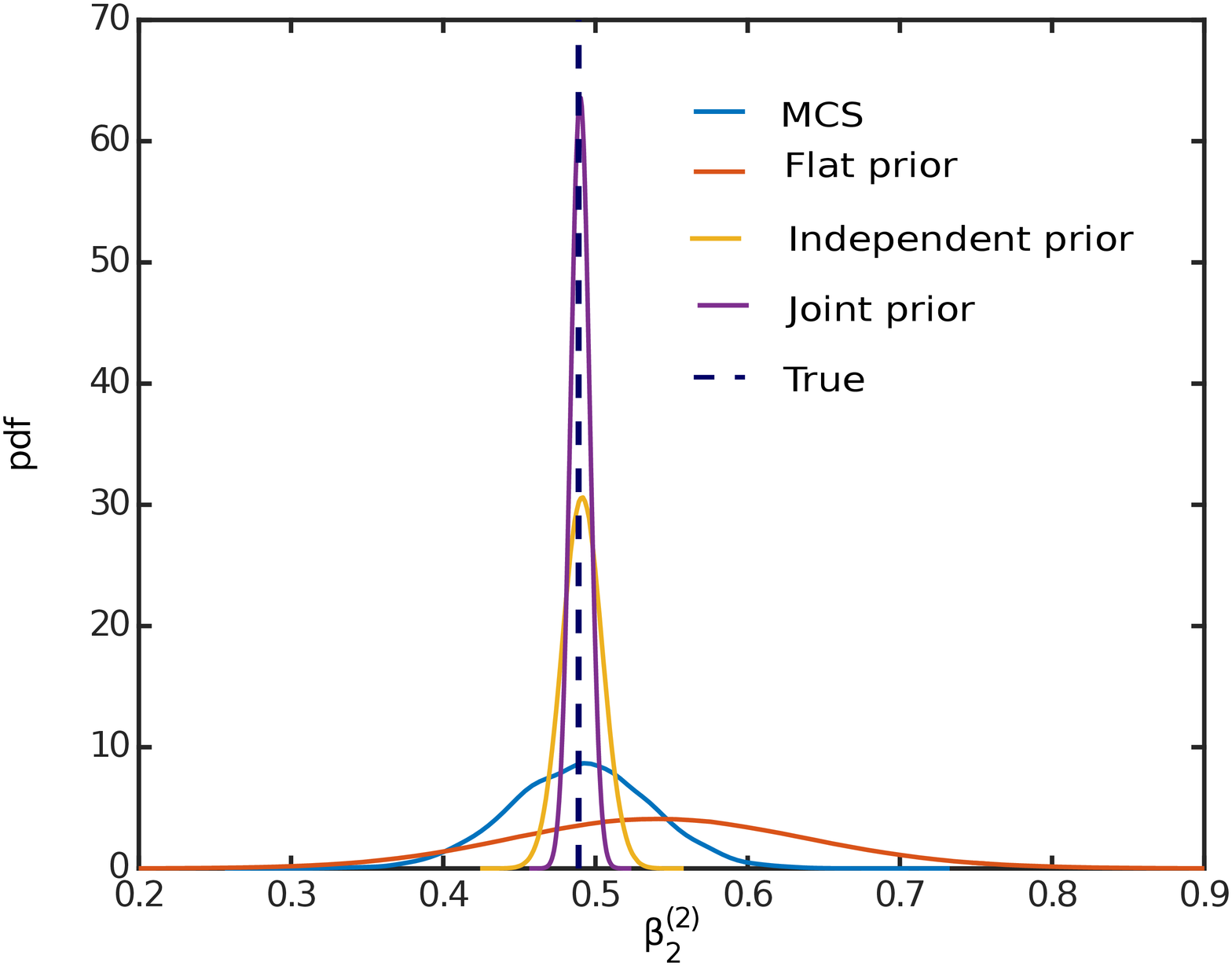}
\caption{Comparison of  posterior of $\beta_2^{(2)}$ for different priors }
\label{3u10cvbeta22}
\end{center}
\end{figure}

\begin{figure}[htbp]
\begin{center}
\includegraphics[width=120mm,height=80mm]{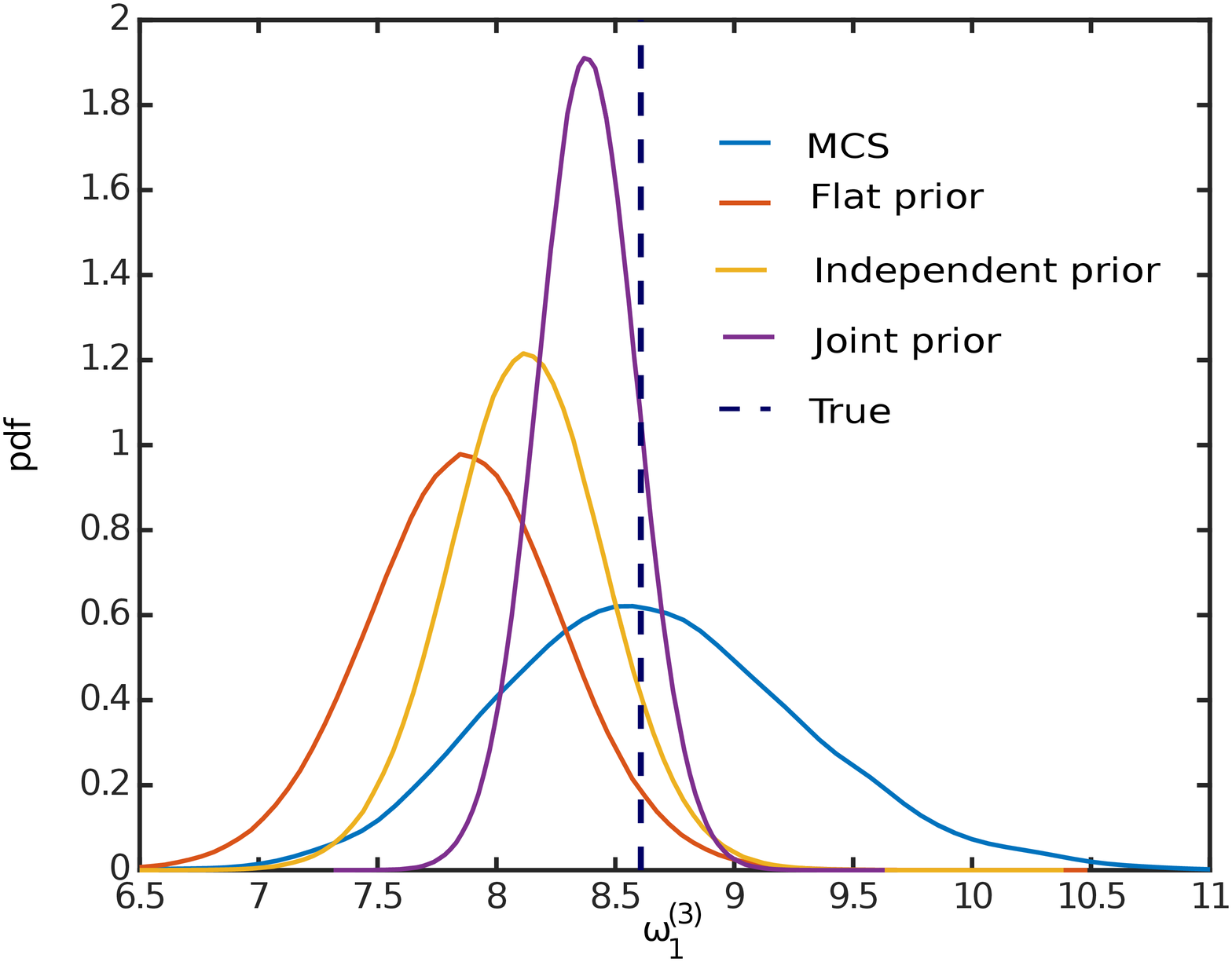}
\caption{Comparison of posterior of $\omega_1^{(3)}$ for different priors }
\label{3u10cvomega31}
\end{center}
\end{figure}
\begin{figure}[htbp]
\begin{center}
\includegraphics[width=120mm,height=80mm]{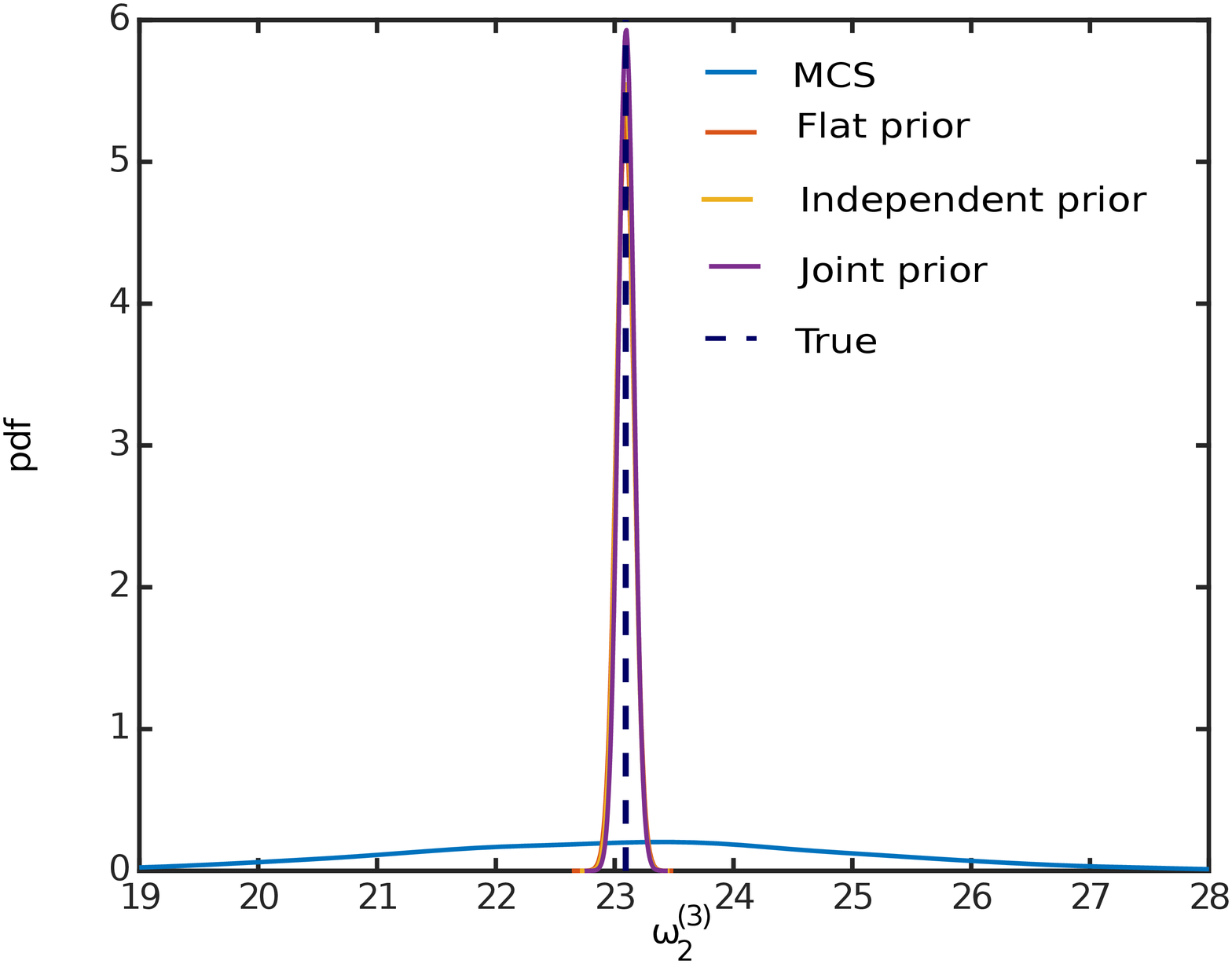}
\caption{Comparison of posterior of $\omega_2^{(3)}$ for different priors }
\label{3u10cvomega32}
\end{center}
\end{figure}
\begin{figure}[htbp]
\begin{center}
\includegraphics[width=120mm,height=80mm]{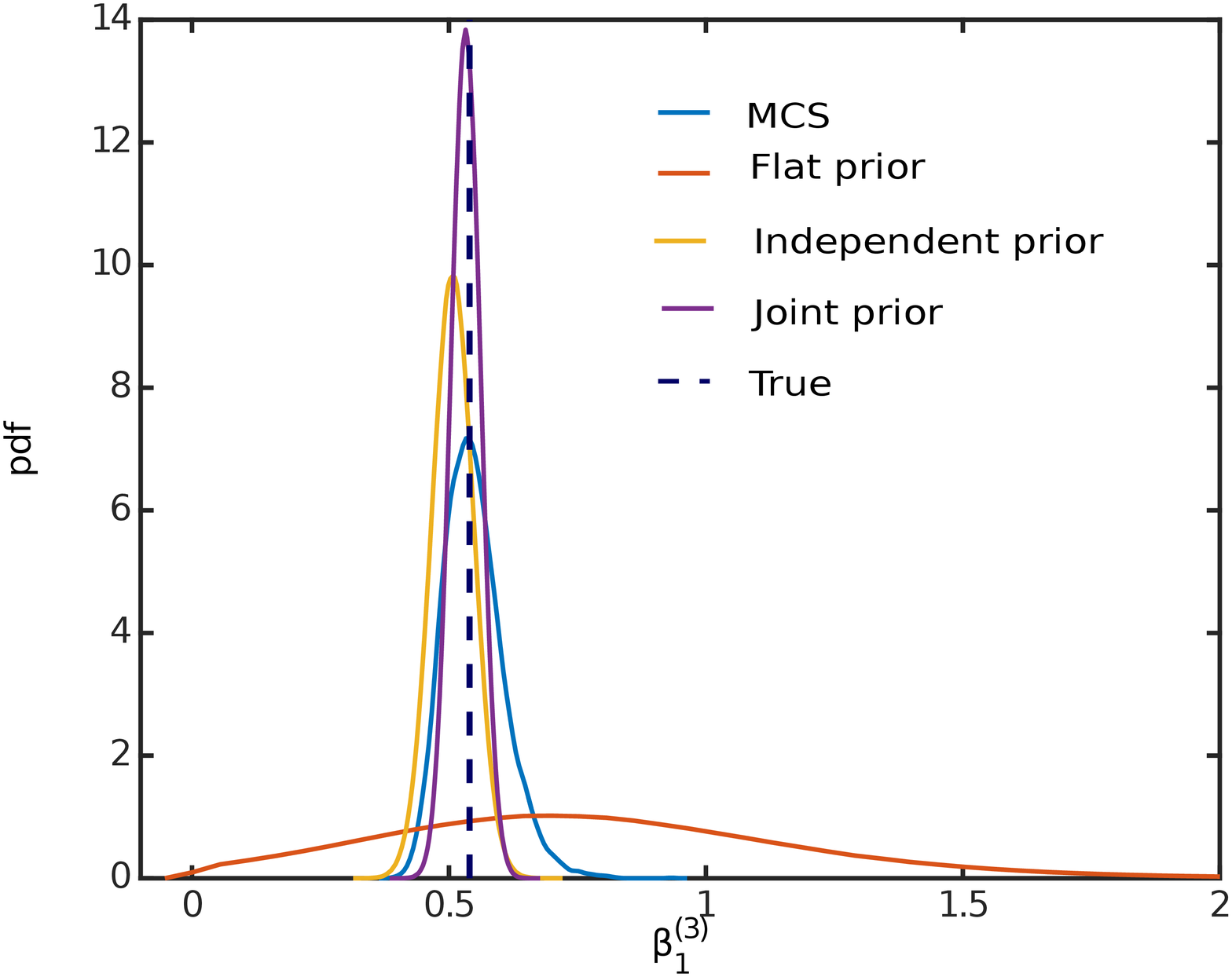}
\caption{Comparison of  posterior of $\beta_1^{(3)}$ for different priors }
\label{3u10cvbeta31}
\end{center}
\end{figure}
\begin{figure}[htbp]
\begin{center}
\includegraphics[width=120mm,height=80mm]{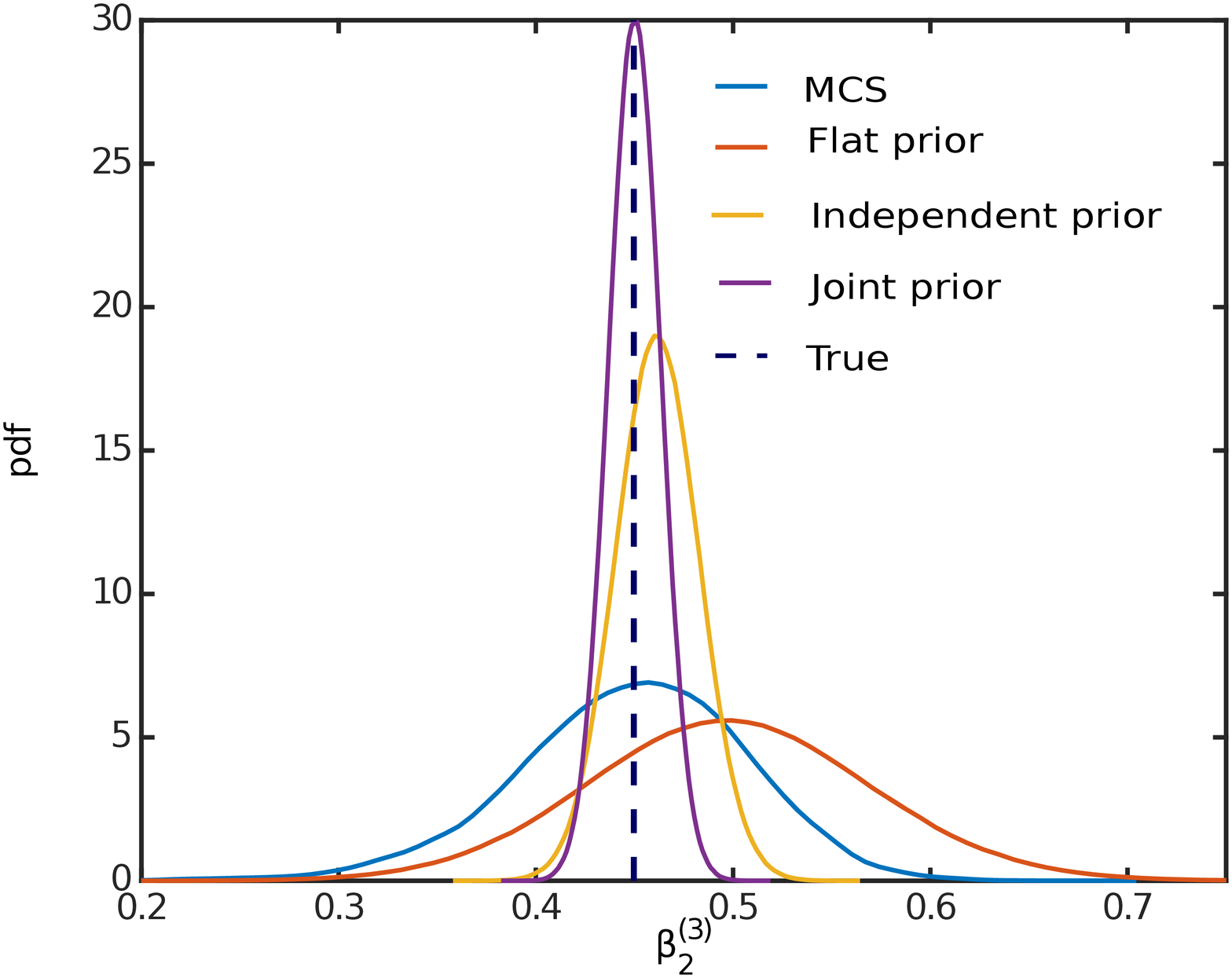}
\caption{Comparison of  posterior of $\beta_2^{(3)}$ for different priors }
\label{3u10cvbeta32}
\end{center}
\end{figure}

\begin{figure}[h!]
\centering
\begin{tabular}{ccc}
\subf{\includegraphics[width=45mm]{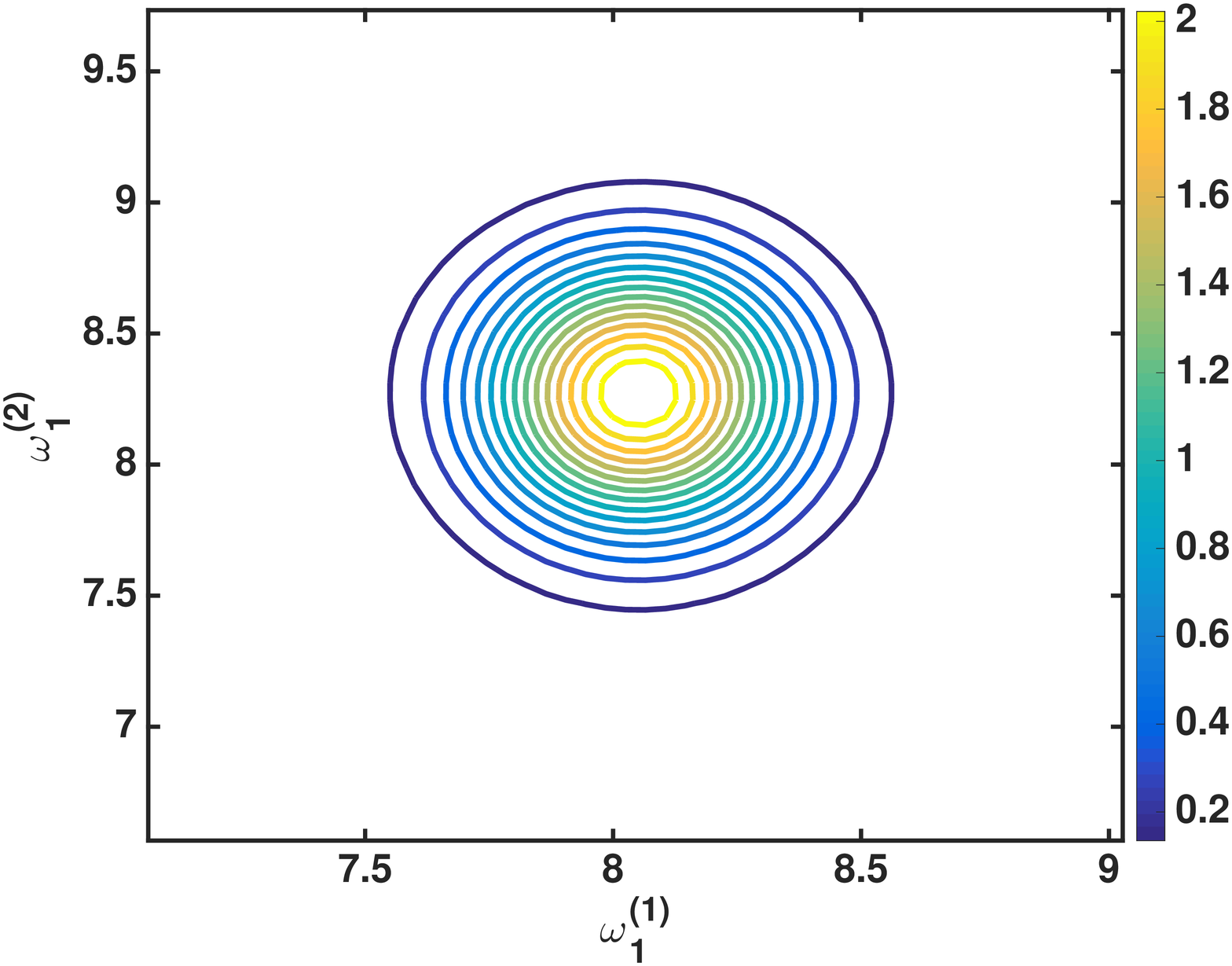}}
     {$$}
&
\subf{\includegraphics[width=45mm]{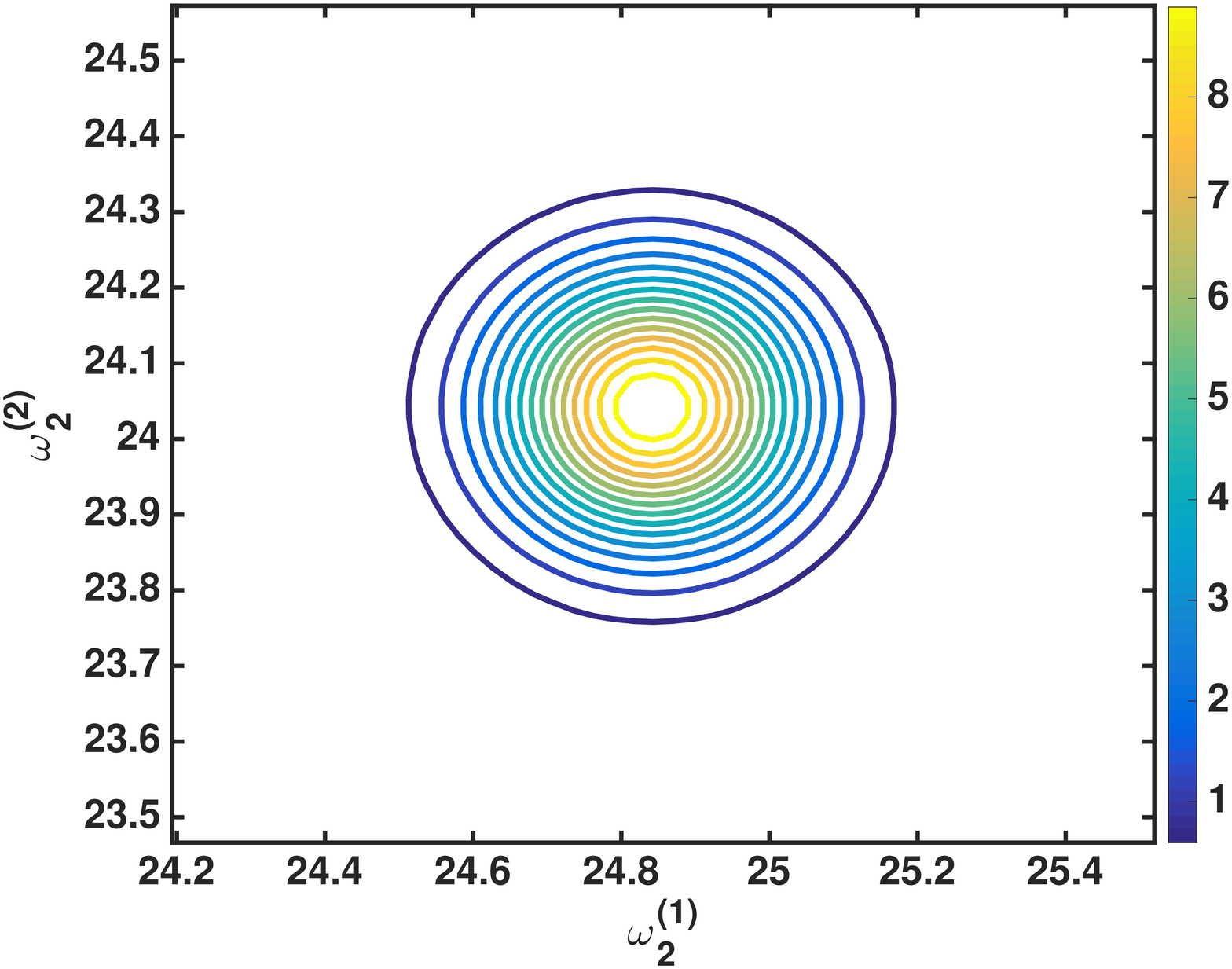}}
     {$$}
&
\subf{\includegraphics[width=45mm]{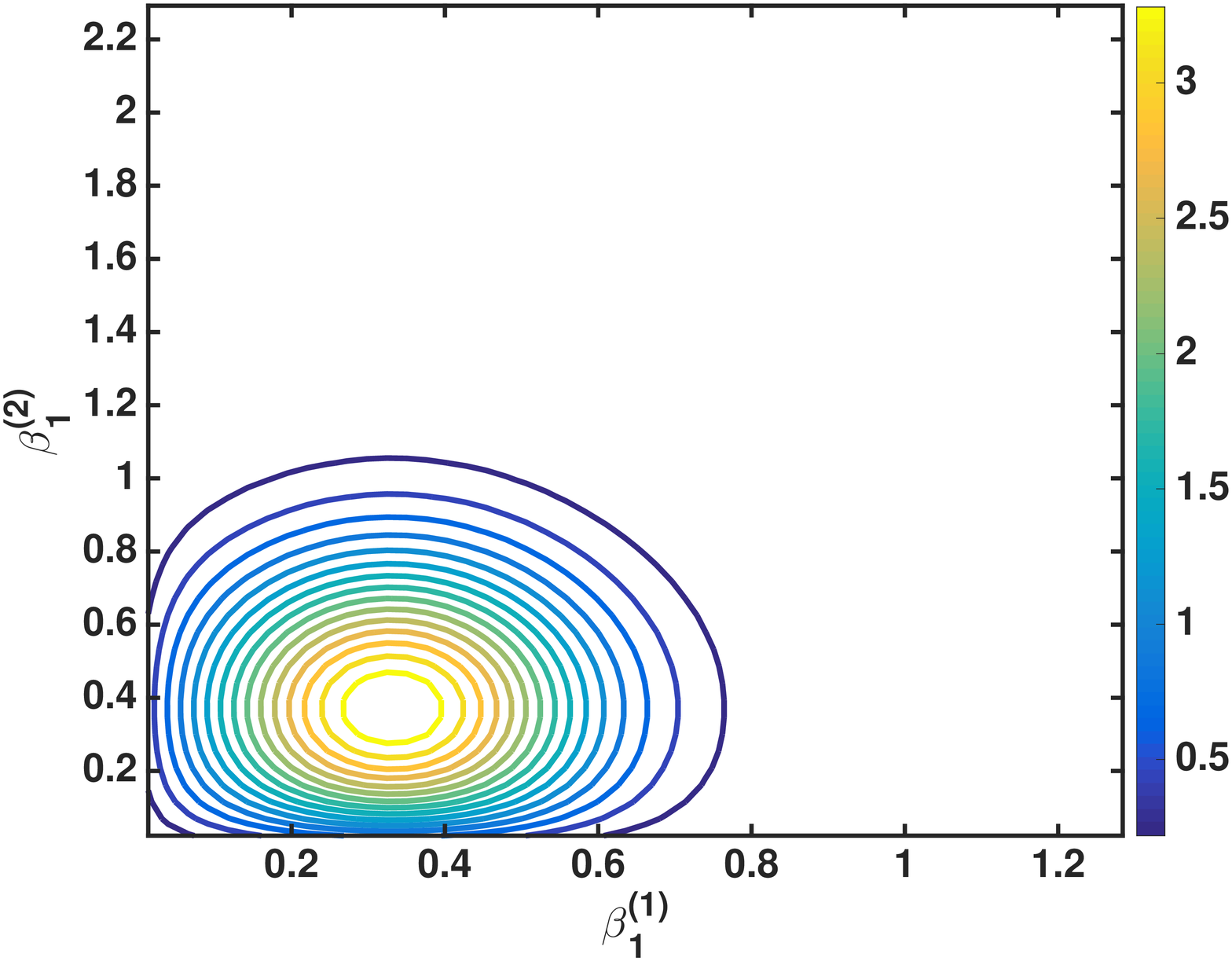}}
     {$$}
\\
\subf{\includegraphics[width=45mm]{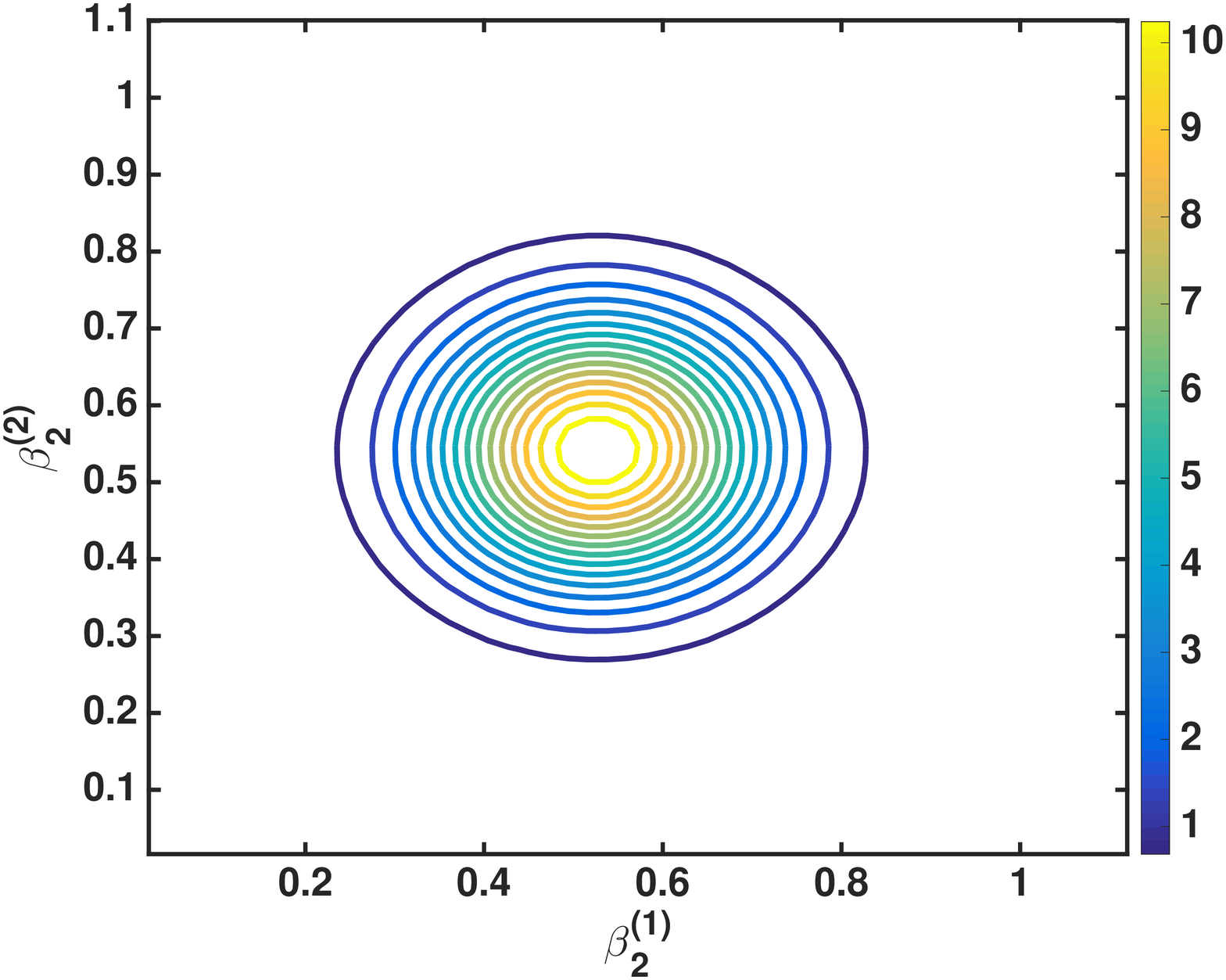}}
     {$$}
&
\subf{\includegraphics[width=45mm]{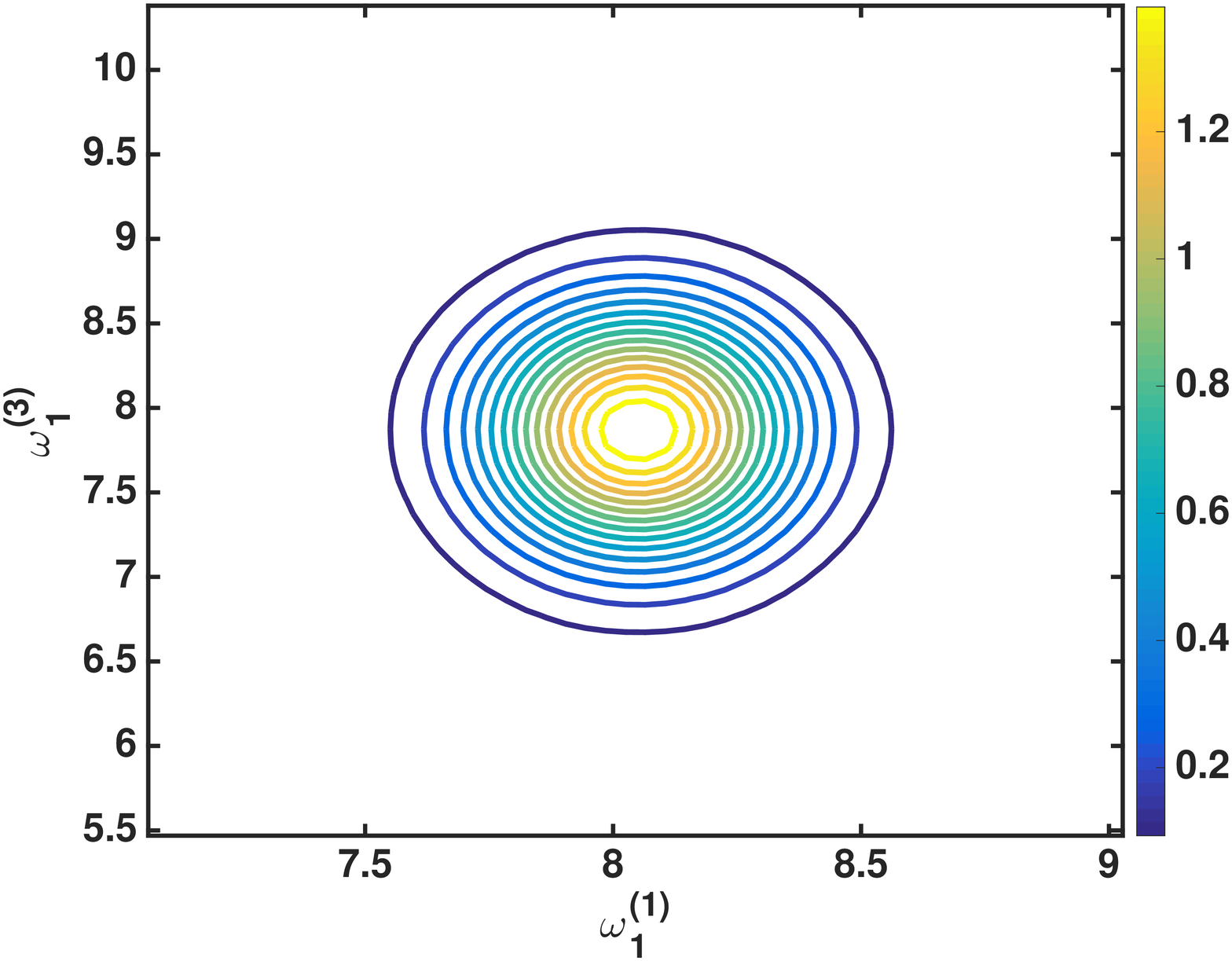}}
     {$$}
&
\subf{\includegraphics[width=45mm]{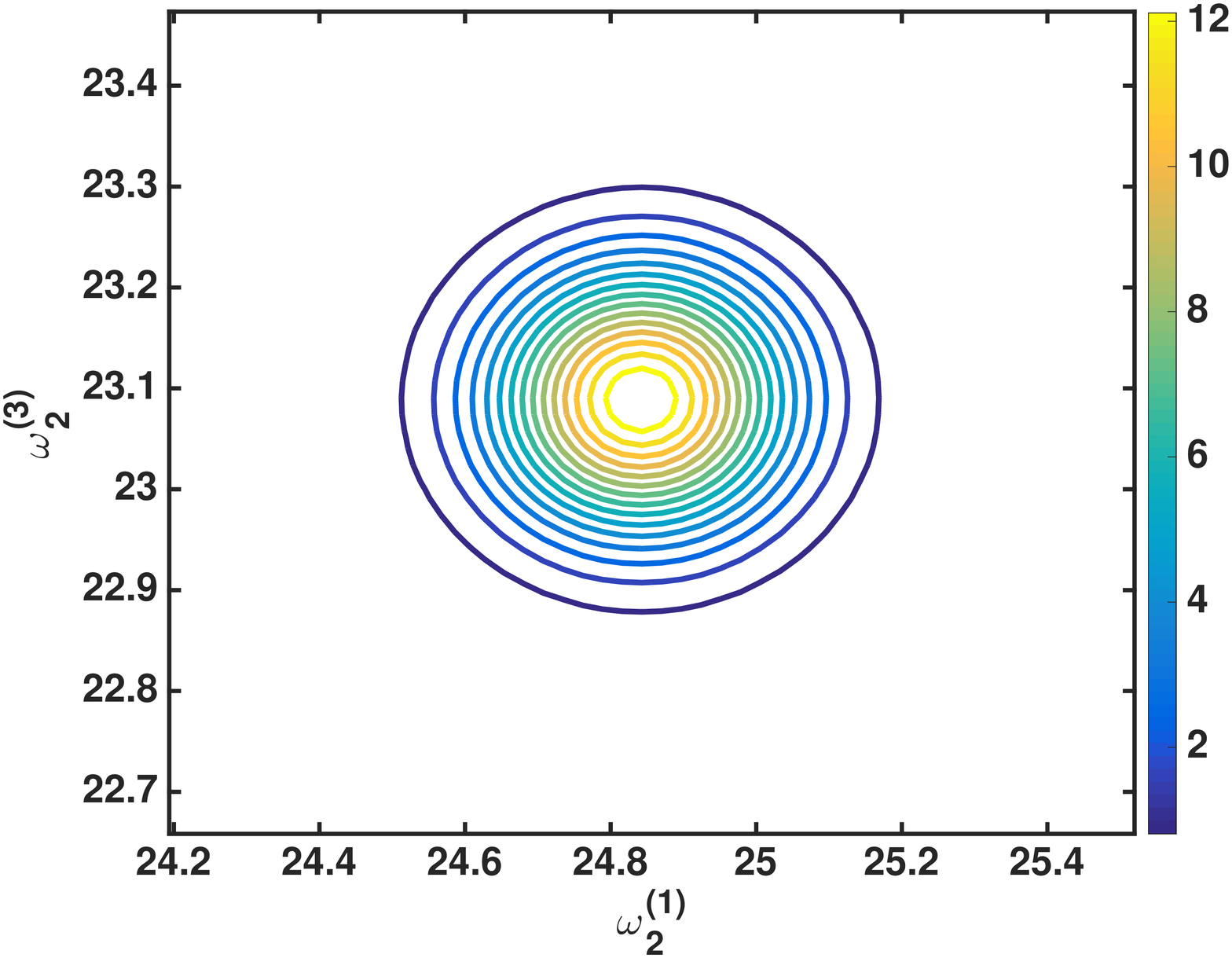}}
     {$$}
\\
\subf{\includegraphics[width=45mm]{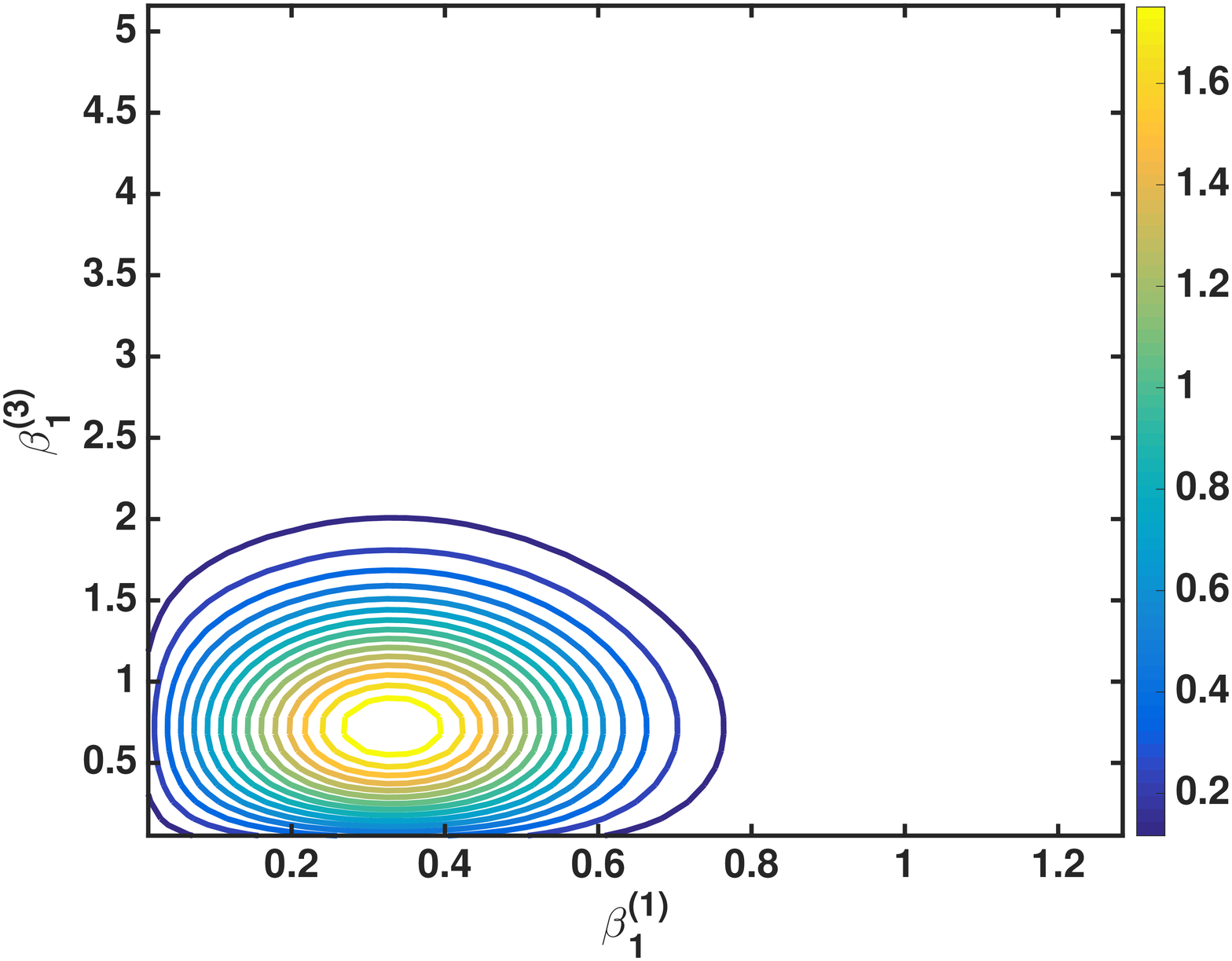}}
     {$$}
&
\subf{\includegraphics[width=45mm]{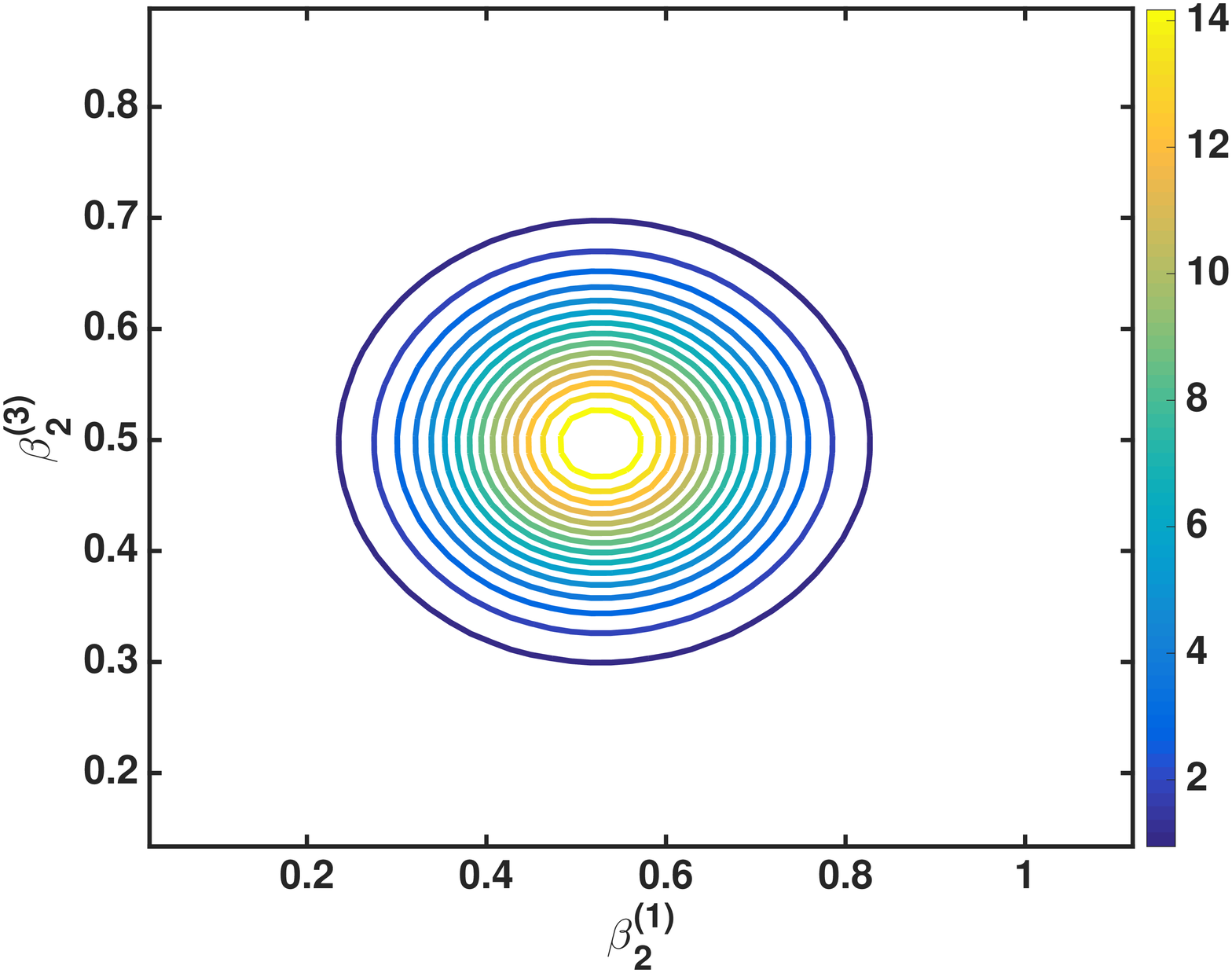}}
     {$$}
&
\subf{\includegraphics[width=45mm]{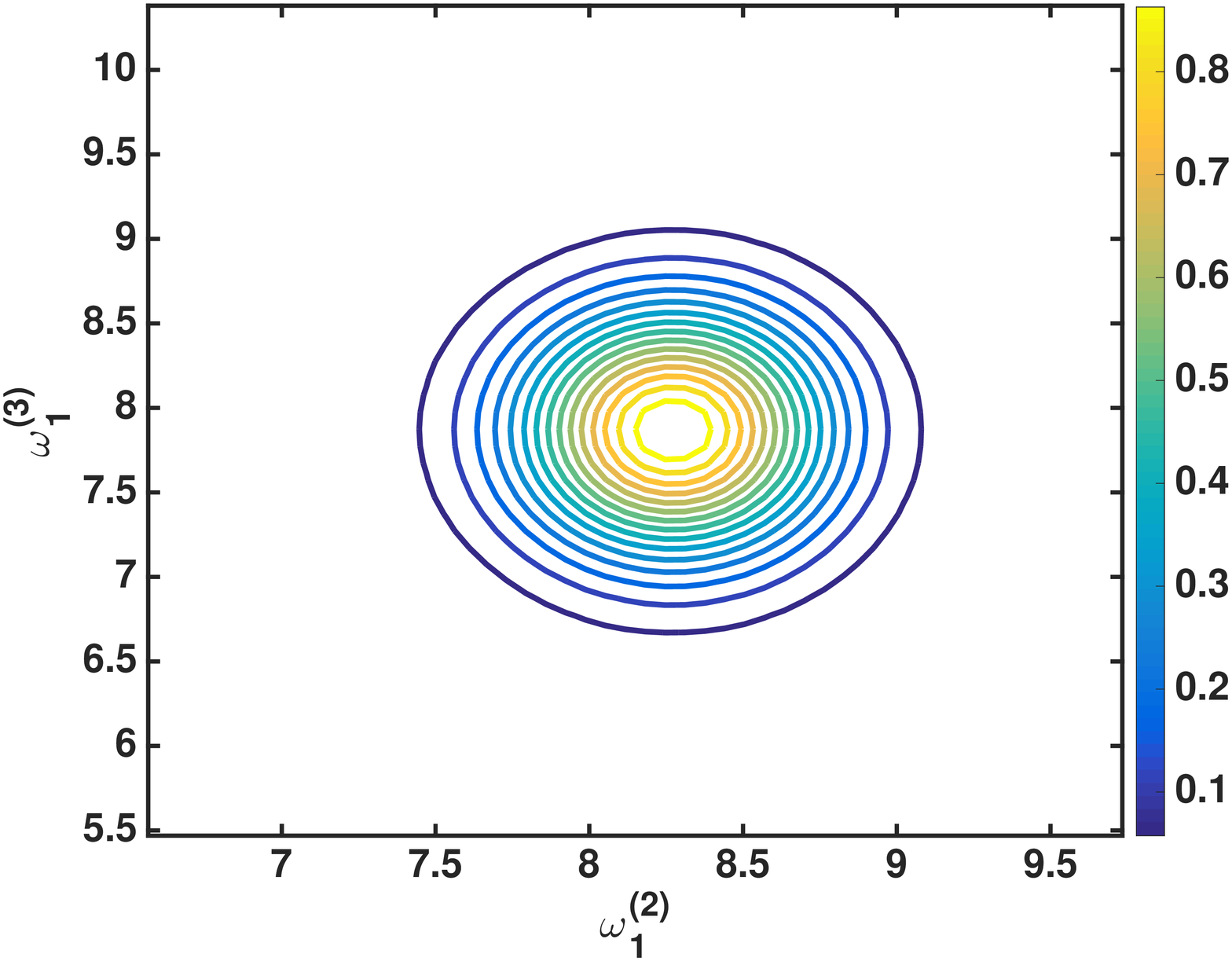}}
     {$$}
\\
\subf{\includegraphics[width=45mm]{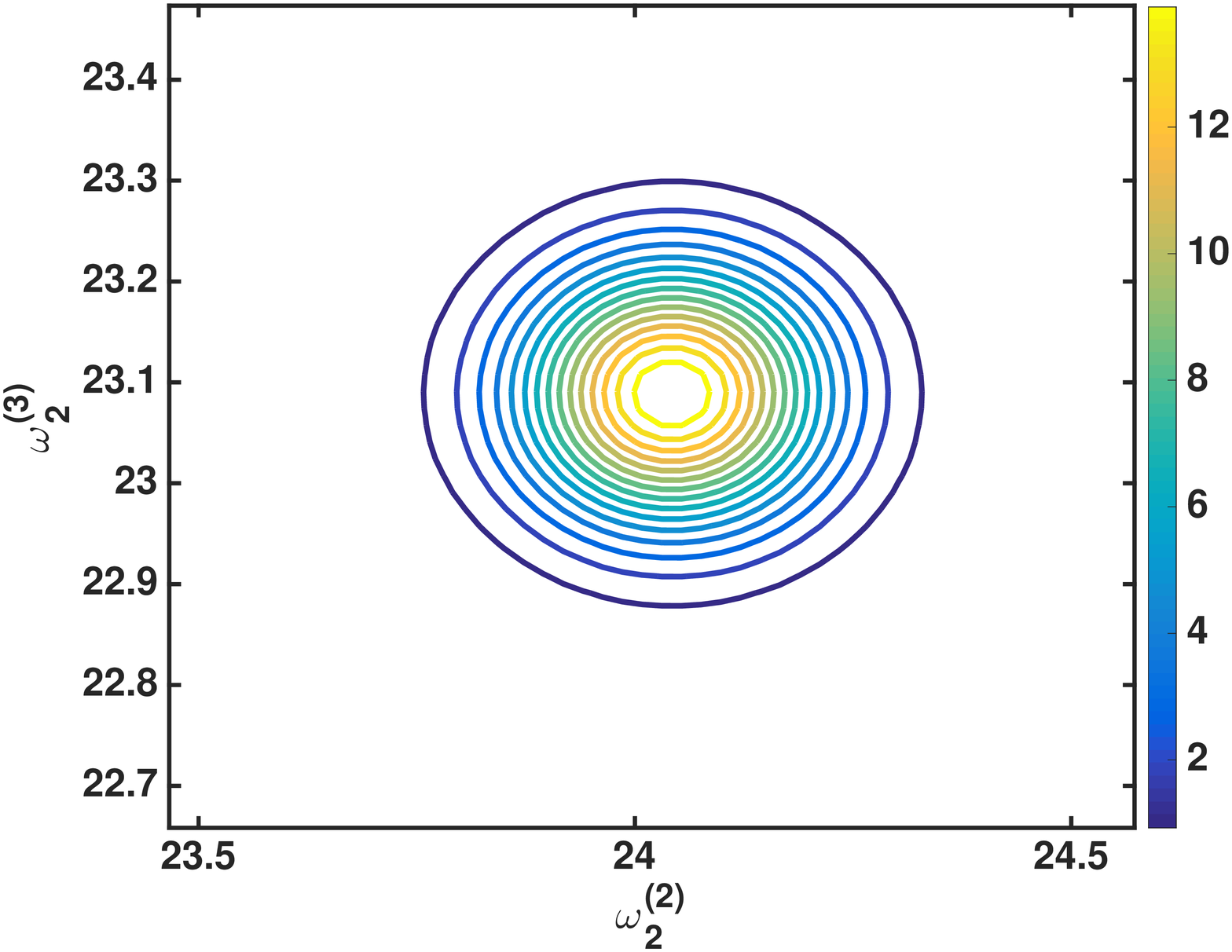}}
     {$$}
&
\subf{\includegraphics[width=45mm]{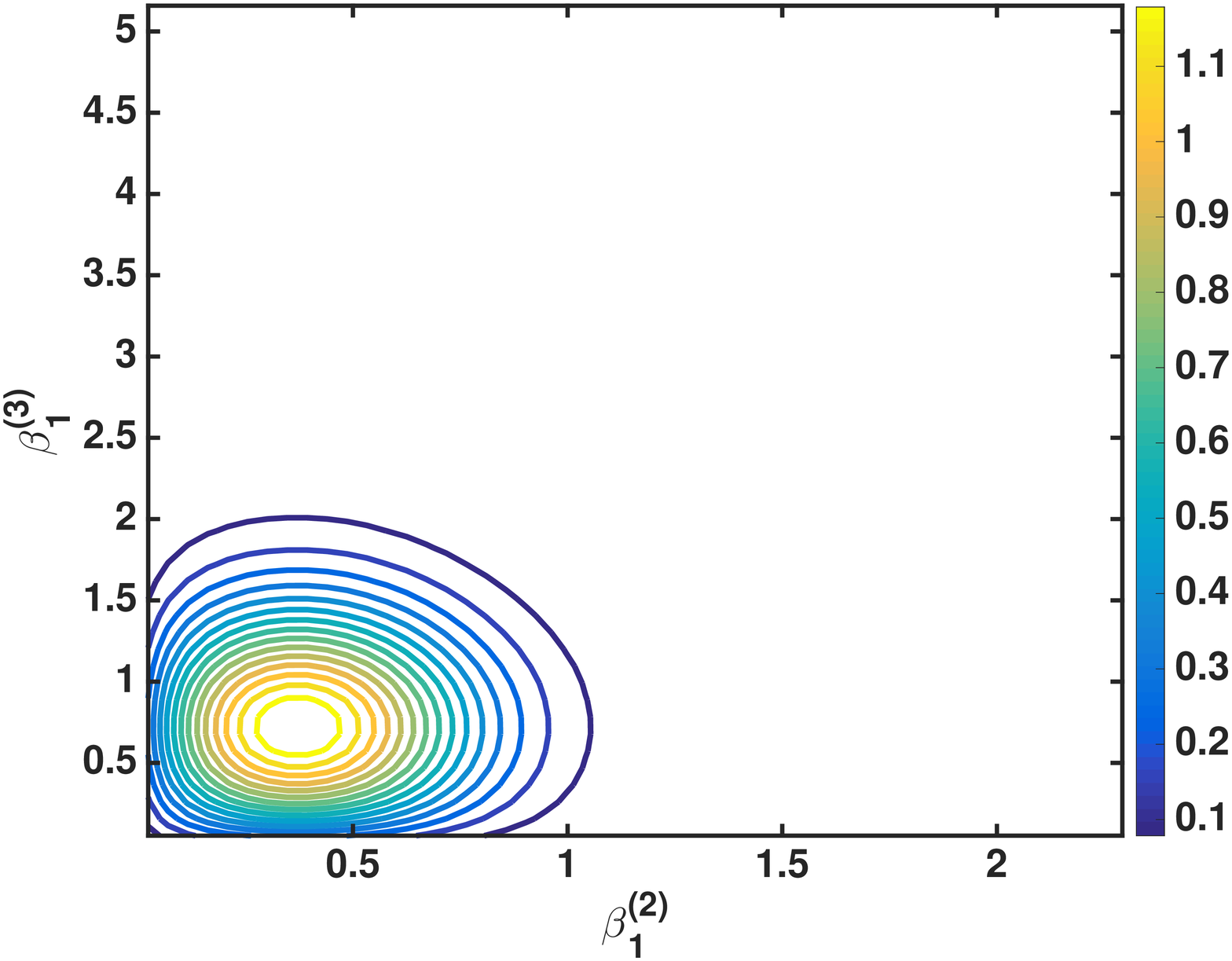}}
     {$$}
&
\subf{\includegraphics[width=45mm]{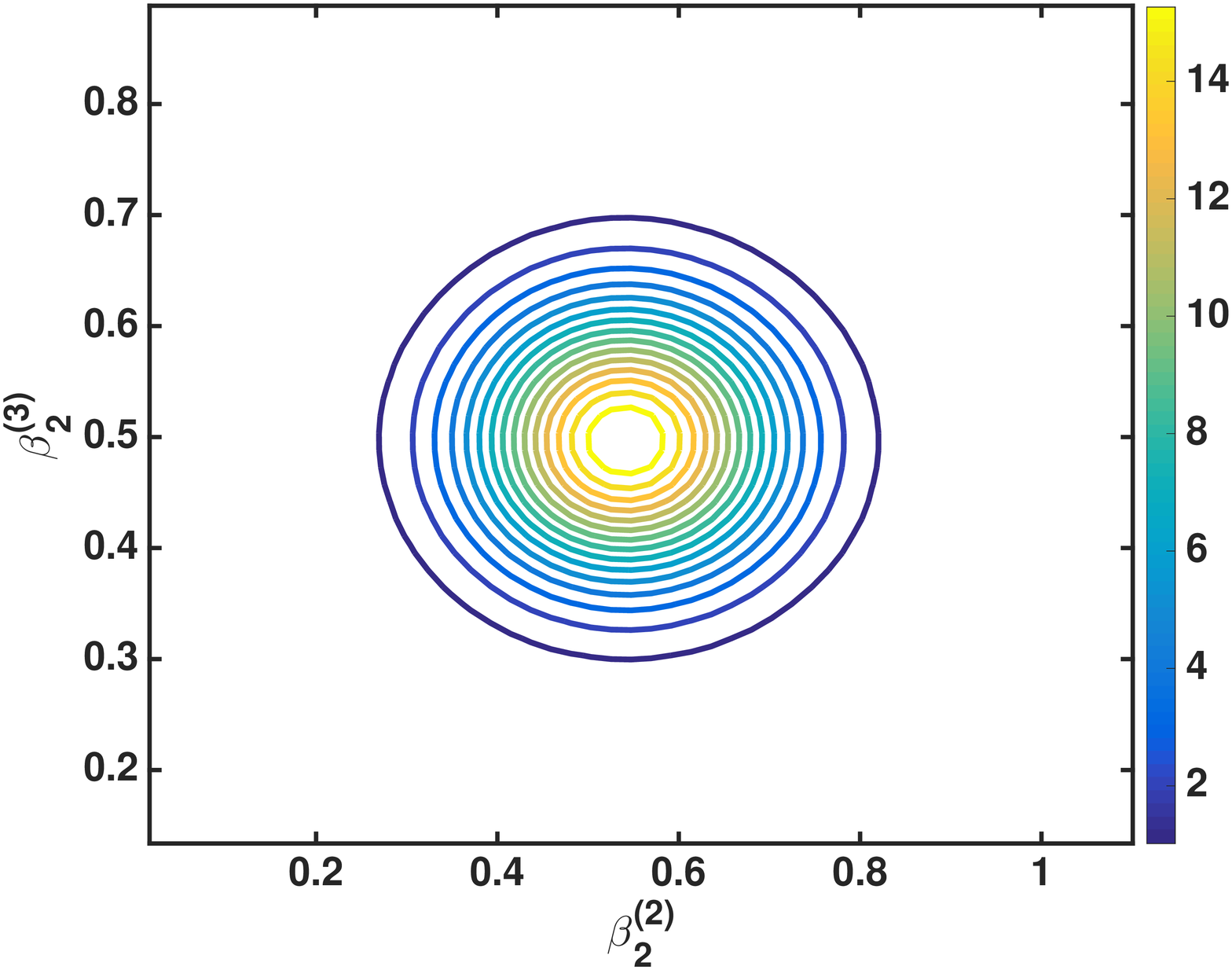}}
     {$$}
\\
\end{tabular}
\caption{Joint posterior pdfs of parameters $\omega_1, \beta_1, \omega_2$ and $\beta_2$ using the flat prior among three airspeeds}
\label{3u10cvjointpostflat}
\end{figure}

\begin{figure}[ht!]
\centering
\begin{tabular}{ccc}
\subf{\includegraphics[width=45mm]{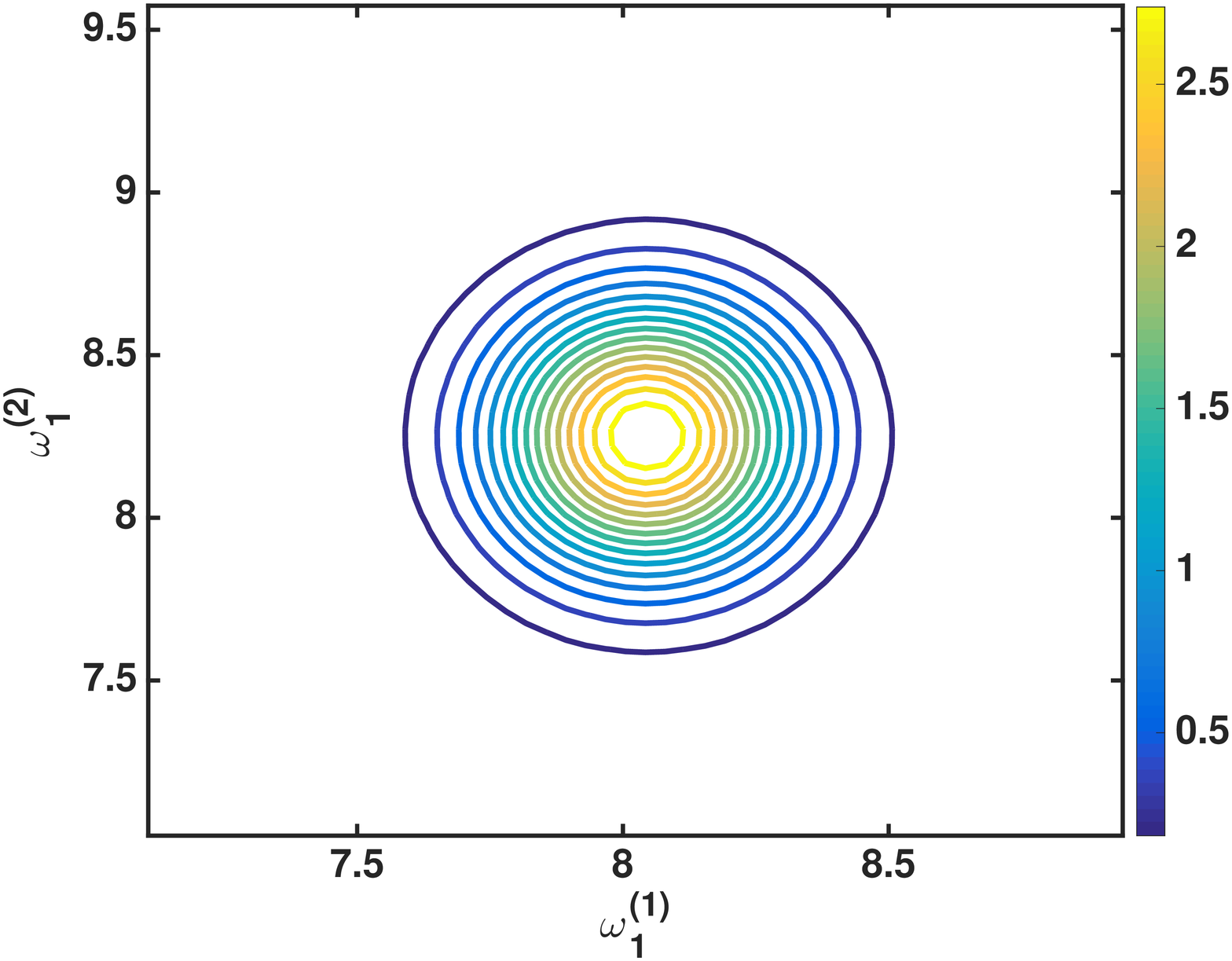}}
     {$$}
&
\subf{\includegraphics[width=45mm]{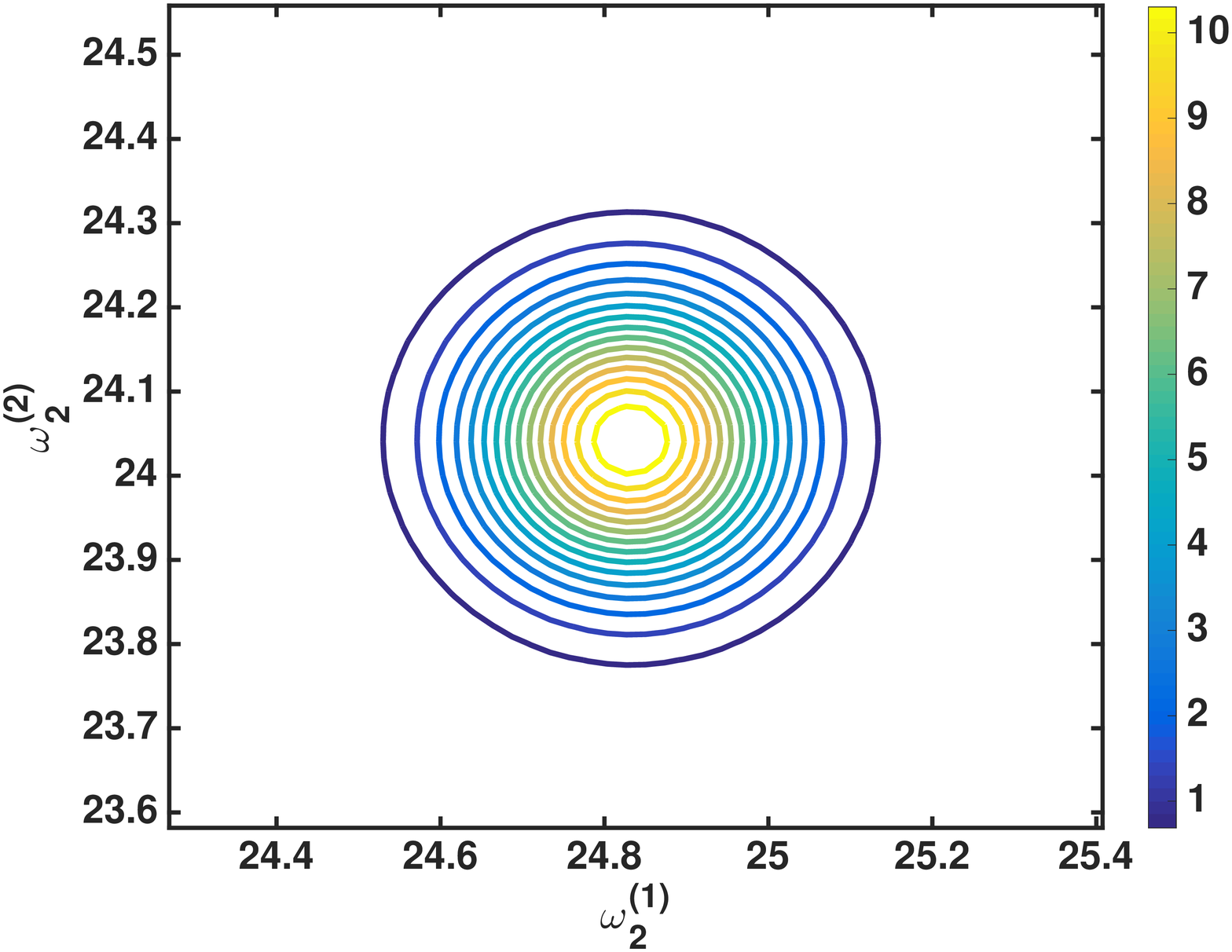}}
     {$$}
&
\subf{\includegraphics[width=45mm]{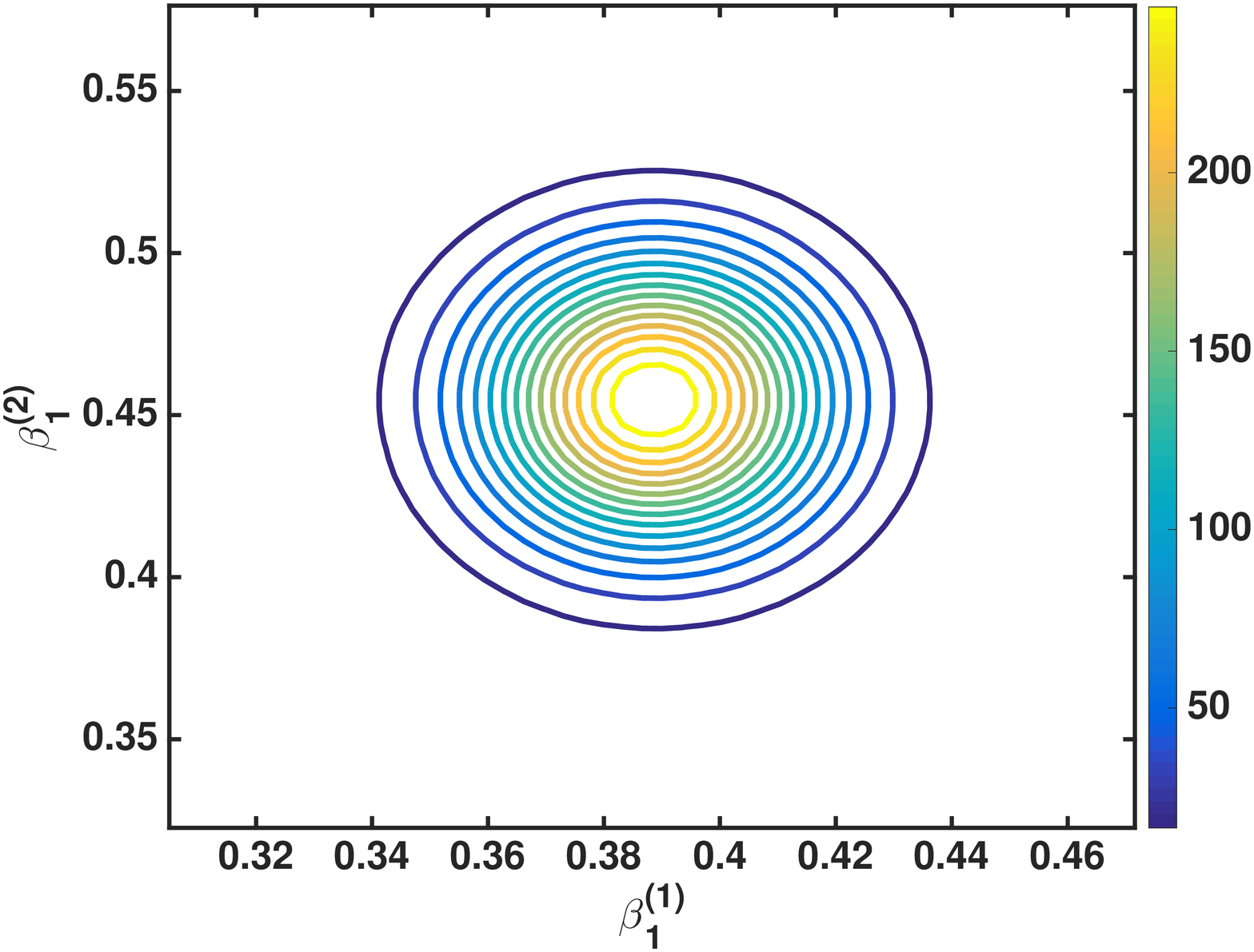}}
     {$$}
\\
\subf{\includegraphics[width=45mm]{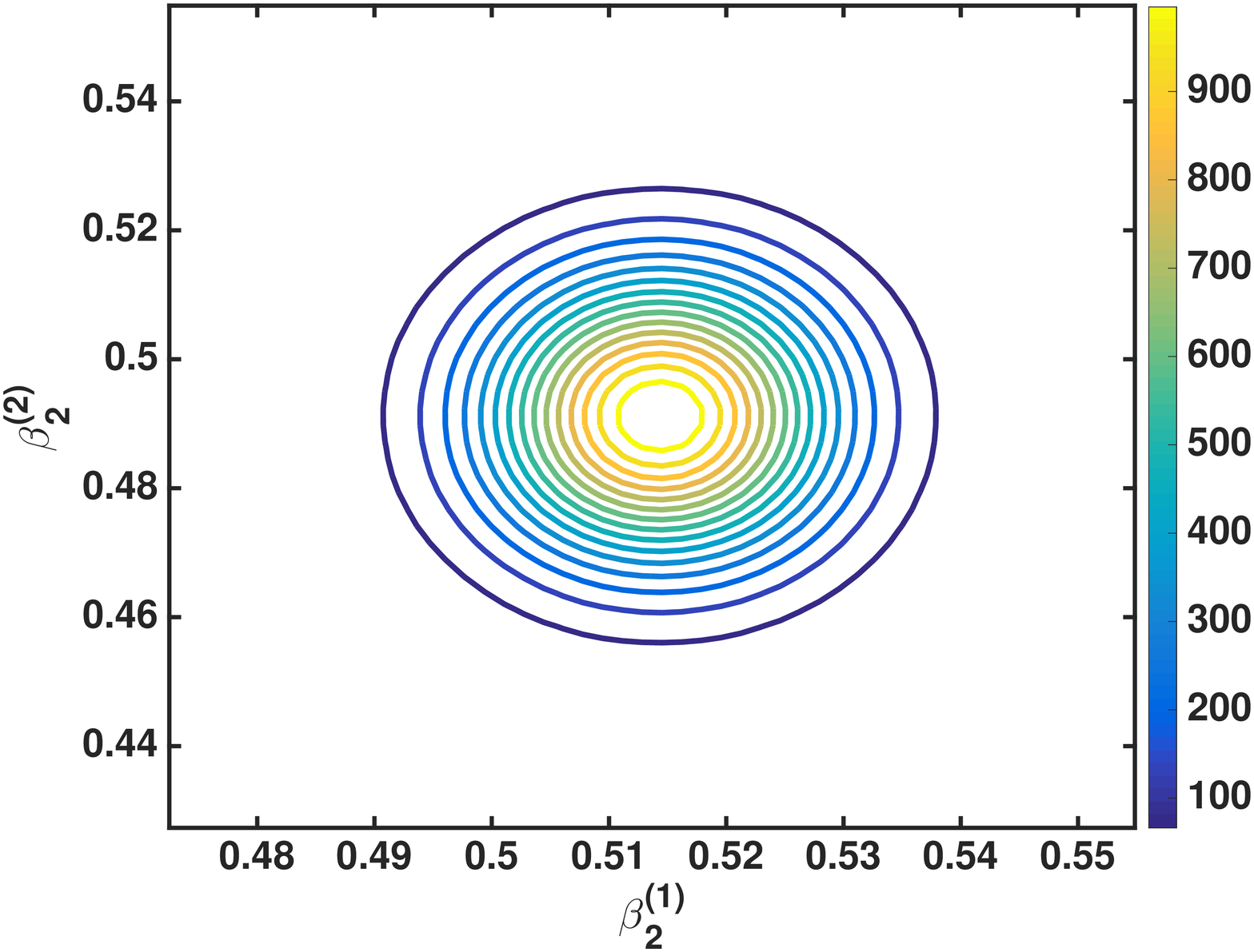}}
     {$$}
&
\subf{\includegraphics[width=45mm]{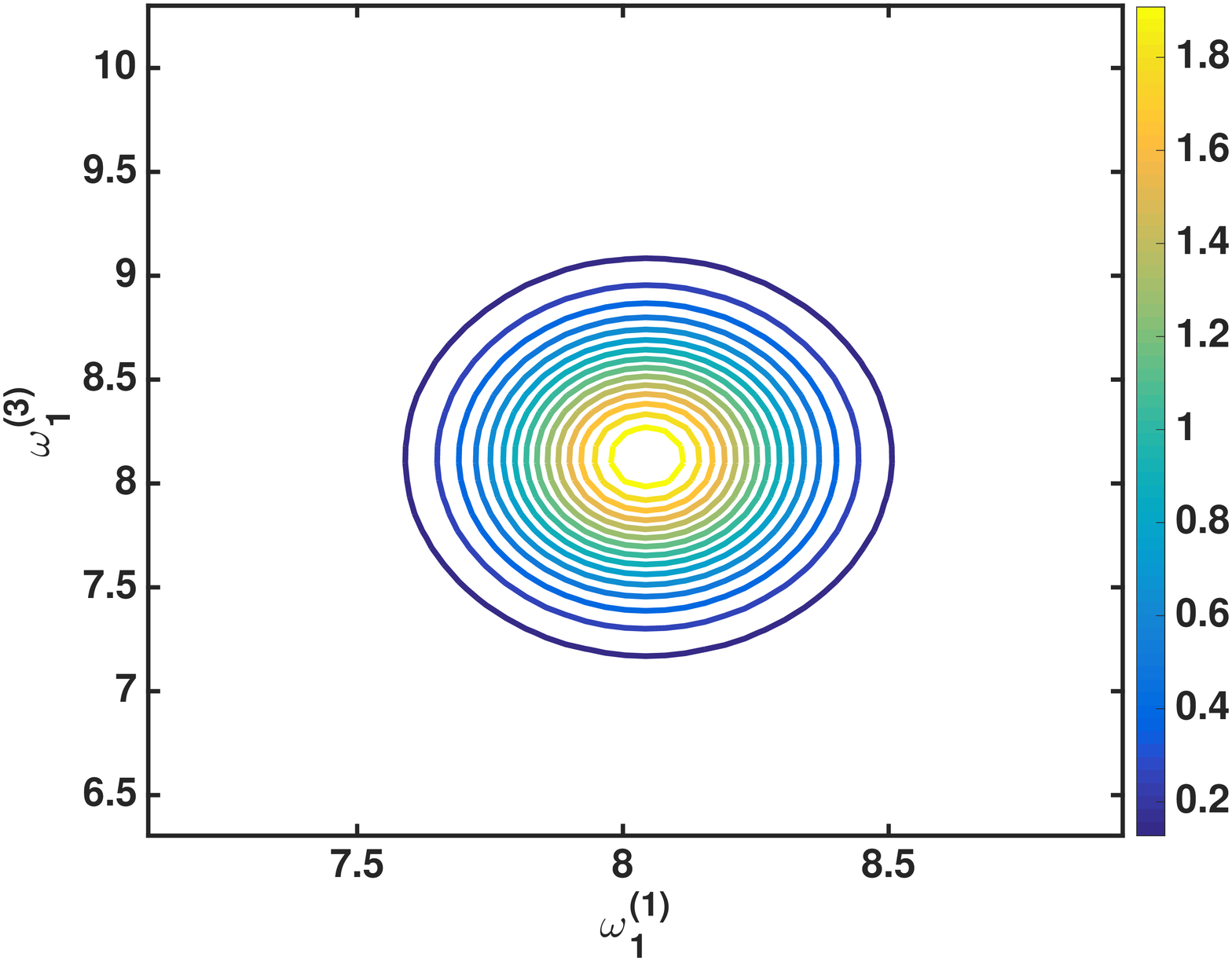}}
     {$$}
&
\subf{\includegraphics[width=45mm]{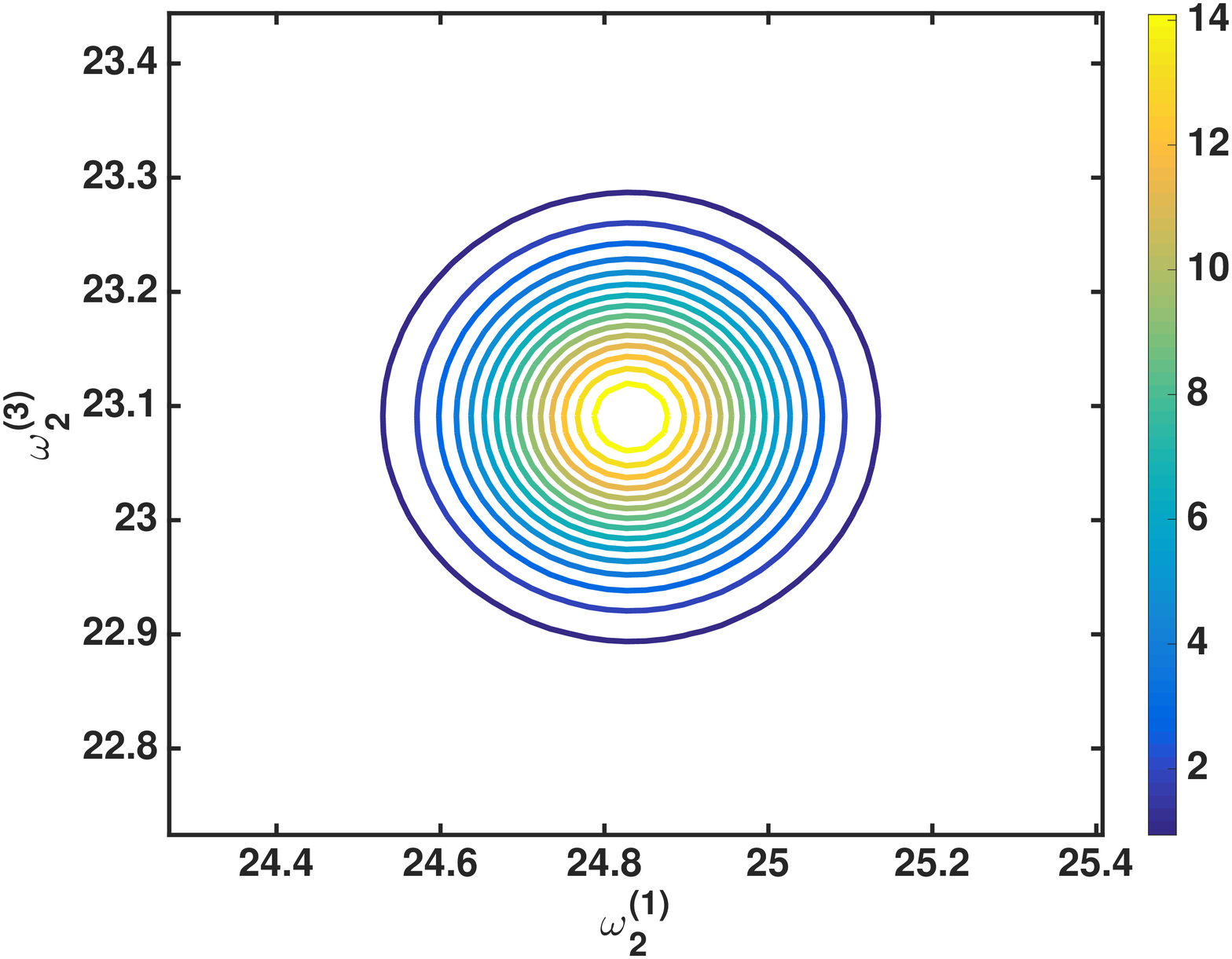}}
     {$$}
\\
\subf{\includegraphics[width=45mm]{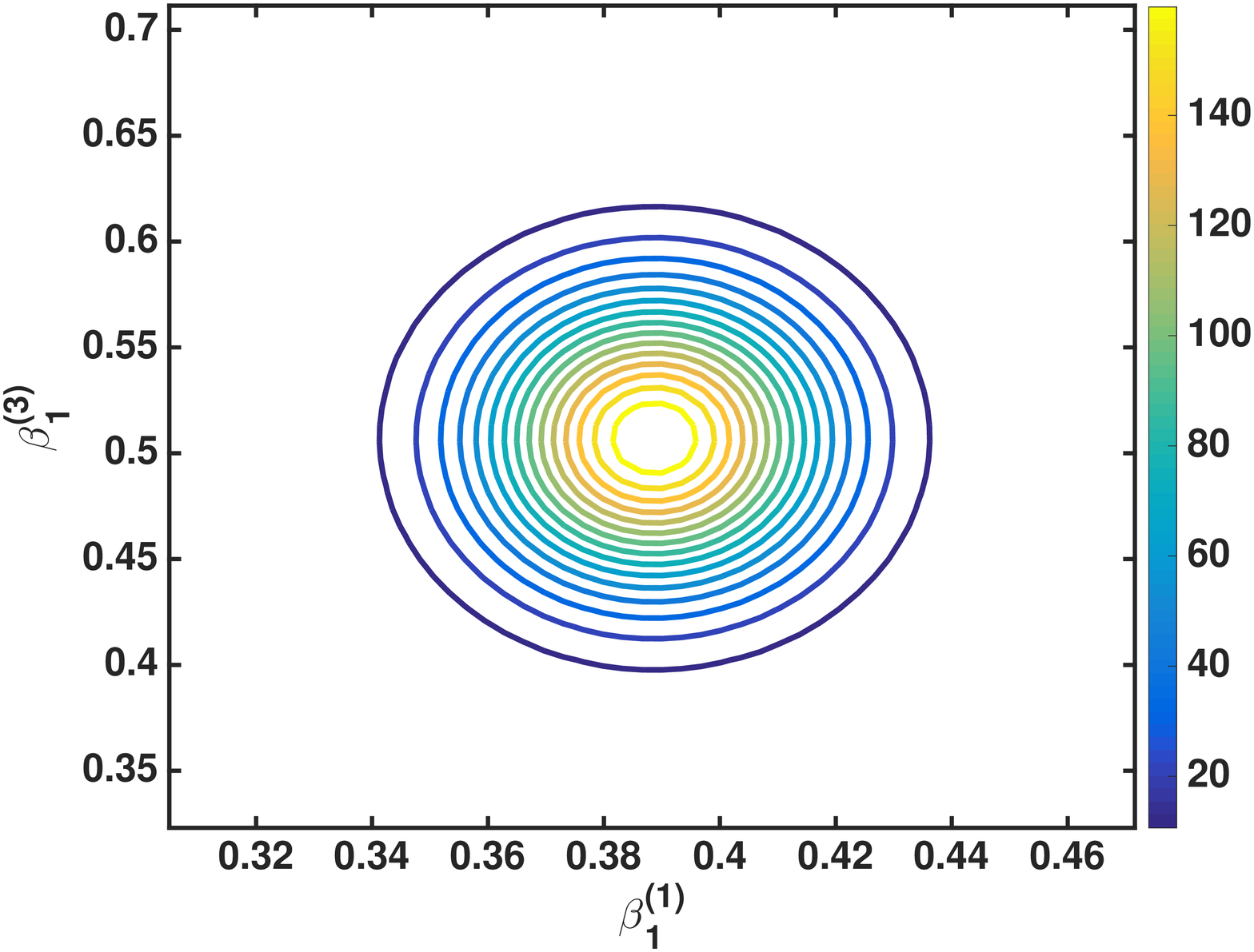}}
     {$$}
&
\subf{\includegraphics[width=45mm]{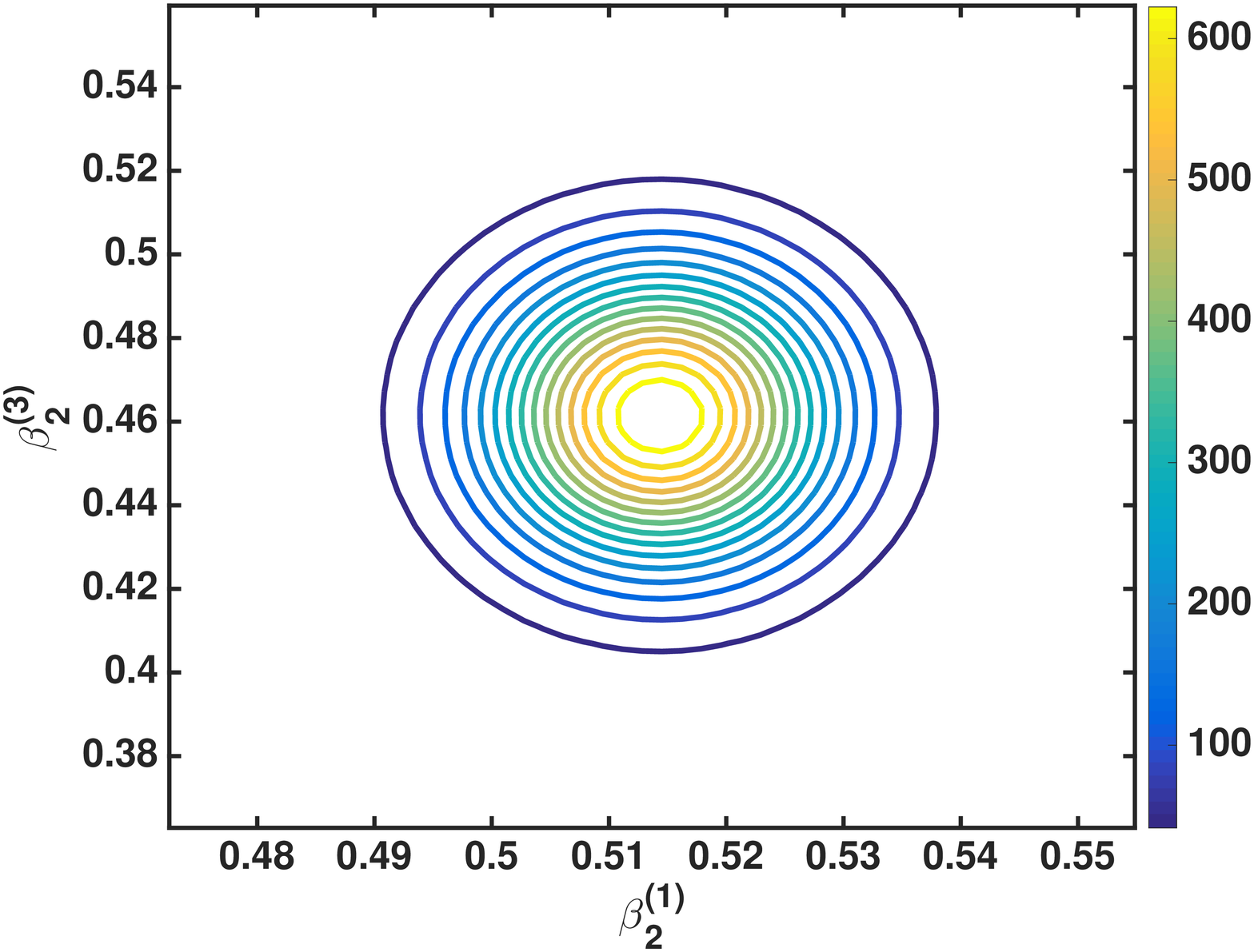}}
     {$$}
&
\subf{\includegraphics[width=45mm]{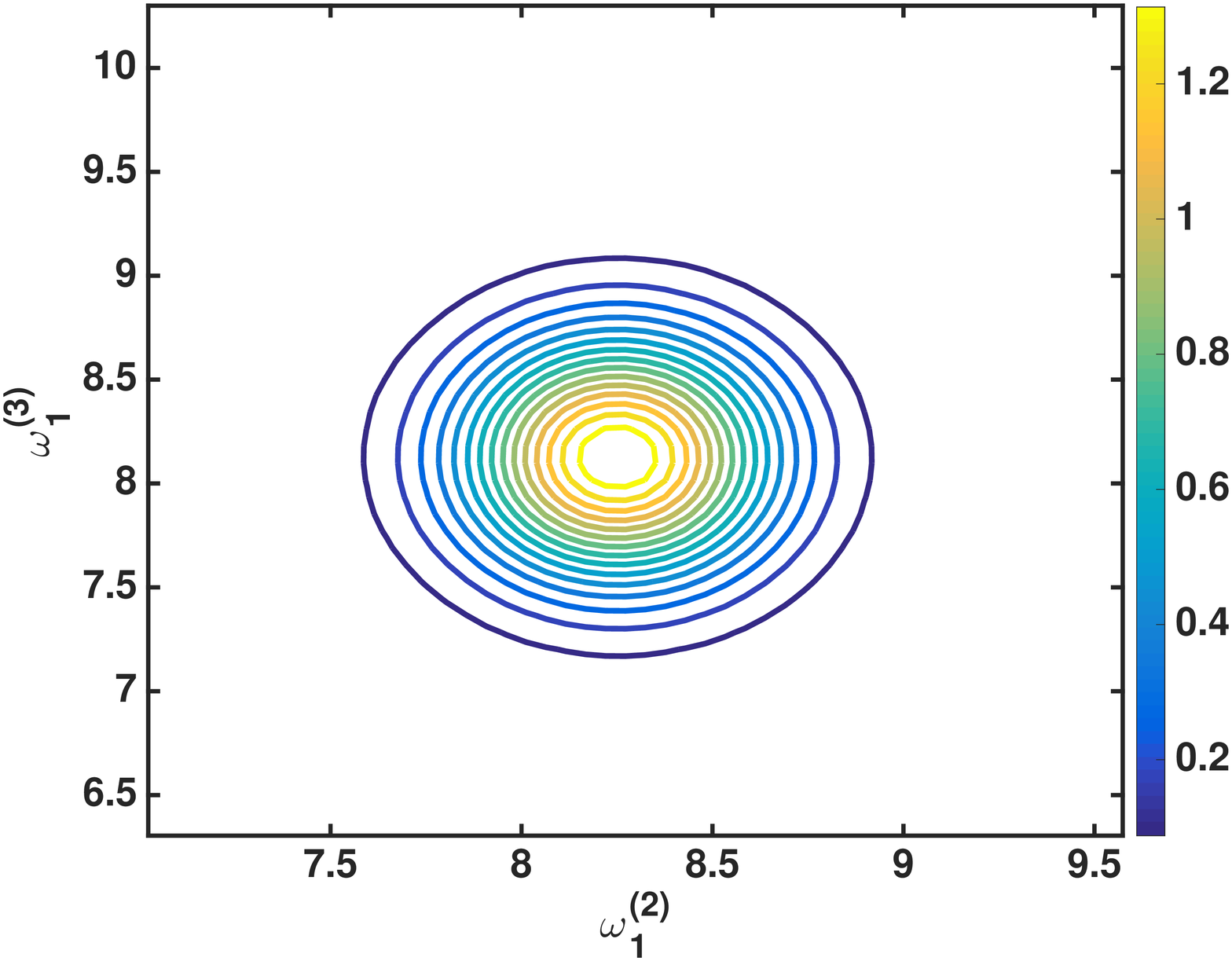}}
     {$$}
\\
\subf{\includegraphics[width=45mm]{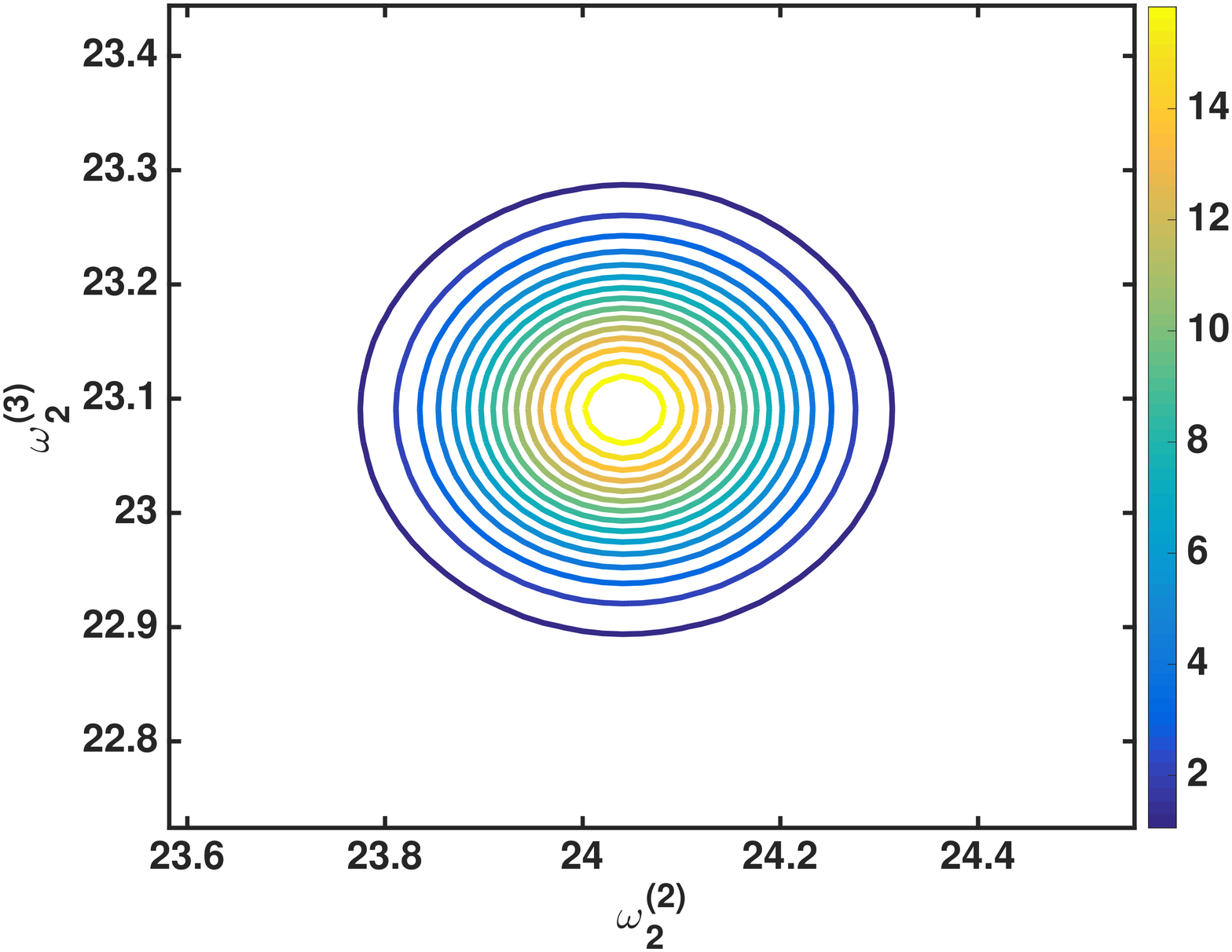}}
     {$$}
&
\subf{\includegraphics[width=45mm]{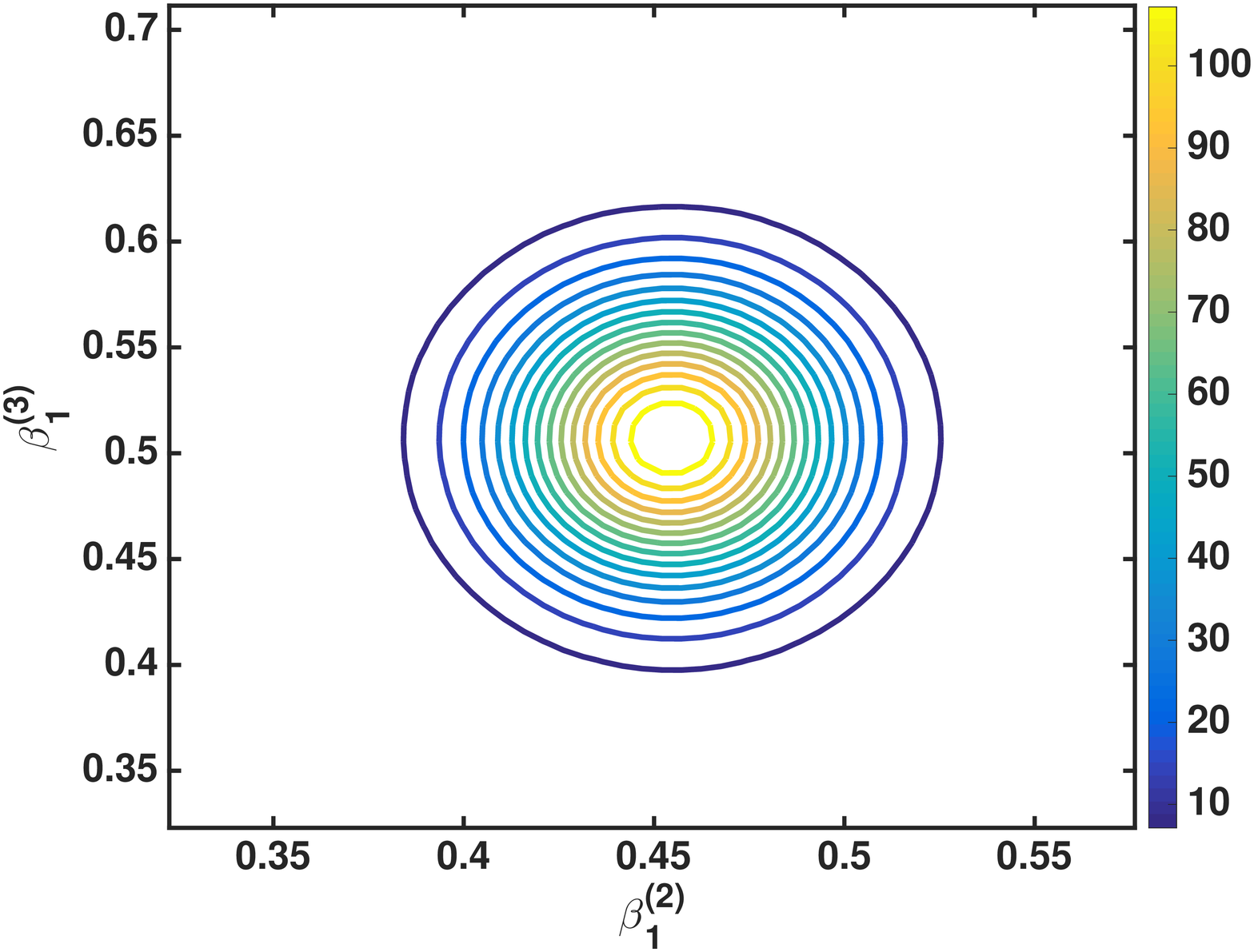}}
     {$$}
&
\subf{\includegraphics[width=45mm]{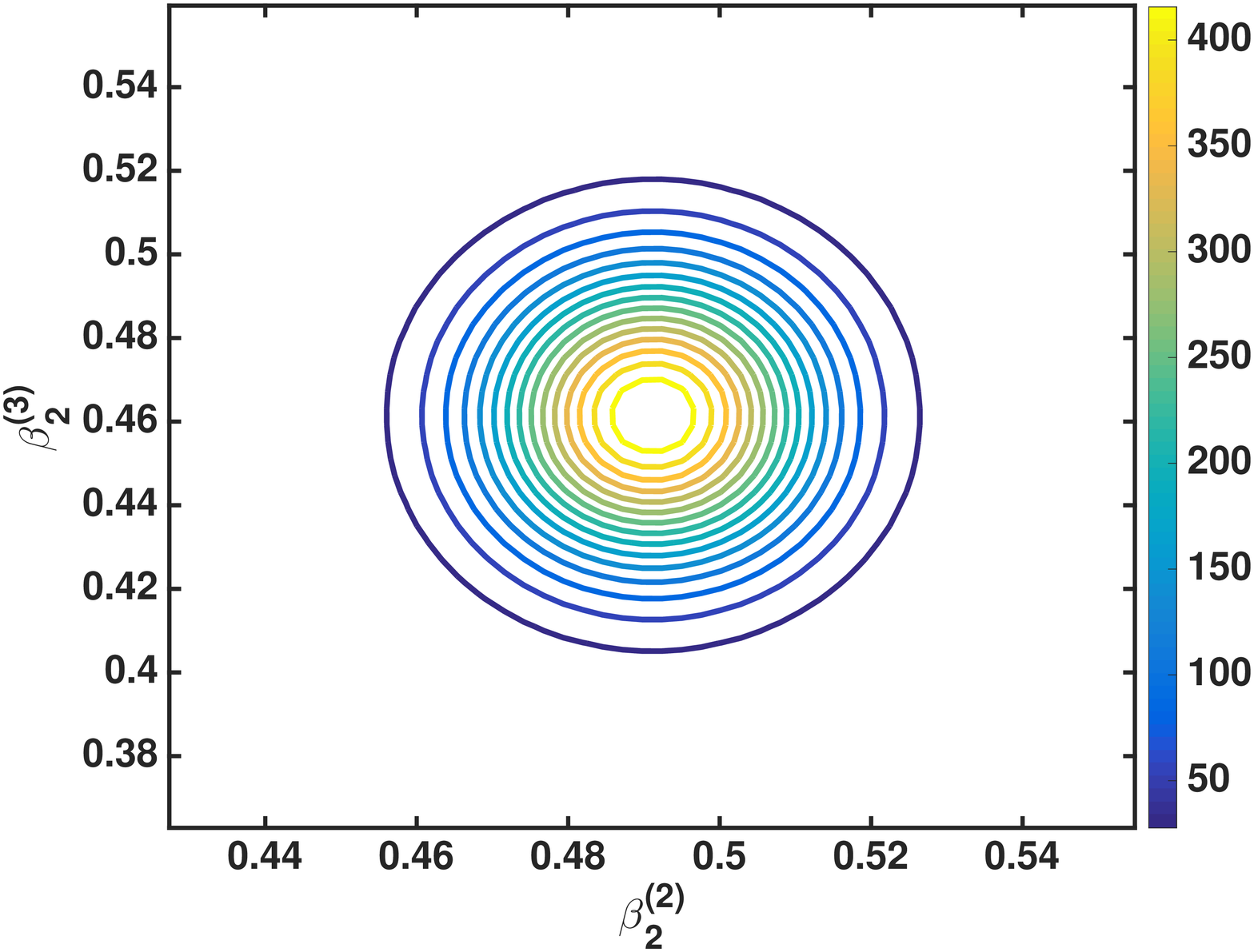}}
     {$$}
\\
\end{tabular}
\caption{Joint posterior pdfs of parameters $\omega_1, \beta_1, \omega_2$ and $\beta_2$ using the independent prior among three airspeeds}
\label{3u10cvjointpostindependent}
\end{figure}

\begin{figure}[ht!]
\centering
\begin{tabular}{ccc}
\subf{\includegraphics[width=45mm]{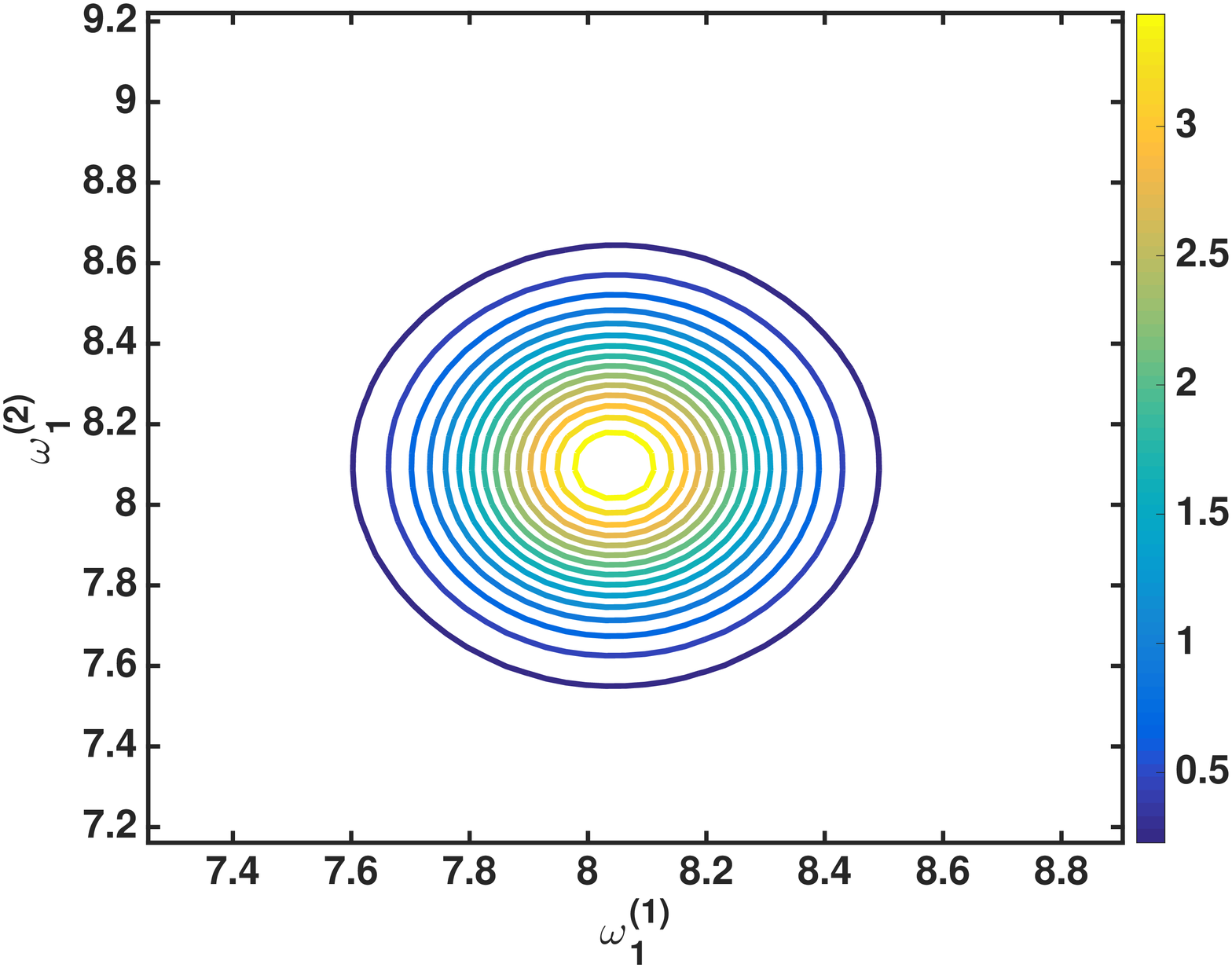}}
     {$$}
&
\subf{\includegraphics[width=45mm]{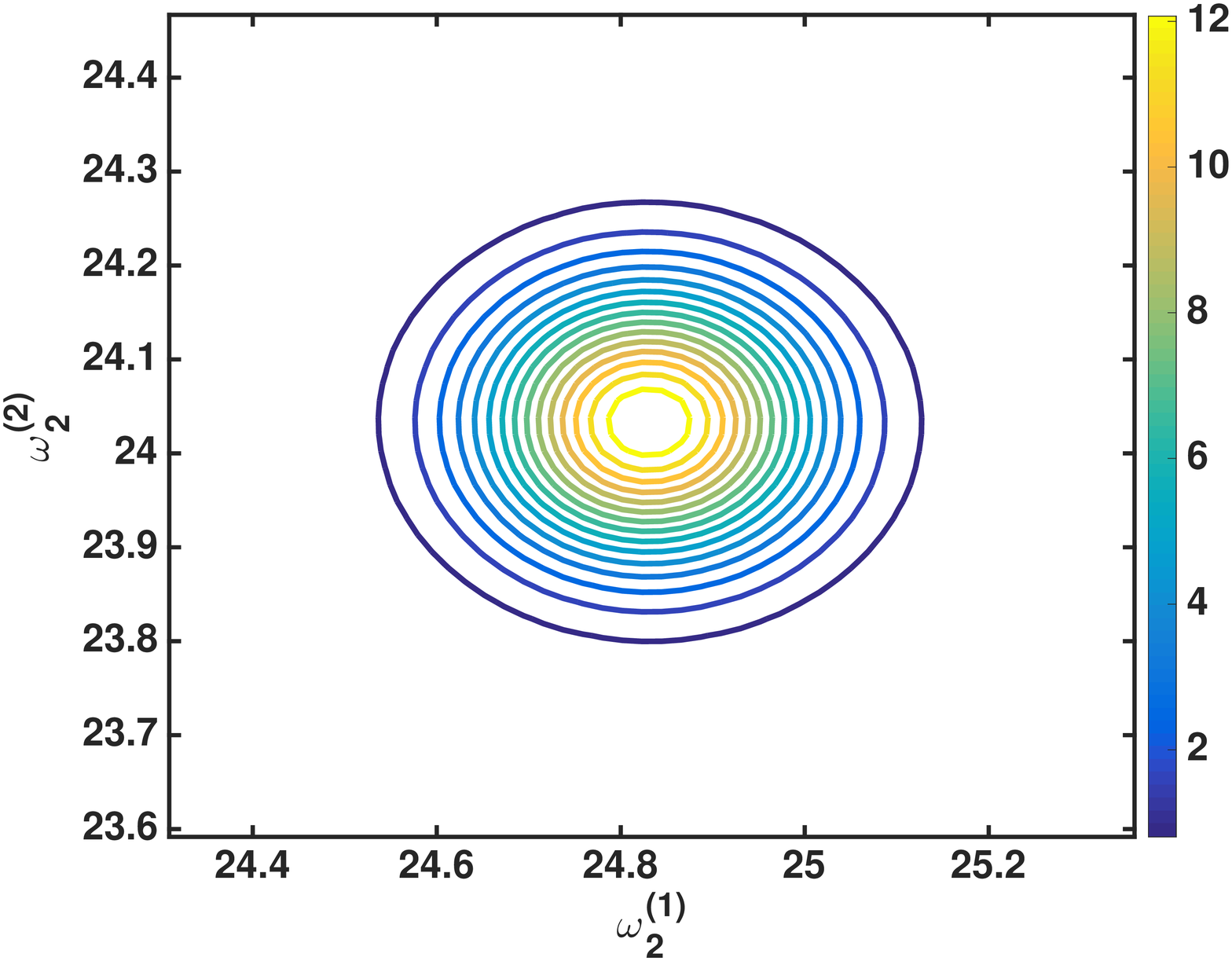}}
     {$$}
&
\subf{\includegraphics[width=45mm]{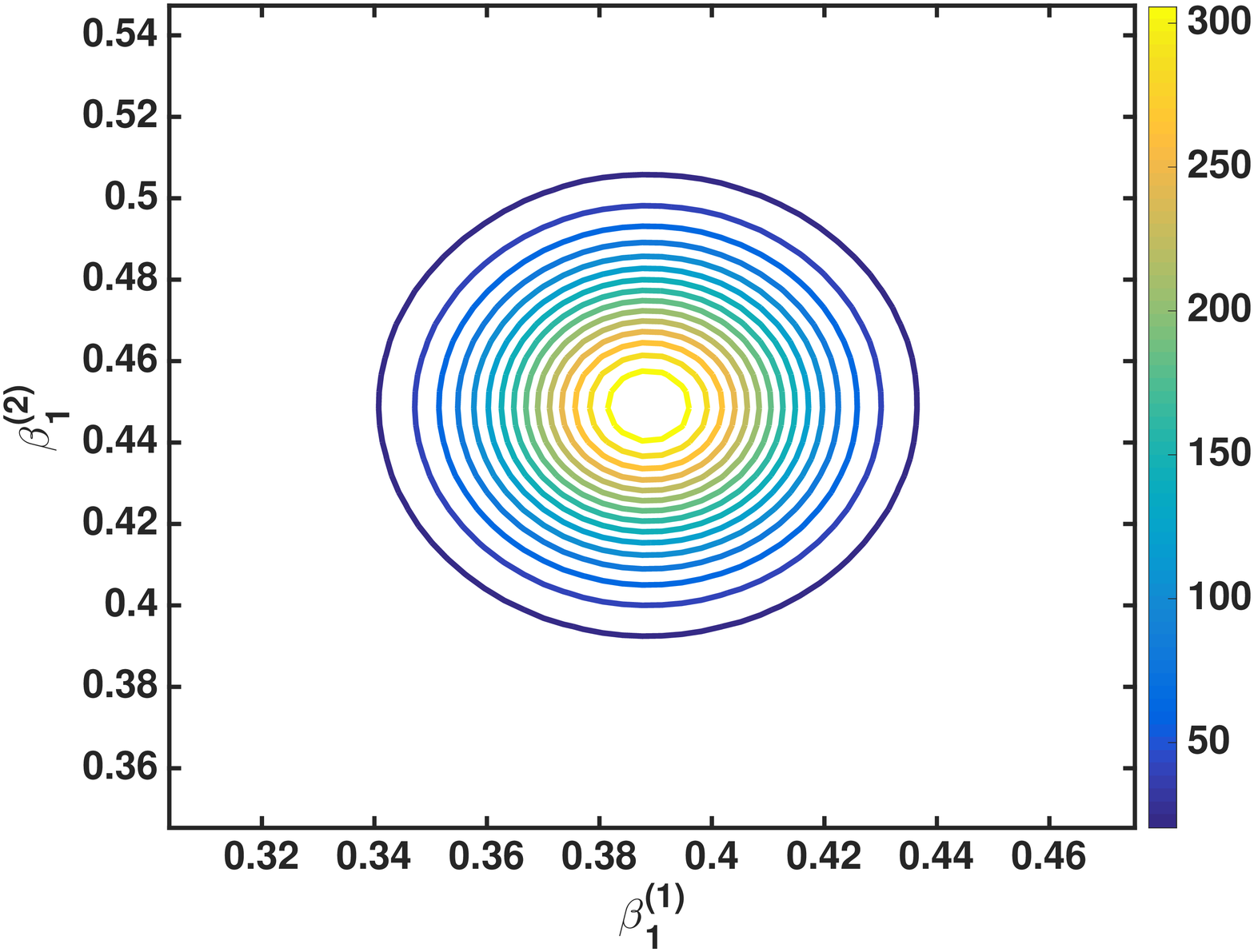}}
     {$$}
\\
\subf{\includegraphics[width=45mm]{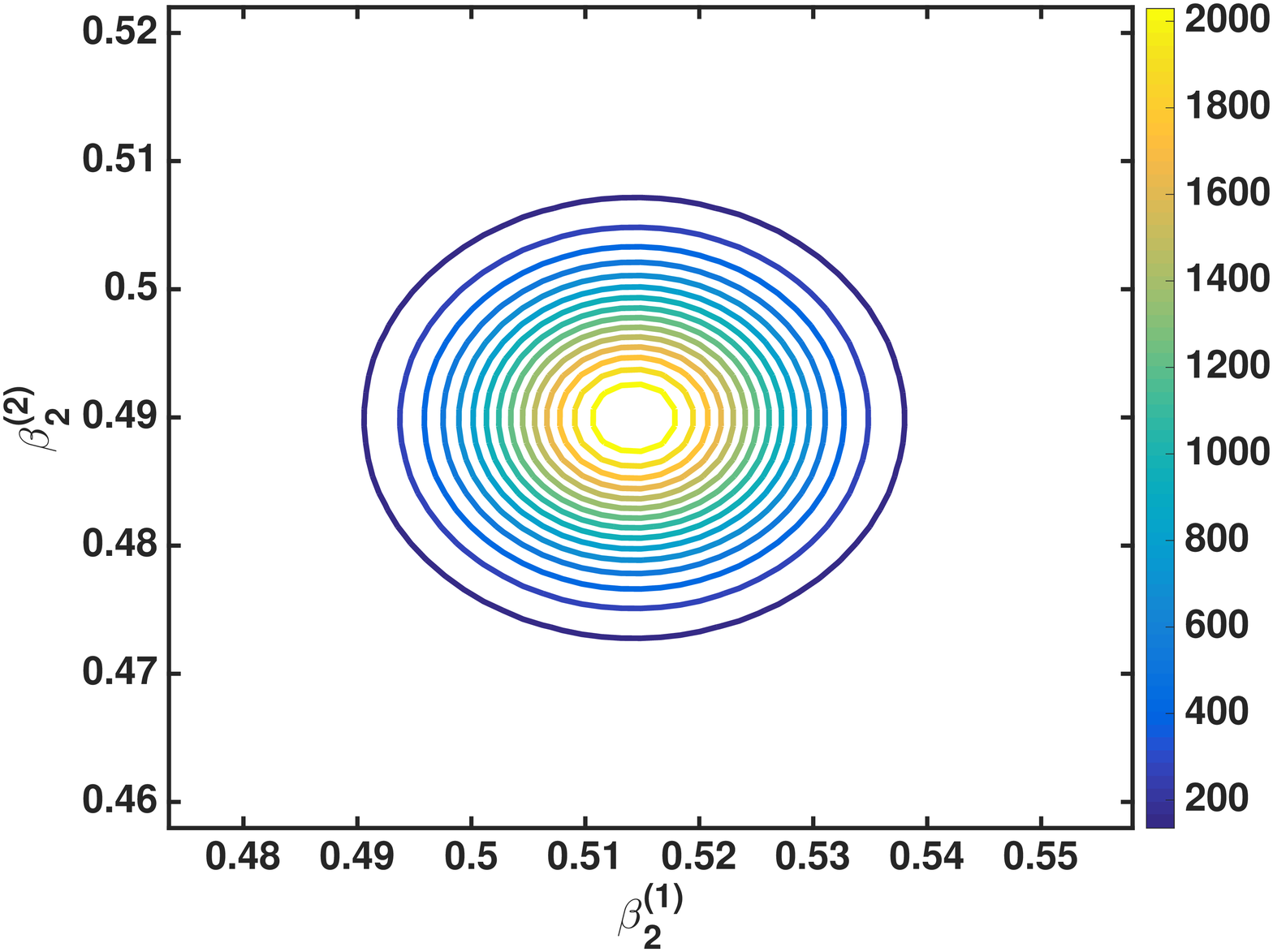}}
     {$$}
&
\subf{\includegraphics[width=45mm]{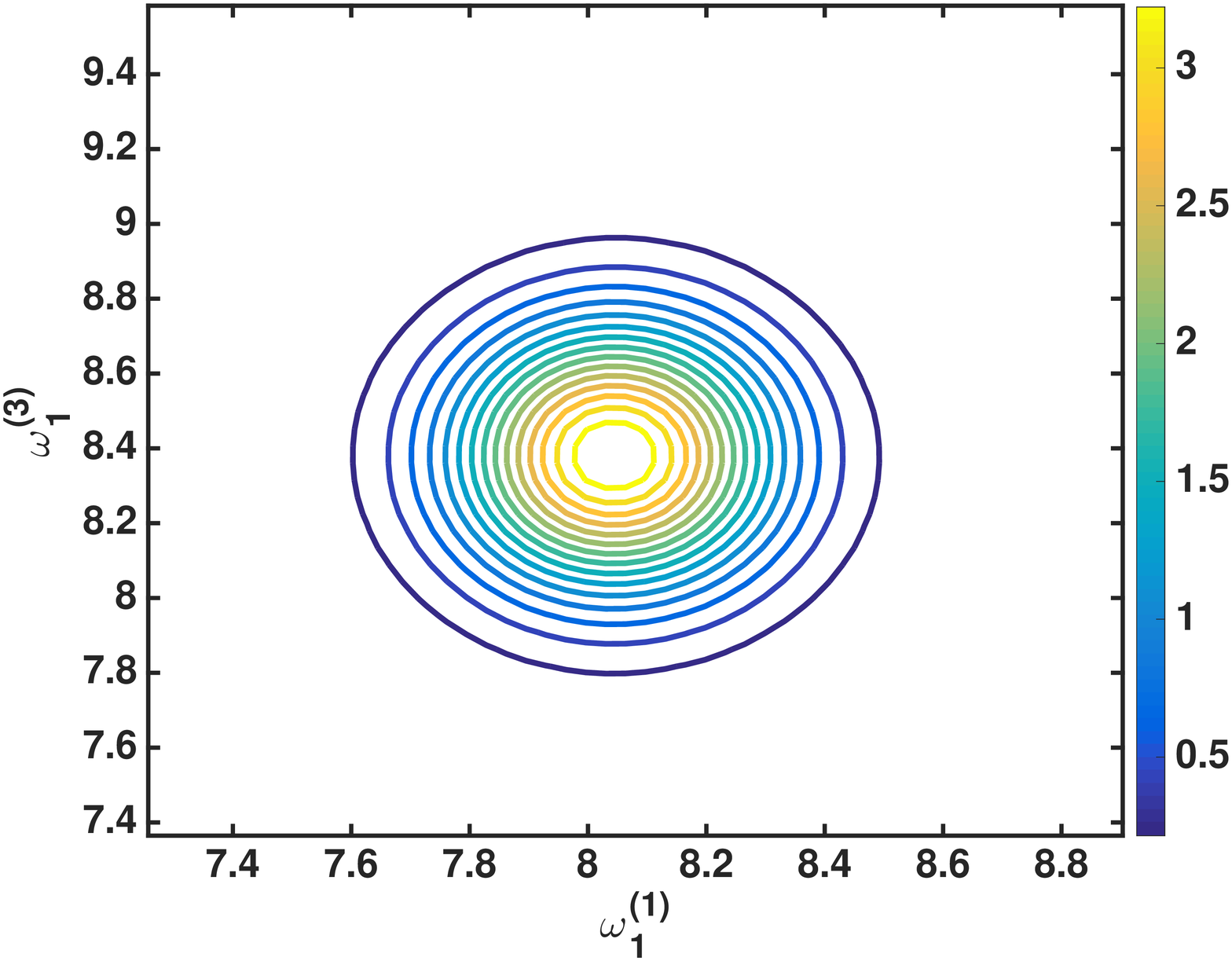}}
     {$$}
&
\subf{\includegraphics[width=45mm]{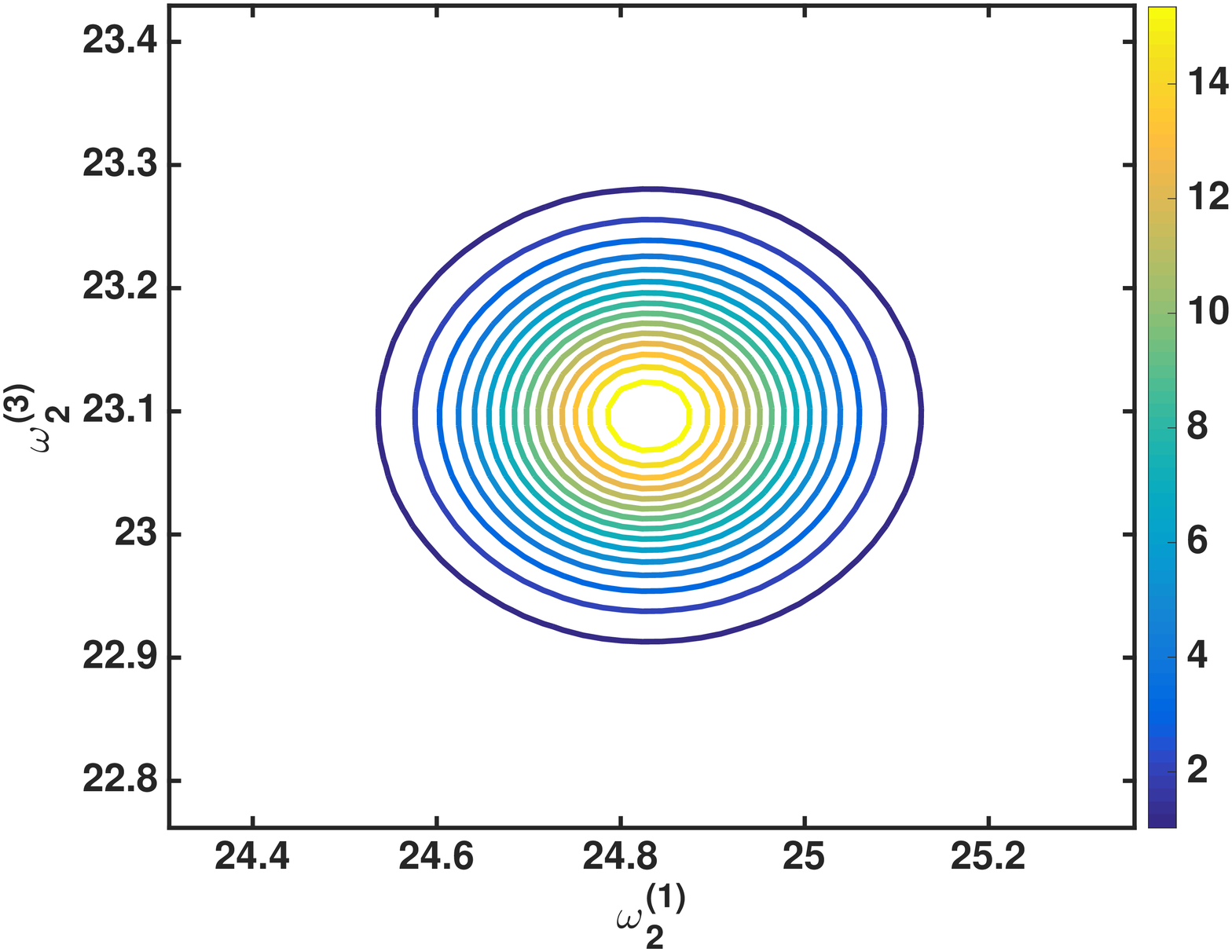}}
     {$$}
\\
\subf{\includegraphics[width=45mm]{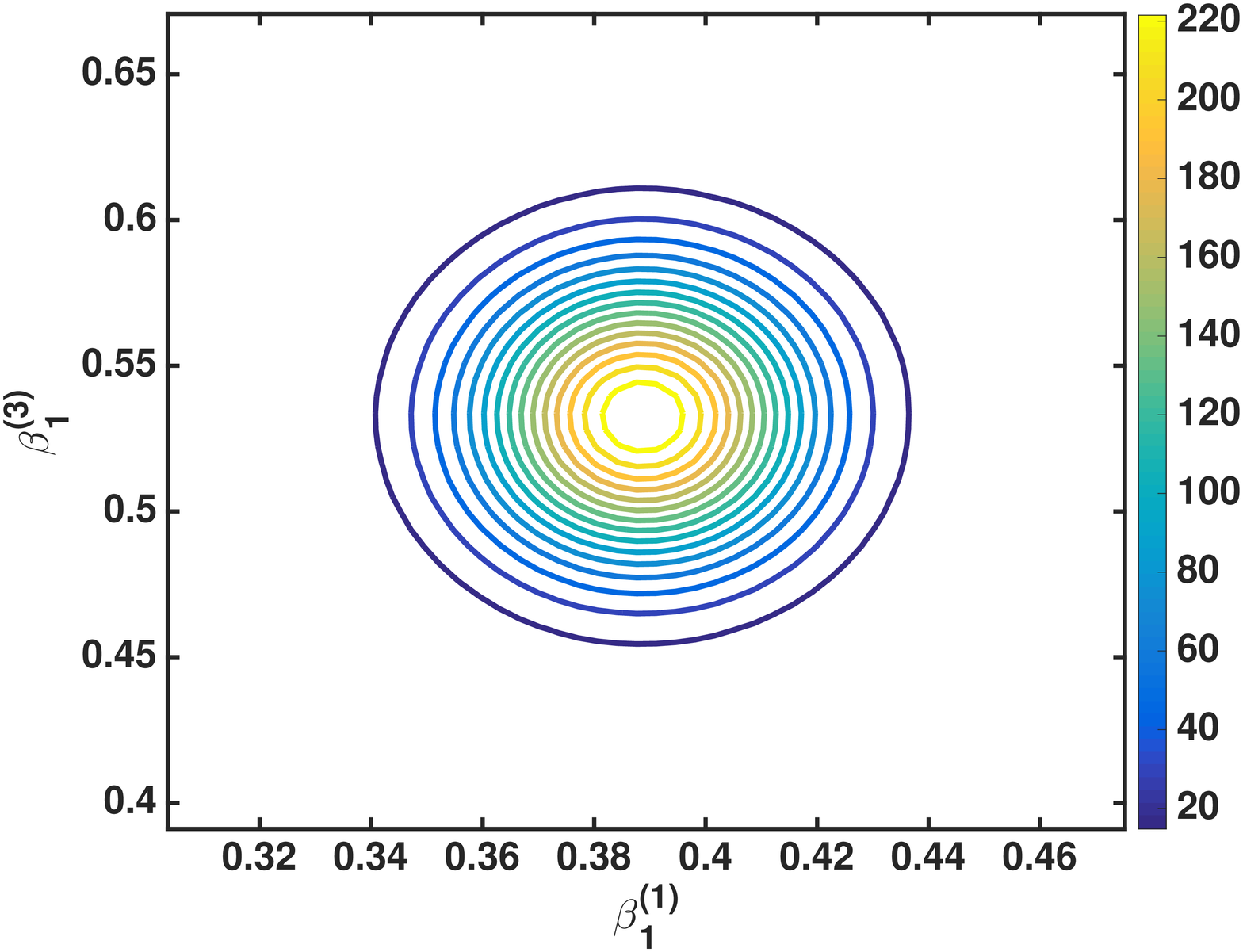}}
     {$$}
&
\subf{\includegraphics[width=45mm]{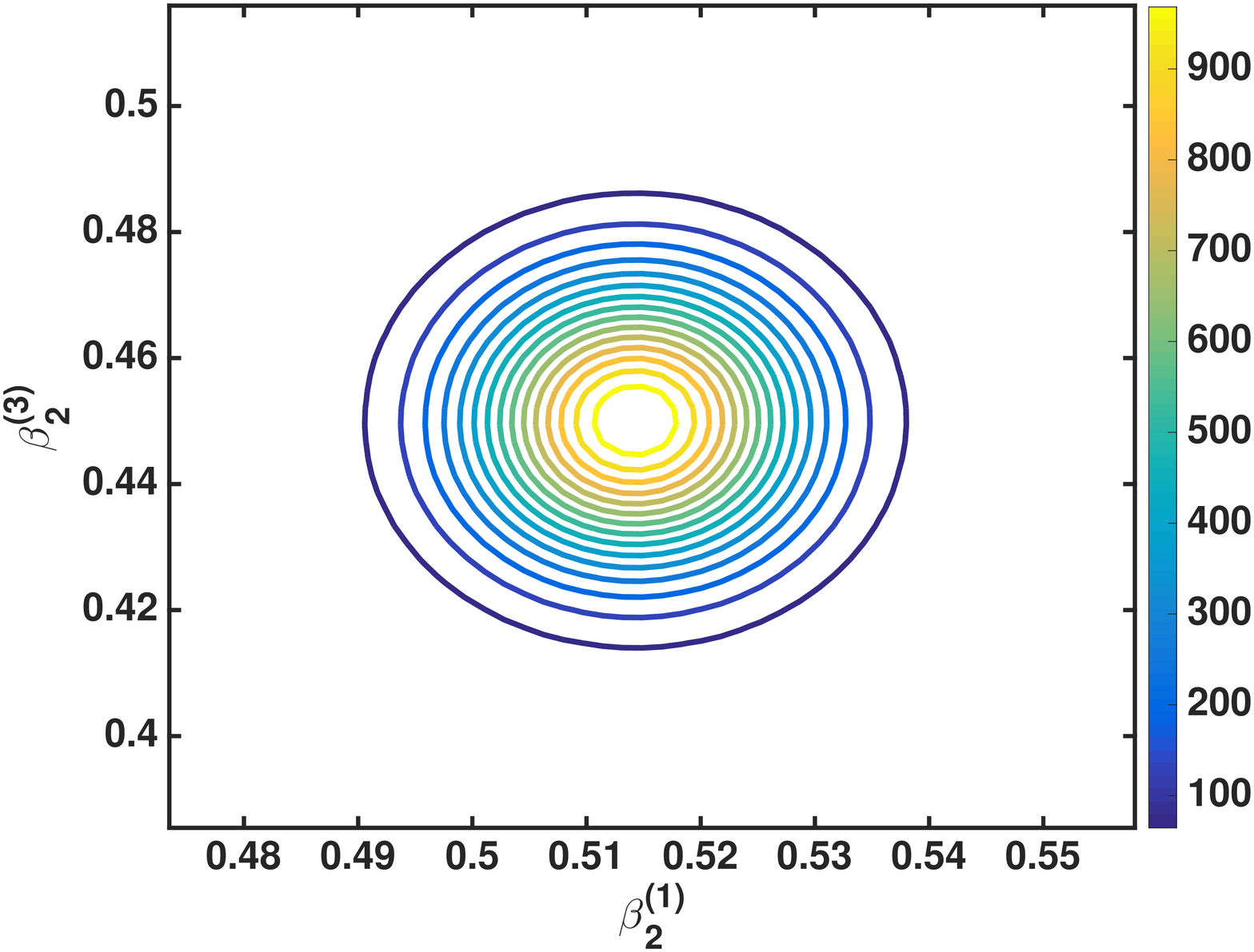}}
     {$$}
&
\subf{\includegraphics[width=45mm]{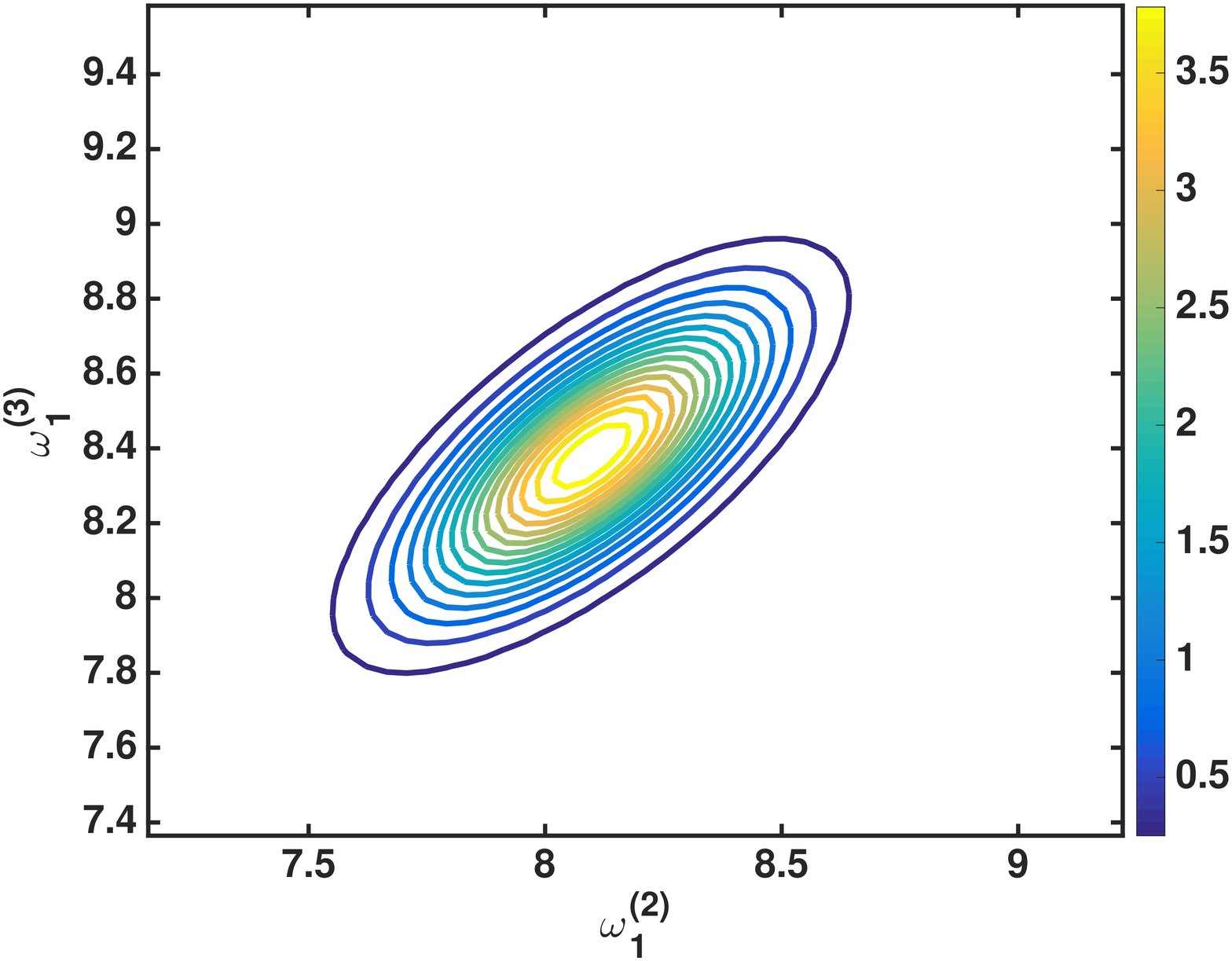}}
     {$$}
\\
\subf{\includegraphics[width=45mm]{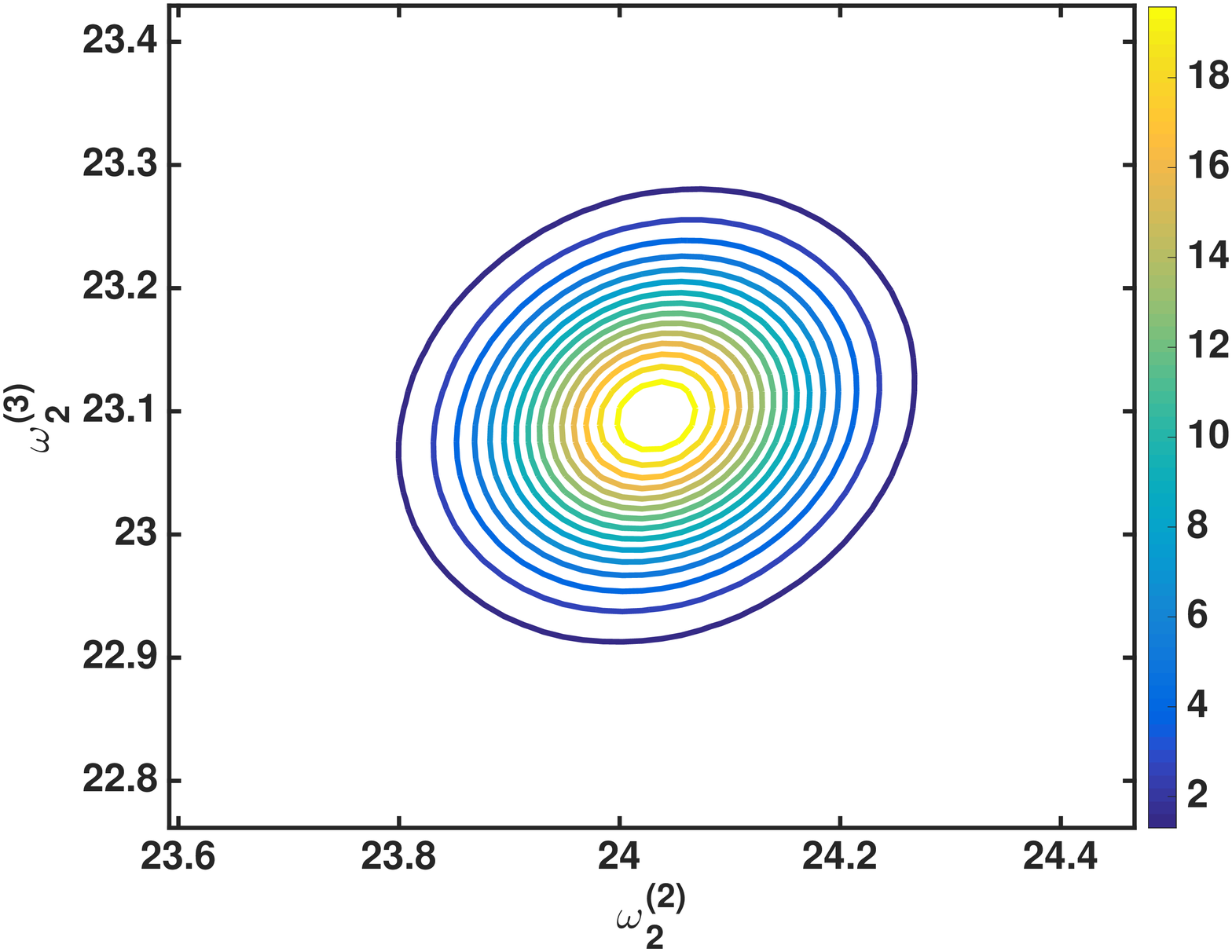}}
     {$$}
&
\subf{\includegraphics[width=45mm]{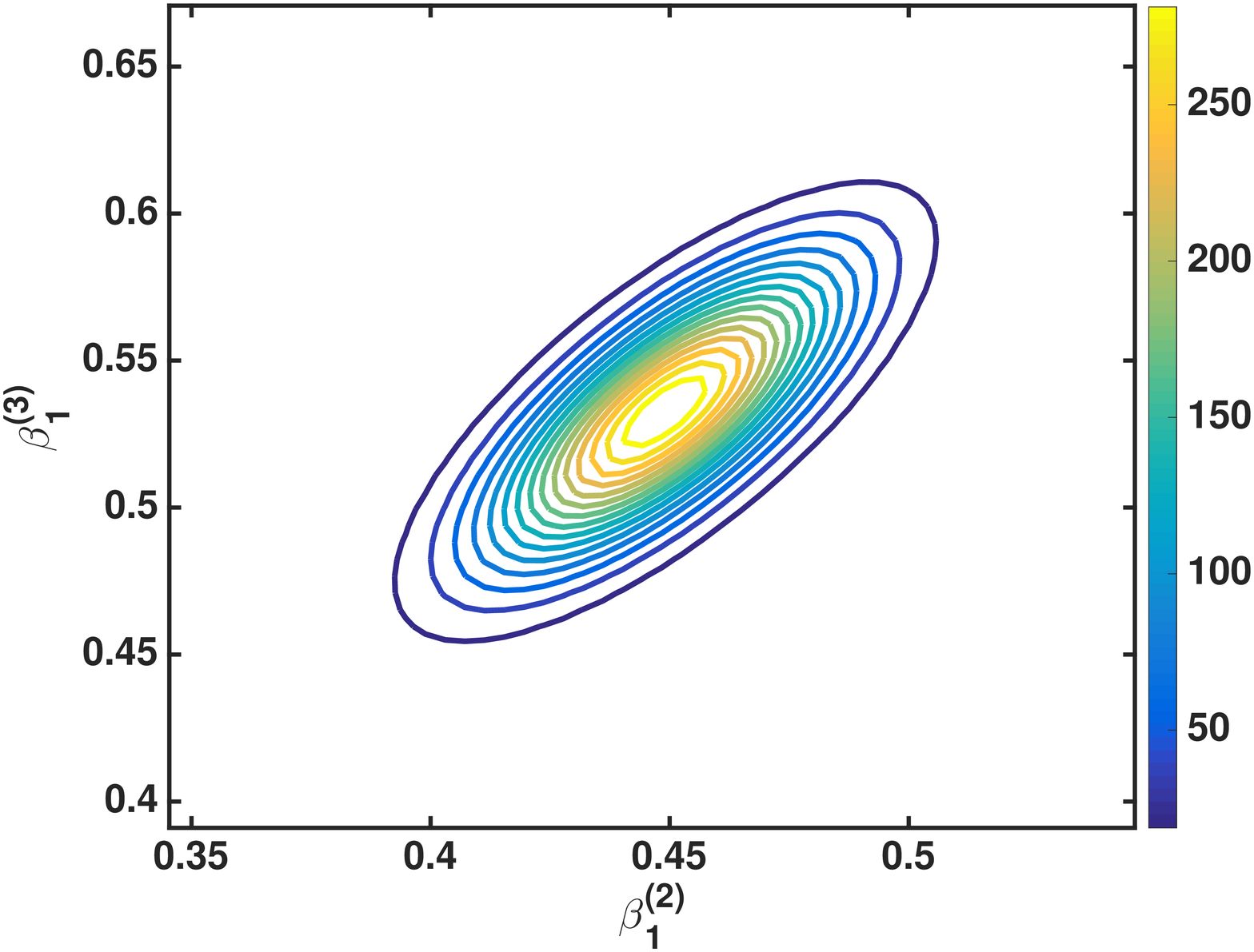}}
     {$$}
&
\subf{\includegraphics[width=45mm]{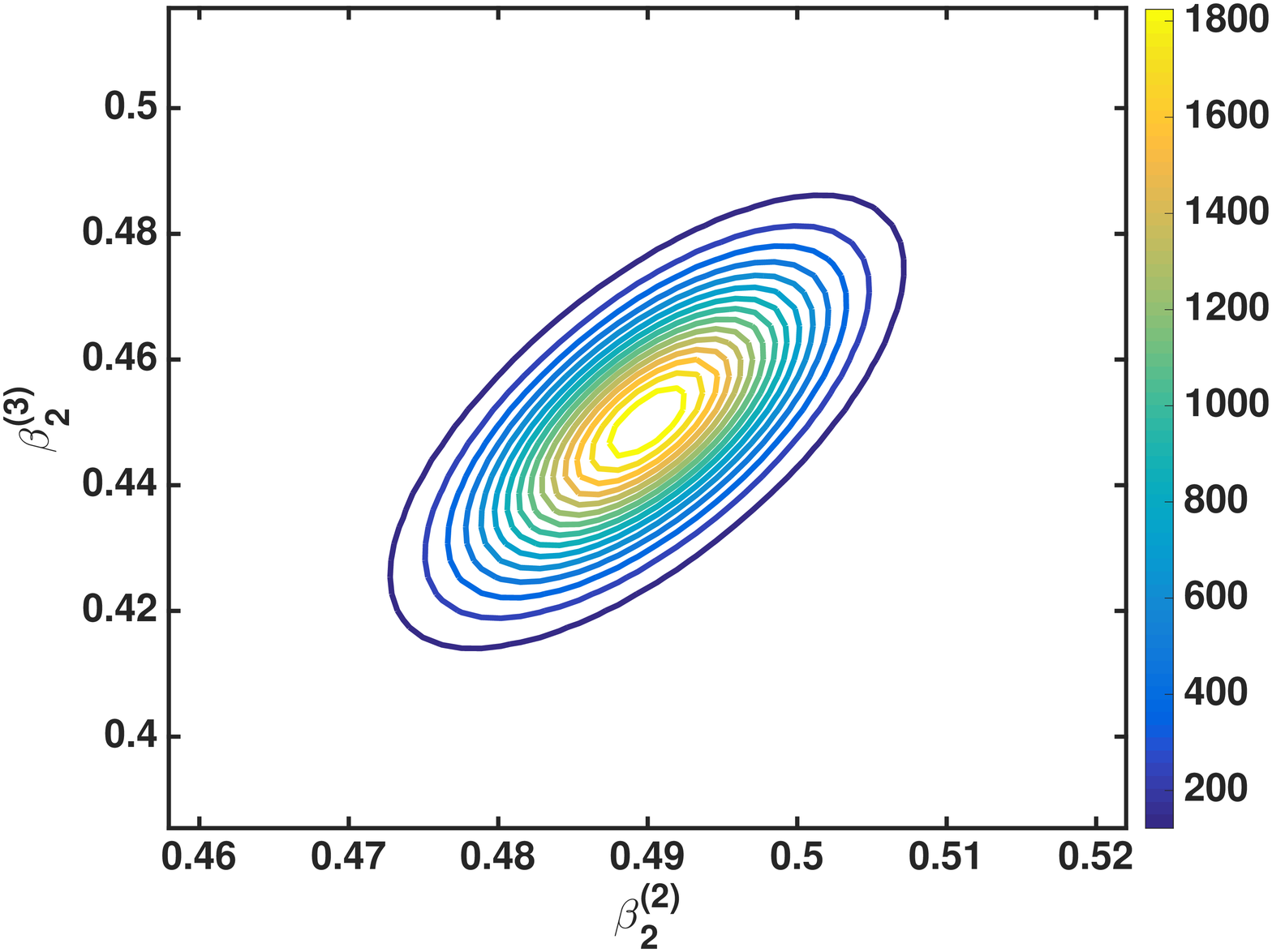}}
     {$$}
\\
\end{tabular}
\caption{Joint posterior pdfs of parameters $\omega_1, \beta_1, \omega_2$ and $\beta_2$ using joint prior among three airspeeds}
\label{3u10cvjointpostjoint}
\end{figure}

\begin{figure}[ht!]
\centering
\begin{tabular}{ccc}
\subf{\includegraphics[width=45mm]{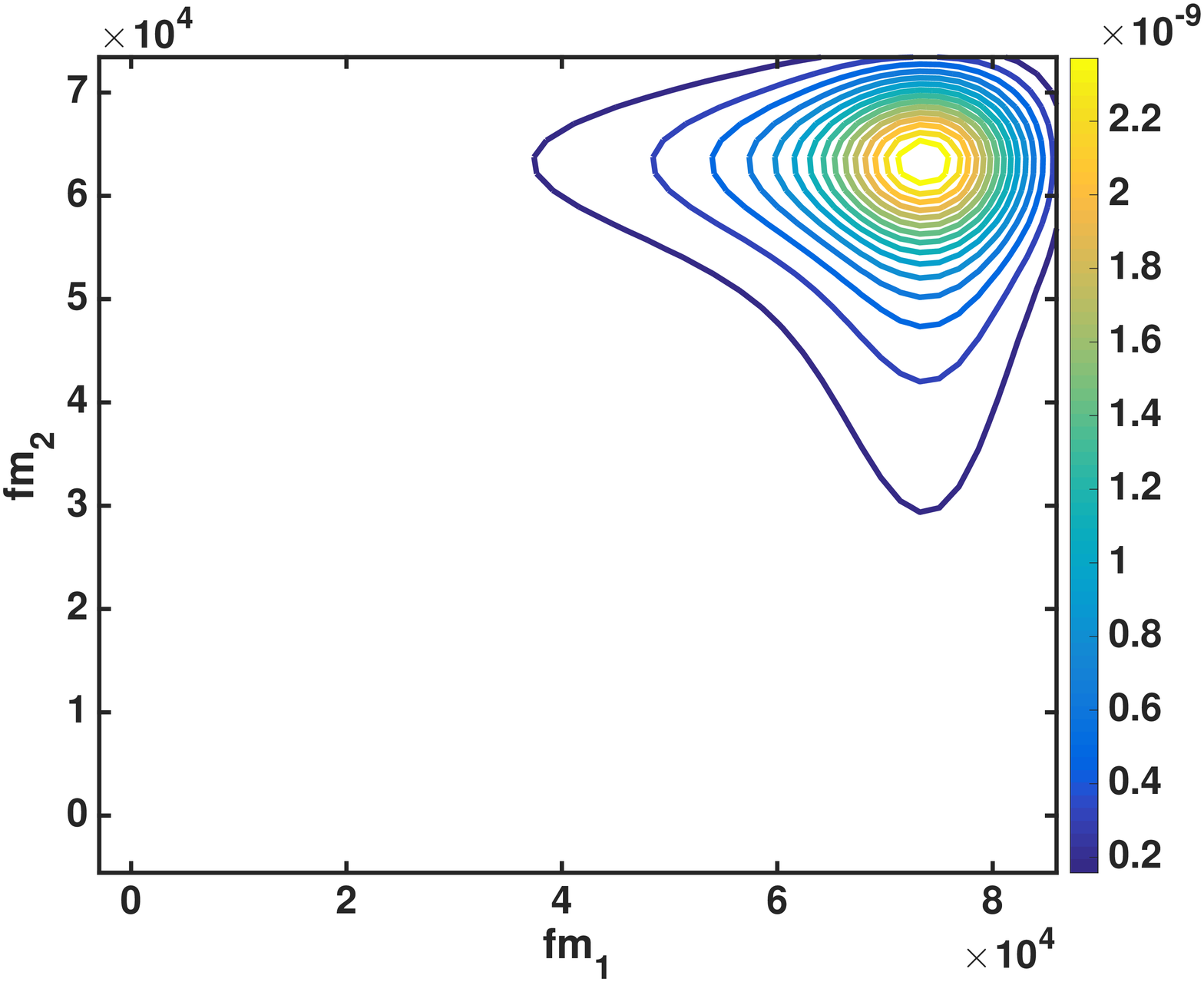}}
     {$$}
&
\subf{\includegraphics[width=45mm]{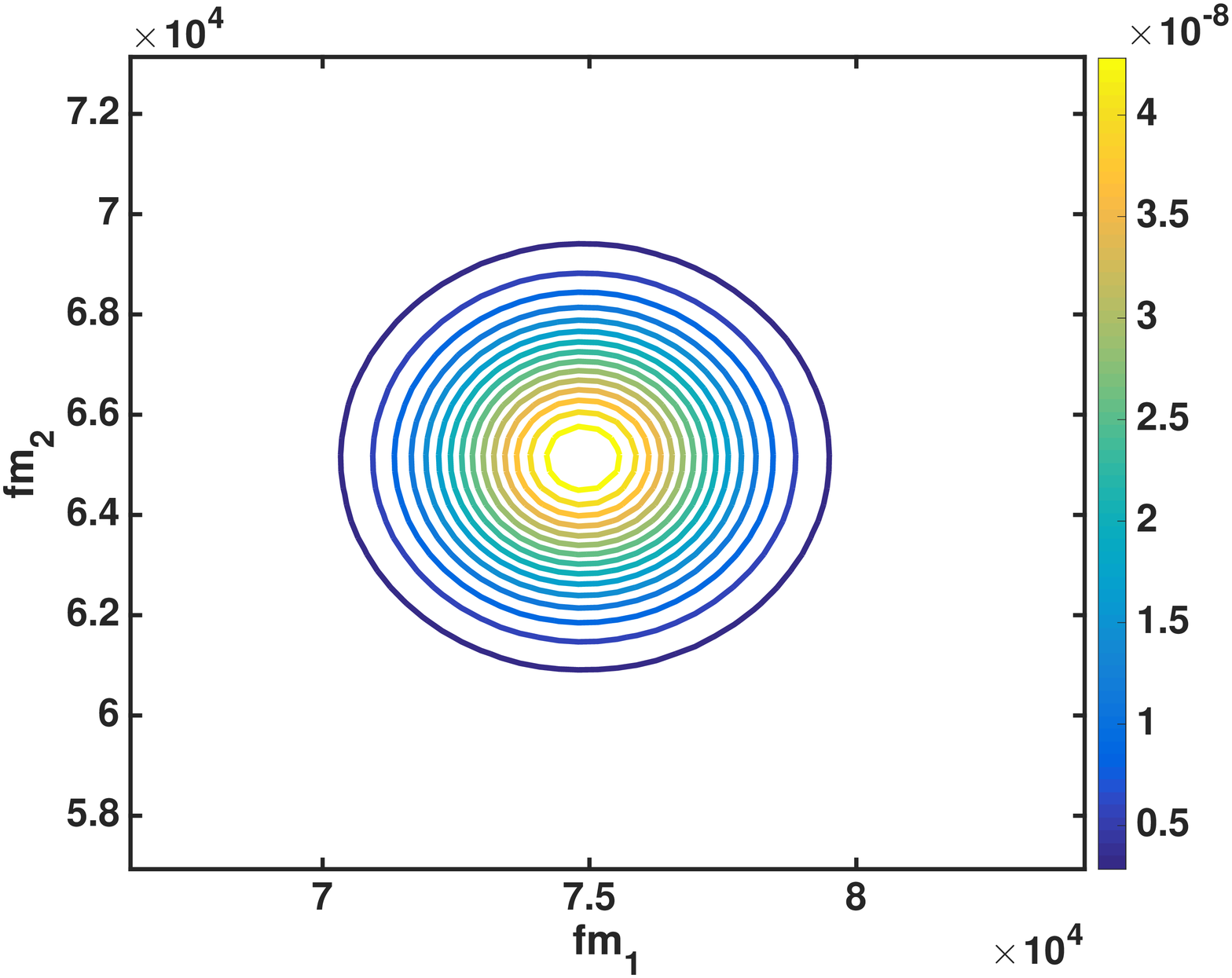}}
     {$$}
&
\subf{\includegraphics[width=45mm]{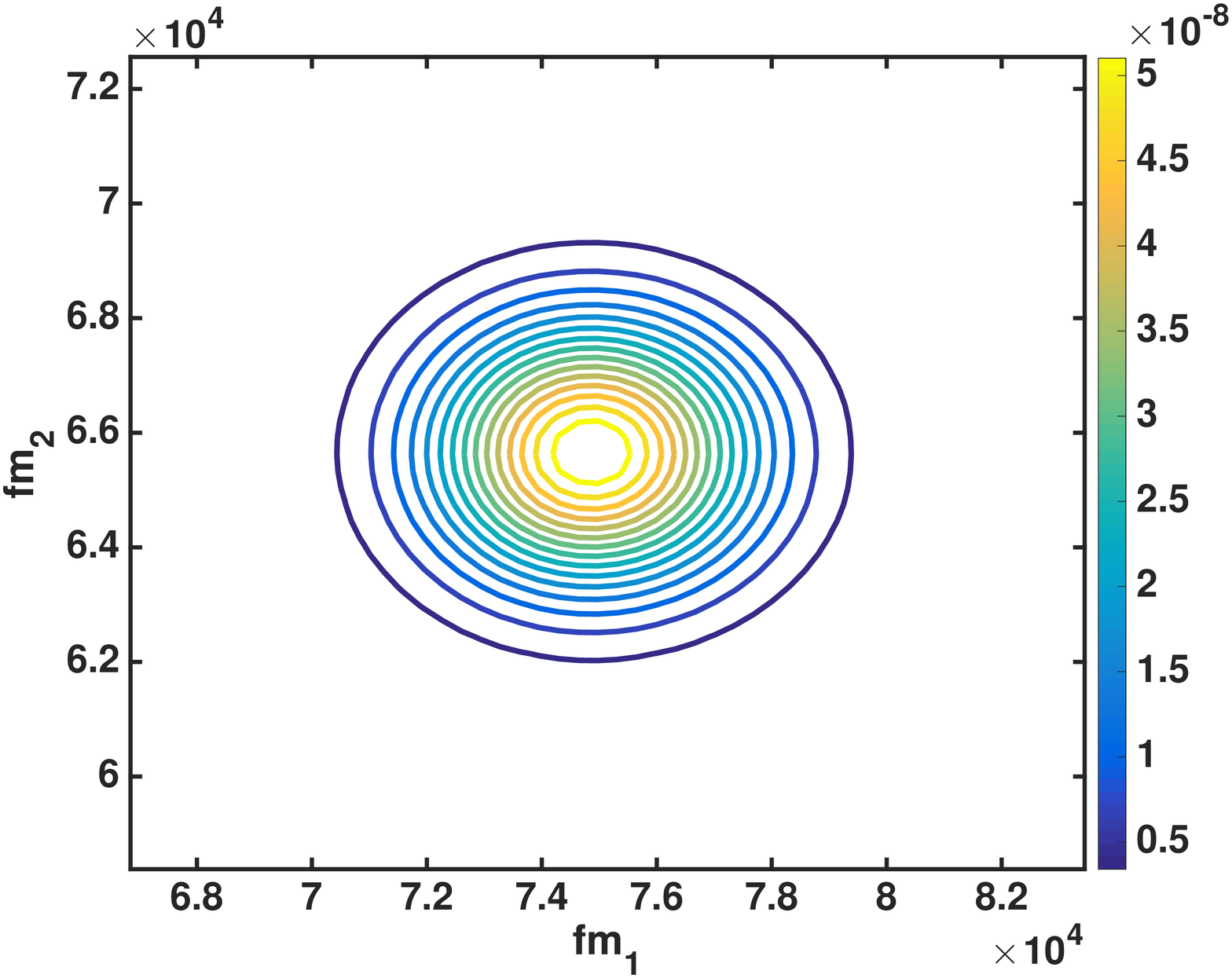}}
     {$$}
\\
\subf{\includegraphics[width=45mm]{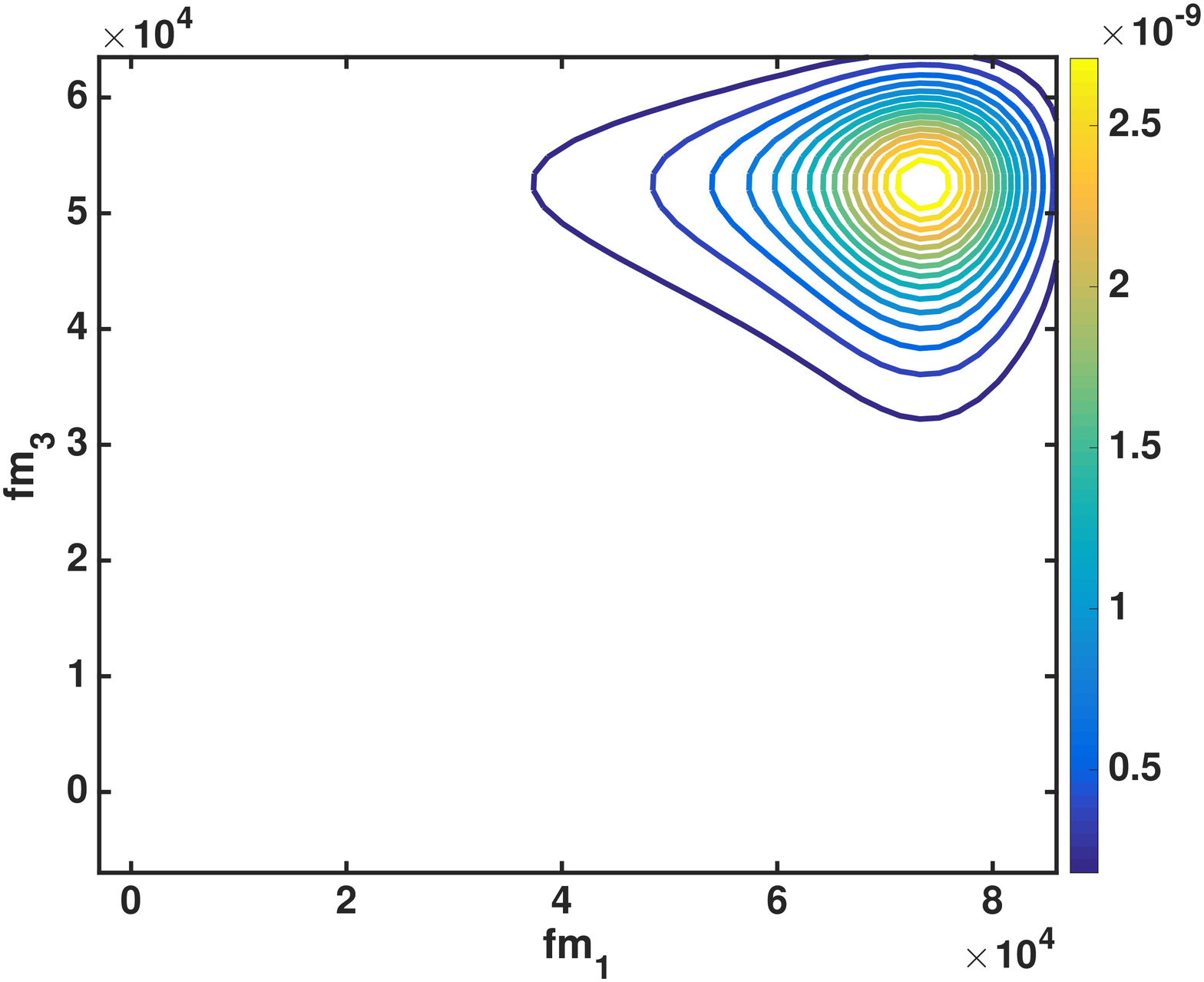}}
     {$$}
&
\subf{\includegraphics[width=45mm]{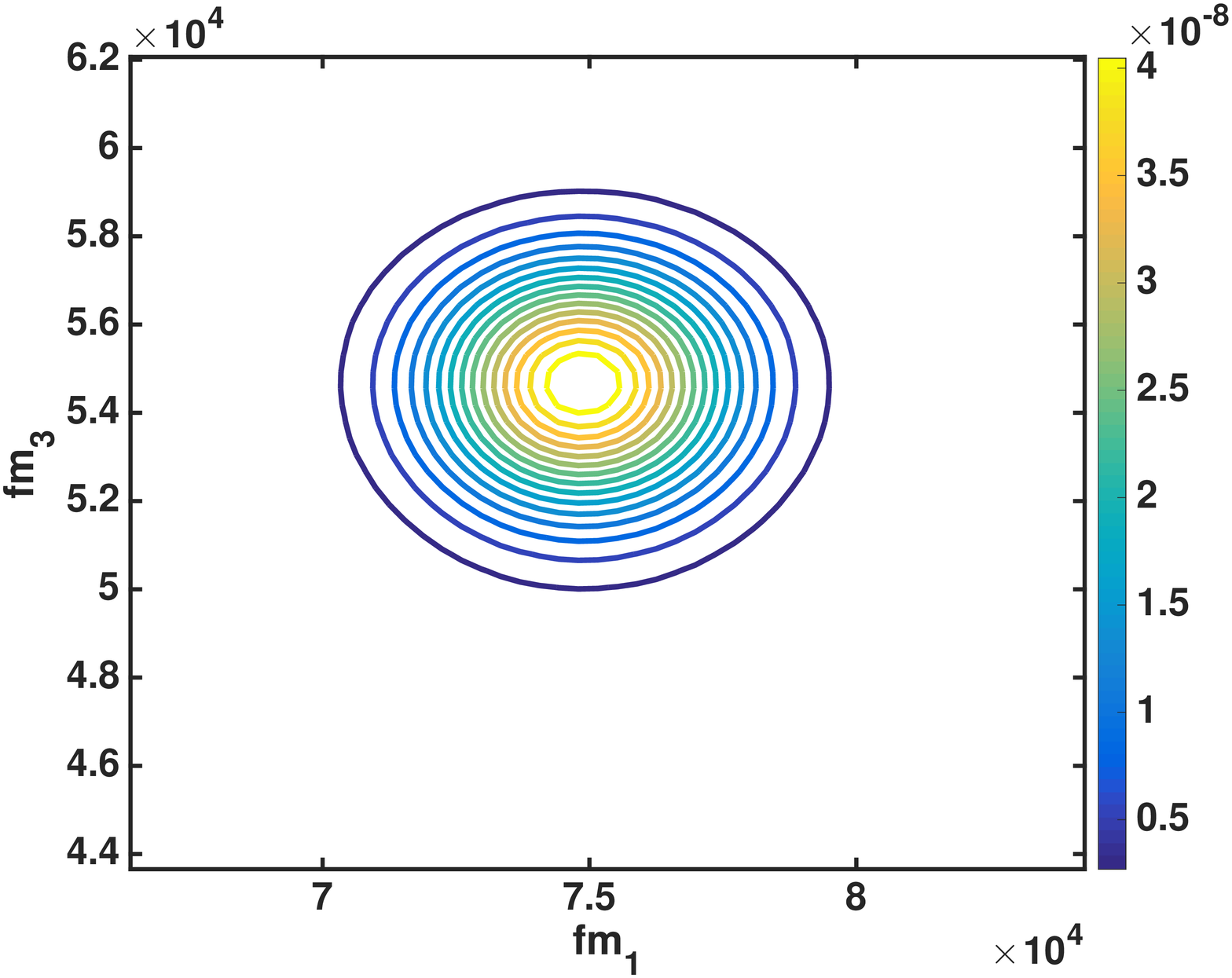}}
     {$$}
&
\subf{\includegraphics[width=45mm]{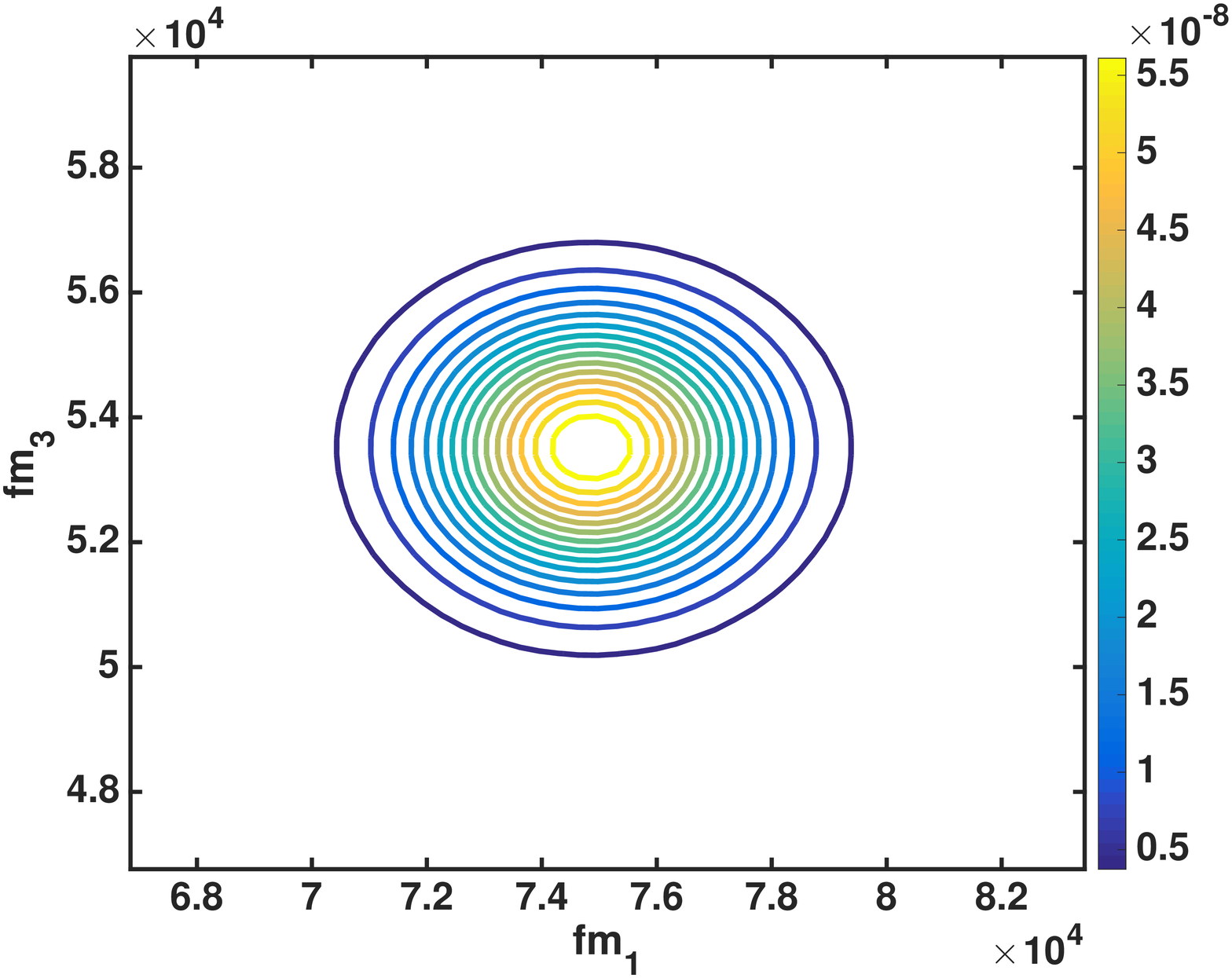}}
     {$$}
\\
\subf{\includegraphics[width=45mm]{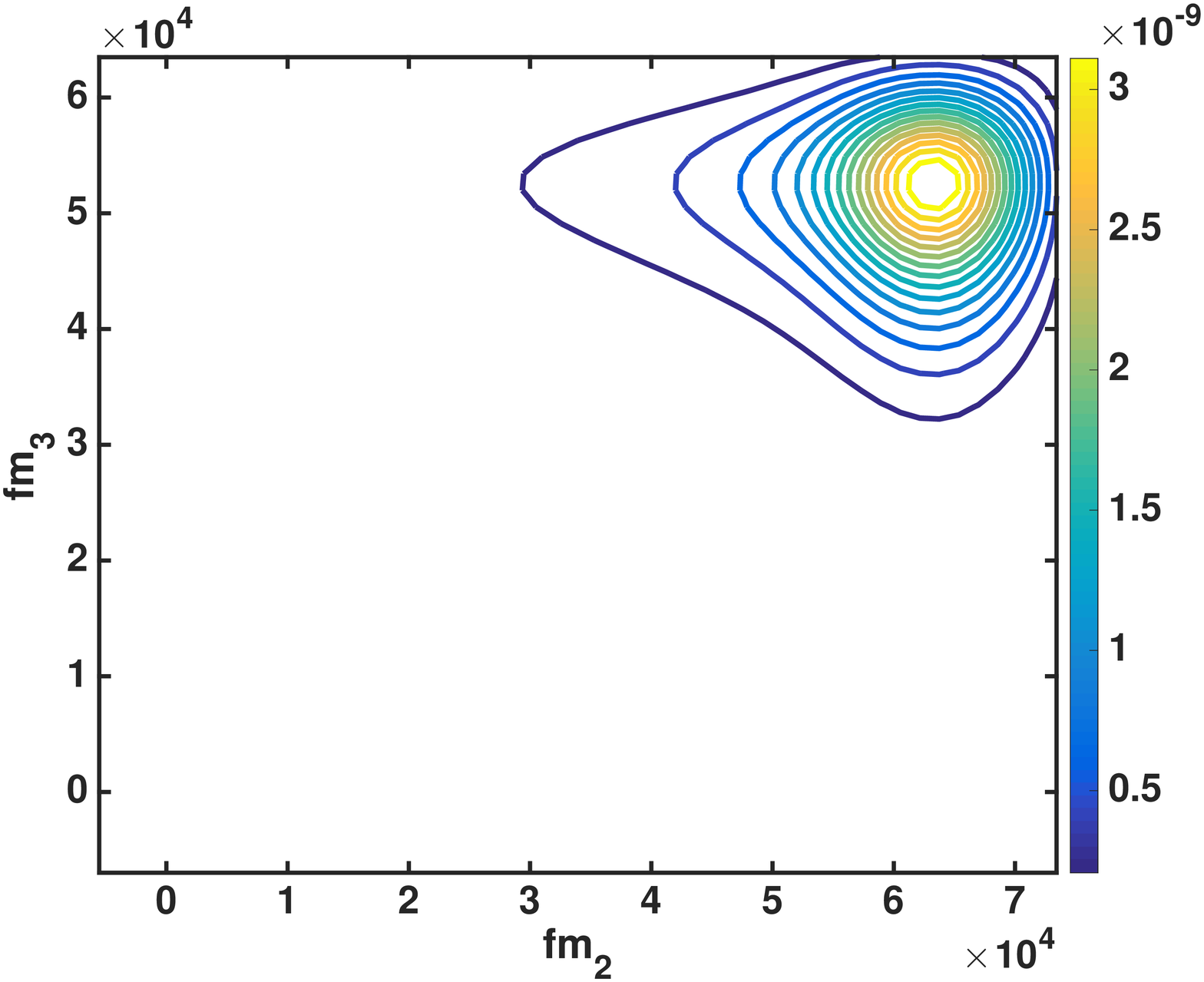}}
     {$$}
&
\subf{\includegraphics[width=45mm]{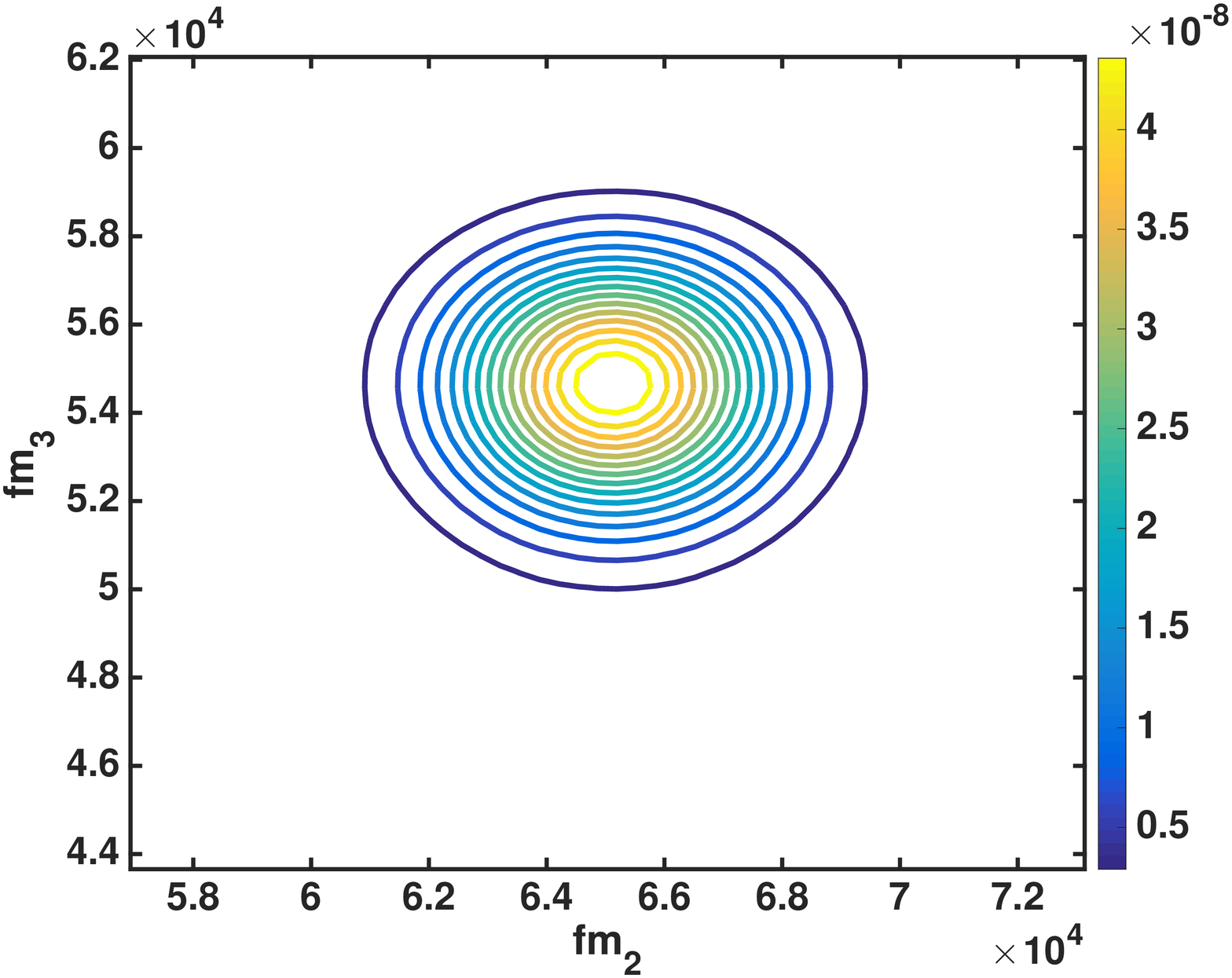}}
     {$$}
&
\subf{\includegraphics[width=45mm]{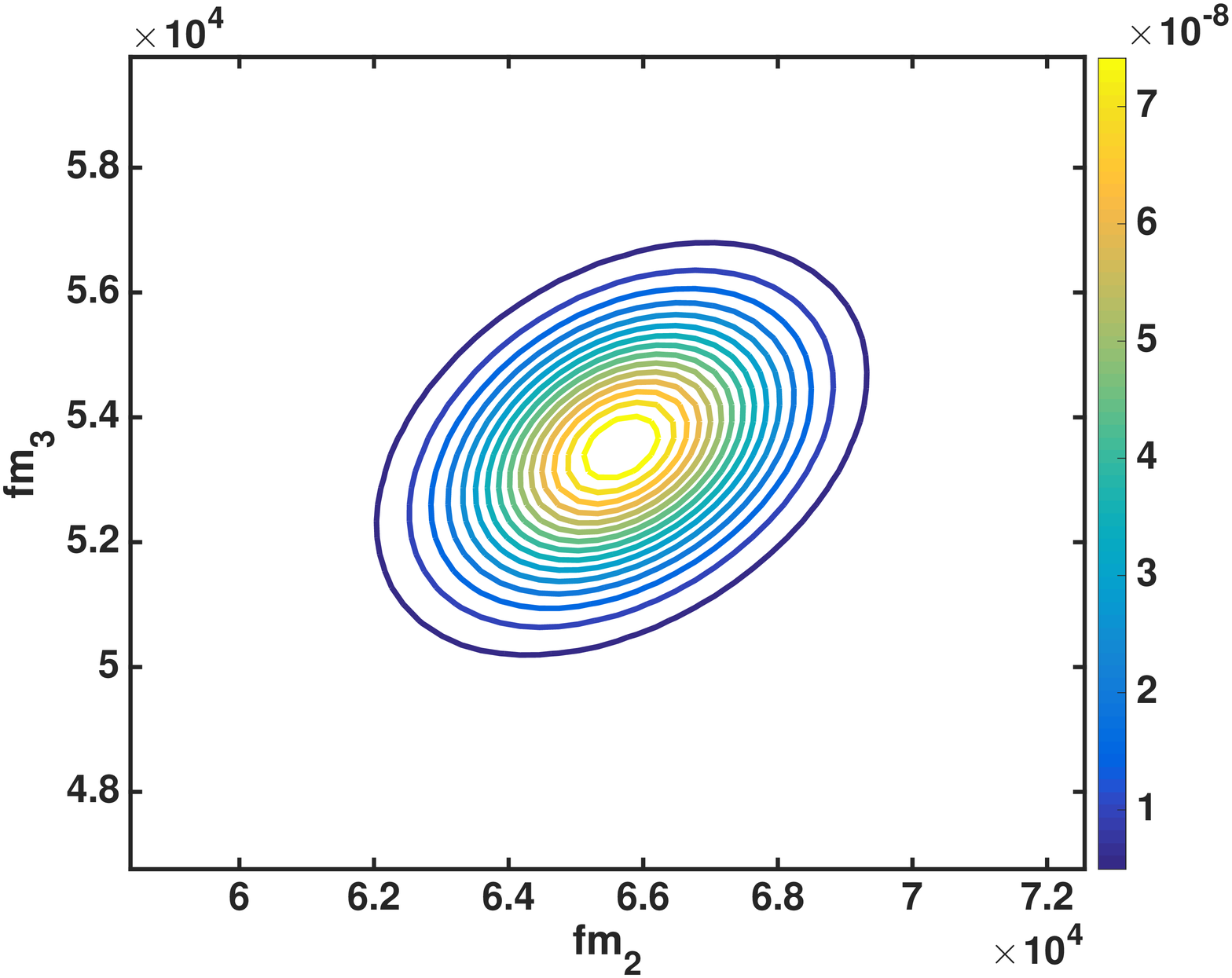}}
     {$$}
\\
\end{tabular}
\caption{Joint posterior pdfs of flutter margins using the flat prior: column 1, 
independent prior: column 2 and joint prior: column 3, among three airspeeds}
\label{3u10cvjointpostfm}
\end{figure}

\begin{figure}[ht!]
\centering
\begin{tabular}{ccc}
\subf{\includegraphics[width=35mm]{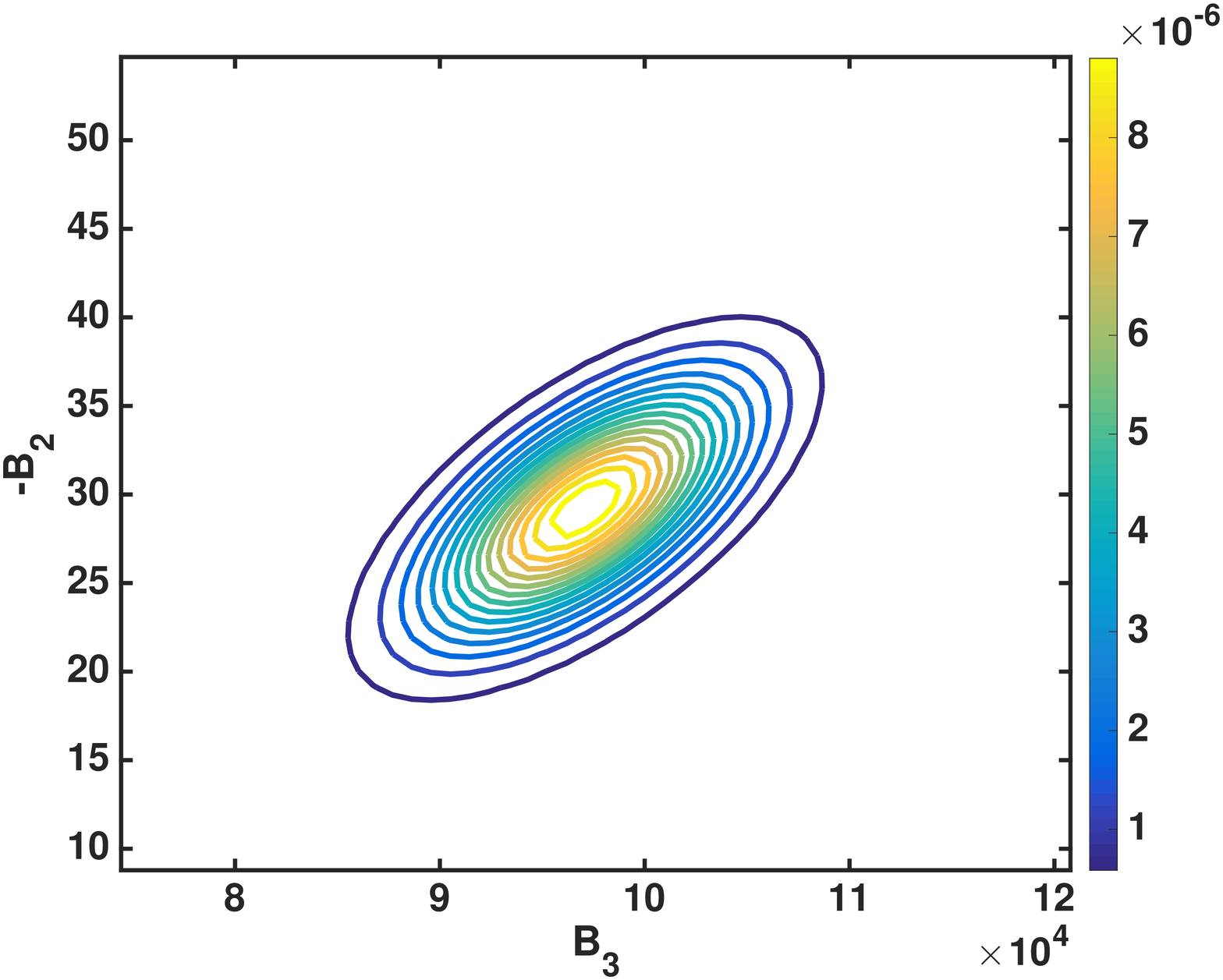}}
     {$$}
&
\subf{\includegraphics[width=35mm]{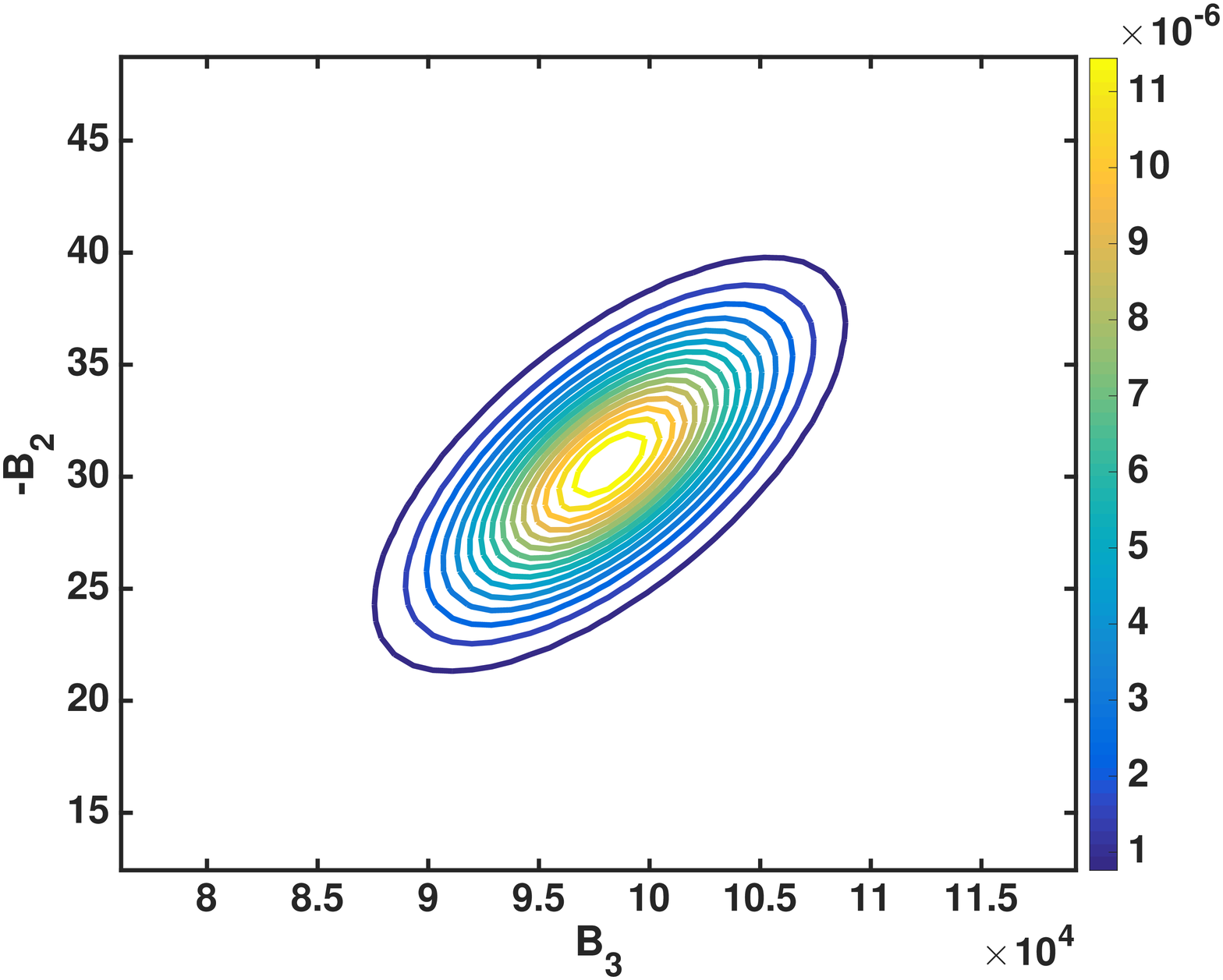}}
     {$$}
&
\subf{\includegraphics[width=35mm]{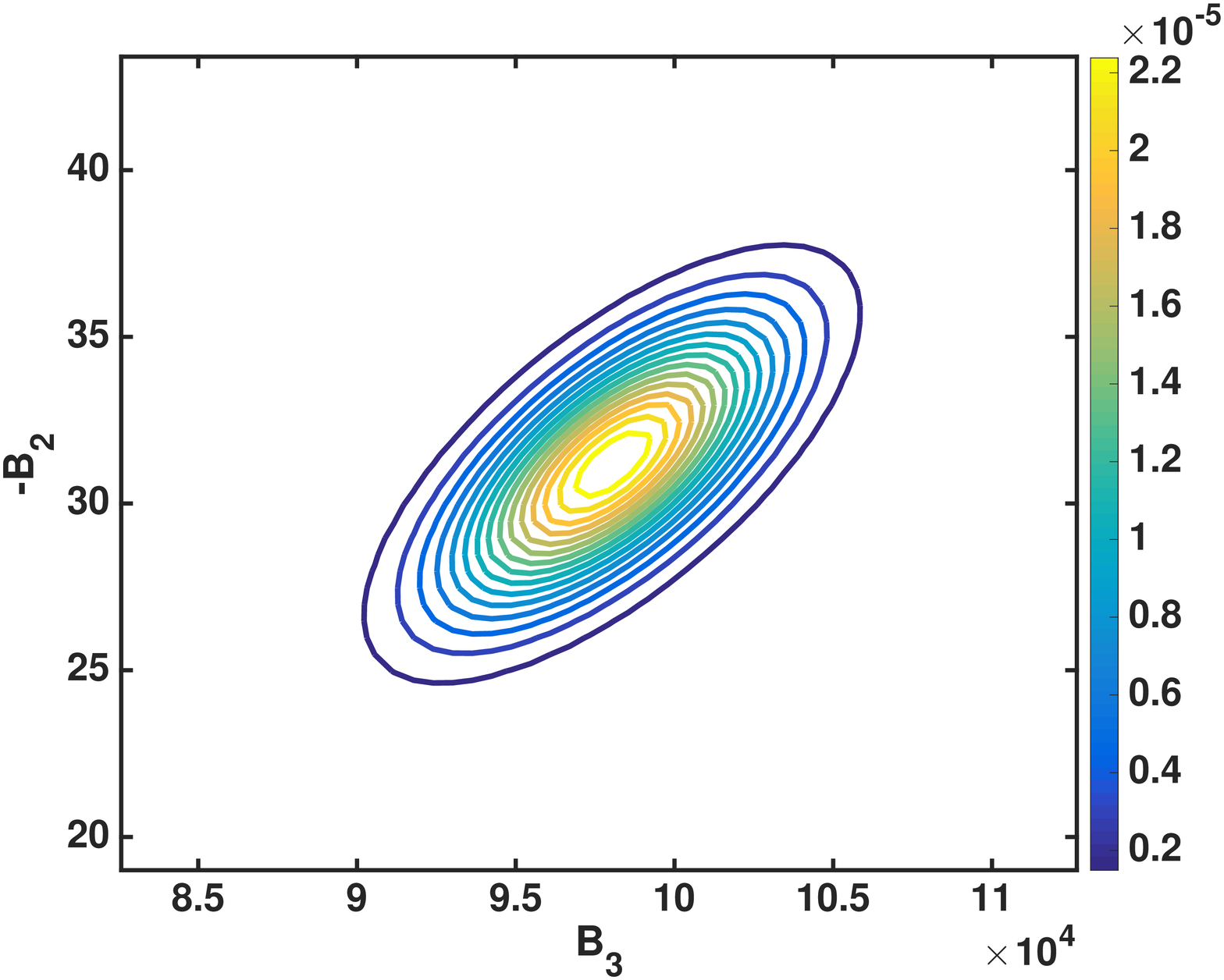}}
     {$$}

\\
\end{tabular}
\caption{Joint posterior pdf  $B_2$ and $B_3$ parameters: 1) the flat prior, 2) independent prior and 3) joint prior}
\label{3u10cvjointpostB1B2}
\end{figure}

\begin{figure}[htbp]
\begin{center}
\includegraphics[width=120mm,height=80mm]{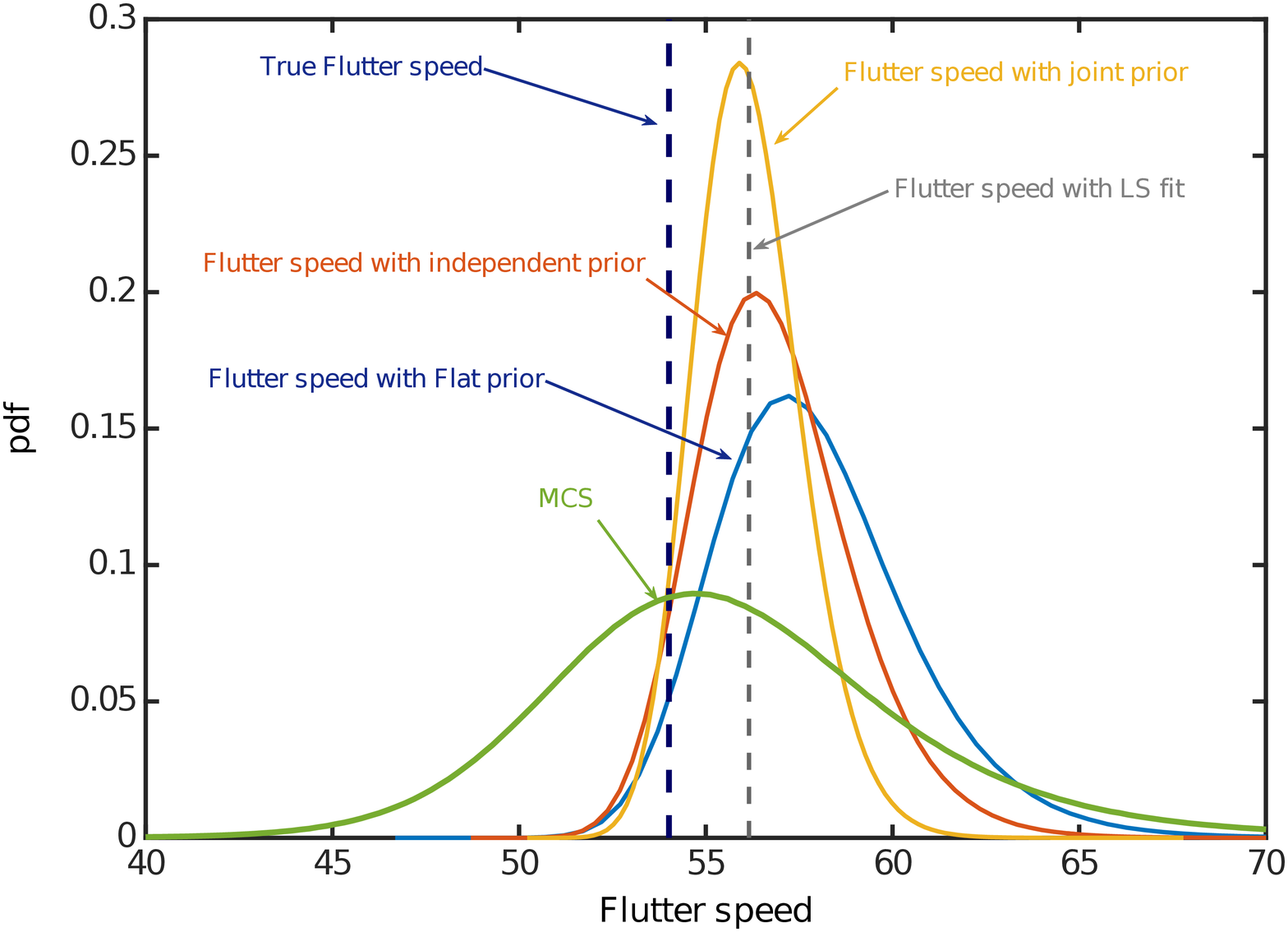}
\end{center}
\caption{ Flutter speed for different priors }
\label{Fs3U10cov}
\end{figure}

\begin{figure}[htbp]
\begin{center}
\includegraphics[width=120mm,height=80mm]{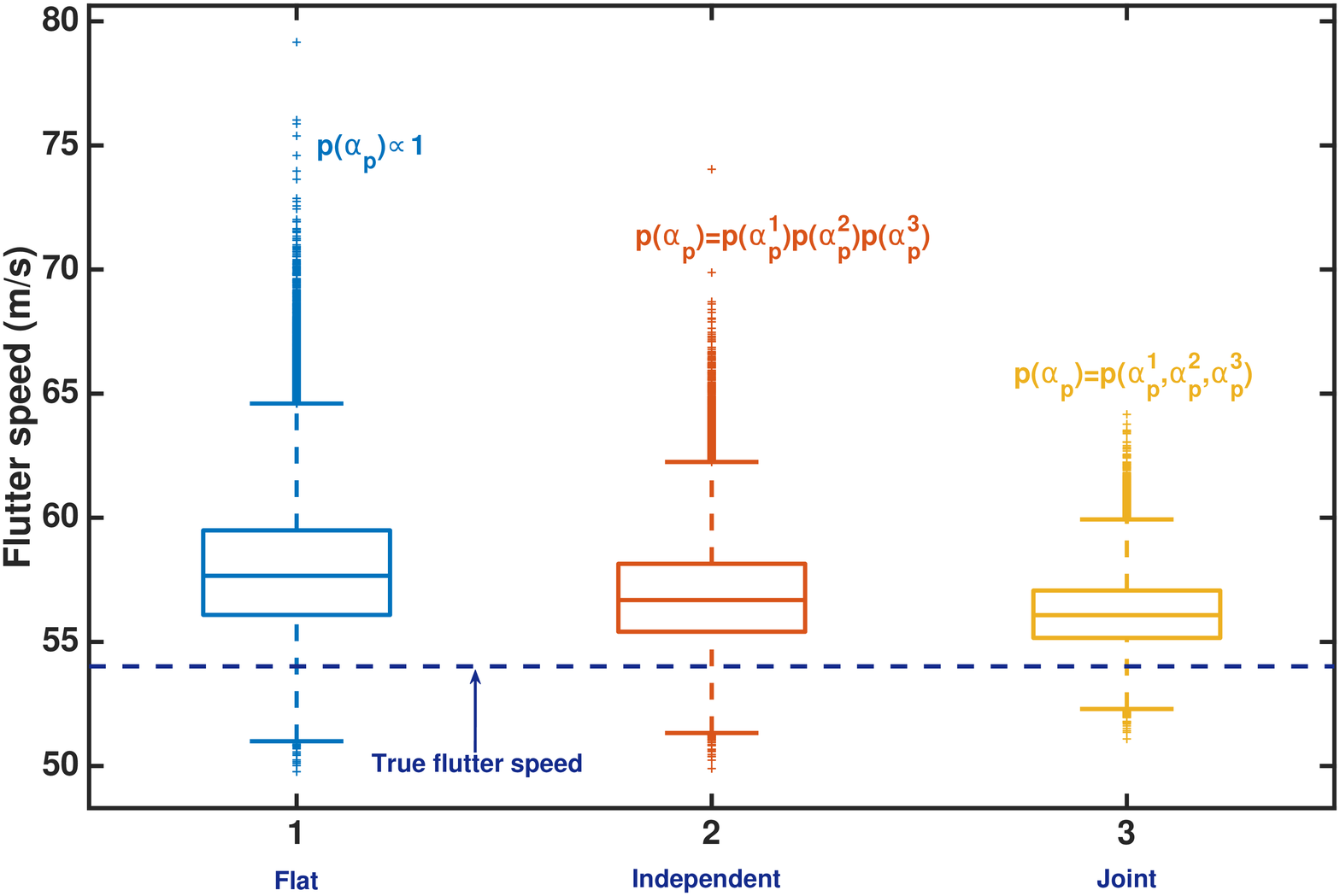}
\end{center}
\caption{Flutter speed for different priors }
\label{Fsboxplot3U10cov}
\end{figure}

\begin{table}[ht!]
\caption{COV in the estimated flutter speed}
\begin{center}
\begin{tabular}{ | c | c | c | c | }
\hline   Flat prior & Independent prior & Joint prior \\ 
\hline $p(\alpha)\propto1$ 
 &$p(\alpha)=p(\alpha^1)p(\alpha^2)p(\alpha^3)$&$p(\alpha)=p(\alpha^1,\alpha^2,\alpha^3)$ \\ \hline
4.5903&	3.7115&	2.5625\\\hline
\end{tabular}\end{center}
\label{Case3FScv}
\end{table}

While quasi-steady aerodynamic model are used in this paper, the Bayesian formulation of the 
flutter margin method is directly applicable to the case when unsteady aerodynamic effects are significant (characterized by the reduced frequency near flutter speed). Even if the quasi-steady model is used when unsteady effects are significant, the modeling error introduced through this assumption can be partly addressed through $\eta_k$ term in \eref{eq:11}. While the Bayesian framework is illustrated using the classical 2-dof pitch-plunge model, the methodology is readily applicable to the continuous systems (three-dimensional wings) using the finite element method and aeroelastic modal anaysis (applicable to  linear dynamical systems with non-symmetric system matrices). The application of the Bayesian flutter margin method to the continuous systems is the subject of future research.

\section{Conclusion}
We report the generalization of the Bayesian flutter margin method which relaxes assumptions made in the previous work by Khalil {\it et al.}. In the previous formulation, no aeroelastic model was used which is required to construct the prior of the modal parameters. Hence a non-informative (flat) prior was used to infer aeroelastic modal parameters at each airspeed. For uncorrelated Gaussian measurement noise, this assumption renders statistically independent modal parameter vectors and flutter
margins across airspeeds. This restriction is relaxed by introducing an aeroelastic model whose  system matrices exhibit random variabilities. The joint probability density function of aeroelastic modal frequencies and decay rates obtained from this stochastic aeroelastic model are statistically dependent  among airspeeds. The joint pdf of the collection of modal frequencies and decay rates for all airspeeds (at which tests are conducted) acts as the prior in order to infer the joint posterior of modal parameters. The flutter margins obtained from these posterior modal parameters exhibit statistical dependence. This statistical dependence among the modal parameters as well as flutter margins across airspeeds can substantially  the uncertainty in the predicted flutter speed. 

\section{Acknowledgements}
A. Sarkar  acknowledges the support of a Discovery Grant from Natural Sciences and Engineering Research Council of Canada and the Department of National Defence, Canada. D. Poirel acknowledges the support of the  Department of National Defence, Canada and a Discovery Grant from Natural Sciences and Engineering Research Council of Canada. The computing infrastructure is supported by the Canada Foundation for Innovation (CFI) and the Ontario Innovation Trust (OIT).

\bibliographystyle{unsrt}
\bibliography{Bibliography}

\end{document}